\documentclass[11pt,french,oldfontcommands,a4paper]{memoir}
\usepackage[latin9]{inputenc}
\usepackage{babel}
\addto\extrasfrench{%
   \providecommand{\og}{\leavevmode\flqq~}%
   \providecommand{\fg}{\ifdim\lastskip>\z@\unskip\fi~\frqq}%
}

\usepackage{verbatim}
\usepackage{float}
\usepackage{units}
\usepackage{textcomp}
\usepackage{amsmath}
\usepackage{amssymb}
\usepackage{graphicx}
\usepackage{esint}
\usepackage{nomencl}
\providecommand{\printnomenclature}{\printglossary}
\providecommand{\makenomenclature}{\makeglossary}
\makenomenclature

\makeatletter

\RequirePackage[sc]{mathpazo}
\RequirePackage[scaled=.95]{helvet}
\RequirePackage{courier}

\RequirePackage{enumerate}

\RequirePackage{lettrine}
\newcommand{\malettrine}[1]{
  \lettrine[lines=2,lhang=0.33,loversize=0.33]{#1}
}

\RequirePackage{caption}
\DeclareCaptionLabelSeparator{mysep}{~--~}
\RequirePackage{geometry}
\newfont{\chapterNumber}{eurb10 scaled 7000}
\renewcommand*{\@makechapterhead}[1]{%
  \thispagestyle{plain}
  \marginpar{\vspace*{1.5em}\flushright\chapterNumber\thechapter}
  \begin{flushleft}\nobreak\Huge\sc#1\end{flushleft}
  \vspace{3cm}
}
\renewcommand*{\@makeschapterhead}[1]{%
  \markboth{#1}{#1}
  \thispagestyle{plain}
  \begin{flushleft}\nobreak\Huge\sc #1\end{flushleft}
  \vspace{3cm}
}

\vfuzz \hfuzz
\newcommand{\colophon}[1]{
  ~\vfill
  \begin{center}
    \small #1
  \end{center}
  \cleardoublepage
}

\RequirePackage{tocbibind}

\@ifundefined{showcaptionsetup}{}{%
 \PassOptionsToPackage{caption=false}{subfig}}
\usepackage{subfig}
\makeatother

\begin{document}

 \thispagestyle{empty}
  \addtocounter{page}{-1}
  \begin{center}     
    \textsc{Service de Recherches de Métallurgie Physique} \\ \bigskip 
    \Huge\textsc{Thèse} \\ \bigskip 
    \large présentée pour obtenir le grade de\\
     \textsc{Docteur de l'Université Paris-Sud 11, Orsay,}\\ spécialité \og Physique du Solide \fg \\ \bigskip     par \\
\bigskip     Maximilien Levesque 
    \vfill     \huge~\textsc{Démixtion et ségrégation superficielle dans les alliages fer--chrome\\}     \Large~\textsc{De la structure électronique aux modèles thermodynamiques}  
  \vfill    
     \normalsize     Thèse soutenue le 26 novembre 2010 devant le jury composé de :\\   
 \vspace{1cm}  
\begin{tabular}{llll}
      M\up{me} & \textsc{Brigitte Beuneu}        & CEA                       &               \\
      M\up{me} & \textsc{Isabelle Demachy-Vacus} & Université Paris-Sud 11 & (Présidente)  \\
      M\up{me} & \textsc{Michèle Gupta}          & Université Paris-Sud 11 & (Directrice)  \\
      M.       & \textsc{Lorenzo Malerba}        & SCK$\cdot$CEN                 & (Rapporteur)  \\
      M.       & \textsc{Pär Olsson}             & EDF R\&D                  & (Rapporteur)  \\
      M.       & \textsc{Frédéric Soisson}       & CEA                     & (Encadrant)   \\
\end{tabular}
 \end{center}   
\cleardoublepage \clearpage

\newpage
\strut
\thispagestyle{empty}
\addtocounter{page}{-1}   
\newpage

  \begin{flushright} 
\thispagestyle{empty}
\addtocounter{page}{-1}
  \large\em\null\vskip1in 
 À mon père.
\vfill 
 \end{flushright} %
\cleardoublepage   \clearpage    
\newpage
\strut
\thispagestyle{empty}
\addtocounter{page}{-1}   
\newpage

\chapter*{Remerciements}

\thispagestyle{empty}\addtocounter{page}{-1}Je tiens d'abord à exprimer
mes remerciements aux personnes sous l'autorité desquelles cette thèse
a été évaluée. Merci à Isabelle Demachy-Vacus d'avoir présidé le jury
ainsi qu'à Lorenzo Malerba et Pär Olsson de m'avoir fait l'honneur
de rapporter ma thèse. Merci également à Brigitte Beuneu d'avoir examiné
ce travail théorique et numérique par son regard d'expérimentatrice.\\

Je remercie François Willaime, chef du Service de Recherches de Métallurgie
Physique du CEA de Saclay pour m'avoir donné l'opportunité de réaliser
ce travail de thèse dans un environnement de grande qualité.\\

Mes plus profonds remerciements vont à Frédéric Soisson, Maylise Nastar
et Chu Chun Fu pour leur aide, leur soutien, leurs conseils, et leur
grande compétence scientifique. Je suis conscient de leur rôle essentiel
dans ce processus qui tente de faire d'un étudiant enthousiaste et
susceptible un chercheur passionné et exigeant.\\

Merci également à Isabelle Drouelle et Clotilde Berdin-Méric pour
leur précieuse introduction au passionnant métier d'enseignant-chercheur.
C'est une chance que d'avoir réalisé mon monitorat sous leur tutorat.
Merci à toute l'équipe pédagogique pour leur patience et leurs conseils.\\

Merci à Bernard Legrand pour sa grande disponibilité et son souci
permanent d'excellence pédagogique. Merci également à Emmanuel Clouet
pour sa présence discrète mais essentielle.\\

Je voudrais aussi remercier Fabrice Legendre pour son admirable bienfaisance.\\

Je souhaite également exprimer ici ma gratitude à l'ensemble du SRMP.
J'ai beaucoup de respect pour la recherche de qualité qui y est faite.
Je ne citerai pas de noms, mais merci à chacun. J'adresse un clin
d'œil particulier (de ce genre~;-)\,) à celles et ceux avec qui
un lien d'amitié sincère s'est tissé au travers de ces trois années.\\

Madame Gupta, tout simplement merci. Comment m'exprimer ici sur votre
exemplarité en seulement quelques lignes ?\\

J'aimerais remercier ma famille. Très bientôt papa, je perçois maintenant
cette subtile éducation, celle qui forme aux plaisirs de l'esprit
mais n'oublie pas la part du cœur. De la \emph{vitesse de la lumière}
à l'\emph{intelligence du c}œ\emph{ur}. N'est-ce-pas Maman ?\\

Enfin, Lucile \ldots{} Les mots sont insuffisants pour qualifier
toute ta patience, ta compréhension, ta douceur, ton écoute \ldots{}
Quant au potentiel immergé, \emph{ab initio} prend un sens nouveau
!\\

\ldots{} chose promise \ldots{} merci aussi à Ravioli, Ratila, Ratchoum
et Rannibal ! \vskip3cm

\begin{flushright} 
\addtocounter{page}{-1}\thispagestyle{empty}
\large\em\null\vskip1in 
L'esprit cherche, le cœur trouve.\\
\rm George Sand
\vfill 
\end{flushright} %

\newpage{}

\tableofcontents{}

\printnomenclature{}

\addcontentsline{toc}{chapter}{Glossaire}
\markboth{Glossaire}{Glossaire}

\nomenclature{SRMP}{Service de Recherches de Métallurgie Physique}\nomenclature{FM}{Ferromagnétique}\nomenclature{NM}{Non-magnétique}\nomenclature{LMTO}{Linear Muffin-Tin Orbital -- Orbitales de "Muffin-Tin"}\nomenclature{LDA}{Local Density Approximation -- Approximation de la densité locale}\nomenclature{PDOS}{Projected Density of States -- Densité d'états projetée}\nomenclature{EXAFS}{Extended X-Ray Absorption Fine Structure -- Spectroscopie d'absorption X de structure fine}\nomenclature{DLM}{Disordered Local Moment -- Moment local désordonné}\nomenclature{DFT}{Density Functional Theory -- Théorie de la fonctionnelle de la densité}\nomenclature{SIESTA}{Spanish Initiative for Electronic Simulations with Thousands of Atoms}\nomenclature{PWSCF}{Plane-Wave Self-Consistent Field}\nomenclature{NC}{Norm-Conserved -- À norme conservée}\nomenclature{US}{Ultra-Soft -- Ultra-doux}\nomenclature{PEEM}{Photoemission electron miscroscopy -- Microscopie de photo-émission d'électron}\nomenclature{XMCD}{X-ray magnetic circular dichroism -- Dichroïsme magnétique de rayons X}\nomenclature{LSDA}{Local Spin Density Approximation -- Approximation de la densité locale polarisée en spin}\nomenclature{AF}{Antiferromagnétique}\nomenclature{PAW}{Projector Augmented Wave -- Ondes augmentées projetées}\nomenclature{GGA}{Generalized Gradient Approximation -- Approximation des gradients généralisés}\nomenclature{KKR}{Korringa-Kohn-Rostoker}\nomenclature{ARPES}{Angle-Resolved PhotoElectron Spectroscopy -- Spectroscopie de photo-électron résolue en angle}\nomenclature{2BM}{Two Band Model -- Modèle à deux bandes}\nomenclature{EMTO}{Exact Muffin-Tin Orbitals -- Orbitales exactes de "Muffin-Tin"}\nomenclature{CDM}{Concentration Dependant Model -- Modèle dépendant de la concentration}\nomenclature{LEED}{Low-Energy Electron Diffraction -- Diffraction d'électrons de basse énergie}\nomenclature{MEIS}{Medium Energy Ion Spectroscopy -- Diffusion des ions de moyenne énergie}\nomenclature{AES}{Auger Electron Spectroscopy -- Spectrométrie d'électron Auger}\nomenclature{STM}{Scanning Tunneling Microscopy -- Microscopie à effet tunnel}\nomenclature{SRO}{Short Range Order -- Ordre à courte distance}

\chapter*{Introduction générale}

\addcontentsline{toc}{chapter}{Introduction générale}
\markboth{Introduction générale}{Introduction générale}

\malettrine{L}{}es aciers ferritiques, probablement renforcés par
dispersion d'oxydes, sont des candidats sérieux comme matériaux de
structure pour les réacteurs nucléaires. Leur utilisation est aujourd'hui
bornée par la limite de solubilité du chrome dans le fer aux températures
basses et intermédiaires d'intérêt technologique. Dans la lacune de
miscibilité a lieu une démixtion en deux solutions solides $\alpha$
et $\alpha$' aux conséquences néfastes sur les propriétés du matériau.
Peu d'observations expérimentales existent pour mesurer la solubilité
limite exacte à ces températures où les durées de mise à l'équilibre
sont de plusieurs années. Des mesures d'ordre à courte distance, des
observations sous irradiation ainsi que des calculs \emph{ab initio}
semblent indiquer que les limites de solubilité admises par extrapolation
depuis les hautes températures sont incorrectes. Afin de mieux comprendre
les propriétés de cet alliage et de ses surfaces, nous avons calculé
\emph{ab initio} de nombreuses propriétés de volume et de surface
de l'alliage modèle fer--chrome puis proposé un modèle énergétique
pour les calculs thermodynamiques de limites de solubilité et d'isothermes
de ségrégation.

Ce travail de thèse s'intègre dans le projet européen collaboratif
GETMAT (Gen IV and Transmutation MATerials). L'un des objectifs de
ce projet est de démontrer l'existence de matériaux suffisamment performants
sous irradiation pour valider des critères de sûreté nucléaire après
des dizaines d'années de service. 

Cette étude a pour but de modéliser l'alliage binaire fer--chrome
et ses surfaces à l'échelle atomique. C'est un projet réalisé au Commissariat
à l'Énergie Atomique (CEA) de Saclay, au sein du Service de Recherche
de Métallurgie Physique (SRMP), du Département des Matériaux pour
le Nucléaire (DMN) intégré lui-même à la Direction de l'Énergie Nucléaire
(DEN). Elle s'inscrit dans un effort de recherche de long terme du
SRMP concernant les propriétés d'équilibre des matériaux d'intérêt
pour le nucléaire. Le système fer--chrome se révèle très singulier
et se situe à la convergence des spécialités du laboratoire. Structure
électronique, thermodynamique chimique et statistique ainsi que modélisation
numérique ont été conjointement mises en œuvre dans cette étude multi-échelle
d'un alliage très usité mais finalement mal connu aux échelles atomiques
et électroniques à basse température.

Ce manuscrit s'articule autour de quatre chapitres.

Dans le premier chapitre, nous passons en revue les propriétés du
système fer--chrome aux échelles électronique, atomique et macroscopique.
Cela permet d'introduire dans un deuxième temps les calculs \emph{ab
initio} et observations sous irradiation à basse température qui remettent
en cause les extrapolations des limites de solubilité depuis les hautes
températures. On commence par présenter succinctement la découverte
historique de la fragilisation des aciers ferritiques et de son lien
avec les limites de miscibilité $\alpha-\alpha$'. On présente également
les propriétés magnétiques de l'alliage, dont l'importance se révélera
tout au long du manuscrit. On discute ensuite les différents modèles
énergétiques proposés pour tenir compte des observations les plus
récentes et pour modéliser toujours plus précisément le système aux
températures d'intérêt pour le nucléaire.

Dans le deuxième chapitre, nous proposons une revue détaillée des
propriétés d'alliage et de surface à température nulle. Après une
introduction méthodologique, nous utilisons la théorie de la fonctionnelle
de la densité électronique afin d'étudier \emph{ab initio} les propriétés
énergétiques, structurales, électroniques et magnétiques du binaire
fer--chrome et de ses surfaces à température nulle. Par la mise en
œuvre de plusieurs approximations de pseudo-potentiels ou de bases,
nous discutons de la validité des différentes méthodes de calcul.
Cela nous permet d'évaluer l'effet des pseudo-potentiels à norme conservée
et des bases d'orbitales localisées pour la mises en œuvre de calculs
à grande dilution. Ces concentrations sont en effet difficilement
accessibles aux approximations plus robustes telle que la méthode
des projecteurs augmentés dans une base d'ondes planes. De plus, ces
approximations numériquement plus légères autorisent des études plus
systématiques des surfaces libres et de la ségrégation du chrome dans
le fer. 

Après une revue de l'ensemble des modèles énergétiques proposés pour
l'alliage binaire fer--chrome cubique centré, nous proposons un hamiltonien
léger et simple ajusté sur nos résultats \emph{ab initio} et sur le
diagramme de phases expérimental à haute température. Par un traitement
statistique de champ moyen validé par des simulations Monte Carlo,
nous déduisons de ce modèle les limites de solubilité $\alpha$--$\alpha$'
dans tout le domaine de température et de composition. Nous terminons
ce chapitre par une comparaison de nos résultats à l'ensemble des
observations expérimentales et propositions théoriques de la littérature.

Afin de pouvoir prédire la concentration surfacique à partir de la
composition volumique, nous adaptons dans le chapitre 4 l'hamiltonien
présenté au chapitre 3 pour tenir compte de l'effet des surfaces sur
les propriétés énergétiques. Pour cela, nous ajustons de nouveaux
paramètres sur les calculs \emph{ab initio} d'énergie de surface et
de ségrégation à dilution infinie. Nous appliquons notre modèle à
l'étude de la ségrégation de surface dans les orientations $\left(100\right)$,
$\left(110\right)$, $\left(111\right)$ et $\left(211\right)$ du
système fer--chrome. À partir d'un traitement statistique de champ
moyen auto-cohérent permettant de décomposer l'énergie de ségrégation
en deux moteurs principaux (effet d'énergie de surface et effet d'alliage),
nous déduisons des isothermes de concentration reliant la concentration
surfacique à la concentration volumique. Nous illustrons également
l'influence du changement de signe de l'énergie de mélange sur les
isothermes de ségrégation de surface.

À l'issu de ces quatre chapitres, nous aurons apporté des éléments
de réponse aux questions suivantes :

Quel est le lien entre les propriétés magnétiques et thermodynamiques
de cet alliage ? En particulier, quel est l'effet du magnétisme sur
les limites de solubilité à basse température et sur les isothermes
de ségrégation ?

Comment tenir compte de ces propriétés magnétiques dans un modèle
énergétique assez léger pour permettre le calcul des limites de solubilité,
des isothermes de ségrégation et envisager les calculs cinétiques
?

\chapter{Les alliages fer-chrome -- usages nucléaires : une modélisation nécessaire\label{cha:Les-alliages-fer-chrome}}

\malettrine{D}{}ans ce premier chapitre, on introduit le contexte
nucléaire et les propriétés physiques de l'alliage binaire fer--chrome.

Nous commençons par une description succincte de la problématique
des matériaux dans un contexte nucléaire. Après une discussion synthétique
et non-exhaustive du cahier des charges lié au choix des matériaux
pour les réacteurs nucléaires du futur, nous indiquons pourquoi les
aciers ferritiques, potentiellement renforcés par exemple par dispersion
d'oxydes, sont des candidats sérieux pour les réacteurs de génération
IV ou de fusion.

L'alliage binaire fer--chrome est le modèle le plus simple des aciers
ferritiques. C'est cet alliage que nous étudions dans la suite de
l'étude. Nous discutons donc dans la suite de ce chapitre des principales
propriétés physico-chimiques de ce binaire. Nous commençons par présenter
son diagramme de phases d'équilibre faisant actuellement référence,
avec une attention particulière aux limites de solubilité des solutions
solides. Le fer et le chrome étant des éléments magnétiques, nous
détaillons les propriétés magnétiques de l'alliage, puis les mesures
d'ordre à courte distance liées au magnétisme. L'effet de ces propriétés
reste pour une part inconnu, en particulier à basse température. C'est
seulement à la lumière de ce chapitre bibliographique que nous présentons
nos résultats dans les chapitres suivants.\vspace{1cm}

\section{Des matériaux de structure pour les réacteurs nucléaires du futur}

 La ressource nucléaire est aujourd'hui la source primaire d'environ
7\,\% de la consommation énergétique mondiale. Plus de 430\,réacteurs
nucléaires à fission sont actuellement répartis dans 31\,pays et
produisent 15\,\% de l'électricité mondiale \cite{zinkle_structural_2009}.
N'émettant pas directement de gaz à effet de serre, cette ressource
aura une place importante dans le bouquet énergétique de demain. En
2001, treize pays dont la France se sont regroupés dans le Forum Génération\,IV
\cite{GEN_IV_homepage} afin d'identifier et de développer dans un
effort conjoint les systèmes nucléaires du futur.

\section{Quels matériaux choisir ?}

Les réacteurs nucléaires sont des environnements très exigeants pour
leurs composants. Le choix des matériaux de structure pour les réacteurs
des futures générations est un défi car la dégradation des matériaux
dans ces environnements peut abaisser le niveau de performance du
réacteur et, dans le pire des cas, induire des incidents potentiellement
graves. Aux États-Unis, deux tiers des 47 responsables de sites nucléaires
considèrent la dégradation des matériaux à long terme comme principale
source d'inquiétude du fait de leur difficile anticipation \cite{zinkle_structural_2009}. 

Soumise à des conditions expérimentales constantes, une pièce peut
être remplacée après une durée de fonctionnement prédéterminée. Les
conditions de fonctionnement varient cependant au cours du temps.
La chimie du milieu primaire, par exemple, est affinée durant les
40 à 60\,années de vie de la centrale. Ces modifications rendent
fois les retours d'expérience difficiles à transférer d'un milieu
ou d'un matériau à l'autre. Une approche physique permettant de comprendre
les phénomènes à l'origine de la dégradation est donc nécessaire car
transférable. C'est une approche qui permet de modéliser ces phénomènes,
donc de les prédire.

\subsection{La dégradation des matériaux}

De nombreux phénomènes modifient les propriétés des matériaux dans
une centrale nucléaire. Nous nous penchons ici plus particulièrement
sur le vieillissement thermique, qui est au cœur de notre étude, et
sur les dommages d'irradiation qui sont à l'origine du choix des aciers
ferritiques. Nous ne traitons cependant pas des dommages d'irradiation
plus avant dans ce manuscrit.

\subsubsection*{Le vieillissement thermique}

Les matériaux utilisés ne sont pas toujours à l'équilibre thermodynamique.
Ce peut être volontaire du fait de propriétés d'intérêt technologique
de phases métastables ou involontaire si le diagramme de phases n'est
pas exactement connu ou si le procédé de fabrication est mal maîtrisé.
L'exposition de ces matériaux pendant 40 à 60\,ans à des températures
de fonctionnement de l'ordre de quelque centaines de Kelvin peut alors
induire des transformations de phases susceptibles d'altérer les propriétés
de ces matériaux.

\subsubsection*{Les dommages d'irradiation\label{sub:les_dommages_dirradiation}}

Les réactions de fission ou de fusion nucléaire ayant lieu dans le
réacteur produisent des neutrons dont l'énergie cinétique est parfois
suffisante pour créer des défauts dans les matériaux \cite{martin_bellon_driven_1996}.
Après 40\,ans de fonctionnement, les matériaux du cœur des réacteurs
nucléaires du futur auront été exposés à des fluences de l'ordre de
$10^{22}$ à $10^{23}$\,neutrons par cm$^{2}$. Il est ainsi envisagé
que les collisions balistiques induise un nombre de déplacement par
atome (dpa) supérieur à 100 durant sa vie.

Un des verrous technologiques clés des réacteurs du futur réside en
conséquence dans le développement de matériaux résistants aux dommages
d'irradiation, c'est-à-dire éliminant rapidement les défauts créés
par ces dommages \cite{yvon_materials_gen_ferritic_aust_2009,zinkle_structural_2009}.

\subsection{Les aciers ferritiques}

Les aciers ferritiques, potentiellement renforcés par dispersion d'oxydes,
remplissent une grande partie du cahier des charges présenté ci-dessus.
Ces aciers, à base de ferrite (fer cubique centré ou fer--$\alpha$)
contiennent un grand nombre d'additifs (Cr, W, Mo, C, N, P, Si \ldots{}).
En particulier, une teneur en chrome de 11 à 13\,\% leur confèrent
une très bonne résistance à la corrosion. Dans les réacteurs nucléaires
actuels, la composition est actuellement limitée à 9\,\% de chrome
à cause d'une perte drastique des propriétés mécaniques au-dessus
de 10\,\%. Une meilleure connaissance des phénomènes induisant la
perte des propriétés mécaniques permettrait d'envisager d'augmenter
la teneur en chrome \cite{yvon_materials_gen_ferritic_aust_2009,zinkle_structural_2009}.

\begin{figure}[h]
\begin{centering}
\includegraphics[scale=0.24]{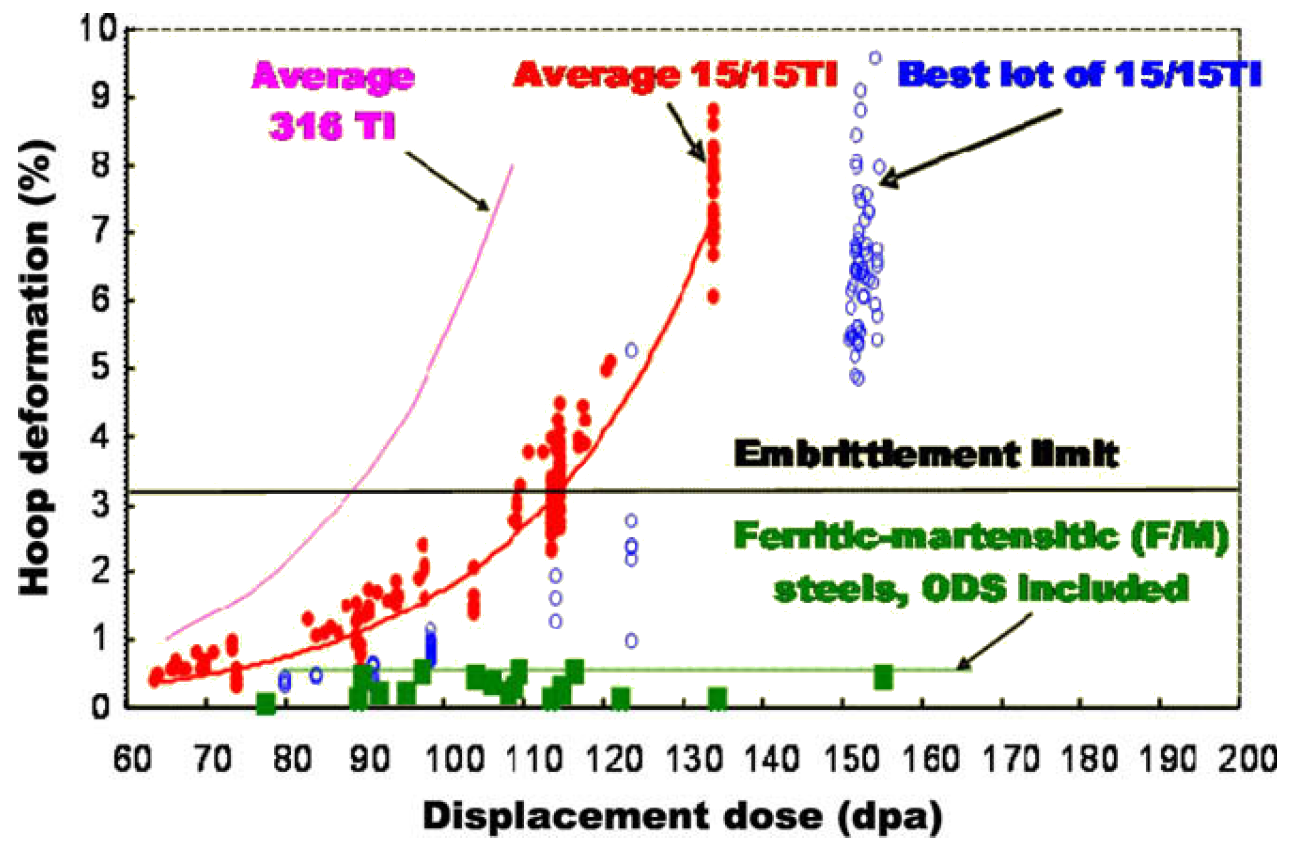}
\par\end{centering}

\caption{Gonflement des aciers austénitiques et ferritiques-martensitiques
en fonction de la dose à différentes températures entre 675 et 825\,K.
Publié par Yvon et Carré \cite{yvon_materials_gen_ferritic_aust_2009}.\label{fig:yvon}}

\end{figure}

Ces aciers sont des candidats sérieux comme matériaux de structure
pour les réacteurs refroidis au sodium ou au plomb, pour les chaudières
sous pression des réacteurs à gaz et pour les matériaux de cœur des
réacteurs à fusion \cite{yvon_materials_gen_ferritic_aust_2009,mansur_materials_2004,klueh_ferritic/martensitic_2007}.
Ils permettent des températures de fonctionnement élevées ($<875$\,K)
qui améliorent les rendements. Comme le montre la figure \ref{fig:yvon},
les aciers ferritiques ont l'avantage par rapport aux austénitiques
et aux alliages base nickel de présenter un faible gonflement induit
par irradiation. En effet, le gonflement des matériaux austénitiques
croît très rapidement après un temps d'incubation, ce qui n'est pas
le cas des ferritiques-martensitiques, même à fortes doses ($>150$\,dpa).
Cette figure illustre également qu'il est nécessaire d'étudier très
précisément (temps longs, fortes doses) les candidats car des effets
de seuil en temps ou dose, par exemple, peuvent apparaître.

Les observations sur les matériaux industriels permettent de construire
une base de données expérimentale et d'en dégager des tendances. On
peut alors affiner le choix des matériaux à étudier. Notre étude se
situe dans le cadre d'une recherche de base sur ces aciers ferritiques.
Il s'agit de comprendre, brique après brique, les phénomènes physiques
qui sous-tendent les propriétés de matériaux complexes. C'est la raison
pour laquelle nous approximons les alliages ferritiques par un alliage
binaire fer--chrome. Ceci permet d'étudier à l'échelle atomique les
principales caractéristiques électroniques, magnétiques, structurelles
et énergétiques des aciers ferritiques, indépendamment des effets
d'impuretés.

\section{L'alliage binaire fer--chrome, modèle des aciers ferritiques}

Les premières études sur cet alliage correspondaient à une volonté
de comprendre la fragilisation observée à basse température. C'est
donc ce phénomène que nous décrivons en premier. Nous discutons ensuite
du diagramme de phases du système, avec une attention particulière
aux limites de solubilité des solutions solides et des effets d'ordre
à courte distance. C'est sur le magnétisme que nous portons ensuite
notre attention, aux échelles macroscopiques et atomiques. On termine
par l'effet des surfaces sur ces propriétés et sur l'introduction
succincte des modèles énergétiques proposés pour l'alliage et ses
surfaces.

\subsection{La fragilisation des aciers ferritiques à basse température}

La perte des propriétés mécaniques des aciers ferritiques au court
du vieillissement thermique à basse température a entraîné le besoin
d'étudier ces matériaux. Dès les années 1930, Becket observe une augmentation
de la dureté et une diminution de la ductilité des aciers ferritiques
à 15\,\% de chrome entre 640 et 813\,\textdegree{}C \cite{becket__1938}.
Il nomme ce phénomène \og 885\,\textdegree{}F embrittlement \fg{}
en anglais, soit littéralement \og la fragilisation à 750\,K \fg{}.
Il montre ainsi que cette propriété générale des aciers est déjà présente
dans l'alliage binaire. Il est le précurseur d'une longue série d'études
sur le système binaire Fe--Cr. Trois hypothèses sont proposées pour
expliquer la fragilisation :
\begin{itemize}
\item la précipitation de la phase $\sigma$ fragilisante déjà observée
à plus haute température \cite{heger_formation_1950},
\item l'existence dans ce domaine de température d'un intermétallique Fe$_{\text{3}}$Cr
fragilisant \cite{tagaya_Fe3Cr_1951,masumoto_chaleurspe_1953,imai1_Fe3Cr_1953,imai2_Fe3Cr_1953},
\item une transformation de phase durcissante : la démixtion $\alpha$--$\alpha'$
entre les solutions solides cubiques centrées $\alpha$ riche en fer
et $\alpha'$ riche en chrome.
\end{itemize}
Fisher et al. \cite{fisher_identification_1953} valident dès 1953
cette dernière hypothèse. Il cœxiste, sous la phase $\sigma$, deux
phases cubiques centrées. L'une est riche en fer (solution solide
$\alpha$) tandis que l'autre est riche en chrome (solution solide
$\alpha'$).

\subsection{Le diagramme de phases de référence}

\begin{figure}[H]
\begin{centering}
\includegraphics[scale=0.35]{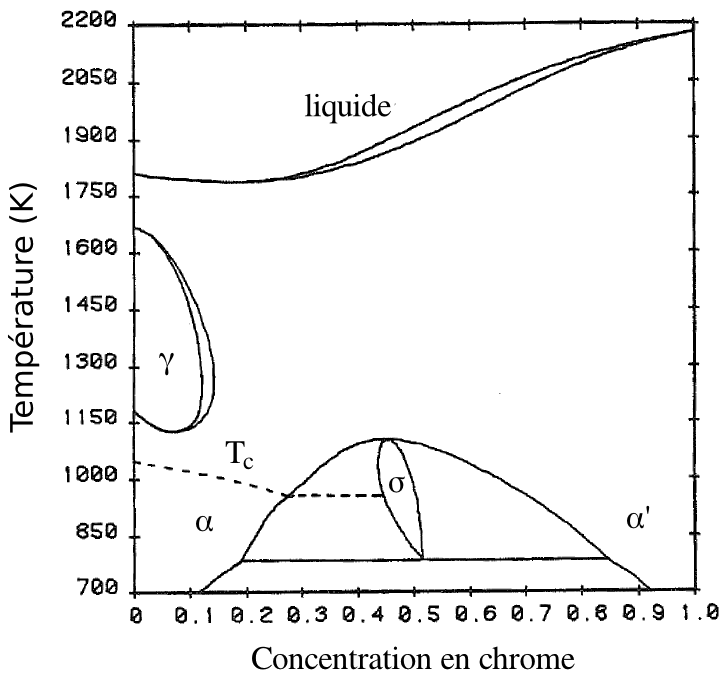}
\par\end{centering}

\caption{Diagramme de phases du système binaire fer--chrome publié par Andersson
et Sundman en 1987 \cite{andersson_thermodynamic_1987}. La température
critique $T_{c}$ est indiquée en tirets pointillés. \label{fig:ddp_biblio_REF_SUNDMAN}}

\end{figure}

De nombreux diagrammes de phases ont été proposés pour le système
binaire fer--chrome. Nous reproduisons sur la figure \ref{fig:ddp_biblio_REF_SUNDMAN}
le diagramme de phases publié par Andersson et Sundman en 1987.

Le diagramme de phases d'équilibre du système fer--chrome comporte
cinq phases. La phase $\alpha$ est la solution solide riche en fer.
La phase $\alpha'$ est la solution solide riche en chrome. Ces deux
solutions solides ont un réseau cubique centré.

La phase $\gamma$ est allotrope de la phase $\alpha$. Il s'agit
d'une solution solide de chrome dans le fer, mais de réseau cubique
à faces centrées. La transformation $\alpha$--$\gamma$, qui a lieu
à 1185\,K, est liée aux moments magnétiques atomiques des atomes
de fer. La modélisation de cette transformation reste un challenge
\cite{dudarev_magnetic_2005,lavrentiev_dudi_magnetic_bccfcc_2010}. 

Le composé intermétallique $\sigma$ est une phase complexe, de groupe
d'espace P$4_{2}$/mnm contenant 30\,atomes par maille élémentaire.
C'est une phase fragilisante qui se forme très lentement, ce qui la
rend technologiquement importante à éviter. Joubert a publié une revue
sur la phase $\sigma$ très récemment \cite{joubert_sigma_2008}.
Dans l'alliage Fe--Cr, la phase $\sigma$ existe entre la température
eutectoïde d'environ 773--783\,K et la température critique de 1093\,K.

La phase liquide apparaît dès 1800\,K.

\subsection{La lacune de miscibilité $\alpha$--$\alpha'$}

\begin{figure}[H]
\begin{centering}
\includegraphics[scale=0.43]{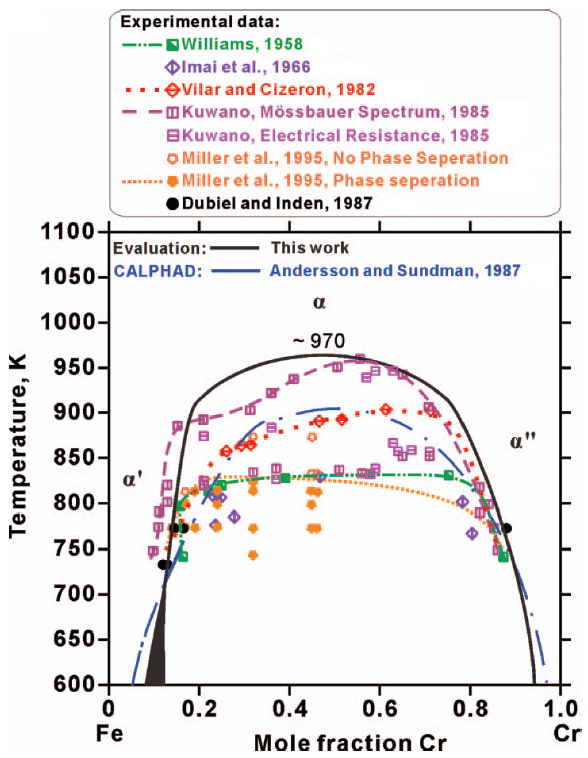}
\par\end{centering}

\caption{Compilation des limites de solubilité expérimentales de Williams \cite{williams_further_1958},
Imai et al. \cite{imai_bcc_1966}, Vilar and Cizeron \cite{Vilar_bcc_1982},
Kuwano \cite{Kuwano_bcc_1985}, Miller et al. \cite{miller_spinodal_1995},
et Dubiel et Inden \cite{dubiel_mossbauer_longtermannealed_1987}.
Ces observations sont comparées aux limites de solubilité proposées
par Andersson et Sundman dont on a extrait la figure \ref{fig:ddp_biblio_REF_SUNDMAN}
et par Xiong et al. \cite{xiong_grrrrbonny_2010}. Publié par Xiong
et al. \cite{xiong_grrrrbonny_2010}.\label{fig:Compilation-des-limites_sol_Xiong}}

\end{figure}

C'est à la lacune de miscibilité $\alpha$--$\alpha'$ que l'on s'intéresse
dans ce travail pour les raisons évoquées ci-dessus. De plus, la position
de la limite de solubilité a un effet important sur les cinétiques
de décomposition car elle contrôle la force motrice de la précipitation.
Bonny et al. puis Xiong et al. ont publié très récemment une revue
détaillée compilant tous les résultats expérimentaux sur la lacune
de miscibilité $\alpha$--$\alpha'$ \cite{bonny_onthe_aa_demixtion_2008,xiong_grrrrbonny_2010}.
On reproduit la compilation de Xiong et al. sur la figure \ref{fig:Compilation-des-limites_sol_Xiong}.
On peut y lire les limites de solubilité observées expérimentalement
\cite{williams_further_1958,imai_bcc_1966,Vilar_bcc_1982,Kuwano_bcc_1985,miller_spinodal_1995,dubiel_mossbauer_longtermannealed_1987},
et les comparer aux limites de solubilité proposées par Andersson
et Sundman \cite{andersson_thermodynamic_1987} (dont on a déduit
la figure \ref{fig:ddp_biblio_REF_SUNDMAN}) et par Xiong et al. \cite{xiong_grrrrbonny_2010}.

Les limites de solubilité stables et métastables à haute température
sont déterminées par des mesures chimiques, mécaniques et de diffusion
de neutrons \cite{williams__1957,williams_further_1958,marcinkowski__1964}.
La transition entre les régimes de nucléation--germination--croissance
et de décomposition spinodale est continue, mais certains auteurs
déterminent les limites spinodales. Marcinkowski \cite{marcinkowski__1964}
et Chandra et al. \cite{chandra_mssbauer_1971} montrent par exemple
qu'à 475\,\textdegree{}C entre 12 et 30\,\%\,Cr, la décomposition
$\alpha$--$\alpha'$ se fait via nucléation et croissance de précipités,
puis via décomposition spinodale \cite{clouet_ASMbook_2009}.

La cinétique de formation de la phase $\sigma$ est plus lente que
la cinétique de formation du domaine biphasé métastable $\alpha-\alpha$'.
On peut déterminer la température critique de la lacune de miscibilité.
Celle-ci se situe entre 900 et 970\,K pour une concentration en Cr
de 40 à 45\,\%. Les limites de solubilité ne sont donc pas symétriques
par rapport à 50\,\%\,Cr. 

Le diagramme de phases n'est pas ou peu connu en-dessous de 700\,K
environ, car les transformations de phases demandent alors plusieurs
années pour se terminer. Ces durées sont difficilement accessibles
à l'expérimentateur. Les limites de solubilité sont parfois extrapolées
des hautes températures vers les basses températures. Il n'y a alors
aucune prise en compte des phénomènes ayant lieu uniquement à basse
température comme les interactions magnétiques. C'est la raison pour
laquelle il est utile d'observer le vieillissement sous irradiation,
cette dernière accélérant la diffusion en première approximation.
Bonny et al. proposent ainsi une revue des résultats expérimentaux
sous irradiation \cite{bonny_onthe_aa_demixtion_2008}. Cependant,
l'irradiation neutronique n'est pas un phénomène linéaire. Comme on
peut l'observer sur la figure \ref{fig:yvon}, des effets de seuil
peuvent apparaître. Les dommages balistiques participent au re-mélange
des précipités, en plus de créer des défauts qui accélèrent le vieillissement.
Il est donc difficile de déduire des limites de solubilité à basse
température d'expériences d'irradiation neutronique.

Les basses températures ne sont pas les seules difficultés singulières
de ce système. Aux températures hautes, où l'alliage est paramagnétique,
Fultz et al. \cite{fultz_phonon_Fe_Cr_1995} étudient les modes de
vibration des éléments purs et de trois alliages de compositions intermédiaires
par diffusion inélastique de neutrons. Ils montrent qu'à 70\,\% de
Cr l'entropie de vibration de l'alliage est très importante. Elle
équivaut à près d'un tiers de l'entropie de mélange maximum à haute
température. L'entropie de vibration qu'ils proposent est asymétrique.
Elle est relativement constante quand la concentration en chrome augmente,
puis varie rapidement dans la solution solide $\alpha'$ riche en
chrome.

\subsection{L'ordre à courte distance\label{sub:L'ordre-a-courte}}

L'existence d'un ordre à courte distance (SRO) (défini dans l'annexe
\ref{sec:SRO}) dans la solution solide $\alpha$ de l'alliage Fe--Cr
a été observée pour la première fois par Vintaykin et Loshmanov par
diffusion de neutrons \cite{vintaykin_475_1966}. En 1983, par la
méthode de perturbation généralisée aux alliages magnétiques, Hennion
\cite{Hennion_perturbation_1983} déduit des potentiels d'interaction
de paires aux premiers voisins fortement dépendants de la concentration
d'impureté. À 0\,K, le potentiel de paire hétéro-atomique serait
négatif en dessous de 25\,\%\,Cr, puis positif au-delà. L'alliage
passerait alors d'une tendance à l'ordre à une tendance à la démixtion
avec l'augmentation de la concentration \cite{clappmoss_I_1966,clappmoss_II_1968,clappmoss_III_1968}.

\begin{figure}[h]
\begin{centering}
\includegraphics[scale=0.3]{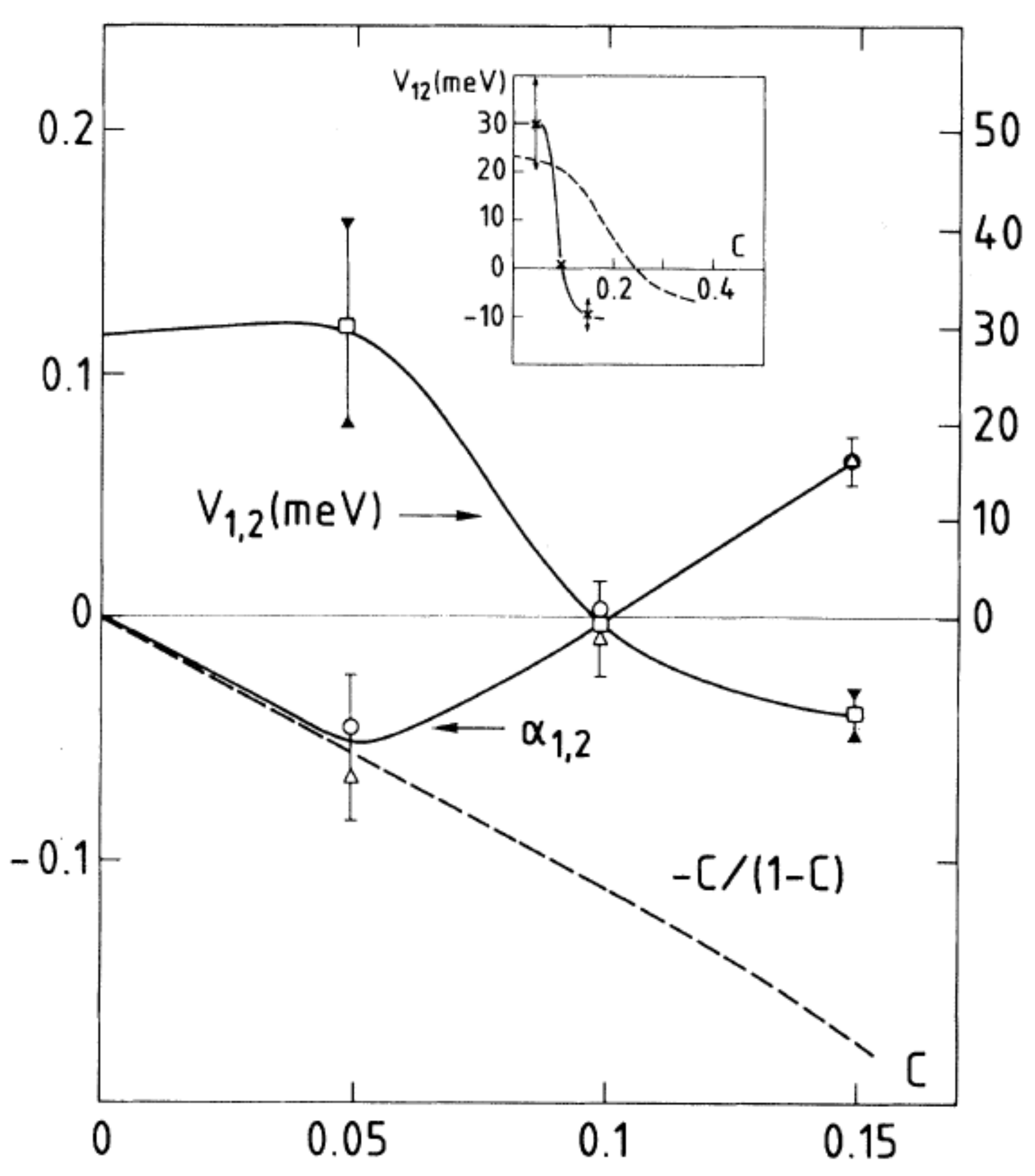}
\par\end{centering}

\caption{Variation du paramètre d'ordre à courte distance de Warren-Cowley
aux premiers et seconds voisins $\alpha_{1,2}$ défini en annexe \ref{sec:SRO}
et du potentiel de paire $V_{1,2}$ en meV en fonction de la concentration
en chrome $C$ à 703\,K. Les triangles et les ronds correspondent
au paramètre d'ordre calculé selon deux hypothèses de calcul. Dans
l'insert, le potentiel de paire proposé par Hennion \cite{Hennion_perturbation_1983}
en pointillés est comparé à celui déduit par Mirebeau et al. en traits
pleins. Publié par Mirebeau et al. \cite{mirebeau_PRL_1984}.\label{fig:variation_SRO_mirebeau}}
\end{figure}

Ces prédictions sont confirmées qualitativement par des observations
de Mirebeau et al. \cite{mirebeau_PRL_1984} par mesures de résistivité
et de diffusion diffuse de neutrons. Ces auteurs ajustent la section
efficace d'absorption neutronique sur les paramètres d'ordre des quatre
premières sphères de coordination \cite{Cowley_SRO_1950} (défini
en annexe \ref{sec:SRO}) en fonction de la concentration en chrome,
en faisant l'hypothèse d'une contribution égale des premiers et deuxièmes
voisins. Ils déduisent ensuite des potentiels de paires à partir des
paramètres d'ordre. On représente sur la figure \ref{fig:variation_SRO_mirebeau}
le paramètre d'ordre $\alpha_{1,2}$ défini en annexe \ref{sec:SRO}
ainsi que les potentiels de paires déduits. Ces derniers sont comparés
en insert de la figure \ref{fig:variation_SRO_mirebeau} aux potentiels
de paires proposés par Hennion.

Le paramètre d'ordre $\alpha_{1,2}$ change de signe aux alentours
de 10\,\%\,Cr à 703\,K. Il y a effectivement une tendance à l'ordre
puis à la démixtion dans l'alliage fer--chrome à cette température
correspondant à 70\,\% de la température critique.

\subsection{Le magnétisme}

\begin{figure}[h]
\begin{centering}
\includegraphics[scale=0.4]{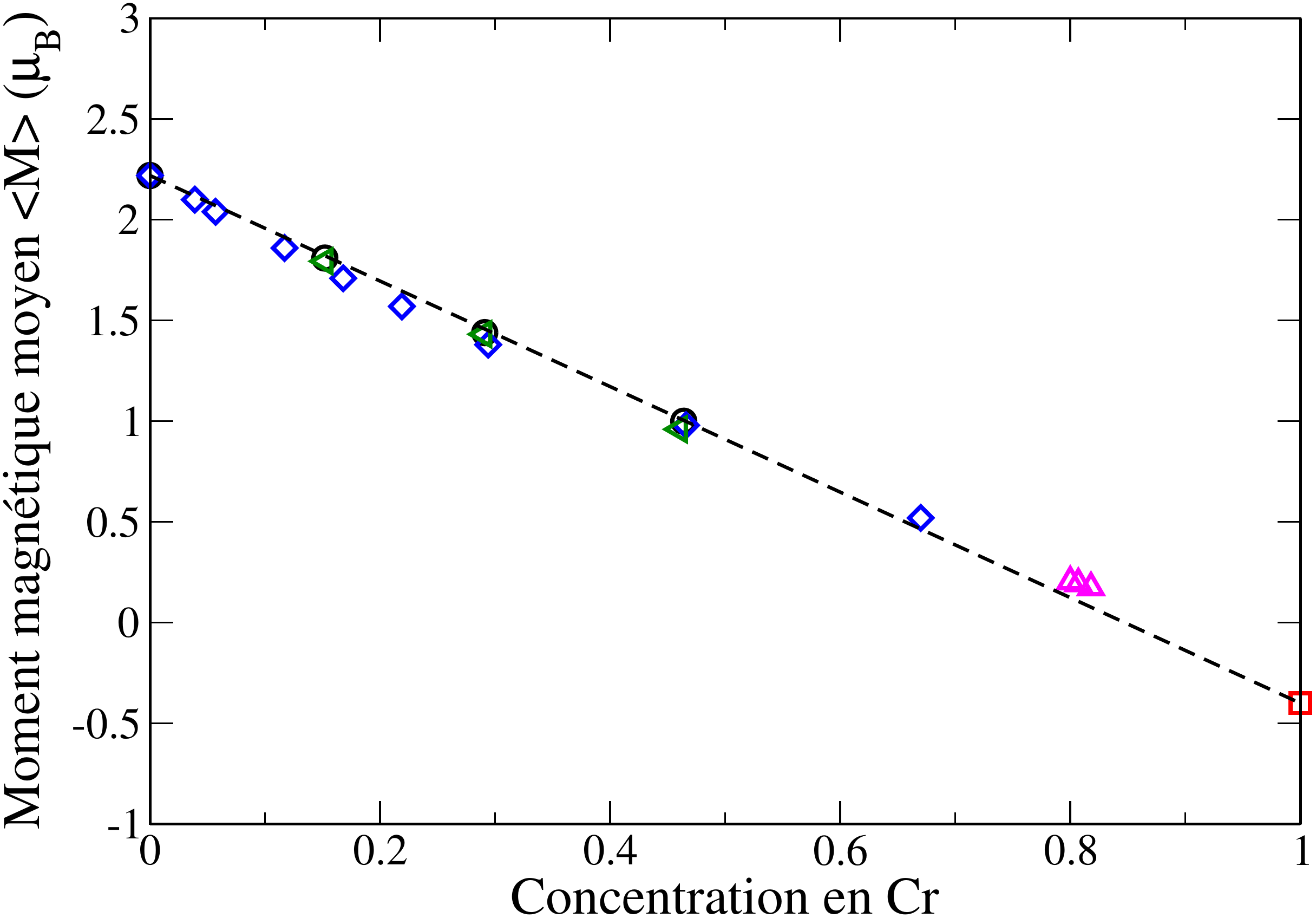}
\par\end{centering}

\caption{Variation du moment magnétique moyen avec la concentration de l'alliage
selon la relation \ref{eq:slater_pauling} de SlaEter-Pauling en traits
pointillés, et comparaison avec les mesures expérimentales de Shull
et Wilkinson \cite{shull_neutron_1955} (ronds noirs), Shull et Wilkinson
\cite{shull_neutron3_1953} (carrés rouges), Fallot \cite{fallot_mag_fecr_1944}
(losanges bleu), Aldred \cite{Aldred_bulkmag_1976} (triangles violets)
et Shull \cite{shull_meanM_c_1956} (triangles verts).\label{fig:Variation-du-moment_avec_c}}
\end{figure}

Nous discutons du magnétisme des éléments purs en détail dans le chapitre
\emph{ab initio}. Dès le développement de la théorie des bandes rigides,
Slater et Pauling \cite{slater_structelec_1937,Pauling_metalforces_1938}
expriment une relation très simple entre le nombre d'électrons $d$
par atome de l'alliage binaire et le moment magnétique atomique moyen
$\left\langle M\right\rangle $. Appliquée à l'alliage Fe$_{x_{Fe}}$Cr$_{x_{Cr}}$,
cette relation conduit à une dépendance linéaire entre $\left\langle M\right\rangle $
et la concentration en chrome $x_{Cr}$ : 
\begin{equation}
\left\langle M\right\rangle =\left(2.2-2.6x_{Cr}\right)\mu_{B}\label{eq:slater_pauling}
\end{equation}
où $\mu_{B}$ est le magnéton de Bohr. Nous représentons l'évolution
du moment magnétique moyen de l'alliage en fonction de la concentration
en chrome selon la relation \ref{eq:slater_pauling} sur la figure
\ref{fig:Variation-du-moment_avec_c}.

Les éléments Cr et Fe ont des numéros atomiques très proches (24 et
26). Les figures de diffractions X et électronique sont en conséquence
très peu contrastées, au contraire des figures de diffraction de neutrons.
Cette dernière méthode expérimentale devient alors la référence pour
l'étude des alliages fer--chrome dès les années 1950 \cite{shull_neutron3_1953,shull_neutron_1955,ishikawa_Fe_imp_dans_Cr_1967,arrott_diffneutrons_first_muCrSDW_1967}.
De plus, les neutrons sont des fermions : ils portent un moment magnétique.
Leur diffusion possède donc une susceptibilité magnétique intéressante
pour notre alliage de deux éléments magnétiques. Les mesures de diffusion
de neutrons de Shull, Fallot, Aldred et al. \cite{shull_neutron_1955,fallot_mag_fecr_1944,Aldred_bulkmag_1976},
que nous reportons sur la figure \ref{fig:Variation-du-moment_avec_c},
confirment la relation de Slater-Pauling. Dans une deuxième étude,
Aldred et al. \cite{Aldred_neutron_1976} montrent cependant de légères
déviations à la relation \ref{eq:slater_pauling} de Slater-Pauling
par diffusion diffuse de neutrons aux grandes longueurs d'ondes. Cette
déviation est le signe du lien très complexe entre les moments magnétiques
de chacun des atomes de l'alliage et leur environnement chimique local.

\begin{figure}[H]
\begin{centering}
\includegraphics[scale=0.45]{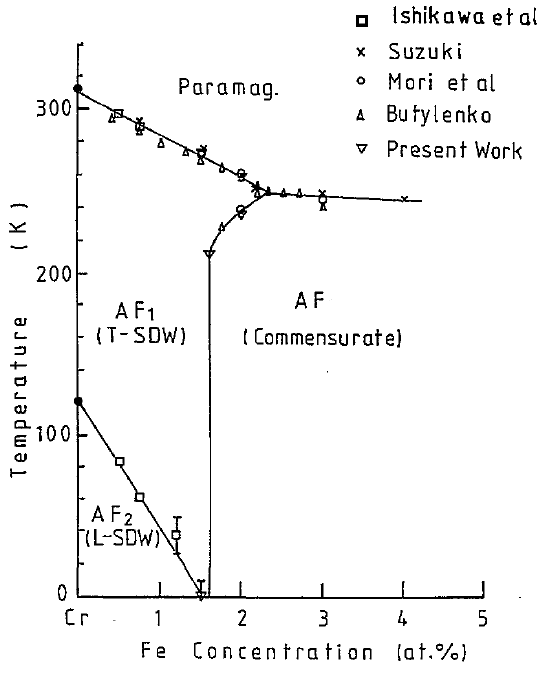}
\par\end{centering}

\caption{Diagramme de phases magnétiques dans le domaine riche en Cr déterminé
par une revue critique des résultats expérimentaux publié par Tsunoda
\cite{tsunoda_FeCr_magnetic_phase_diagram_1994}. La température de
Curie du fer est de 1043\,K. Pour le détail expérimental, se référer
à la publication originale.\label{fig:ddp_magnetique}}

\end{figure}

Ces mesures, associées à des mesures de magnétisme global, permettent
de tracer le diagramme de phases magnétiques du système riche en Cr
qu'on représente sur la figure \ref{fig:ddp_magnetique} d'après une
revue des mesures expérimentales par Tsunoda \cite{tsunoda_FeCr_magnetic_phase_diagram_1994}. 

La température de Néel est de 311\,K pour le chrome pur. Au-delà,
le Cr devient paramagnétique. Pour des teneurs en fer faibles, inférieures
à 1.5\,\%, l'alliage Fe-Cr présente une onde de spin incommensurable
transversale (T-SDW) ou longitudinale (L-SDW). Au-delà, l'alliage
est antiferromagnétique (AF) de période $a$ le paramètre de la maille
cubique centrée. On discute plus précisément des ondes de spin dans
le chapitre 2.

Pour toutes les concentrations étudiées, le moment magnétique atomique
des Cr est anti-aligné à ceux des Fe \cite{shull_neutron_1955,fallot_mag_fecr_1944,Aldred_bulkmag_1976}.
En particulier, cette observation est vérifiée dans les alliages ne
contenant que 1 à 2\,\%\,Cr par Collins et Low \cite{Collins_mudistri_1965}.
Dans ces alliages, le moment magnétique du fer est augmenté sur plusieurs
sphères de coordination autour de l'impureté (plus de 10\,\AA{}).
Cependant, Collins et Low montrent que la portée est encore supérieure
dans les alliages plus riches en chrome.

\subsection{Mesures thermodynamiques et estimations du diagramme de phases à
basse température}

Les premières mesures thermodynamiques datent de la fin des années
50 \cite{backhurst__1958,martens_heat_1956}. Dès les années 60, des
mesures d'enthalpie de mélange de l'alliage sont réalisées entre 10
et 90\,\%\,Cr entre 1190 et 1570\,K \cite{dench_adiabatic_1963,cook_jones_1943}.
Elles sont positives sur tout le domaine de concentration, comme
dans une classique solution régulière. Ces mesures, complétées par
des mesures de pression de vapeur de Cr et de coefficients d'activité,
sont utilisées par Kubaschewski, Heymer et Chart \cite{kubaschewski_thermodynamics_1960,kubaschewski_calc_miscibily_1965}
pour proposer une première prédiction du diagramme de phases de l'alliage.
La température critique de la lacune de miscibilité est raffinée
\cite{muller_thermodynamic_1969,rao_computed_miscib_1972}.  Jusqu'ici,
toutes les mesures thermodynamiques, réalisées à hautes températures,
sont en accord relatif avec le modèle classique de solution régulière.
Cependant, en 1953, Masumoto et al. \cite{masumoto_chaleurspe_1953}
montrent que la chaleur spécifique de l'alliage dépend de la concentration,
ce qui est en désaccord avec le modèle de solution régulière. En particulier,
un alliage à moins de 24\,\%\,Cr montre un comportement spécifique
qu'ils attribuent à une transformation liée au magnétisme. La température
critique de référence est mesurée par Kuwano et Hamaguchi \cite{kuwano_mssbauer_1988}
par spectroscopie Mössbauer à 950\,\,K. 

\begin{figure}[h]
\begin{centering}
\includegraphics[scale=0.32]{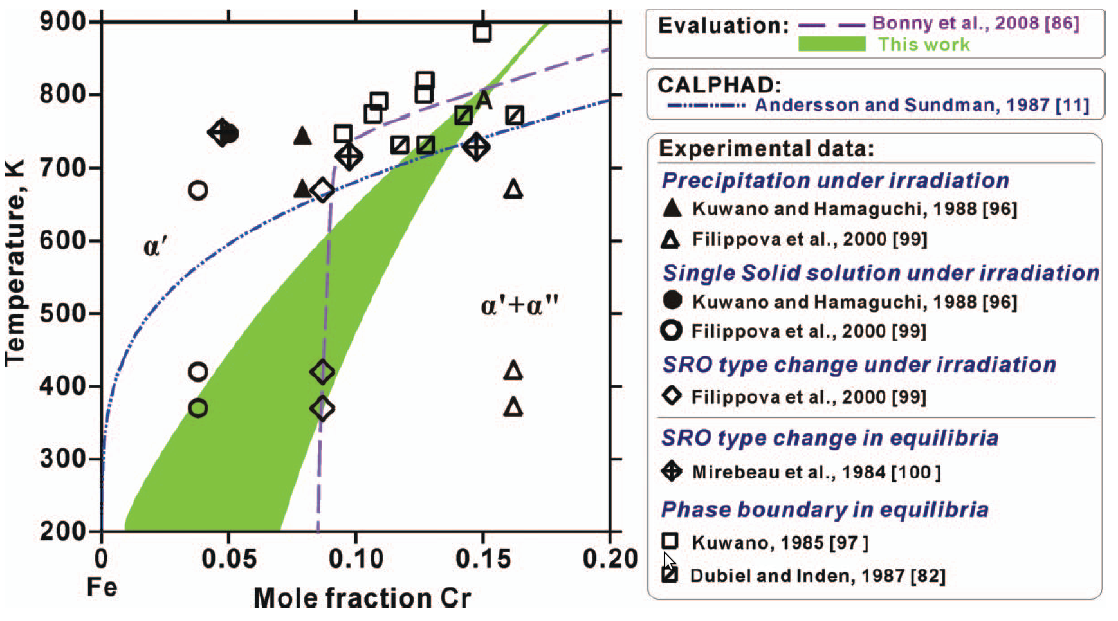}
\par\end{centering}

\caption{Comparaison des limites de solubilité $\alpha$--$\alpha'$ expérimentales,
calculées  par Andersson et Sundman \cite{andersson_thermodynamic_1987}
et estimées par Bonny et al. \cite{bonny_onthe_aa_demixtion_2008}
et Xiong et al. \cite{xiong_grrrrbonny_2010}. Publié par Xiong et
al. \cite{xiong_grrrrbonny_2010}. Se référer à l'article de Xiong
et al. pour le détail des points expérimentaux. \label{fig:nouvelles_limites}}
\end{figure}

Bonny et al. \cite{bonny_onthe_aa_demixtion_2008} puis Xiong et
al. compilent les résultats expérimentaux les plus récents et proposent
de nouvelles limites de solubilité qui remettent en cause les extrapolation
à partir des hautes températures. On représente la compilation de
Xiong et al. sur la figure \ref{fig:nouvelles_limites}. L'évaluation
de Bonny et al. est indiquée en tirets mauves.

Bonny et al. et Xiong et al., malgré des interprétations expérimentales
parfois divergentes, proposent des limites de solubilité très similaires.
Les limites de solubilité de Andersson et Sundman faisant jusqu'ici
référence tendent vers zéro à température nulle. Au contraire, Bonny
et al. et Xiong et al. proposent une solubilité assez large de Cr
dans Fe à basse température, de l'ordre de quelques pourcents atomiques.
Alors que Xiong et al. proposent un domaine de limites possibles,
Bonny et al. proposent une solubilité limitée de Cr dans Fe de $8 pc$ Cr
à 0 K. Ces hypothèses, qui ne font pas encore consensus \cite{xiong_grrrrbonny_2010},
relancent le développement de modèles de cohésion qui permettraient
de calculer (et non plus extrapoler) les limites de solubilité à basse
température.

\subsection{Les modèles énergétiques}

D'excellentes revues ont été publiées très récemment par Bonny, Malerba
ou Xiong et al. sur les modèles énergétiques pour les calculs thermodynamiques
sur le système fer--chrome \cite{bonny_stateofart_thermo_numeric_2009,malerba_modelling_2007,Malerba_revue_2008,caro_implications_2006,dudarev_eu_2009,xiong_grrrrbonny_2010}.

De nombreux modèles classiques ont tenté de reproduire les propriétés
de l'alliage Fe--Cr ferritique : ajustement sur la solution régulière
par Williams \cite{williams__1957,williams_further_1958}, ajustements
CALPHAD ou encore développements en amas par Lavrentiev et al. \cite{lavrentiev_classicalCE_FeCr_2007,nguyen-manh_classicalCE2_FeCr_2008}.
Ces modèles classiques sont mis en défaut par les calculs \emph{ab
initio} qui montrent que le moment magnétique des atomes de Cr dépend
de la composition chimique locale, induisant des propriétés de mélange
singulières \cite{klaver_magnetism_2006,olsson_ab_2003,Olsson_defautsdft_2007,olsson_electronic_2006}.
De nouveaux modèles physiques ont ainsi été développés pour reproduire
ces propriétés en étendant les modèles d'Ising \cite{ackland_magnetically_2006,ackland_orderedsigma_2009,inden_thermodynamicFeCr_2008}
ou les développements en amas \cite{lavrentiev_CEmag2_2010,lavrentiev_dudi_magnetic_bccfcc_2010,lavrentiev_magnetic_CE_2009}
à un nouveau degré de liberté : le moment magnétique atomique. Ces
modèles sont discutés en détail dans le chapitre 3 dédié à la thermodynamique
du système, dans une section introductive à notre modèle.

Le traitement explicite de la dépendance du moment magnétique des
atomes de Cr en la composition chimique locale est une première difficulté
numérique. L'autre difficulté concerne la prise en compte de l'entropie
non-configurationnelle. La modélisation de ce système est une entreprise
difficile mais une large communauté y travaille conjointement.

\section{Les surfaces libres de l'alliage}

Les surfaces libres des alliages ferritiques industriels ont été largement
étudiées à l'échelle macroscopique du fait de leur importance technologique
due à leur résistance à la corrosion pour un coût faible. Leur réactivité
à l'oxygène de l'environnement rend les études expérimentales sur
les surfaces libres difficiles car elles nécessitent un ultra-vide.
De plus, leur magnétisme rend les observations au microscope en transmission
également difficiles. Les études atomistiques des surfaces libres
des éléments Fe et Cr purs et de l'alliage modèle Fe--Cr sont donc
plus rares.

\subsubsection*{Lien avec l'oxydation}

De façon très générale, c'est la formation d'une couche d'oxyde de
chrome Cr$_{\text{2}}$O$_{\text{3}}$ très stable, qui empêche la
diffusion de l'oxygène de la surface vers le volume, qui rend les
aciers à plus de 7--9\,\%\,Cr passifs face à la corrosion.

L'oxydation est un phénomène de nature principalement électrochimique,
lié au potentiel de la surface. Dans cette étude, nous ne tenons pas
compte de ces phénomènes. Il ne s'agit donc pas ici de modéliser l'oxydation
des matériaux. Pour une revue détaillée de l'oxydation des matériaux
de transition et sur les phénomènes de corrosion, on peut se référer
à l'ouvrage généraliste de Landolt \cite{landolt_1997}. On peut tout
de même espérer qu'une meilleure compréhension des propriétés thermodynamiques
de l'alliage Fe--Cr sera utile pour comprendre leur oxydation.

\subsubsection*{Magnétisme, surfaces libres et modèles énergétiques}

Qualitativement, l'effet du magnétisme sur les surfaces libres de
métaux de transition a été largement étudié \cite{andrieu_surfaces_solides_livre_2005}.
Les atomes en surface ont une coordinence plus faible qu'en volume.
Leur bande électronique $d$ est plus étroite, ce qui a pour effet
d'augmenter leur moment magnétique selon le critère de Stoner \cite{stoner_1939}. 

De tous les modèles introduits dans la sous-section \ref{sub:Les-mod=0000E8les-de_coh=0000E9sion},
aucun ne permet de construire le diagramme de phases complet du système
fer--chrome tout en reproduisant la dépendance en environnement chimique
local du moment magnétique local. Ackland et Ropo et al. \cite{ackland_orderedsigma_2009,ropo_Crseg_2007}
ont proposé les premiers modèles permettant de prédire la concentration
surfacique de l'alliage en fonction de la concentration volumique.
Cependant, aucun de ces deux modèles ne reproduit de façon satisfaisante
l'effet des surfaces sur les propriétés magnétiques et énergétiques.
Nous les discutons en détail dans le chapitre 4.

\section{Conclusion}

Les aciers ferritiques sont des candidats sérieux comme matériaux
de structure pour les centrales nucléaires du futur. C'est le système
binaire fer--chrome qui est le modèle d'étude de ces matériaux. Son
diagramme de phases montre aux températures d'intérêt pour le nucléaire
une coexistence entre les solutions solides $\alpha$ riche en fer
et $\alpha'$ riche en chrome. Les limites de solubilité de ces solutions
solides expérimentées seulement aux température élevées restent mal-connues
à basse température. À haute température, l'entropie de vibration
au moins est très importante. À des températures plus basses, l'existence
d'une transition entre une tendance à l'ordre et une tendance à la
démixtion aux alentours de 10\,\% en chrome pourrait indiquer des
limites de solubilité inattendues. Cela semble confirmé par des calculs
\emph{ab initio} récents qui montrent que le moment magnétique atomique
des atomes de chrome est particulièrement sensible à son environnement
chimique local. Cela rend le mélange à basse température complexe,
et les conséquences sur les limites de solubilité restent hypothétiques.
Il semblerait que la solubilité du chrome dans le fer-$\alpha$ soit
supérieure à la limite extrapolée depuis les hautes températures.

\section{Notre contribution}

Cette méconnaissance des limites de solubilité à basse température
explique pourquoi de nombreux modèles énergétiques ont été développés
ces dernières années. Il s'agit de développer un modèle énergétique
qui puisse être utilisé pour le calcul de propriétés en température
par des méthodes statistiques (simulations Monte Carlo, champs moyens
...). La physique de ces matériaux est de plus largement modifiée
en présence de surfaces qui sont le lieu des phénomènes de corrosion.
Jusqu'à aujour'hui, aucun modèle présenté n'est satifaisant à la fois
en volume et en surface.

C'est ce dernier point qui a motivé notre étude. Comment construire
un modèle énergétique qui se révèle juste à basse température malgré
le peu d'indices expérimentaux, tout en restant valide à haute température
? Comment tenir compte de la présence des surfaces ? Peut-on prévoir
la ségrégation de l'un des éléments en surface ?

Pour tenter de répondre à ces questions, nous avons choisi de compléter
les connaissances des propriétés énergétiques, magnétiques et structurales
de l'alliage binaire fer-chrome cubique centré à très basse température
par des calculs \emph{ab initio}. Nous utilisons dans un deuxième
temps ces propriétés pour la construction d'un modèle énergétique
dont nous extrayons les limites de solubilité à toutes températures
par différents traitements statistiques. Enfin, en lien avec les phénomènes
de corrosion, nous apportons des éléments quant à la ségrégation du
chrome dans le fer en présence de surfaces:
\begin{enumerate}
\item On complète nos connaissances de l'alliage à très basse température
par des calculs \emph{ab initio ;}
\item On utilise ces résultats ainsi que ceux de la littérature pour construire
un modèle énergétique ;
\item Par un traitement statistique de champ moyen et des simulations Monte
Carlo, on déduit les limites de solubilité ;
\item On développe le modèle pour qu'il tienne compte des surfaces dont
on modélise l'enrichissement en l'un des éléments par un traitement
de champ moyen sur site.
\end{enumerate}

\chapter{Calculs DFT et propriétés énergétiques\label{cha:DFT}}

\malettrine{L}{}e système fer-chrome reste mal-connu à basse température.
En particulier, les surfaces de l'alliage sont très difficiles à observer
expérimentalement.

C'est la raison pour laquelle nous commençons cette étude par la construction
d'une base de données \emph{ab initio} des propriétés énergétiques,
électroniques, magnétiques et structurales de l'alliage à 0\,K à
l'échelle atomique.

Dans un premier temps, nous introduisons la théorie de la fonctionnelle
de la densité électronique que nous avons mise en œuvre. Dans ce travail,
nous avons utilisé deux codes de calculs DFT pour évaluer les approximations
des fonctionnelles d'échange et corrélation, des pseudo-potentiels
et des bases. 

Ensuite, nous présentons nos résultats à 0\,K à partir d'une étude
des éléments purs pour valider notre méthodologie. Nous étudions dans
un deuxième temps les propriétés de mélange des deux éléments dans
l'ensemble du domaine de concentration en commençant par l'introduction
d'impuretés puis en étendant la concentration en éléments substitutionnels.
Nous détaillons l'effet de l'évolution de la concentration en chrome
sur les propriétés structurales (paramètre de maille, dilatations
ou compressions locales), énergétiques (énergies de mise en solution
et de mélange) et magnétiques (moment magnétique atomique local et
moyen) de l'alliage, en lien avec sa structure électronique.

Dans une dernière section, nous conservons la démarche décrite précédemment
pour l'étude des surfaces libres de l'alliage, que nous comparons
à la littérature. Nous discutons de l'effet des surfaces sur la structure
électronique de chacun des éléments, puis étudions la ségrégation
de Cr dans Fe à différentes concentrations volumiques et surfaciques.\vspace{1cm}

\section{Description succincte de la théorie de la fonctionnelle de la densité
mise en œuvre}

Dans la première partie de notre travail, nous utilisons la théorie
de la fonctionnelle de la densité électronique (\og density functionnal
theory \fg{} ou DFT en anglais) comme un outil pour le calcul de
propriétés à l'échelle atomique du système fer-chrome. Dans ce sous-chapitre,
nous rappelons succinctement les fondements de cette théorie, ses
approximations et ses limites.

La résolution des équations issues de la DFT nécessite des hypothèses
physiques et des approximations que nous rappelons. Nous comparons
alors les deux codes de calcul que nous utilisons en mettant en lumière
les approximations différentes implémentées dans chacun de ces codes.

Pour en savoir plus sur la DFT et ses implémentations, nous recommandons
le livre de revue de Martin \cite{martin_electronic_2004} et les
ouvrages particulièrement pédagogiques de Springborg \cite{springborg_2000}
et Koch \cite{koch_chemistDFT_2001}.

\subsection{La densité électronique comme unique variable : le théorème de Hohenberg
et Kohn}

L'objet de la DFT est de remplacer le problème insoluble des interactions
à $N$ corps (ici des électrons) par un problème à une unique variable
d'intégration : la densité électronique $n\left(\mathbf{r}\right)$
\cite{thomas_1927,fermi_1927}. En 1964, Hohenberg  et Kohn établissent
\cite{hohenberg_inhomogeneous_1964} \emph{ad absurdo} que toutes
les propriétés à l'état fondamental (souvent appelées \og observables \fg{})
d'un système à $N$ électrons en interaction soumis à un potentiel
extérieur $v_{ext}\left(\mathbf{r}\right)$ sont des fonctionnelles
de la densité électronique de l'état fondamental $n_{0}\left(\mathbf{r}\right)$.
Connaissant $n_{0}\left(\mathbf{r}\right)$, toutes les propriétés
de l'état fondamental sont déterminées. Ainsi, en vertu du principe
variationnel, l'énergie totale de l'état fondamental d'un système
à $N$ électrons en interaction dans un potentiel extérieur peut être
obtenue par minimisation de l'énergie de Hohenberg et Kohn, $E_{HK}\left[n\right]$.

Nos systèmes sont composés de noyaux et d'électrons. L'énergie totale
du système \{électrons + noyaux\} est donc la somme de $E_{HK}$ (due
aux électrons uniquement) et de l'énergie coulombienne classique entre
électrons et noyaux $E_{\acute{e}-N}$ et entre noyaux $E_{N-N}$.
\begin{equation}
E_{tot}=E_{HK}+E_{\acute{e}-N}+E_{N-N}
\end{equation}
Dans l'approximation adiabatique, les noyaux sont des charges fixes.
$E_{\acute{e}-N}$ et $E_{N-N}$ ont donc la forme classique d'une
énergie d'interaction électrostatique. L'énergie de Hohenberg et Kohn
$E_{HK}$ s'exprime : 
\begin{equation}
E_{HK}\left[n\right]=E_{H}\left[n\right]+E_{ext}\left[n\right]+T\left[n\right]+E_{xc}\left[n\right]
\end{equation}
où l'énergie de Hartree $E_{H}\left[n\right]$ est la part électrostatique
(non-quantique) de l'énergie d'interaction électronique :
\begin{equation}
E_{H}=\frac{1}{2}\int d\mathbf{r}d\mathbf{r'}\frac{n\left(\mathbf{r}\right)n\left(\mathbf{r'}\right)}{\left|\mathbf{r}-\mathbf{r'}\right|}.\label{eq:E_hartree}
\end{equation}
$E_{ext}\left[n\right]$ est l'énergie coulombienne (non-quantique)
d'interaction des électrons avec un potentiel électrostatique extérieur
$v_{ext}$:

\[
E_{ext}=\int d\mathbf{r}n\left(\mathbf{r}\right)v_{ext}\left(\mathbf{r}\right)
\]
$E_{xc}\left[n\right]$ est l'énergie d'échange et corrélation, part
non-électrostatique de l'énergie totale due aux interactions entre
électrons, et $T\left[n\right]$ est l'énergie cinétique des électrons.
Nous savons \emph{ad absurdo} que $T\left[n\right]$ et $E_{xc}\left[n\right]$
existent, sans avoir de forme analytique. Ce problème est l'objet
de l'ansatz de Kohn et Sham.

\subsection{L'ansatz de Kohn-Sham}

Kohn et Sham utilisent le théorème de Fermi sur le gaz d'électrons
libres pour minimiser la partie inconnue de $E_{HK}$ \cite{kohn_self-consistent_1965}.
Ils dissocient l'énergie cinétique $T$ des électrons en interaction
en :
\begin{itemize}
\item une partie correspondant à l'énergie cinétique d'un gaz d'électrons
libres soumis à un potentiel. Ce potentiel, dit de Kohn et Sham, $v_{KS}$,
est choisi de sorte que les électrons libres aient la même densité
à l'état fondamental que les électrons en interaction.
\item le reste, correspondant à la partie corrélation de l'énergie cinétique.
Cette dernière partie est dans la suite des écritures intégrée à $E_{xc}$.
\end{itemize}
Cet ansatz est un jeu de réécriture du problème. La nouvelle expression
de l'énergie d'échange et corrélation $E_{xc}$ doit être approximée
: c'est le rôle des fonctionnelles d'échange et corrélation.

\subsection{Les fonctionnelles d'échange et corrélation}

De l'énergie totale, seule l'énergie d'échange et corrélation $E_{xc}$
n'est pas précisément définie et doit être approximée. Kohn et Sham
\cite{kohn_self-consistent_1965} proposent à cette fin l'approximation
de la densité locale (\og Local Density Approximation \fg{}, la
LDA).

\subsubsection*{L'approximation de la densité locale}

Kohn et Sham réécrivent l'énergie d'échange et corrélation comme l'intégrale
d'une densité d'énergie d'échange et corrélation $\epsilon_{xc}\left[n\right]$
\begin{equation}
E_{xc}\left[n\right]=\int d\mathbf{r}n\left(\mathbf{r}\right)\epsilon_{xc}\left[n\right]\left(\mathbf{r}\right)
\end{equation}

en faisant l'hypothèse qu'en chaque point de l'espace \textbf{r},
la densité d'énergie d'échange et corrélation $\epsilon_{xc}\left[n\right]\left(\mathbf{r}\right)$
est la même que celle d'un gaz d'électrons homogène de même densité
$n\left(\mathbf{r}\right)$. La densité d'énergie d'échange et corrélation
$\epsilon_{xc}$ d'un gaz d'électrons homogène a été estimée par simulations
Monte Carlo quantiques à haute et basse densité électronique \cite{ceperley_alder_1980,ceperley_1978}.
L'interpolation analytique entre ces deux limites est l'expression
de la densité d'énergie d'échange et corrélation $\epsilon_{xc}$
comme fonction de la densité électronique $n$. Le choix de la fonction
d'interpolation est un paramètre des calculs LDA.

\subsubsection*{Au delà de la LDA : la GGA et les développements en gradients}

La LDA peut se révéler imprécise (voir tableau \ref{tab:comparaison_fonctionnelles}).
Des développements en gradient permettent de mieux tenir compte des
inhomogénéités de la densité électronique des systèmes réels. Le développement
en gradient le plus simple de $\epsilon_{xc}$ est appelé \og approximation
des gradients généralisée \fg{} (GGA). $\epsilon_{xc}$ est alors
une fonction de la densité locale $n\left(\mathbf{r}\right)$ et de
son gradient $\nabla n\left(\mathbf{r}\right)$. Contrairement à la
LDA, et bien que des approches systématiques existent \cite{perdew_GGAsimple_1996},
plusieurs paramètres ajustables sont nécessaires au développement
d'une fonctionnelle GGA. Celle-ci améliore cependant souvent la qualité
des résultats par rapport à la LDA, même si elle sur-corrige parfois
cette dernière \cite{martin_electronic_2004}. Les informations compilées
dans le tableau \ref{tab:comparaison_fonctionnelles}, limitées car
très génériques, permettent toutefois de comparer les effets des fonctionnelles
d'échange et corrélation LDA et GGA sur les propriétés calculées des
éléments.

\begin{table}[h]
\begin{centering}
\begin{tabular}{|c|c|c|}
\hline 
 & LDA & GGA \\
\hline 
\hline 
paramètre de maille & sous-estime & surestime \\
\hline 
module de compressibilité & surestime & pas de tendance \\
\hline 
énergie de cohésion & surestime & pas de tendance \\
\hline 
énergie de formation des lacunes & raisonnable & parfois imprécise \\
\hline 
\end{tabular}
\par\end{centering}

\caption{Comparaison des tendances générales de l'effet des fonctionnelles
d'échange et corrélation LDA et GGA sur les propriétés physiques calculées
dans les métaux de transition. \label{tab:comparaison_fonctionnelles}}
\end{table}

Le tableau \ref{tab:comparaison_fonctionnelles} est indicatif mais
trop général pour justifier le choix du type de fonctionnelle d'échange
et corrélation pour l'étude d'un système précis. Nous verrons au paragraphe
\ref{sub:choix_GGA_sur_Fe} que ce choix a des conséquences importantes
sur les propriétés calculées, comme l'état structural et magnétique
de plus basse énergie \ldots{} Le choix de la fonctionnelle LDA ou
GGA se fait donc en fonction du ou des éléments que l'on veut étudier.

Notons qu'il existe des développements en gradients de plus haut degré.
Par exemple, la méta-GGA de Perdew et al. \cite{perdew_metagga_1999}
propose de tenir compte du laplacien de la densité locale. Plus lourde
numériquement, la méta-GGA n'améliore pas toujours la qualité des
résultats.

Dans cette étude, nous utilisons des fonctionnelles d'échange et corrélation
LDA et GGA.

\subsubsection*{La polarisation en spin}

Les premiers calculs polarisés en spin datent des années 1970 \cite{von_barth_LSD_1972}.
Dans ces calculs, la variable de densité électronique est subdivisée
en deux densités scalaires ($\uparrow$ et $\downarrow$) dont la
somme est la densité électronique dont nous discutions jusqu'alors
: 
\begin{equation}
n\left(\mathbf{r}\right)=n_{\uparrow}\left(\mathbf{r}\right)+n_{\downarrow}\left(\mathbf{r}\right)
\end{equation}
La somme de leurs intégrales :
\begin{eqnarray}
\int d\mathbf{r}n\left(\mathbf{r}\right) & = & \int d\mathbf{r}n_{\uparrow}\left(\mathbf{r}\right)+\int d\mathbf{r}n_{\downarrow}\left(\mathbf{r}\right)\\
 & = & N_{\uparrow}+N_{\downarrow}\\
 & = & N
\end{eqnarray}
où $N$ est le nombre total d'électrons du système. $N_{\uparrow}$
($N_{\downarrow}$) est le nombre total d'électrons de moment $\uparrow$
($\downarrow$).

Le moment magnétique $M$ de chaque atome (le spin atomique) est la
différence entre les deux densités électroniques : 
\begin{equation}
M=N_{\uparrow}-N_{\downarrow}
\end{equation}

Dans cette approximation de polarisation en spin scalaire, l'énergie
d'échange et corrélation s'écrit : 
\begin{equation}
E_{xc}\left[n_{\uparrow},n_{\downarrow}\right]=\int d\mathbf{r}n\left(\mathbf{r}\right)\epsilon_{xc}\left[n_{\uparrow},n_{\downarrow}\right]\left(\mathbf{r}\right)
\end{equation}

À la fin des années 1980, la DFT est étendue aux systèmes de spins
non-colinéaires \cite{kubler_DFT_spinnoncolineair_1988} : le moment
magnétique atomique $M$ devient vectoriel.

Dans cette étude, nous utilisons : 
\begin{itemize}
\item la DFT non-polarisée en spin
\item la DFT polarisée en spin, exclusivement dans l'approximation des spins
colinéaires.
\end{itemize}

\subsection{Les codes SIESTA et PWSCF}

Deux codes ont été utilisés pour cette étude : SIESTA (Spanish Initiative
for Electronic Simulations with Thousands of Atoms) \cite{ordejon_self-consistent_1996,anglada_systematic_2002,artacho_linear-scaling_1999,junquera_numerical_2001,sanchez-portal_density-functional_1997,soler_siesta_2002,siesta_homepage}
et PWSCF (Plane-Wave Self-Consistent Field) \cite{quantumespresso_homepage,giannozzi_quantumespresso_2009}.
Ces deux codes utilisent l'approximation des pseudo-potentiels et
résolvent les équations de Kohn et Sham dans des bases différentes.
Le tableau \ref{tab:SIESTA-et-PWSCF.} compare les propriétés principales
de SIESTA et PWSCF ainsi que leurs avantages et inconvénients. Le
détail de cette comparaison fait l'objet de la suite de cette section.

\begin{table}[h]
\begin{centering}
\begin{tabular}{|c|c|c|}
\hline 
 & SIESTA & PWSCF \\
\hline 
\hline 
pseudo-potentiels & NC & PAW / NC / US \\
\hline 
base & orbitales localisées & ondes planes \\
\hline 
résultats & moins robuste & très robuste \\
\hline 
calculs en volume & très efficace & efficace$^{(1)}$ \\
\hline 
calculs en surface & très efficace & moins efficace \\
\hline 
\end{tabular}
\par\end{centering}

\caption{Comparaison des propriétés principales des codes SIESTA et PWSCF et
de leurs points faibles et forts. Les pseudo-potentiels NC sont à
norme conservée quand les PAW et US sont des Projector Augmented Wave
et Ultra-Soft\protect \\
$^{(1)}$ en fonction des symétries.\label{tab:SIESTA-et-PWSCF.}}
\end{table}

\subsection{Les pseudo-potentiels}

\begin{figure}[h]
\begin{centering}
\includegraphics[scale=0.3]{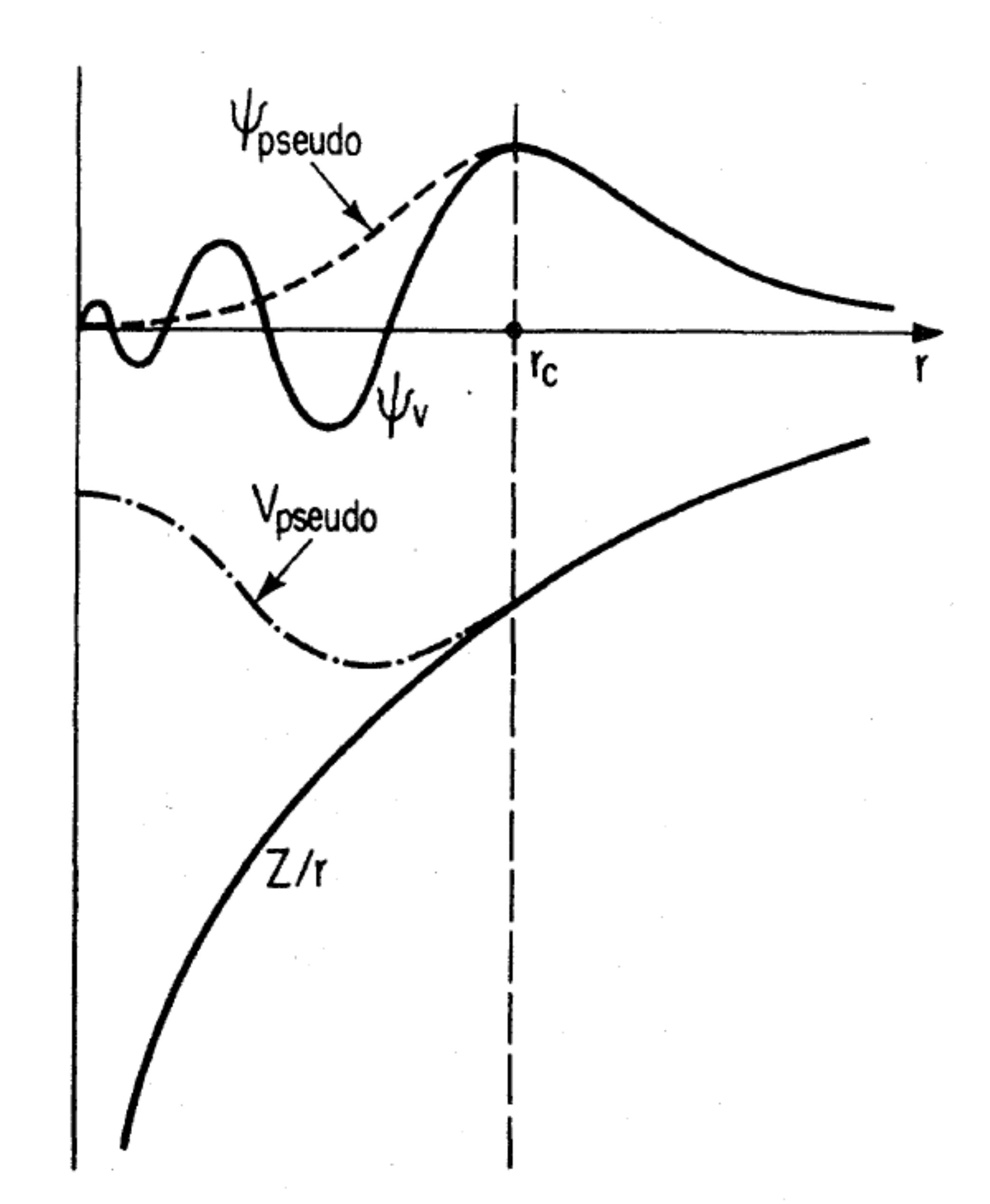}
\par\end{centering}

\caption{Comparaison schématique d'un potentiel \og tous électrons \fg{}
(trait continu) avec un pseudo-potentiel (trait pointillé), d'une
part, et des fonctions d'ondes correspondantes d'autre part. À partir
du rayon de coupure $r_{c}$, le potentiel \og tous électrons \fg{}
et le pseudo-potentiel sont indiscernables. Publié par Payne et al.
\cite{payne_abinitio_1992}. \label{fig:schema_pseudopotentiel}}

\end{figure}

Les électrons d'un atome peuvent être divisés en deux groupes : les
électrons de cœur et les électrons de valence. Les électrons de cœur
sont les électrons des couches électroniques les plus basses en énergie.
Très localisés, ils sont fortement liés au noyau de l'atome. Ils sont
en conséquence peu affectés par l'environnement chimique de l'atome
et ne participent pas ou peu aux liaisons chimiques ou à la cohésion
du solide. Ils sont par contre responsables de la plus grande partie
du coût de calcul. Dans l'approximation des pseudo-potentiels, le
potentiel coulombien du noyau et des électrons de cœur est fixé (c'est
l'approximation des cœurs gelés) et remplacé par un potentiel effectif
qui agit sur les électrons de valence. La figure \ref{fig:schema_pseudopotentiel}
illustre schématiquement l'intérêt de remplacer un potentiel \og tous
électrons \fg{} par un pseudo-potentiel : les oscillations, coûteuses
en temps de calcul, sont atténuées. Un pseudo-potentiel qui oscille
peu (beaucoup) est dit doux (dur).

Nous avons utilisé et comparé trois types de pseudo-potentiels dans
ce travail.

\subsubsection*{Les pseudo-potentiels à norme conservée}

L'approche des pseudo-potentiels à norme conservée a été développée
par Hamann, Schlüter et Chiang \cite{hamann_NC_pseudo_1979}. Elle
consiste à construire une pseudo-fonction d'onde qui répond à une
liste de critères :
\begin{itemize}
\item les énergies propres obtenues par un calcul tous-électrons ($\epsilon^{TE}$)
et les pseudo-énergies ($\epsilon^{PS}$) sont identiques :
\begin{equation}
\epsilon^{PS}=\epsilon^{TE}
\end{equation}

\item les pseudo-fonctions d'onde et les fonctions d'onde tous-électrons
sont identiques au-delà d'un rayon de cœur $r_{c}$ choisi arbitrairement
:
\begin{equation}
\Phi^{PS}\left(r\right)=\Phi^{TE}\left(r\right)\,\forall r>r_{c}
\end{equation}

\item l'intégrale de 0 à $r_{c}$ de la densité de probabilité de présence
associée aux pseudo-fonctions d'onde et à la fonction d'onde tous
électrons est identique (conservation de la norme) :
\begin{equation}
\int_{0}^{r_{c}}\left|\Phi^{PS}\left(r\right)\right|^{2}r^{2}dr=\int_{0}^{r_{c}}\left|\Phi^{TE}\left(r\right)\right|^{2}r^{2}dr.\label{eq:norme_conservee}
\end{equation}

\end{itemize}
Bien que l'implémentation de la méthode PAW soit en cours \cite{postdoc_Tristana},
l'approche à norme conservée est la seule implémentée dans SIESTA.
Il existe des critères normalisés de validation de ce type de pseudo-potentiels
qui permettent une transférabilité satisfaisante \cite{vanderbilt_pseudo_transferability_1995}.

\subsubsection*{Les pseudo-potentiels ultra-softs}

L'objectif des pseudo-potentiels est de créer des pseudo-fonctions
aussi douces que possible \cite{vanderbilt_ultrasoft_1990}. Afin
de lisser davantage les pseudo-potentiels, la condition \ref{eq:norme_conservee}
de conservation de la norme peut être relâchée : les densités électroniques
sont encore plus douces et un gain substantiel en temps de calcul
est réalisé. Des problèmes de transférabilité et de charge électronique
totale peuvent résulter. Le calcul peut alors converger vers un mauvais
état fondamental \cite{martin_electronic_2004}.

\subsubsection*{La méthode PAW}

La méthode \og Projector Augmented Wave \fg{} (PAW), proposée par
Blöchl en 1994 \cite{blochl_PAW__1994,kresse_PAW_1999}, est une méthode
\og tous-électrons \fg{} qui s'appuie sur l'idée de base des pseudo-potentiels.
Elle consiste à décomposer la vraie fonction d'onde $\Psi$ en plusieurs
parties. Chaque partie est transformée, avec tous ses nœuds, en une
fonction d'onde $\Psi_{PAW}$ qui se développe aisément dans la base
choisie.

Dans l'approximation du cœur gelé, et uniquement dans ce cas, une
correspondance exacte peut être faite entre les pseudo-potentiels
ultra-softs et le formalisme PAW. Proche des calculs tous-électrons,
la méthode PAW est plus robuste que l'approximation des pseudo-potentiels
ultra-softs ou à norme conservée.

\subsection{Les bases}

Afin de résoudre les équations de Kohn-Sham à un électron, les fonctions
d'onde $\Psi_{i}$ sont développées linéairement dans une base d'orbitales
$\left|\alpha_{i}\right\rangle $ :
\begin{equation}
\left|\Psi_{i}\right\rangle =\sum_{p=1}^{n}c_{i,p}\left|\alpha_{i}\right\rangle 
\end{equation}

Les orbitales $\left|\alpha_{i}\right\rangle $ de la base peuvent
être localisées (gaussiennes, orbitales pseudo-atomiques, ...), délocalisées
(ondes planes ou sphériques, ...) ou mixtes.

\subsubsection*{Les bases localisées}

SIESTA \cite{ordejon_self-consistent_1996,sanchez-portal_density-functional_1997,artacho_linear-scaling_1999,junquera_numerical_2001,soler_siesta_2002,anglada_systematic_2002,siesta_homepage}
utilise des bases localisées. Ce sont des orbitales numériques pseudo-atomiques
qui s'annulent au-delà d'un rayon de coupure $R$. La fonction d'onde
$\Psi$ de $i$ atomes à $q$ orbitales pseudo-atomiques $\psi_{i}$
de rayon de coupure $\mathbf{R}_{i}$.
\begin{equation}
\Psi=\sum_{i}\sum_{p=1}^{q}c_{i,p}\psi_{i,p}\left(\mathbf{r}_{i,p}-\mathbf{R}_{i}\right)
\end{equation}

Les bases d'orbitales pseudo-atomiques sont efficaces, mais plus
dépendantes du système, contrairement par exemple aux ondes planes.
Mathématiquement, ce n'est pas un ensemble complet, contrairement
aux bases d'ondes planes. La convergence n'est donc pas systématique
quand on augmente la taille de la base. Elles sont particulièrement
adaptées aux systèmes moléculaires ou aux solides présentant des défauts
comme par exemple des surfaces. Il faut cependant prendre soin de
bien décrire l'évanescence de la densité électronique à la surface.

\subsubsection*{Les bases d'ondes planes}

PWSCF utilise une base d'ondes planes. À chaque énergie est associée
l'énergie cinétique d'une onde plane. C'est une base mathématiquement
complète : théoriquement, une infinité d'ondes planes reproduit l'état
fondamental de n'importe quel système. Une énergie de coupure $E_{cutoff}$
est définie. Elle correspond à l'énergie cinétique maximum des ondes
planes à considérer.

\subsubsection*{Projections de la densité électronique sur les bases (localisées
et d'ondes planes)}

Dans cette étude, nous faisons souvent appel à la notion de moment
magnétique atomique, qui est une grandeur calculée à partir des densités
électroniques de spin $\uparrow$ et $\downarrow$ résultant de la
boucle d'auto-cohérence électronique. La densité est projetée sur
les bases puis la charge totale et le moment magnétique atomique sont
calculés. Dans une base localisée, le calcul du moment magnétique
sur chaque atome se fait directement (à quelques éléments non-diagonaux
de la matrice de densité électronique près) par une projection de
la densité électronique sur chacune des orbitales pseudo-atomiques.
C'est l'analyse classique de Mulliken \cite{mulliken_1955}. Dans
les bases d'ondes planes, la densité électronique est projetée sur
des sphères centrées sur les noyaux atomiques. Le rayon des sphères
reste au choix de l'utilisateur. Des recouvrements ou au contraire
des zones de l'espace qui n'appartiennent à aucune sphère et ne sont
pas projetées peuvent exister. On peut également choisir de diviser
l'espace en cellules de Voronoï \cite{voronoi_1907} ou projeter sur
des orbitales pseudo-atomiques. La méthode de projection en ondes
planes est plus arbitraire qu'en bases localisées.

\subsubsection*{Avantages et inconvénients des bases localisées et des bases d'ondes
planes}

Les avantages et inconvénients des bases d'orbitales pseudo-atomiques
et des bases d'ondes planes sont comparés dans le tableau \ref{tab:Comparaison_bases}.

\begin{table}[h]
\begin{centering}
\begin{tabular}{|c|c|c|}
\hline 
 & bases localisées & ondes planes \\
\hline 
\hline 
taille de la base & faible & importante \\
\hline 
convergence systématique & non  & oui \\
\hline 
précision & moins précis & plus précis \\
\hline 
description du vide & très adaptées & peu efficace \\
\hline 
représentation & intuitives & peu intuitives \\
\hline 
projections (charge, $M$) & directes & indirectes \\
\hline 
\end{tabular}
\par\end{centering}

\caption{Comparaison qualitative des bases d'orbitales localisées pseudo-atomiques
utilisées par SIESTA et des bases d'ondes planes utilisées par PWSCF.\label{tab:Comparaison_bases}}
\end{table}

La taille de la base à manipuler par exemple lors de la diagonalisation
de la matrice hamiltonien est beaucoup plus faible dans le cas des
bases localisées qui ne contiennent au plus que quelques dizaines
d'orbitales par atome.

Comme nous l'indiquions au chapitre précédent, les bases localisées
ne forment pas une base complète au sens mathématique. Il n'y a donc
pas de convergence systématique en augmentant la taille de la base,
contrairement aux bases d'ondes planes.

Les bases localisées ont cependant l'avantage de l'efficience de calcul.

\subsection{Bilan sur les pseudo-potentiels et les bases}

Pour des raisons d'efficacité numérique de calcul développées dans
la suite du manuscrit, nous utilisons dans cette étude majoritairement
le code SIESTA utilisant des pseudo-potentiels à norme conservée et
des bases d'orbitales pseudo-atomiques. Cela permet la réalisation
de calculs systématiques dans des super-cellules contenant plusieurs
centaines d'atomes et des défauts étendus, par exemple pour l'étude
de systèmes dilués et des surfaces libres de l'alliage. Ponctuellement,
nous confrontons ces résultats aux calculs PWSCF utilisant la méthode
PAW et US pour qualifier et, dans la mesure du possible, évaluer l'influence
de l'approximation des pseudo-potentiels et/ou de la base.

Dans un effort de clarté, dans la suite de ce manuscrit, les calculs
réalisés avec SIESTA ou PWSCF seront notés SIESTA-NC, PWSCF-PAW et
PWSCF-US, selon le type de pseudo-potentiel mis en œuvre.

\section{L'alliage Fe--Cr volumique : calculs \emph{ab initio} antérieurs}

De nombreux calculs DFT ont déjà été réalisés pour l'alliage Fe--Cr.
L'objet de cette deuxième section est d'en faire une revue afin d'éclairer
les résultats présentés dans la section suivante.

\subsection{L'impureté Cr substitutionnelle dans Fe}

Le lien entre structure électronique et stabilité des impuretés substitutionnelles
$3d$ et $4d$ dans les métaux de transition est étudié par Drittler
et al. \cite{drittler_green_1989}. Ces auteurs calculent en KKR-LDA
par la méthode des fonctions de Green la structure électronique des
impuretés de métaux de transition dans le fer. 
\begin{itemize}
\item la bande de spin $\downarrow$ (qu'on appellera parfois minoritaire)
de l'impureté V, Cr, Mn, Co dépend peu de la nature de l'élément chimique
: elle est divisée en deux pics séparés par un minimum (décrit Drittler
et al. \cite{drittler_green_1989} comme un pseudo-gap) dans lequel
se situe le niveau de Fermi $E_{F}$ qui sépare les états liants ($E<E_{F}$)
des états antiliants ($E>E_{F})$.
\item selon la nature de l'impureté, la bande de spin $\uparrow$ (qu'on
appellera parfois majoritaire) se déplace autour du niveau de Fermi.
Cette bande est le principal contributeur à la (dé)stabilisation de
l'élément substitutionnel : l'énergie du système est directement liée
à la position des extrema de densité électronique par rapport au niveau
de Fermi.
\end{itemize}

Dans le fer pur, l'énergie de Fermi se situe dans le pic antiliant
de la densité d'états du canal majoritaire, il contient 0.2 à 0.3
états vides. Cela fait du fer un ferromagnétique faible.

L'énergie de Fermi de l'impureté Cr dans Fe est dans le pseudo-gap
du canal majoritaire. La bande $\uparrow$ est décalée vers les plus
hautes énergies, le pic antiliant est essentiellement vide. Cela induit
un moment magnétique faible.

\subsection{L'alliage Fe--Cr}

\subsubsection*{L'alliage non-magnétique\label{sub:Les-calculs-non-magn=0000E9tiques}}

Les premiers calculs de structure électronique sont non-polarisés,
puis polarisés uniquement pour Fe \cite{moroni_electronic_1993,turchi_stabilitySROCPA_1994},
dans l'approximation du potentiel cohérent (CPA)\nomenclature{CPA}{Coherent Potential Approximation -- Approximation du potentiel cohérent}
\cite{soven_CPA_1967,velicky_kirkpatrick_ehrenreich_1968,velicky_1969}.
L'idée développée par Soven \cite{soven_CPA_1967} puis par Velicky,
Kirkpatrick et Ehrenreich \cite{velicky_kirkpatrick_ehrenreich_1968,velicky_1969}
est de remplacer l'environnement ordonné d'un site par un potentiel
effectif moyen. Cela permet de simuler les propriétés électroniques
de l'alliage. La solution solide Fe--Cr cubique centrée se comporte
comme une solution solide régulière. Les interactions de paires effectives
déduites de ces premiers résultats permettent de calculer en Monte
Carlo des limites de solubilité $\alpha-\alpha'$ et des limites spinodales
quasi-symétriques. Elles ressemblent à celles proposées plusieurs
années plus tôt par minimisation de l'enthalpie libre déduite des
mesures à haute température \cite{williams_miscibility_1974}.

\subsubsection*{L'alliage magnétique}

\paragraph{Vers une prise en compte du magnétisme}

La méthode DLM (Disordered Local Moment) permet de reproduire le désordre
magnétique de la phase paramagnétique. Le système binaire $\mathrm{Fe_{x_{Fe}}Cr_{x_{Cr}}}$
est décrit comme un quaternaire $\mathrm{\left(Fe_{\downarrow}Fe_{\uparrow}\right)_{x_{Fe}}\left(Cr_{\downarrow}Cr_{\uparrow}\right)_{x_{Cr}}}$.
Cette approximation a l'avantage de mieux simuler l'état paramagnétique
de l'alliage au-dessus de $T_{Curie}$ que la suppression artificielle
des moments magnétiques atomiques par des calculs non-polarisés en
spin, car elle tient compte des couplages magnétiques. Cette prise
en compte du magnétisme stabilise la phase paramagnétique de plusieurs
dixièmes d'eV \cite{akai_CPA_1993}, ce qui remet en cause les conclusions
basées sur les calculs non-magnétiques présentés ci-dessus : en phase
diluée, les moments magnétiques des atomes de chrome sont anti-alignés
aux moments des atomes de fer \cite{Kulikov_CPA_1997}.

\paragraph{La densité d'états électroniques projetée (PDOS) moyenne de l'alliage
Fe--Cr}

Lorsque la proportion de Fe augmente, le canal majoritaire se remplit
alors que la bande minoritaire ne bouge pas \cite{ghosh_electronic_2001}.
Le moment magnétique moyen $\left\langle M\right\rangle =M_{\uparrow}-M_{\downarrow}$
augmente ainsi, et de manière quasi linéaire, en accord avec Slater
et Pauling \cite{slater_structelec_1937,Pauling_metalforces_1938}.
Ce remplissage d'états hauts en énergie, passant par un maximum de
densité d'état pour $x_{Cr}\thickapprox0.5$, est corrélé à un mélange
énergétiquement défavorable des éléments Fe et Cr \cite{olsson_electronic_2006}.

\paragraph{L'énergie de mélange négative du côté riche en Fe}

En calculant \emph{ab initio} l'énergie de mélange de l'alliage désordonné
dans tout le domaine de concentration, Olsson et al. \cite{olsson_ab_2003,kissavos_emix_2006,olsson_electronic_2006,olsson_erratum:_2006,klaver_magnetism_2006}
observent une énergie de mélange négative pour $x_{Cr}\lessapprox6$\,\%.
Les auteurs émettent l'hypothèse d'une transition topologique de la
surface de Fermi, dont on peut trouver une description par Smirnova
et al. dans Al--Zn \cite{Smirnova_ETT_2001}, qui relie la dérivée
seconde de la contribution enthalpique de la structure de bande $H_{BS}$
à l'inverse du nombre d'états au niveau de Fermi de l'alliage $n\left(E_{F}\right)$
et aux numéros atomiques de Fe (Cr) $Z_{Fe(Cr)}$: 
\begin{equation}
\dfrac{\partial^{2}H_{BS}}{\partial x^{2}}\propto\dfrac{\left(Z_{Cr}-Z_{Fe}\right)^{2}}{n\left(E_{F}\right)}
\end{equation}

$n\left(E_{F}\right)$ varie rapidement avec la concentration, ce
qui influe directement sur la stabilité de l'alliage. On peut concevoir
cette relation comme un écart à la théorie de la solution solide régulière
dans laquelle la dérivée seconde de l'énergie de mélange par rapport
à $x$ est constante.

Les mêmes calculs réalisés dans un système sans magnétisme macroscopique
(DLM ou NM) \cite{kissavos_emix_2006} reproduisent les résultats
expérimentaux à haute température \cite{dench_adiabatic_1963} dans
la phase paramagnétique. L'alliage s'y comporte comme une solution
régulière. C'est bien le magnétisme qui est responsable du changement
de signe de l'énergie de mélange du côté riche en Fe.

Olsson et al. \cite{olsson_electronic_2006} lient ce changement de
signe à la position du niveau de Fermi. Ce qui se passe cependant
entre 0 et 10\,\%\,Cr est très particulier. Contrairement à la tendance
globale de dépeuplement du pic antiliant du canal majoritaire quand
$x_{Cr}$ augmente (décrite ci-dessus), on observe le phénomène inverse
de 0 à 10\,\%\,Cr, causé par des variations très locales de la densité
d'état projetée. Dans ce même domaine de concentration, $E_{F}$ se
rapproche encore du minimum du pseudo-gap du canal minoritaire. Ces
deux effets stabilisent l'alliage jusqu'à 10\,\%\,Cr : $\Delta H_{mix}<0$.
Par des arguments de population de la densité d'états au niveau de
Fermi, Olsson et al. parviennent également à estimer la limite entre
les régimes de germination--nucléation--croissance et de décomposition
spinodale à 17\,\%\,Cr à 0\,K. Les PDOS de l'alliage \og paramagnétique \fg{}
ne montrent pas ces variations locales à basse concentration en Cr.

\paragraph{Du point de vue atomistique : énergie, structure et magnétisme}

Des arguments atomistiques sont mis en avant par Klaver et al. \cite{klaver_magnetism_2006}
pour expliquer le changement de signe de l'énergie de mélange. Les
énergies de liaison Cr--Cr calculées avec des méthodes de référence
(PAW) sont répulsives.

À faible concentration, les atomes de Cr sont trop éloignés pour
interagir : la mise en solution favorable de l'impureté stabilise
le système. Quand $x_{Cr}$ augmente, des interactions Cr--Cr répulsives
apparaissent et augmentent l'énergie totale du système.

De la microscopie électronique de photo-émission (PEEM), de l'XMCD
et de l'EXAFS  couplés à des calculs DFT \og tous-électrons \fg{}
sont utilisés pour décrire la structure locale d'un atome de Cr dans
le fer à 6.2 et 12.5\,\%\,Cr \cite{froideval_magnetic_2007}. À
6.2\,\%\,Cr, l'atome de Cr a un effet structural négligeable. À
12.5\,\%\,Cr, l'atome de chrome induit une légère contraction des
deux premières sphères de voisinage.

\paragraph{Magnétisme et ordre local}

Le lien entre changement de signe de l'énergie de mélange, magnétisme
et tendance à l'ordre local est proposé par Paxton et Finnis en 2008
\cite{paxton_magnetic_2008}. Ces auteurs introduisent un paramètre
de Stoner dans un potentiel liaisons-fortes qui leur permet de contrôler
le moment magnétique des atomes et d'en déduire son rôle dans la physique
de l'alliage. Le moment magnétique des atomes de Fe est très stable,
même en présence de Cr. Au contraire, le moment magnétique d'une impureté
Cr dans le fer est très dépendante de l'environnement chimique local.
Dans la phase diluée, le moment de l'impureté Cr s'anti-aligne avec
les moments des atomes de la matrice Fe. Leur structure électronique
est alors très différente de celle de l'élément pur. Quand deux Cr
sont premiers voisins, la répulsion entre impuretés Cr modifie leur
structure électronique.

Pour que les Cr ne se repoussent plus, il faut qu'ils soient assez
nombreux et proches voisins pour retrouver la structure électronique
de la phase cubique centrée antiferromagnétique. La coopération entre
Cr de plusieurs mailles est nécessaire. Si le nombre de Cr voisins
n'est pas suffisant, la structure électronique la plus stable correspond
à l'atome isolé dans Fe. Quand la concentration augmente, les Cr précipitent
donc pour retrouver la structure électronique de l'élément pur. À
faible concentration, ils préfèrent s'isoler les uns des autres :
c'est la naissance d'une forme d'ordre local discutée au paragraphe
\ref{sub:L'ordre-a-courte}.

\section{Notre étude de l'alliage Fe--Cr}

À la lumière de cet état de l'art, cette section présente l'ensemble
des résultats \emph{ab initio} sur le volume de l'alliage fer-chrome.
L'objectif est dans un premier temps de valider la méthodologie sur
des paramètres structuraux connus, puis de comprendre les phénomènes
structuraux, magnétiques et énergétiques clés conférant à l'alliage
ses propriétés singulières.

\subsection{Validation méthodologique : les éléments purs\label{sub:choix_GGA_sur_Fe}}

Les éléments Fe et Cr ont déjà été très largement étudiés par le passé.
Dans cette sous-section, nous commençons par présenter chacun des
deux éléments du point de vue structural, énergétique et magnétique.
En comparant nos résultats ou ceux de la littérature aux observations
expérimentales, on choisit une fonctionnelle d'échange et corrélation.
On réalise ensuite une étude des deux éléments en utilisant les logiciels
SIESTA et PWSCF dans les approximations des pseudo-potentiels à norme
conservée (SIESTA-NC), ultra-softs (PWSCF-US) et PWSCF-PAW. Cela permet
d'évaluer l'effet des approximations des pseudo-potentiels et des
bases sur chacun des deux éléments, et de définir des références structurales
et énergétiques. Rappelons que cette étude se limite aux alliages
sans défauts.

\subsubsection*{Le fer}

\paragraph{Propriétés énergétiques et structurales}

À basses pression et température, le fer est expérimentalement ferromagnétique
de moment magnétique atomique $M_{Fe}=2.22$\,$\textrm{\ensuremath{\mu_{B}}}$.
Il a un réseau cristallin cubique centré de paramètre de maille $a_{Fe}=2.86$\,Å.
Sa température de Curie $T_{Curie}$ est 1043\,K \cite{kittel_introduction_1958}.
Une revue détaillée des propriétés magnétiques a été écrite par Autès
et al. \cite{autes_magnetism_2006}.

Wang et al. ont montré que la LDA polarisée en spin (LSDA) décrit
intrinsèquement mal le magnétisme dans les métaux de transition $3d$
\cite{wang_FeLDAGGA_1985}. En LDA, Fe est non-magnétique ($M_{Fe}=0$ $\mu_{B}$),
de structure hexagonale. En GGA, Fe est ferromagnétique ($2.2<M_{Fe}<2.8$ $\mu_{B}$),
de structure cubique centrée de paramètre de maille $a_{Fe}$ compris
entre 2.83 et 2.88\,Å. Le choix d'une fonctionnelle d'échange et
corrélation de type GGA semble donc s'imposer pour l'étude d'un alliage
à base de fer.

\begin{figure}[h]
\begin{centering}
\includegraphics[scale=0.4]{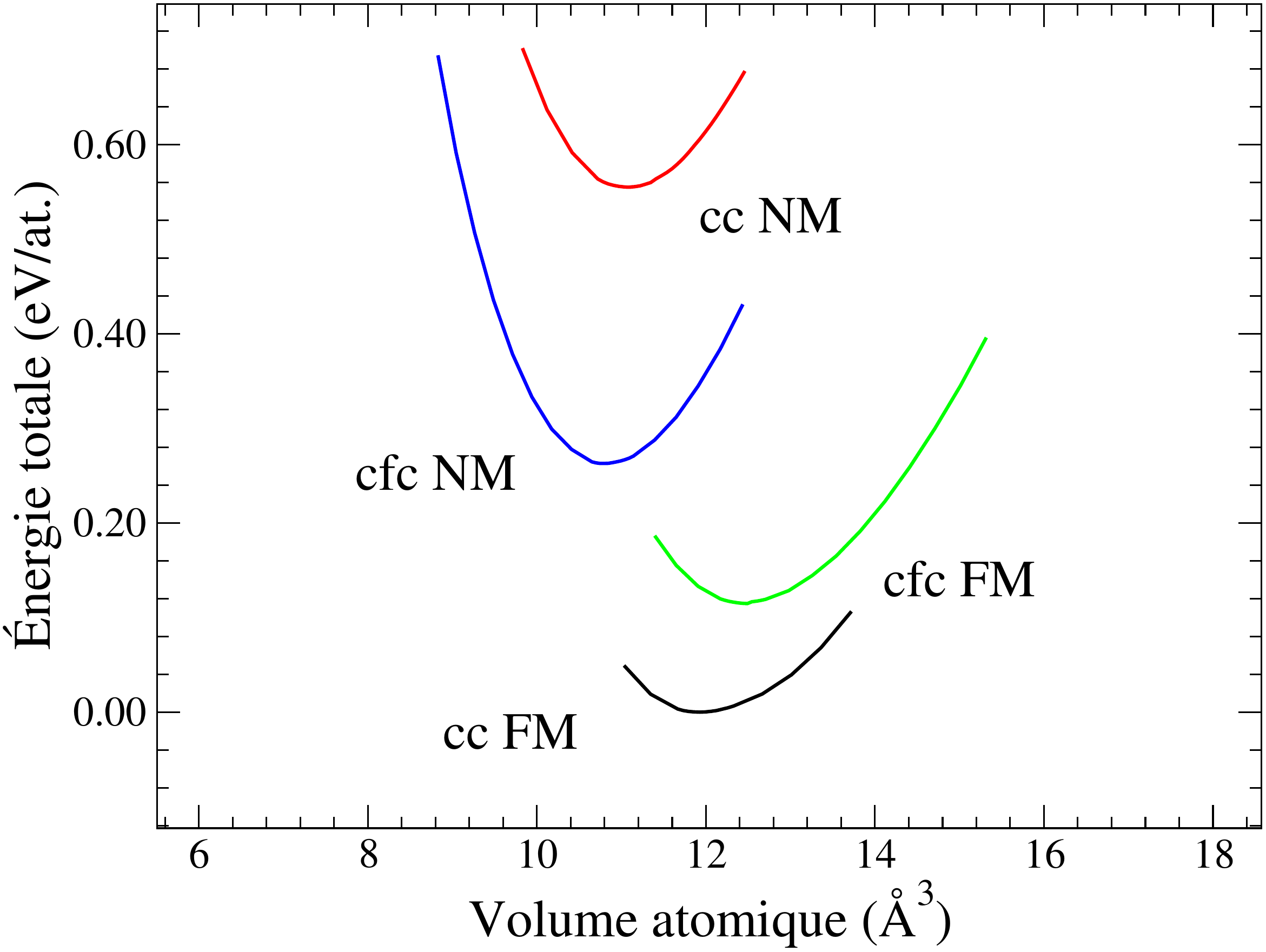}
\par\end{centering}

\caption{Énergies relatives calculées avec SIESTA-NC des phases cubique centrée
(cc) et cubique faces-centrées (cfc) ferromagnétique (FM) et non-magnétique
(NM) du fer en fonction du volume atomique. La phase cubique centrée
de volume atomique 11.93\,Å$^{\text{3}}$ sert de référence en énergie.\label{fig:parametredemaille_fe}}
\end{figure}

Les figures \ref{fig:parametredemaille_fe} et \ref{fig:etot_Cr_alat}
présentent les énergies totales calculée avec SIESTA-NC des phases
cubique centrée (cc) et cubique à faces centrées (cfc) ferromagnétiques
(FM), antiferromagnétique (AF) et non-magnétiques (NM) en fonction
du volume atomique. Le paramètre de maille d'équilibre ainsi que le
module de compressibilité de l'élément sont déduits d'une régression
des couples paramètre de maille -- énergie totale par l'équation d'état
de Murnaghan \cite{murnaghan_1944}.

Le paramètre de maille d'équilibre calculé avec SIESTA-NC est 2.879\,Å
(5.44\,Bohr). Fe NM est plus énergétique que Fe cc de 0.555\,eV,
c'est-à-dire 6641\,K plus énergétique. Cette température est très
différente de la température de Curie $T_{C}$ (1043\,K) car nous
rappelons qu'un calcul non-polarisé en spin n'est pas un calcul dans
la phase paramagnétique. Dans le premier cas, le moment magnétique
atomique est nul. Dans le deuxième cas, le moment est non-nul, mais
la moyenne des moments sur tous les atomes du système est nulle. Cette
différence d'énergie interne entre structures magnétiques et non-magnétiques
est en accord avec les résultats de Klaver et al. (0.546\,eV/atome)
\cite{klaver_magnetism_2006} et de Jiang et al. (0.56\,eV/atome)
\cite{jiang_sqsfecr_2004}. La différence d'énergie entre les phases
magnétiques cubique centrée et cubique à faces centrées correspond
à une température de 1320\,K. C'est une première approximation, mais
on retrouve l'ordre de grandeur de la température de transition de
phases ferritique--austénitique du fer (expérimentalement 1185\,K).

\paragraph{Le moment magnétique du fer}

Le moment magnétique de chaque atome a été calculé avec SIESTA-NC.
Nous présentons ces résultats ainsi que ceux issus d'autres méthodes
\emph{ab initio} dans les tableaux \ref{tab:Moments-magn=0000E9tiques-Fe}
et \ref{tab:mu_Cr_pur}.

Avec SIESTA-NC le moment magnétique atomique du fer $M_{Fe}$ est
2.32\,$\mu_{B}$, en accord avec les $2.22$\,$\mu_{B}$ expérimentaux.
Comme le proposaient Wang et al. \cite{wang_FeLDAGGA_1985}, les calculs
LDA ne sont pas en accord avec l'expérience. Tous les calculs GGA
sont globalement équivalents : seuls les calculs VASP-US surestiment
nettement $M_{Fe}$ de 13\,\%. On ne peut cependant associer cette
surestimation à un effet du pseudo-potentiel ou de la base car le
même type de calcul (ondes planes et pseudo-potentiels US) donne un
résultat satisfaisant (PWSCF-US : 2.23\,$\mu_{B}$). Il s'agit peut-être
d'un problème lié au pseudo-potentiel ultrasoft fourni avec le code
VASP et qu'on ne peut modifier.

\begin{table}[h]
\begin{centering}
\begin{tabular}{|c|c|c|}
\hline 
Référence & Méthodologie & $M_{Fe}$ ($\mu_{B}$) \\
\hline 
\hline 
Ce travail & SIESTA-NC GGA & 2.32 \\
\hline 
Ce travail  & PWSCF-US GGA & 2.23 \\
\hline 
\cite{Olsson_defautsdft_2007} & VASP-US GGA & 2.52 \\
\hline 
Ce travail & PWSCF-PAW GGA & 2.20 \\
\hline 
\cite{Olsson_defautsdft_2007} & VASP-PAW GGA & 2.21 \\
\hline 
\cite{klaver_magnetism_2006} & VASP-PAW GGA & 2.2 \\
\hline 
\cite{wang_FeLDAGGA_1985} & LAPW LDA & 2.08 \\
\hline 
\cite{kittel_introduction_1958} & Exp. & 2.22 \\
\hline 
\end{tabular}
\par\end{centering}

\caption{Comparaison du moment magnétique atomique $M_{Fe}$ ($\mu_{B}$) du
fer cubique centré ferromagnétique calculé avec SIESTA-NC et PWSCF-PAW
avec des calculs ab initio antérieurs. \label{tab:Moments-magn=0000E9tiques-Fe}}
\end{table}

Une première conclusion pour Fe : les résultats SIESTA-NC sont en
accord satisfaisant à la fois avec les calculs réputés plus robustes
PWSCF-PAW et avec l'expérience.

\subsubsection*{Le chrome\label{par:Cr_SDW}}

\paragraph{Propriétés énergétiques et structurales}

\begin{figure}[h]
\begin{centering}
\includegraphics[scale=0.5]{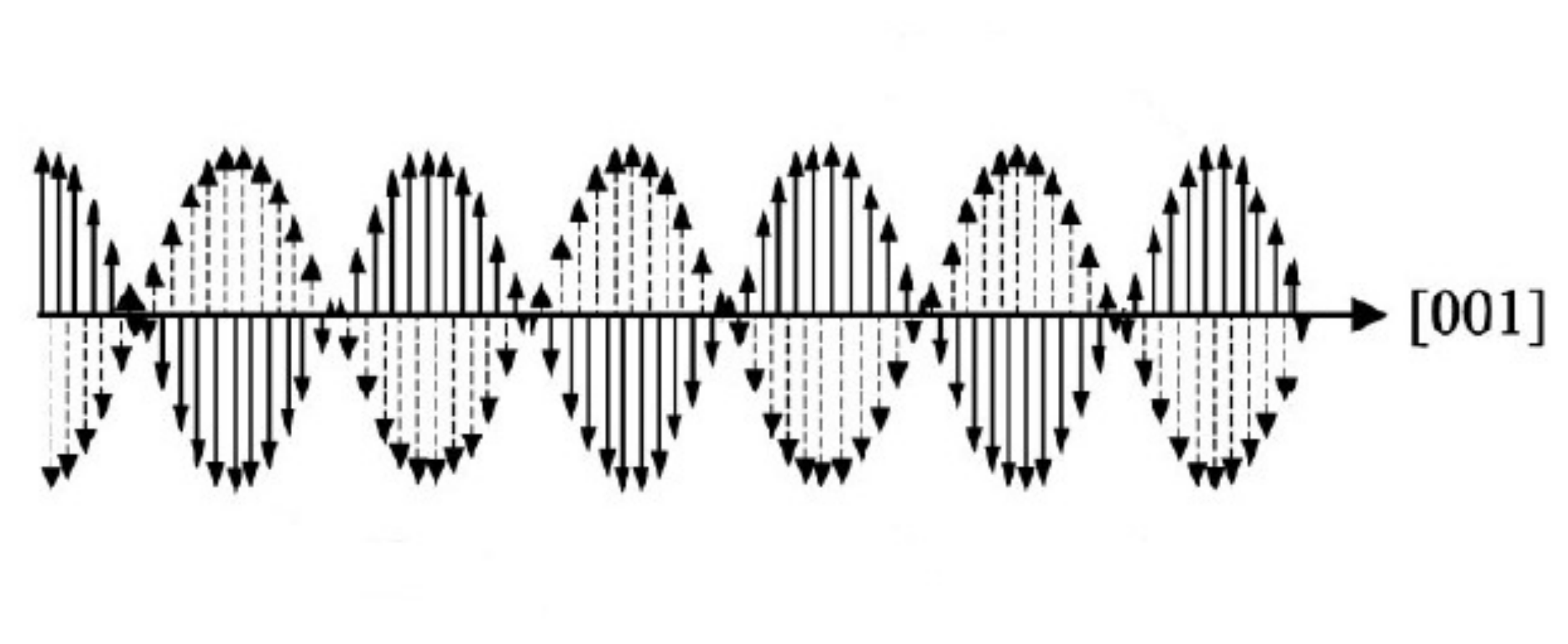}
\par\end{centering}

\caption{Représentation schématique d'une onde de densité de spin (SDW\nomenclature{SDW}{Onde de densité de spin})
de vecteur d'onde dans la direction $\left[100\right]$. L'amplitude
des moments magnétiques atomiques représentés par les flèches varie
de façon sinusoïdale dans la direction de propagation. Figure de Zabel
\cite{zabel_Crmag_1999}.\label{fig:Onde de spin}}
\end{figure}

Le chrome est un élément complexe. Fawcett a écrit une revue très
détaillée de ses propriétés énergétiques et structurales expérimentales
\cite{fawcett_Crpur_1988}. On trouvera plus d'informations quant
à ses propriétés magnétiques dans la revue de Zabel \cite{zabel_Crmag_1999}.
À l'état fondamental, il a un réseau cubique centré de paramètre de
maille $a_{Cr}=2.884$\,Å. C'est un élément antiferromagnétique très
singulier, car il présente une onde de spin \cite{corliss_antiphaseCr_1959},
ce qui signifie que l'amplitude de son moment magnétique local est
modulée de façon sinusoïdale. Le schéma d'une onde de spin est proposé
dans la figure \ref{fig:Onde de spin}. En l'absence de défauts (lacunes,
impuretés, surfaces, dislocations, \ldots{}), cette onde de spin
est incommensurable (ISDW\nomenclature{ISDW}{Incommensurate Spin Density Wave -- Onde de spin incommensurable}),
c'est-à-dire que le rapport de la période du cristal sur celle de
l'onde de spin n'est pas entier. Le vecteur de propagation de l'onde
de spin $\vec{Q}$ est dans la direction $\left[100\right]$ du cristal
\cite{zabel_Crmag_1999}. Sa periode varie de 60\,Å à température
nulle jusqu'à environ 78\,Å à la température de Néel $T_{N\acute{e}el}$
de 311\,K. C'est à cette température relativement élevée en comparaison
des températures de Néel des autres éléments antiferromagnétiques
que le chrome devient paramagnétique (pour rappel, $T_{Curie}(Fe)=1043$\,K).

\begin{figure}[h]
\begin{centering}
\includegraphics[clip,scale=0.5]{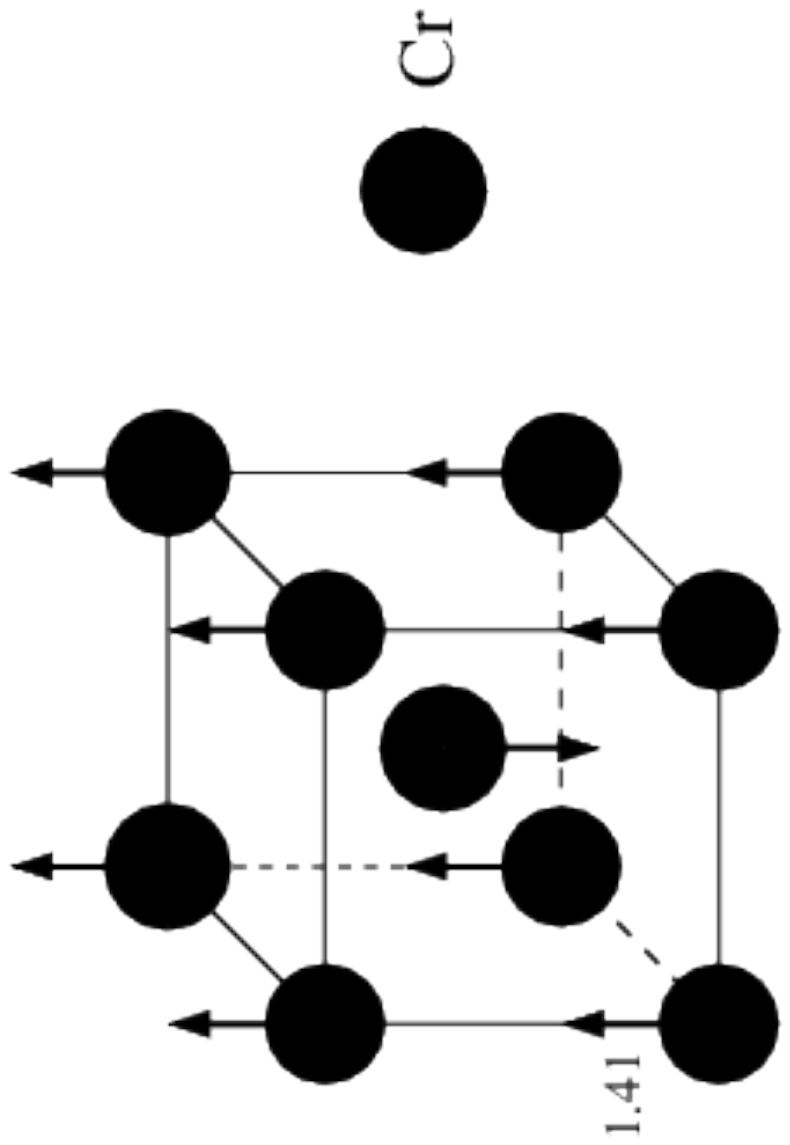}
\par\end{centering}

\caption{Maille élémentaire de Cr dans la phase antiferromagnétique AF. Les
moments magnétiques atomiques sont représentés par des vecteurs anti-parallèles
entre plus proches voisins. Cette phase est une approximation de la
phase ISDW présentée dans la figure \ref{fig:Onde de spin}. \label{fig:Structure-antiferromagn=0000E9tique-(AFM)}}
\end{figure}

Cette onde de densité de spin incommensurable (ISDW) n'est pas l'état
magnétique fondamental calculé par les fonctionnelles d'échange et
corrélation LDA et GGA \cite{hafner_SDWCr_2002,cottenier_SDWCr_2002,marcus_SDWCr_1998}.
Hafner, Cottenier ou Marcus et al., par exemple, montrent que ces
fonctionnelles d'échange et corrélation conduisent pour Cr à un état
fondamental antiferromagnétique commensurable (AF) représente sur
la figure \ref{fig:Structure-antiferromagn=0000E9tique-(AFM)}. Dans
cette phase magnétique, une maille cubique centrée contient un Cr
de spin \og + \fg{} et un Cr de spin \og $-$ \fg{}. Les états
ISDW et AF sont en fait quasi-dégénérés. C'est la raison pour laquelle
l'ensemble des calculs \emph{ab initio} de la littérature se font
dans l'approximation AF.

\begin{figure}[h]
\begin{centering}
\includegraphics[scale=0.4]{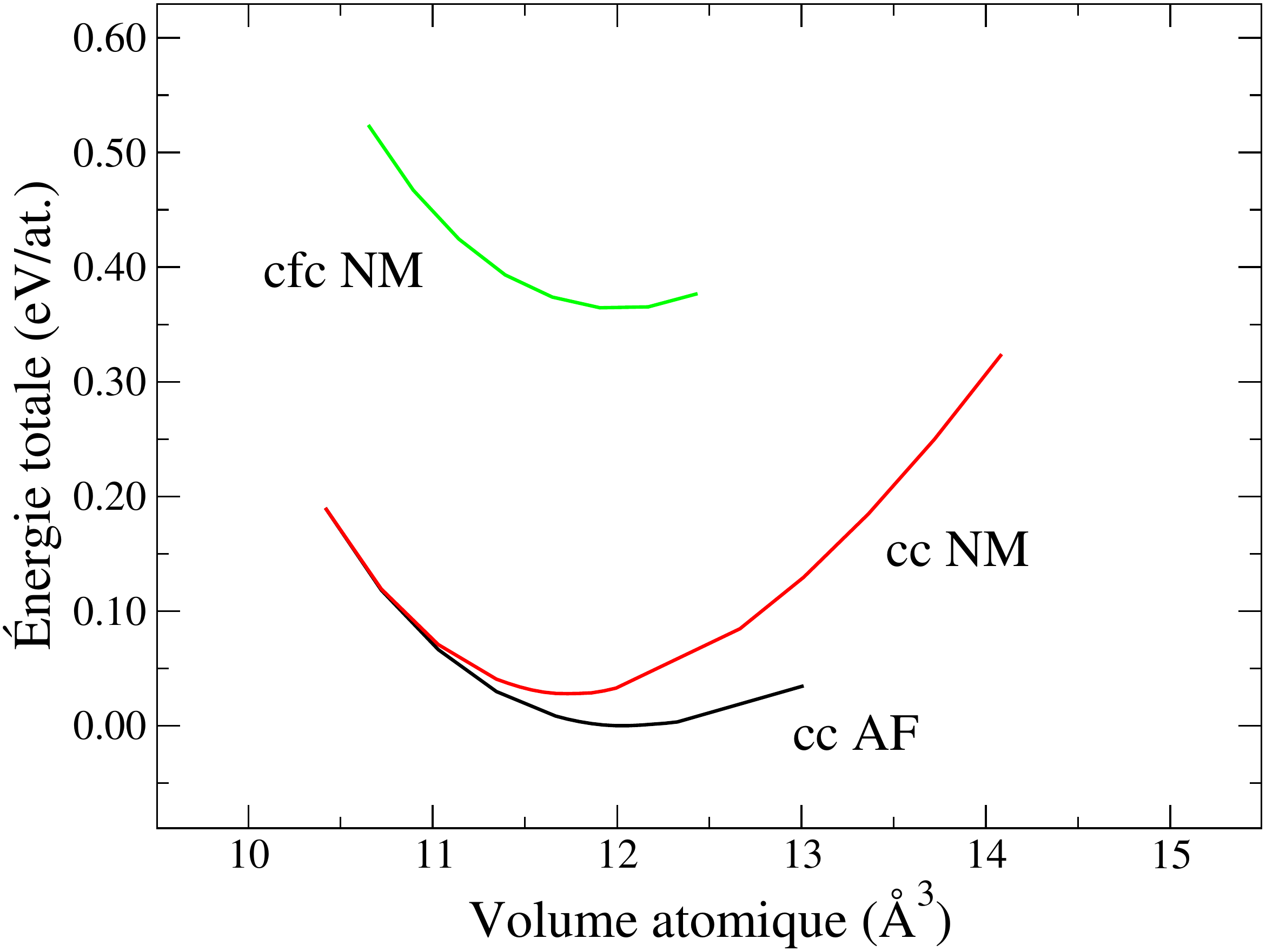}
\par\end{centering}

\caption{Énergies relatives calculées avec SIESTA-NC des phases cubique centrée
(cc) et cubique faces-centrées (cfc) antiferromagnétique (AF) et non-magnétique
(NM) du chrome en fonction du volume atomique. La référence en énergie
est la phase cc AF de paramètre de maille $a_{Cr}=\unit[2.88]{\AA}$.
Les calculs polarisés en spin dans la structure cfc tombent dans un
minimum d'énergie qui correspond à un état non-magnétique.\label{fig:etot_Cr_alat}}
\end{figure}

Comme pour Fe, nous avons calculé l'énergie totale de Cr dans les
réseaux cubique centré et cubique à faces centrées pour différents
volumes atomiques, que nous présentons dans la figure \ref{fig:etot_Cr_alat}.
Le chrome magnétique est instable dans la phase cubique à faces centrées. 

Le paramètre de maille d'équilibre avec SIESTA-NC est $a_{Cr}=2.89$\,Å
(5.445\,Bohr). La structure non-magnétique (NM) est 0.028\,eV, c'est-à-dire
324\,K plus énergétique que l'état fondamental cc antiferromagnétique,
en accord relatif avec Klaver et al. (0.014\,eV/atome) \cite{klaver_magnetism_2006}.
Rappelons que la température de Néel de Cr est $T_{N}=311$\,K. Pour
Marcus et al. \cite{marcus_SDWCr_1998}, cette faible différence d'énergie
entre les phases AF et NM est responsable de l'inversion de stabilité
ISDW--AF en GGA.

\paragraph{Le moment magnétique du chrome}

Comme le montre le tableau \ref{tab:mu_Cr_pur}, le moment magnétique
local $M_{Cr}$ du chrome antiferromagnétique AF dépend beaucoup de
la méthode de calcul. 

\begin{table}[h]
\begin{centering}
\begin{tabular}{|c|c|c|}
\hline 
Référence & Technique & $M_{Cr}$ ($\mu_{B}$) \\
\hline 
\hline 
Ce travail & SIESTA-NC GGA & 1.41 \\
\hline 
Ce travail & PWSCF-US GGA & 1.30 \\
\hline 
\cite{Olsson_defautsdft_2007} & VASP-US GGA & 1.97 \\
\hline 
Ce travail & PWSCF-PAW GGA & 1.20 \\
\hline 
\cite{Olsson_defautsdft_2007} & VASP-PAW GGA & 0.92 \\
\hline 
\cite{klaver_magnetism_2006} & VASP-PAW GGA & 1.1 \\
\hline 
\cite{hafner_SDWCr_2002} & VASP-PAW GGA & 1.19 \\
\hline 
\cite{mishra_FeinCr_2008} & FLAPW GGA & 1.19 \\
\hline 
\cite{hashemifar_FLAPW_Fe_in_Cr_2006} & FLAPW GGA & 1.00 \\
\hline 
\cite{cottenier_SDWCr_2002} & FLAPW GGA & 1.08 \\
\hline 
\cite{singh_magnetismeGGA_1992} & FLAPW GGA & 1.40 \\
\hline 
\cite{bihlmayer_Cr100V100_2000} & FLAPW GGA & 0.99 \\
\hline 
\cite{hafner_SDWCr_2002} & LMTO GGA & 0.99 \\
\hline 
\cite{bihlmayer_Cr100V100_2000} & FLAPW LDA & 0.44 \\
\hline 
\cite{chen_LDACr_1988} & FLAPW LDA & $0.67$, 0.70, 1.39$^{a}$ \\
\hline 
\cite{shull_neutron_1955} & diffraction de neutrons & 0.40 \\
\hline 
\cite{arrott_diffneutrons_first_muCrSDW_1967} & diffusion de neutrons & 0.43$^{b}$, 0.62$^{c}$ \\
\hline 
\end{tabular}
\par\end{centering}

\caption{Moment magnétique calculé avec différentes méthodes \emph{ab initio}
et résultats expérimentaux pour le chrome cubique centré \label{tab:mu_Cr_pur}.\protect \\
$^{a}$ selon la fonctionnelle d'échange et corrélation LSDA utilisée\protect \\
$^{b}$ valeur moyenne de l'onde de spin à 4.2\,K\protect \\
$^{c}$ valeur maximum de l'onde de spin à 4.2\,K.}
\end{table}

Toutes les valeurs calculées avec une fonctionnelle d'échange et corrélation
GGA sont supérieures aux valeurs expérimentales d'un facteur deux
ou plus. Chen et al. montrent que cette propriétés de Cr dépend fortement
de la paramétrisation LDA choisie \cite{chen_LDACr_1988}. Les calculs
LSDA sont généralement plus proches des résultats expérimentaux que
les calculs GGA. Le moment magnétique de Cr est en effet très sensible
à la fonctionnelle d'échange et corrélation utilisée \cite{cottenier_SDWCr_2002}.
Le moment calculé avec SIESTA-NC est dans la moyenne supérieure des
résultats \emph{ab initio}. Les résultats PAW sont inférieurs aux
résultats NC, eux-même inférieurs aux résultats US. En particulier,
le pseudo-potentiel US fourni avec le code VASP très largement répandu
surestime très largement le moment magnétique des atomes de Cr. Les
résultats NC sont surestimés par rapport aux PAW. On n'observe pas
d'effet des bases. Une fonctionnelle d'échange et corrélation GGA
surestime le moment magnétique de Cr, mais la LDA n'est pas capable
de trouver le bon état fondamental cubique centré \cite{chen_LDACr_1988,marcus_SDWCr_1998,hafner_SDWCr_2002,moruzzi_AFCr_1992,cottenier_SDWCr_2002}. 

Nos calculs sont faits dans l'approximation des spins colinéaires
: les moments magnétiques atomiques sont des scalaires. Ce travail
est, de fait, plus particulièrement orienté vers l'alliage riche en
fer. Or, le diagramme de phases magnétiques du système montre que
dès  1.5 à 1.75\,\%\,Fe, l'alliage à 0\,K est AF \cite{tsunoda_FeCr_magnetic_phase_diagram_1994,arrott_diffneutrons_first_muCrSDW_1967,ishikawa_Fe_imp_dans_Cr_1967}.
La non-prise en compte de l'ISDW dans notre étude est donc raisonnable.

\subsubsection*{Lien entre volume atomique et moment magnétique}

Nous récapitulons les paramètres de maille calculés pour Fe et Cr
cubiques centrés dans les états magnétiques et non-magnétiques dans
le tableau \ref{tab:Param=0000E8tres-de-maille}. Tous les résultats
présentés sont issus de calculs GGA dans l'approximation des pseudo-potentiels
NC, US ou PAW. Seule la dernière ligne correspond à un calcul tous
électrons en LDA, pour indication. Les paramètres de maille de Fe
et de Cr calculés en PAW et GGA sont légèrement sous-estimés (d'environ
1\,\%). Au contraire, les calculs SIESTA et US surestiment de moins
de 1\,\% le paramètre de maille.

\begin{table}[h]
\begin{centering}
\begin{tabular}{|c|c|c|c|c|c|}
\hline 
 &  & \multicolumn{2}{c|}{$a_{Fe}$ (Å)} & \multicolumn{2}{c|}{$a_{Cr}$ (Å)} \\
\hline 
Référence & Méthodologie & FM & NM & AF & NM \\
\hline 
\hline 
Ce travail & SIESTA NC GGA & 2.879 & 2.81 & 2.886 & 2.86 \\
\hline 
\cite{Olsson_defautsdft_2007} & VASP US GGA & 2.879 & - & 2.918 & 2.851 \\
\hline 
Ce travail & PWSCF PAW GGA & 2.836 & 2.82 & 2.87 & 2.85 \\
\hline 
\cite{Olsson_defautsdft_2007} & VASP PAW GGA & 2.831 & - & 2.850 & 2.836 \\
\hline 
\cite{klaver_magnetism_2006} & VASP PAW GGA & 2.829 & - & 2.863 & - \\
\hline 
\cite{wang_FeLDAGGA_1985} & LAPW LDA & 2.758 & 2.70 & - & - \\
\hline 
\cite{kittel_introduction_1958} & Exp. & 2.87 & - & 2.88 & - \\
\hline 
Pearson selon \cite{olsson_electronic_2006} & Exp. & 2.86 & - & - & - \\
\hline 
\cite{wang_FeLDAGGA_1985} & Exp. & 2.861 & - & - & - \\
\hline 
\end{tabular}
\par\end{centering}

\caption{Paramètres de maille calculés et expérimentaux en Ångströms (Å). \label{tab:Param=0000E8tres-de-maille}}
\end{table}

L'approximation des pseudo-potentiels n'a pas d'effet notable sur
le paramètre de maille.

\begin{figure}[h]
\begin{centering}
\includegraphics[scale=0.4]{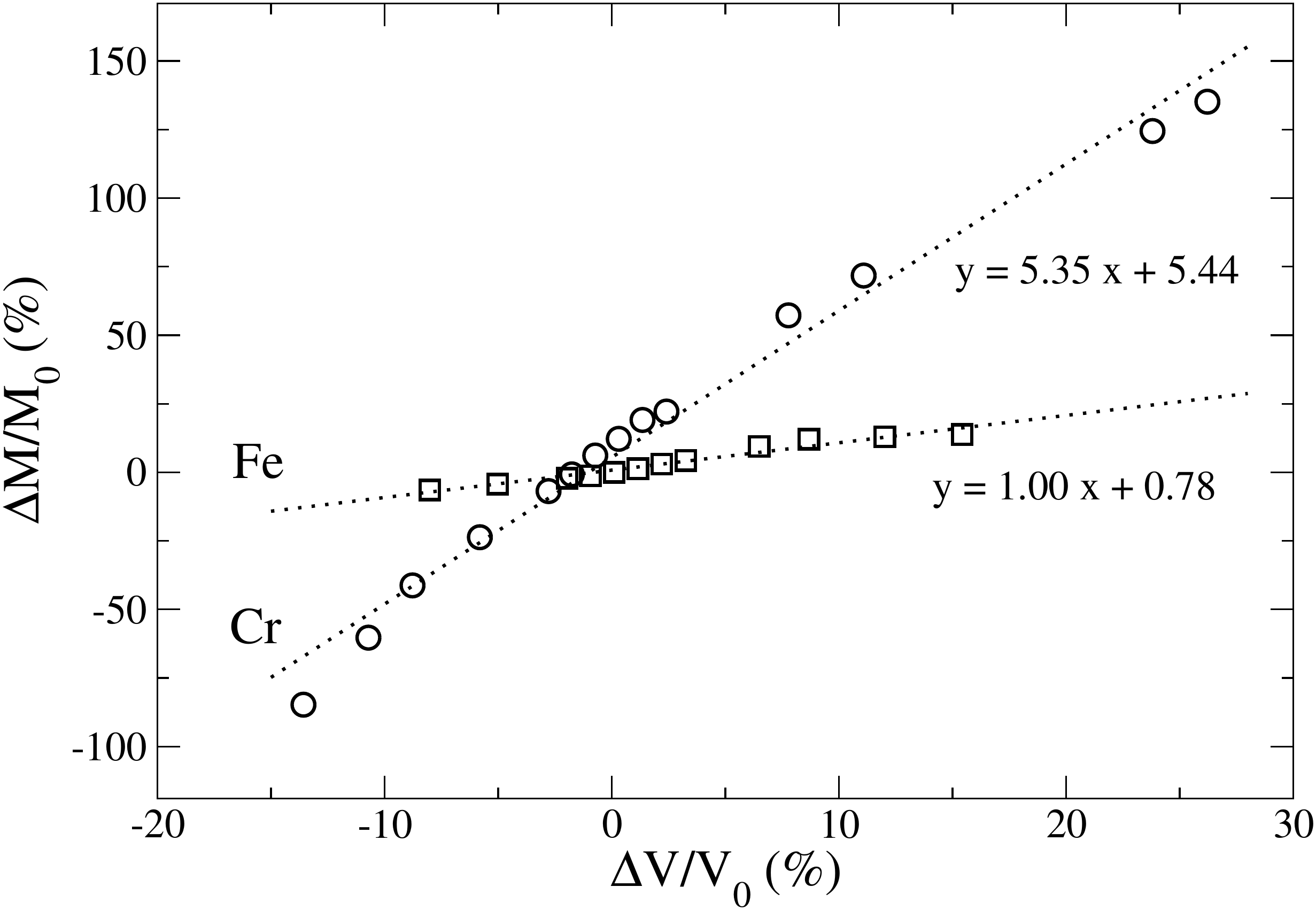}
\par\end{centering}

\caption{Évolution relative du moment magnétique des atomes de Fe et Cr cc
avec le volume atomique dans l'approximation des pseudo-potentiels
à norme conservée (calculs SIESTA-NC).\label{fig:=0000C9cart-relatif-du_moment_a}}
\end{figure}

Nous avons calculé le moment magnétique du fer et du chrome en imposant
différents paramètres de maille. Nous représentons sur la figure \ref{fig:=0000C9cart-relatif-du_moment_a}
l'écart relatif du moment magnétique local en fonction de l'écart
relatif du volume atomique. Les calculs sont ici réalisés en SIESTA-NC.

On observe sur la figure \ref{fig:=0000C9cart-relatif-du_moment_a}
que l'augmentation relative de moment magnétique est quasi-linéaire
avec l'augmentation relative de volume atomique. Ces calculs montrent
que la sensibilité du moment magnétique au volume atomique est environ
5.3 fois supérieure dans le chrome que dans le fer. Pour des variations
de volume de l'ordre de $\unit[2]{\%}$ par rapport au volume d'équilibre,
le moment magnétique local du fer varie de moins de 5\,\%. C'est
une variation de 25\,\% que subit le moment magnétique du chrome
pour une même dilatation de la maille. On peut alors expliquer pourquoi
le moment magnétique atomique de Cr est sensible à l'approximation
des pseudo-potentiels. De faibles variations de paramètres de maille,
qui ne remettent pas en cause l'approximation des pseudo-potentiels,
ont de grandes conséquence sur le moment magnétique calculé.

\subsubsection*{Densités électroniques du fer et du chrome non-polarisés en spin}

\begin{figure}[h]
\begin{centering}
\includegraphics[scale=0.4]{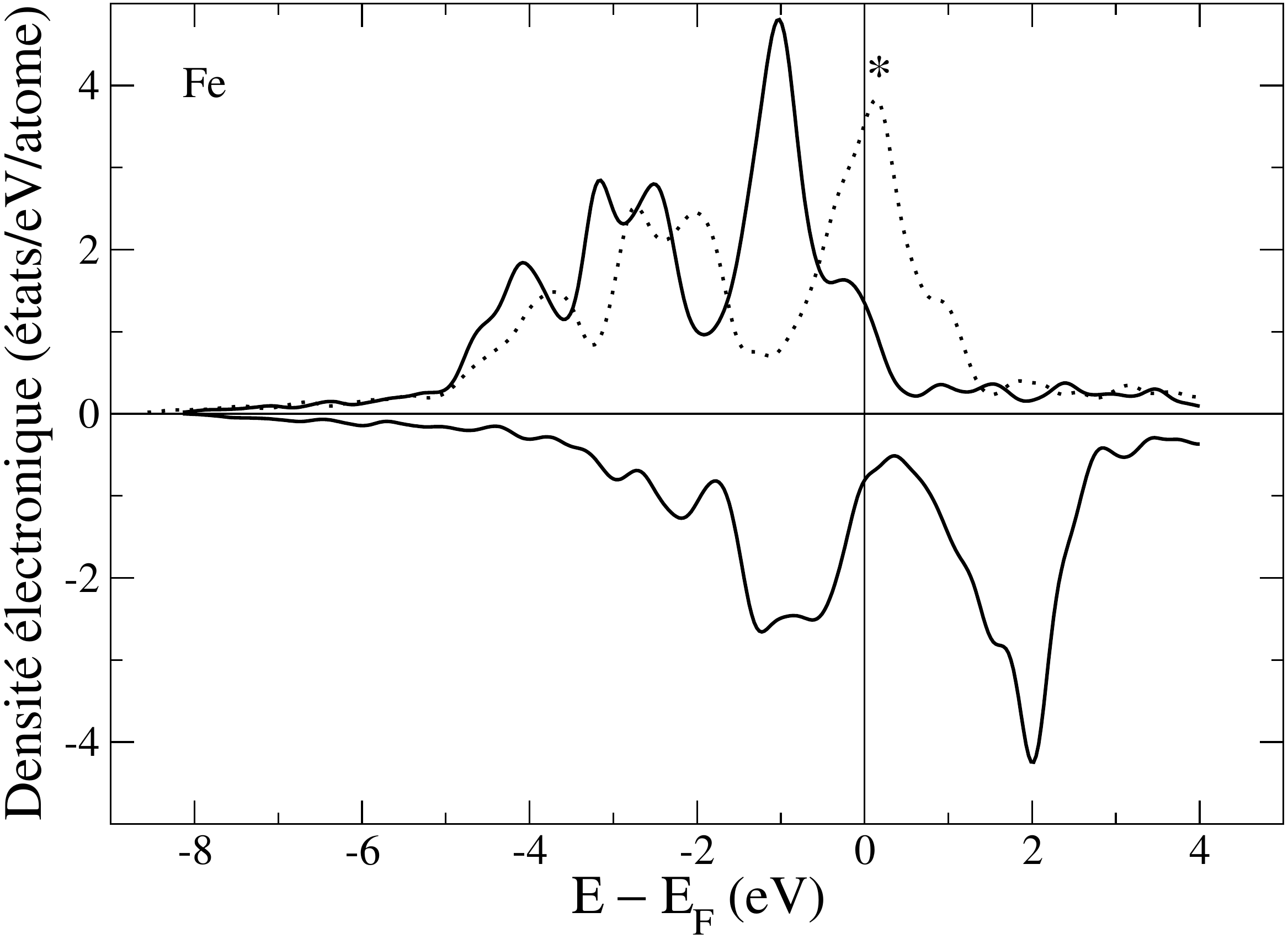}
\par\end{centering}

\begin{centering}
\includegraphics[scale=0.4]{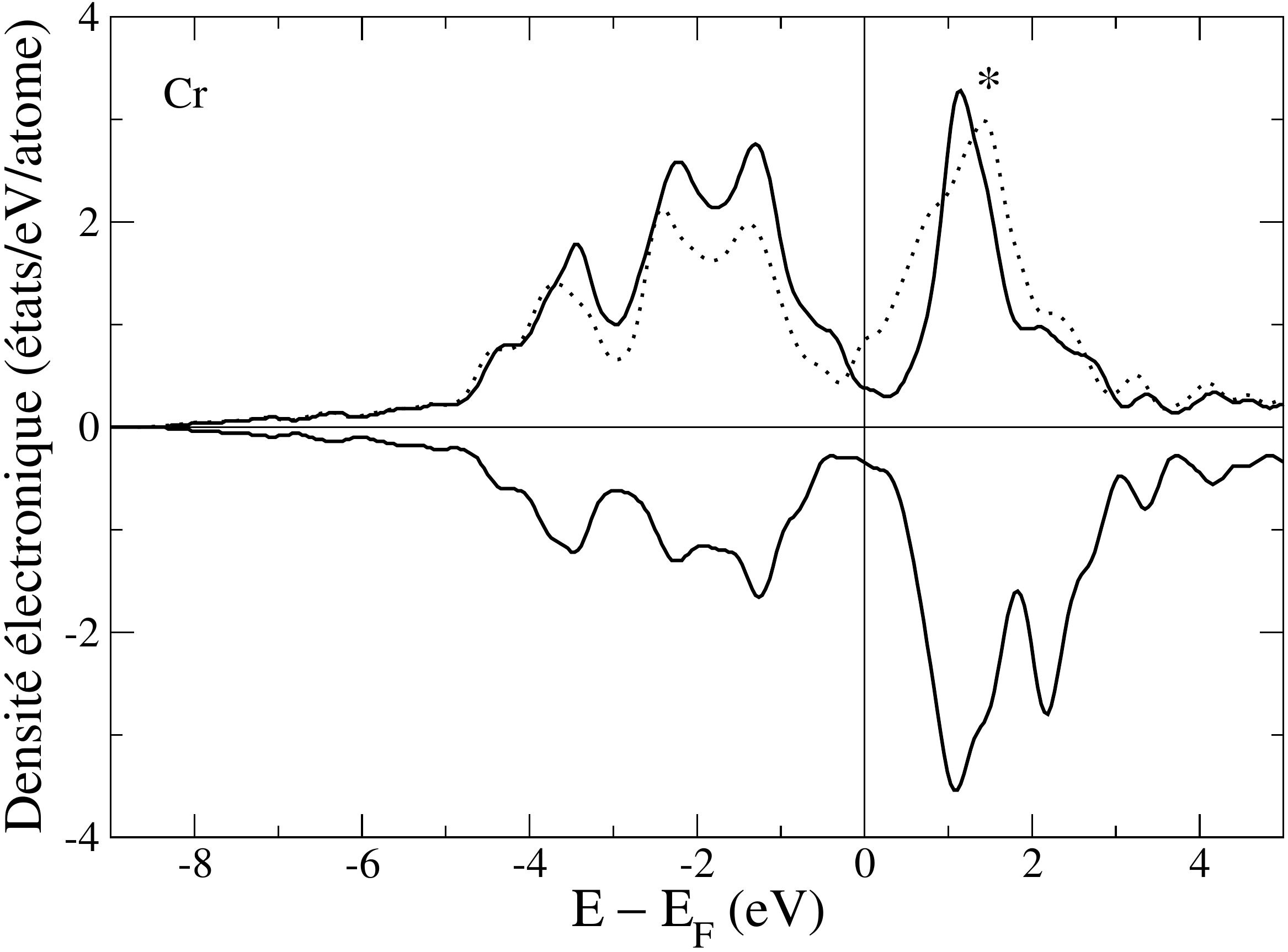}
\par\end{centering}

\caption{Densité électronique projetée des éléments purs Fe et Cr cubiques
centrés magnétiques (trait plein) et non-magnétiques (trait pointillé)
calculés en SIESTA-NC. Le pic antiliant des éléments non-magnétiques
est repéré par une étoile. \label{fig:PDOS_elements_purs_M}}
\end{figure}

Nous représentons dans la figure \ref{fig:PDOS_elements_purs_M} les
densités d'états électroniques projetées des éléments Fe (en haut)
et Cr (en bas) magnétiques et non-magnétiques. Les PDOS non-magnétiques
permettent de mieux appréhender les conséquences du magnétisme sur
les éléments.

Pour les deux éléments, nous voyons sur la figure \ref{fig:PDOS_elements_purs_M}
que la structure électronique non-magnétique est très classique :
on retrouve la structure typique à 3 bandes (liante et non-liante
difficilement discernables puis anti-liante) d'un métal cubique centré.

Dans Fe, le niveau de Fermi tombe dans un maximum de densité d'états.
C'est une position peu stable.

Dans Cr, le niveau de Fermi tombe dans le deuxième pseudo-gap (un
minimum) de la densité d'états. C'est une position beaucoup plus stable.

\subsubsection*{Densités électroniques du fer et du chrome magnétiques}

Les densités électroniques du fer et du chrome magnétiques calculées
avec SIESTA-NC sont également représentées sur la figure \ref{fig:PDOS_elements_purs_M}.

La densité électronique projetée de Cr magnétique est similaire au
Cr non-magnétique : le niveau de Fermi se situe entre les pics des
états liants et antiliants, dans la bande majoritaire ($\uparrow$)
ou minoritaire ($\downarrow$). 

Pour le fer, la densité électronique est très différente. On peut
la lire dans le cadre théorique de Stoner, dans lequel les bandes
$\uparrow$ et $\downarrow$ sont décalées par rapport au niveau de
Fermi à l'état magnétique dans des sens différents. Le canal majoritaire
($\uparrow$) est décalé vers les basses énergies, tandis que le canal
minoritaire est décalé vers les hautes énergies. On peuple ainsi un
pic de plus basse énergie dans le canal majoritaire, la densité d'état
au niveau de Fermi est plus faible. La bande ($\downarrow$) est décalée
vers les énergies plus hautes. Le pic antiliant est ainsi complètement
vidé, de sorte que le niveau de Fermi tombe dans un pseudo-gap.

\subsubsection*{Bilan sur les éléments purs}

Cette première partie d'étude \emph{ab initio} nous permet cependant
les conclusions suivantes :
\begin{itemize}
\item méthodologiquement d'abord : nos résultats sont cohérents avec les
résultats antérieurs, ce qui valide nos choix. Nous utilisons une
fonctionnelle d'échange et corrélation de type GGA. Après plusieurs
tests, nous choisissons le paramétrage de Perdew, Burke et Ernzerhof
(PBE) \cite{perdew_GGAsimple_1996}.
\item L'influence la plus notable entre les différentes approximations mises
en œuvre concerne le lien entre pseudo-potentiels et moment magnétique
atomique de Cr. Celui-ci est surestimé dans les calculs utilisant
des pseudo-potentiels à norme conservée.
\end{itemize}

\subsection{Propriétés d'alliage}

L'objet de ce paragraphe est de comprendre les phénomènes liés au
mélange du fer et du chrome à température nulle, dans l'objectif de
pouvoir extrapoler ces résultats à température finie par la suite.
Les matériaux d'intérêt pour le nucléaire étant constitués d'une matrice
riche en fer, c'est la dissolution du chrome dans le fer qui retient
plus particulièrement notre attention. Nous étudions dans un premier
temps la mise en solution de l'un des éléments dans l'autre. Dans
un deuxième temps, nous étudions l'interaction de deux impuretés.
Enfin, nous étudions les propriétés de mélange dans l'ensemble du
domaine de définition. Nous portons une attention particulière aux
propriétés structurales, magnétiques et énergétiques de ce mélange.

\subsubsection*{Dissolution d'une impureté de chrome dans une matrice de fer\label{sub:def_mu}}

\paragraph{Propriétés structurales}

Lors de la dissolution d'une impureté de chrome dans le fer, nous
n'observons pas de variation notable des paramètres structuraux. Les
deux éléments ayant des paramètres de maille similaires ($\Delta r/r=\unit[0.28]{\%}$),
la distance entre l'impureté Cr et l'atome de Fe 1$^{\text{er}}$
voisin est augmentée de seulement 0.28\,\%.

\paragraph{Propriétés énergétiques\label{par:dissolutionCrdansFe}}

La variation énergétique due à la substitution d'un atome de matrice
par une impureté est donnée par l'énergie de mise en solution $\Delta E_{sol}^{Cr\, dans\, Fe}$
que nous calculons \emph{ab initio} :

\begin{equation}
\Delta E_{sol}^{Cr\, dans\, Fe}=E\left(Fe_{N}Cr_{1}\right)-\left(NE\left(Fe\right)+E\left(Cr\right)\right)
\end{equation}
où $E\left(Fe_{N}Cr_{1}\right)$ est l'énergie totale d'une super-cellule
contenant un maximum d'atomes de Fe et un seul Cr. $E\left(Fe\right)$
($E\left(Cr\right)$) est l'énergie d'un atome de Fe (Cr) dans la
structure cubique centrée FM (AF) fondamentale. Le paramètre de maille
de la super-cellule Fe$_{\text{N}}$Cr$_{\text{1}}$ est celui de
Fe pur, Cr étant très dilué. Pour tous les résultats présentés, $N$
est testé jusqu'à $249$ (super-cellule $5\times5\times5$ à $2$
atomes par maille), où toutes les propriétés sont convergées.

Une énergie de mise en solution négative signifie que la réaction
est exothermique, c'est-à-dire que la substitution est favorable.

L'énergie de mise en solution d'une impureté Cr dans Fe calculée avec
SIESTA-NC est de $\Delta E_{sol}^{Cr\, dans\, Fe}=-0.466$\,eV et
$-0.20$\,eV en PWSCF-PAW. Ces énergies de mise en solution sont
en accord relatif avec les calculs précédents récapitulés dans le
tableau \ref{tab:Mom_Cr_imp_dans_matrice_de_Fe}. L'énergie de mise
se calcule à la limite infiniment diluée, ce qui est impossible pour
des calculs DFT périodiques. Nous indiquons les concentrations des
super-cellules de calcul pour référence. La différence entre les méthodes
est à relier aux propriétés magnétiques du mélange.

\begin{table}[h]
\begin{centering}
\begin{tabular}{|c|c|c|c|c|}
\hline 
Référence & Méthode & M ($\mu_{B}$) & $\Delta E_{sol}^{Cr\, dans\, Fe}$ (eV) & $x_{Cr}$ (\%) \\
\hline 
\hline 
Ce travail & SIESTA NC & $-2.55$ & $-0.47$ & 0.4 \\
\hline 
Ce travail & PWSCF PAW & $-2.20$ & $-0.20$ & 0.4 \\
\hline 
\cite{Olsson_defautsdft_2007} & VASP US & $-2.40$ & $-0.46$ & 1.85$^{\text{a}}$ \\
\hline 
\cite{Olsson_defautsdft_2007} & VASP PAW & $-1.59$ & $-0.12$ & 1.85$^{\text{a}}$ \\
\hline 
\cite{klaver_magnetism_2006} & VASP PAW & $-1.8$ & - & 1.85 \\
\hline 
\cite{Collins_mudistri_1965} & Exp. à $300$\,K & $-0.7$ & - &  \\
\hline 
\cite{Aldred_neutron_1976} & Exp. à $4.2$\,K & $-1.16$ & - &  \\
\hline 
\end{tabular}
\par\end{centering}

\caption{Moment magnétique porté par un atome de Cr en impureté (concentration
$x_{Cr}$) dans une matrice de Fe et énergie de dissolution de Cr
dans Fe. Les expériences sont de diffusion de neutrons.\protect \\
$^{\text{a}}$ Ce chiffre n'est pas explicite dans l'article de Olsson
et al. \cite{Olsson_defautsdft_2007} \label{tab:Mom_Cr_imp_dans_matrice_de_Fe}}
\end{table}

\paragraph{Propriétés magnétiques}

\begin{figure}[h]
\begin{centering}
\includegraphics[scale=0.5]{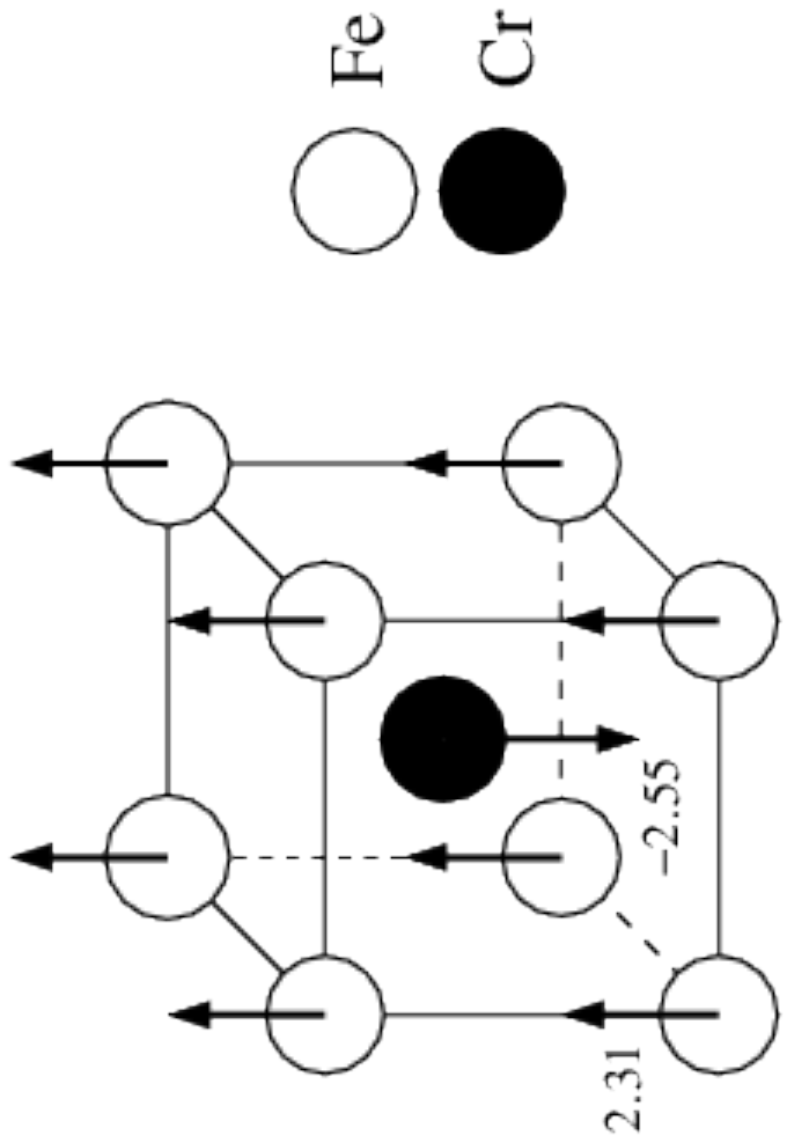}
\par\end{centering}

\caption{Impureté Cr (en noir) et sa première sphère de coordination dans une
matrice Fe (en blanc). Les flèches indiquent le moment magnétique
de chacun des atomes : $+2.31$\,$\mu_{B}$ pour les Fe de la première
sphère de coordination, au lieu de $+2.32$\,$\mu_{B}$ sans impureté,
et $-2.55$\,$\mu_{B}$ pour l'impureté Cr, antiferromagnétique au
lieu de $\pm1.31$\,$\mu_{B}$ dans l'élément pur. \label{fig:impuret=0000E9_Cr_dans_Fe}}
\end{figure}

Nos calculs montrent que $M_{Cr}$ est anti-aligné à $M_{Fe}$. La
figure \ref{fig:impuret=0000E9_Cr_dans_Fe} illustre la configuration
magnétique de Cr (en noir) et de sa première sphère de coordination
(en blanc). Les sphères suivantes ne sont pas modifiées par la mise
en solution.

Le moment du fer reste ainsi quasiment inchangé, alors que celui de
Cr augmente de 95\,\% et s'anti-aligne aux Fe. Les moments magnétiques
des Fe 2$^{\text{èmes}}$ et 3$^{\text{èmes}}$ voisins sont 2.32
et 2.34\,$\mu_{B}$. On observe un transfert de charge de Cr vers
la matrice de Fe. L'impureté perd $0.36$ électron redistribué de
façon homogène sur l'ensemble des Fe de la super-cellule.

Olsson et al. \cite{Olsson_defautsdft_2007} calculent une énergie
de mise en solution similaire en US et 0.3\,eV plus faible en PAW.
Le moment porté par l'impureté calculée en PAW est plus faible ($-1.59$\,$\mu_{B}$).
Le tableau \ref{tab:Mom_Cr_imp_dans_matrice_de_Fe} reprend les différentes
mesures du moment porté par un Cr en impureté dans une matrice de
Fe issues dans la littérature. Collins et Low \cite{Collins_mudistri_1965}
mesuraient un moment antiparallèle de $-0.7$\,$\mu_{B}$ par diffraction
de neutron, mais à température ambiante, température correspondant
au tiers de la température de Curie. Le moment est donc déjà fortement
diminué par rapport à 0\,K. Un peu plus tard, Aldred et al. mesurent
par diffraction de neutrons $-1.16$\,$\mu_{B}$ pour Cr et $+2.174$\,$\mu_{B}$
pour les premières sphères de coordination de nature Fe à 5\,K \cite{Aldred_bulkmag_1976,Aldred_neutron_1976}.
Les résultats DFT sont en accord raisonnable avec ces mesures expérimentales.

\paragraph{Propriétés électroniques}

\begin{figure}[h]
\begin{centering}
\includegraphics[scale=0.4]{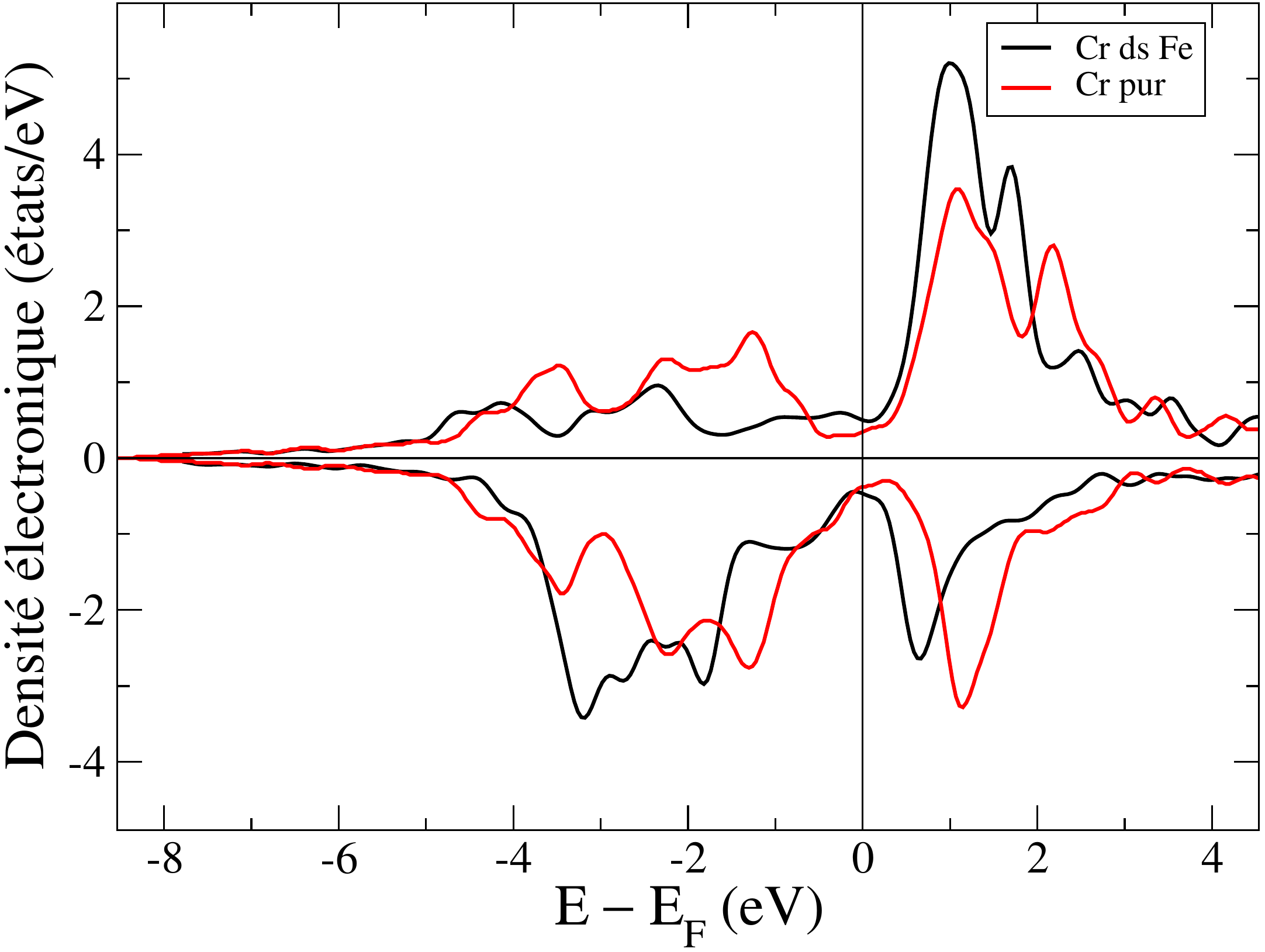}
\par\end{centering}

\caption{Densité électronique projetée d'un atome de Cr dans une matrice de
Fe (noir) ou dans un matrice de Cr (rouge). La courbe rouge correspond
à la figure \ref{fig:PDOS_elements_purs_M}. \label{fig:LDOS_Cr_dans_Fe}}
\end{figure}

Pour comprendre le lien entre le moment anti-aligné du chrome et l'énergie
de mise en solution négative, nous représentons sur la figure \ref{fig:LDOS_Cr_dans_Fe}
la densité électronique projetée d'un atome de Cr dans une matrice
de Fe (en noir) et dans une matrice de Cr (en rouge, correspondant
à la figure \ref{fig:PDOS_elements_purs_M} présentant la PDOS du
chrome pur).

La bande $\uparrow$ au-dessous du niveau de Fermi de l'impureté Cr
dans Fe se dépeuple au profit de la bande $\downarrow$ : le moment
magnétique atomique de Cr augmente. De plus, le pic liant de la bande
$\downarrow$ entre 0 et $-3$\,eV est déplacé vers des énergies
plus basses, entre $-1$ et $-4$\,eV, ce qui minimise l'énergie
de bande de Cr et diminue ainsi l'énergie totale du système.

\begin{figure}[h]
\begin{centering}
\includegraphics[scale=0.4]{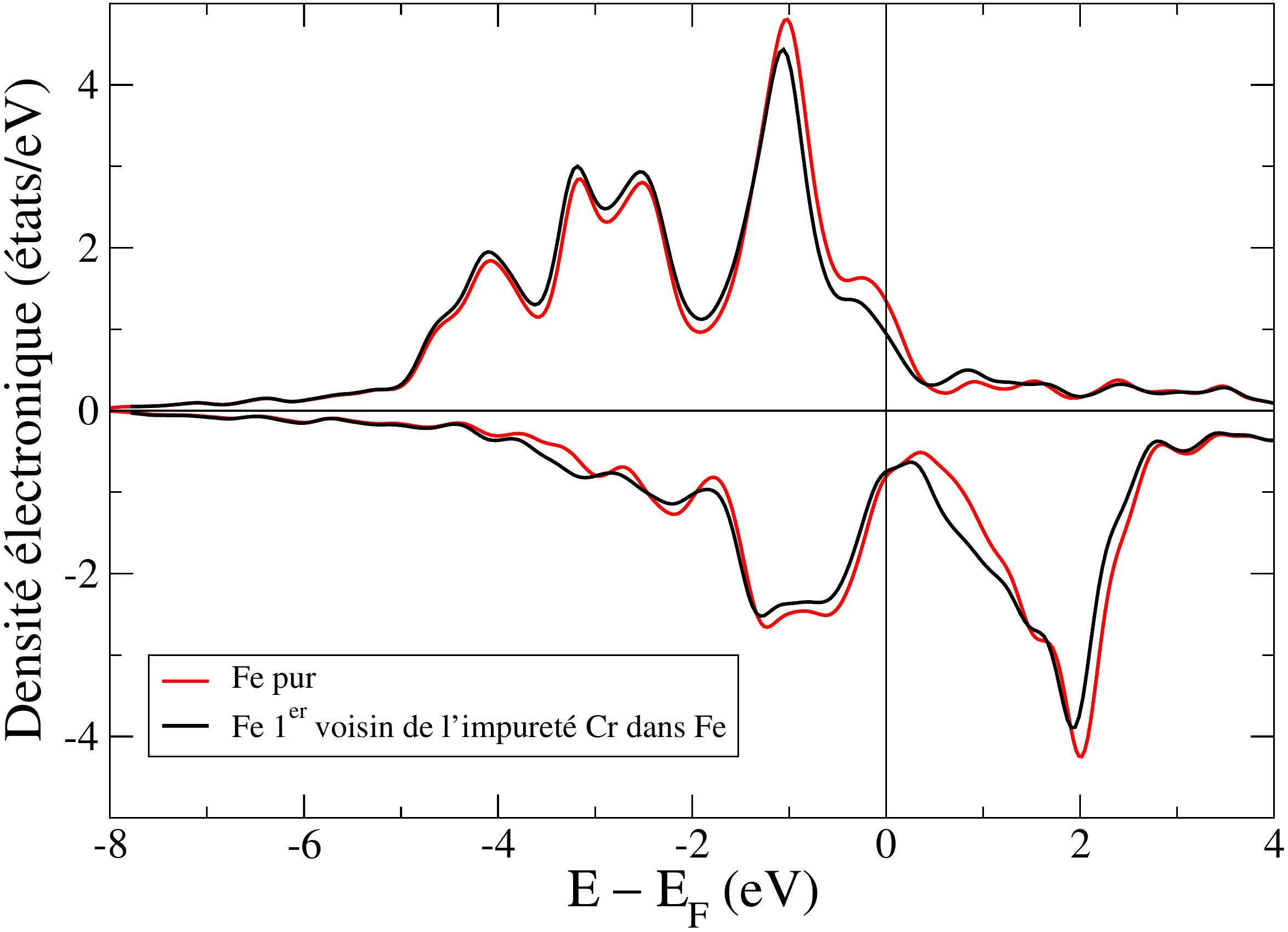}
\par\end{centering}

\caption{Densité électronique projetée sur l'atome de Fe premier voisin de
l'impureté de Cr (noir) et comparaison avec celle de Fe pur (rouge).
Ces résultats sont issus de calculs en SIESTA-NC.\label{fig:LDOS_FeNN_Crimp_dans_Fe}}
\end{figure}

On représente sur la figure \ref{fig:LDOS_FeNN_Crimp_dans_Fe} la
PDOS de Fe dans Fe (en rouge, correspondant à la figure \ref{fig:PDOS_elements_purs_M}
présentant la PDOS du fer pur), et de Fe premier voisin de l'impureté
Cr dans une matrice de fer. 

On observe sur cette figure \ref{fig:LDOS_FeNN_Crimp_dans_Fe} que
la PDOS de Fe est très faiblement modifiée par la présence de l'impureté
Cr. L'énergie de mise en solution de Cr dans Fe est donc une conséquence
de l'effet des atomes Fe sur Cr et non de Cr sur la matrice Fe.

On peut alors expliquer la surestimation de l'énergie de mise en solution
de Cr dans Fe par les méthodes et approximations qui surestiment le
moment magnétique de Cr. Que les pseudo-potentiels soient US, NC ou
PAW, la diminution de l'énergie du système lors de la mise en solution
de Cr dans Fe est due à la diminution de l'énergie de bande de Cr,
c'est-à-dire à l'augmentation du moment magnétique de Cr. Les approximations
US et NC, qui surestiment le moment magnétique de Cr, surestiment
en conséquence la diminution de l'énergie de bande et donc l'énergie
de mise en solution.

En résumé : plus le moment magnétique de Cr dans Fe est fort, plus
la mise en solution est favorable. Les pseudo-potentiels US et NC
surestiment le moment magnétique de Cr, ils surestiment donc la stabilisation
de Cr en solution dans Fe.

\subsubsection*{Dissolution d'un atome de fer dans une matrice de chrome}

À l'autre extrémité du diagramme de phases, Fe est en solution dans
Cr. À seulement 2.8\,\%\,Fe, Moze et al. \cite{moze_CrrichimpFe_1988}
montrent par diffusion diffuse de neutrons en faisceau polarisé que
l'impureté Fe possède un moment antiferromagnétique $M_{imp}=-2.13$\,$\mu_{B}$
couplé à la phase antiferromagnétique commensurable AF de Cr (représentée
sur la figure \ref{fig:Structure-antiferromagn=0000E9tique-(AFM)}).
L'impureté ne perturbe pas le moment magnétique des atomes de matrice
voisins d'amplitude 0.72\,$\mu_{B}$.

La mise en solution d'un Fe dans une super-cellule de taille $5\times5\times5$,
c'est-à-dire de 250 atomes, correspond à une concentration de 0.4\,\%
en Fe inférieure aux 2.8\,\% délimitant la transition de phases ISDW--AF.
À cette concentration, c'est donc une phase ISDW (représentée sur
la figure \ref{fig:Onde de spin}) qui est observée expérimentalement
\cite{moze_CrrichimpFe_1988,fawcett_Crpur_1988,zabel_Crmag_1999}.
Les calculs présentés ici, qui sont dans l'approximation AF, sont
donc moins justifiés que pour des concentrations plus fortes en Fe.
Nous présentons donc ici des résultats succincts.

\begin{figure}[h]
\begin{centering}
\includegraphics[scale=0.5]{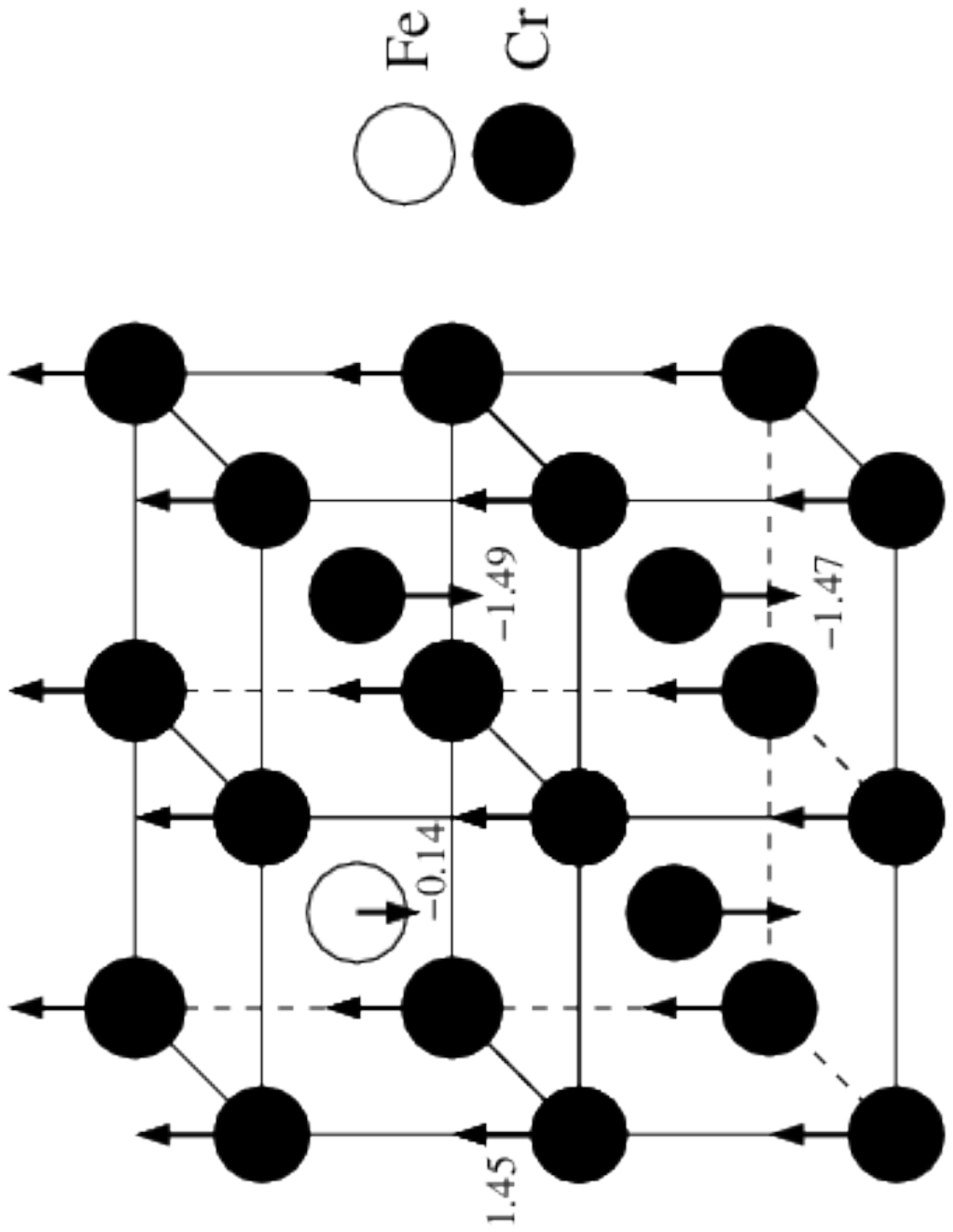}
\par\end{centering}

\caption{Structure magnétique calculée en SIESTA-NC d'une impureté Fe (en blanc)
dans une matrice Cr (en noir). Les moments magnétiques atomiques des
1$^{\mbox{er}}$, 2$^{\mbox{e}}$ et 3$^{\mbox{e}}$ voisins de l'impureté
sont indiqués.\label{fig:Fe_impuret=0000E9_dans_Cr}}
\end{figure}

L'énergie de mise en solution $\Delta H_{sol}^{Fe\, dans\, Cr}$ calculée
avec SIESTA est de $+0.288$\,eV. Il est donc défavorable de substituer
un atome de Cr par un atome de Fe. La structure magnétique de l'impureté
et de ses trois premières sphères de coordinations est représentée
sur la figure \ref{fig:Fe_impuret=0000E9_dans_Cr}. Le moment magnétique
de l'impureté Fe est $-0.14$\,$\mu_{B}$. Il est anti-aligné au
moment magnétique local de ses premiers voisins Cr de moment 1.45\,$\mu_{B}$,
et aligné aux moments des deuxièmes voisins d'amplitude $-1.49$\,$\mu_{B}$.
La matrice, qui reste dans une phase magnétique AF, est très faiblement
perturbée par la présence de Fe : Le moment des troisièmes voisins
est d'amplitude $-1.47$\,$\mu_{B}$, pour converger ensuite à l'amplitude
de Cr pur : $\pm1.41$\,$\mu_{B}$.

Ces résultats sont en désaccord avec les récents calculs de Mishra
\cite{mishra_FeinCr_2008}. Cet auteur mesure l'effet d'un atome d'impureté
Fe dans une matrice de Cr AF en FLAPW\nomenclature{FLAPW ou FPLAPW}{Full Potential Linearized Augmented Plane Wave -- Ondes planes augmentées avec linéarisation}.
Fe y a un moment faible ($0.22$\,$\mu_{B}$) aligné avec ses permiers
voisins. Les moments magnétiques atomiques des premiers, deuxièmes
et troisièmes voisins sont différents de ceux des atomes de volume
de moins de 5\,\%.

Pour trouver l'état fondamental magnétique d'une structure, il faut
tester un grand ensemble de configurations magnétiques initiales,
chacune convergeant vers un minimum local de la surface de potentiel.
Nous avons en effet trouvé une structure très comparable à celle de
Mishra, mais il ne s'agit que d'un minimum local. La structure la
plus basse en énergie est celle présentée ci-dessus et représentée
sur la figure \ref{fig:Fe_impuret=0000E9_dans_Cr}. D'autres calculs
\emph{ab initio} sont en désaccord relatif, mais les super-cellules
sont alors de concentrations supérieures à 6\,\% \cite{hashemifar_FLAPW_Fe_in_Cr_2006}.

\subsubsection*{Interactions entre impuretés}

La dissolution d'une impureté chrome dans le fer est exothermique.
Cette propriété est liée au moment magnétique des atomes de Cr. Nous
voulons maintenant comprendre comment interagissent deux impuretés
dans une matrice de fer.

Nous pouvons, pour comprendre si les impuretés s'attirent ou se repoussent,
calculer l'énergie de liaison $E_{liaison}^{Cr-Cr}\left(i\right)$
entre atomes de Cr $i$$^{\text{èmes}}$ voisins dans une matrice
de Fe : 
\begin{equation}
E_{liaison}^{Cr-Cr}\left(i\right)=E_{tot}^{\left(i\right)}\left(Fe_{N-2}Cr_{2}\right)+NE\left(Fe\right)-2E_{tot}\left(Fe_{N-1}Cr_{1}\right),
\end{equation}
où $E_{tot}^{\left(i\right)}\left(Fe_{N-2}Cr_{2}\right)$ est l'énergie
totale d'une super-cellule de $N$ atomes dans laquelle deux des atomes
sont des Cr $i$$^{\text{èmes}}$ voisins.

\begin{figure}[!h]
\begin{centering}
\includegraphics[scale=0.4]{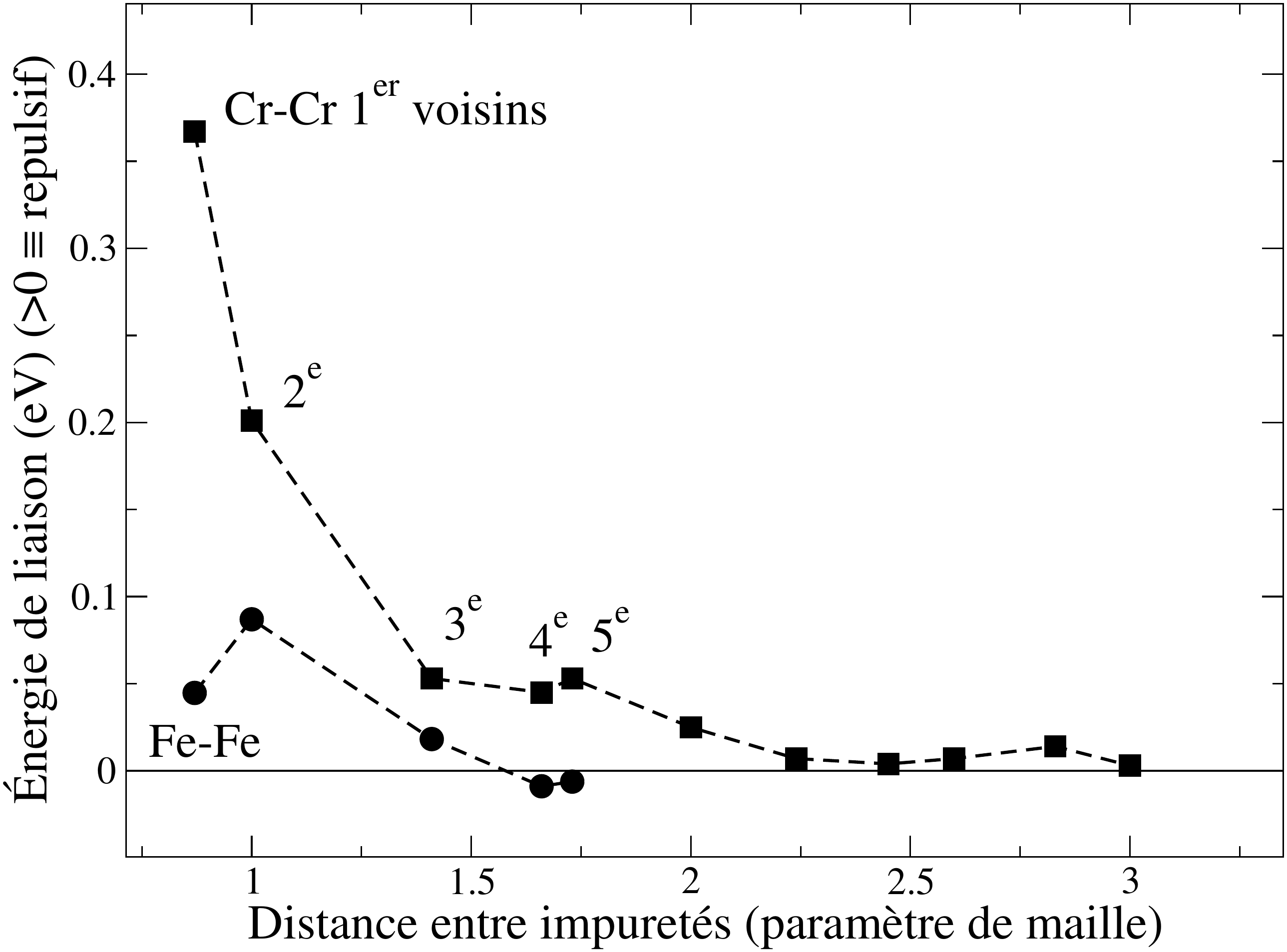}
\par\end{centering}

\begin{centering}
\includegraphics[scale=0.4]{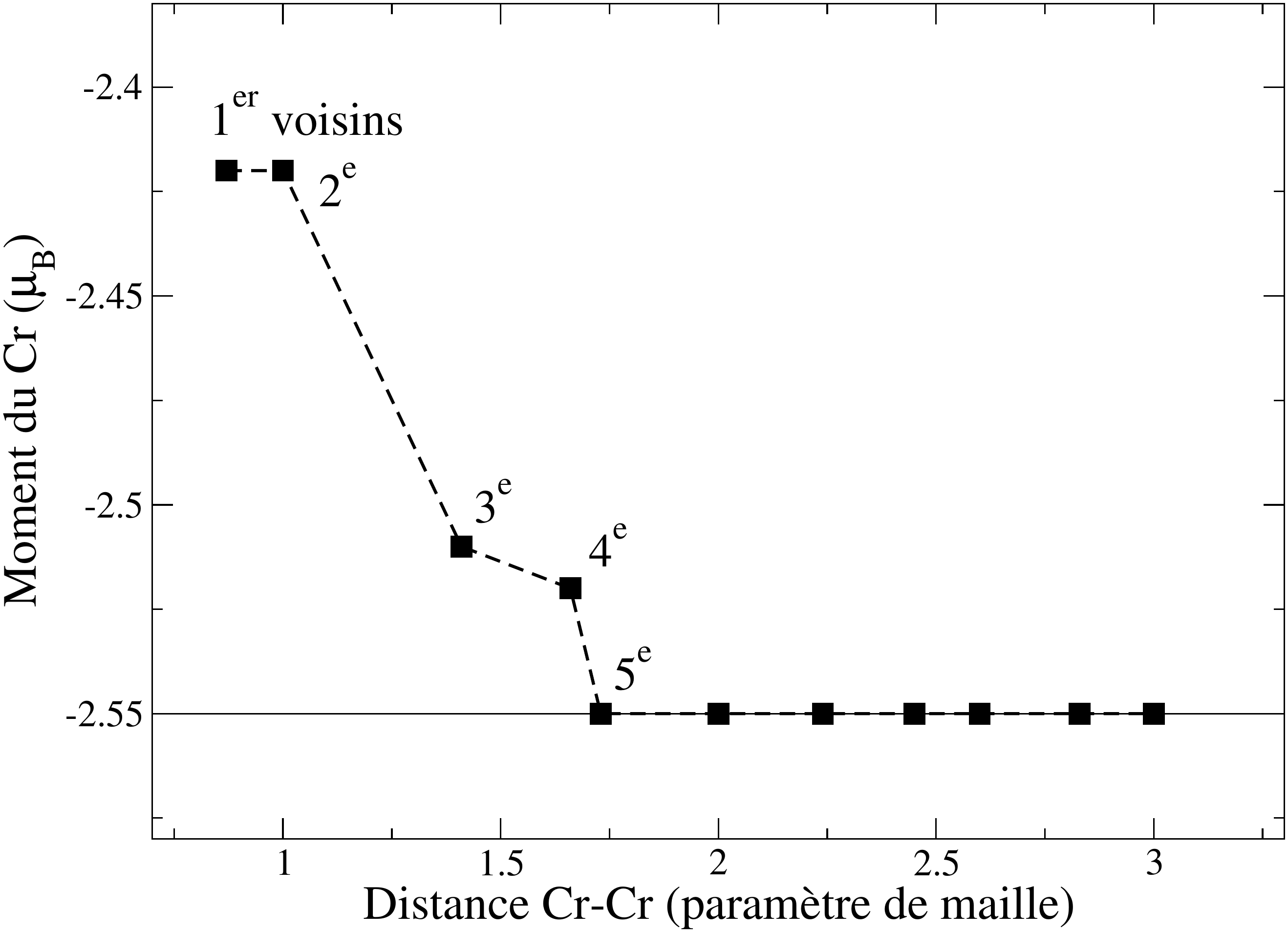}
\par\end{centering}

\caption{(en haut) énergies de liaison Cr--Cr dans Fe (carrés) et Fe--Fe dans
Cr (cercles) calculées avec SIESTA-NC. Les distances sont indiquées
en unités réduites et en voisinage.\protect \\
(en bas) moment magnétique des deux impuretés Cr dans Fe en fonction
de la distance Cr--Cr. Les moments des Fe ne sont pas indiqués car
de nombreuses configurations magnétiques dégénérées existent pour
Fe dans Cr.\label{fig:EbindCrCr}}

\end{figure}

La figure \ref{fig:EbindCrCr} présente l'évolution de l'énergie de
liaison Cr--Cr (Fe--Fe) dans une matrice de Fe (Cr) pour une distance
entre impuretés allant jusqu'aux 13$^{\text{e}}$ voisins. Ces calculs
sont réalisés dans une super-cellule $5\times5\times5$ (250\,atomes)
en SIESTA-NC. Les données brutes sont indiquées dans le tableau \ref{tab:Ebind_CrCr}.
L'énergie de liaison Cr--Cr croît rapidement lorsque les impuretés
se rapprochent. Elle diminue de moitié entre 1$^{\text{ers}}$ et
2$^{\text{e}}$ voisins, puis est divisée par quatre entre 2$^{\text{e}}$
et 3$^{\text{e}}$ voisins.

\begin{table}[h]
\begin{centering}
\begin{tabular}{|c|c|c|c|}
\hline 
$d_{Cr-Cr}$ (voisins) & $d_{Cr-Cr}$ (unités réduites) & $M_{Cr}$ ($\mu_{B}$) & $E_{liaison}^{Cr-Cr}$ (eV) \\
\hline 
\hline 
1 & 0.87 & $-2.42$ & 0.37 \\
\hline 
2 & 1.00 & $-2.42$ & 0.20 \\
\hline 
3 & 1.41 & $-2.51$ & 0.05 \\
\hline 
4 & 1.66 & $-2.52$ & 0.05 \\
\hline 
5 & 1.73 & $-2.55$ & 0.05 \\
\hline 
6 & 2.00 & $-2.55$ & 0.03 \\
\hline 
7 à 13 & 2.18 & $-2.55$ & 0.00 \\
\hline 
\end{tabular}
\par\end{centering}

\caption{Énergies de liaisons Cr--Cr (SIESTA-NC) dans une matrice de Fe pour
des distances entre impuretés de 1 à 13 sphères de coordination, et
moment magnétique $M_{Cr}$ porté par chaque impureté Cr.\label{tab:Ebind_CrCr}}
\end{table}

\paragraph{Lien avec le magnétisme}

Le moment d'un atome de Cr en impureté est anti-aligné aux moments
des atomes de Fe de la matrice. Quand deux atomes de Cr se rapprochent
l'un de l'autre, un phénomène de frustration a lieu . Quelle que
soit la distance entre Cr, leurs moments sont anti-alignés aux Fe.
Les moments des Cr sont donc frustrés. Les moments frustrés portés
par les Cr diminuent lorsqu'ils se rapprochent. L'énergie de liaison
est en conséquence positive (i.e. répulsive).

La convergence n'est atteinte que pour des distances entre impuretés
Cr de 7 sphères de coordinations ou plus. C'est une très longue portée.

\begin{figure}[h]
\begin{centering}
\includegraphics[scale=0.5]{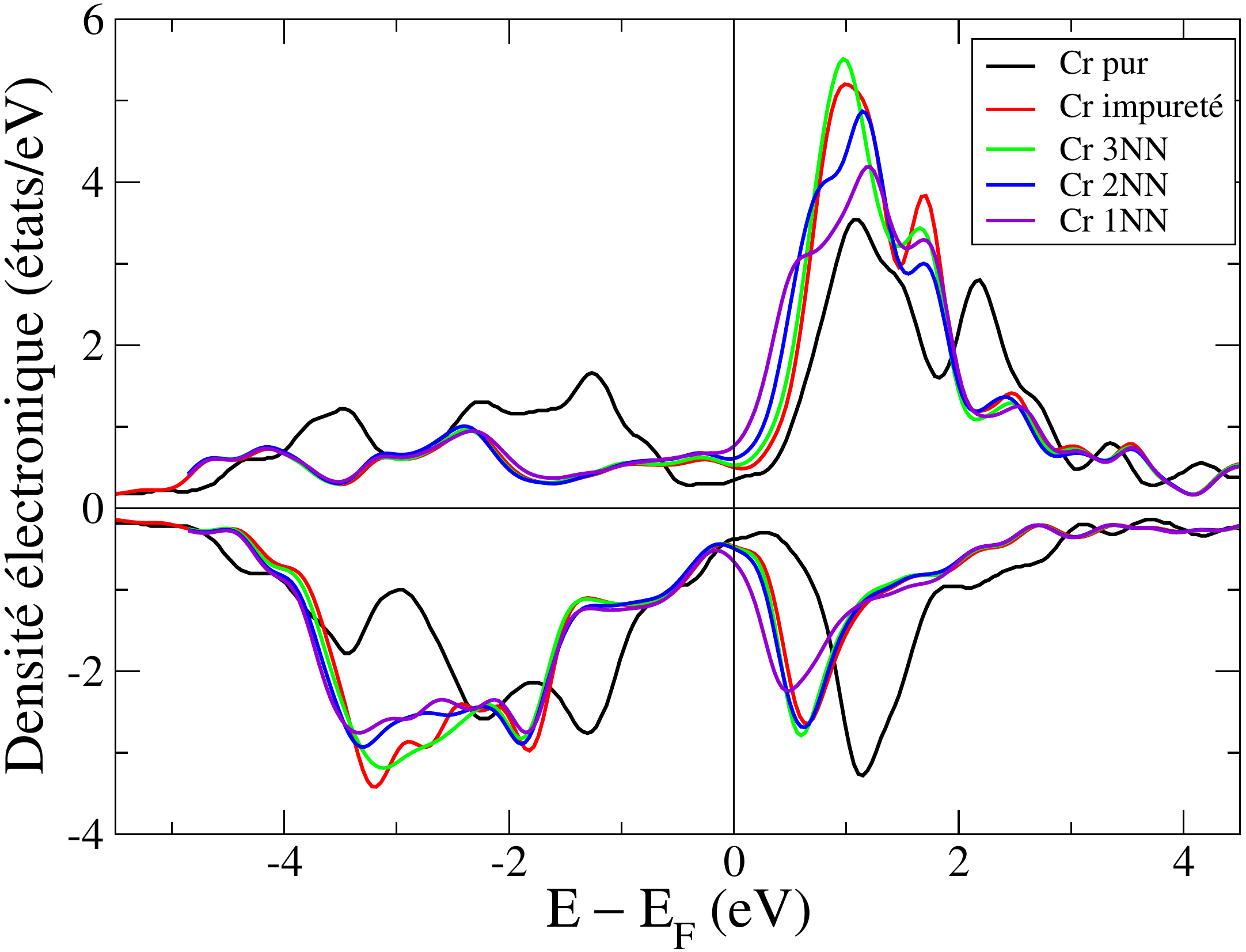}
\par\end{centering}

\caption{PDOS calculées avec SIESTA-NC de Cr pur (en noir), Cr en impureté
dans Fe (en rouge), deux Cr premiers voisins dans Fe (en violet),
et deuxièmes (en bleu) et troisièmes voisins (en vert).\label{fig:LDOS-Cr-Cr}}
\end{figure}

Au contraire, deux impuretés Fe dans Cr ont des moments magnétiques
atomiques quasi-nuls, quelle que soit leur distance. L'énergie de
liaison est faiblement répulsive en 1$^{\text{e}}$ et 2$^{\text{e}}$
voisins d'un ordre de grandeur en-dessous de la répulsion Cr--Cr dans
Fe. En 3$^{\text{e}}$ voisins, les impuretés Fe ne se voient déjà
quasiment plus. Ces interactions sont de beaucoup moins longue portée
que pour Cr--Cr dans Fe.

\subsubsection*{Énergies de mélange\label{sub:Energies-de-m=0000E9lange_DFT}}

L'énergie de mélange $\Delta E_{mix}$ d'une distribution donnée de
$N$ atomes de Fe et $M$ atomes de Cr est calculée \emph{ab initio}
selon la relation suivante :
\begin{equation}
\Delta E_{mix}=\dfrac{E\left(Fe_{N}Cr_{M}\right)-\left(NE\left(Fe\right)+ME\left(Cr\right)\right)}{N+M}
\end{equation}
où $E\left(Fe_{N}Cr_{M}\right)$ est l'énergie totale d'une super-cellule
contenant $N$ atomes de Fe et $M$ atomes de Cr. Les énergies de
référence de Fe et Cr sont calculées \emph{ab initio} dans la phase
cubique centrée, avec le paramètre de maille de l'élément pur.

Le paramètre de maille utilisé pour la super-cellule contenant de
Fe et de Cr est celui de Fe. En effet, les paramètres de maille de
Fe et Cr sont très proches, et nous nous intéressons principalement
à la solution solide riche en Fe. Nous avons comparé les résultats
de la solution solide $\alpha'$ avec des calculs similaires utilisant
le paramètre de maille de Cr $a_{Cr}$, ou celui obtenu par la loi
de Vegard : $a_{Vegard}=x_{Fe}a_{Fe}+x_{Cr}a_{Cr}$. C'est une relation
similaire qui a été obtenue par diffraction de rayons X par Leslie,
Zwell et Speich \cite{leslie_iron_1972,zwell_effects_1973} : $a_{Fe_{1-x}Cr_{x}}=a_{Fe}+(5\pm0.5)\times10^{-4}x$.
La différence d'énergie totale entre les super-cellules avec ces différents
paramètres de maille est inférieure à la précision des calculs.

Les structures Fe$_{N}$Cr$_{M}$ ont été générées suivant deux méthodes.
La plus grande partie de ces structures sont des structures simples
choisies pour donner \emph{a priori} des énergies de mélange faibles
ou correspondent à des structures ordonnées classiques des réseaux
cubiques centrés (B$_{\text{2}}$, DO$_{\text{3}}$, B$_{\text{32}}$,
\ldots{}). Dans ces deux cas, les structures sont ordonnées, au sens
où les fonctions de corrélation sont non-nulles. Le reste des structures
sont des structures spéciales quasi-aléatoires (SQS). Le détail théorique
et pratique de la génération de ces structures par Enrique Martinez
avec le code ATAT \cite{ATAT_homepage} est donné en annexe \ref{sec:annexe_SQS}.
L'intérêt des structures SQS est d'approximer une solution solide
idéale en générant des structures ayant des fonctions de corrélation
nulles.

\begin{figure}[h]
\begin{centering}
\includegraphics[scale=0.4]{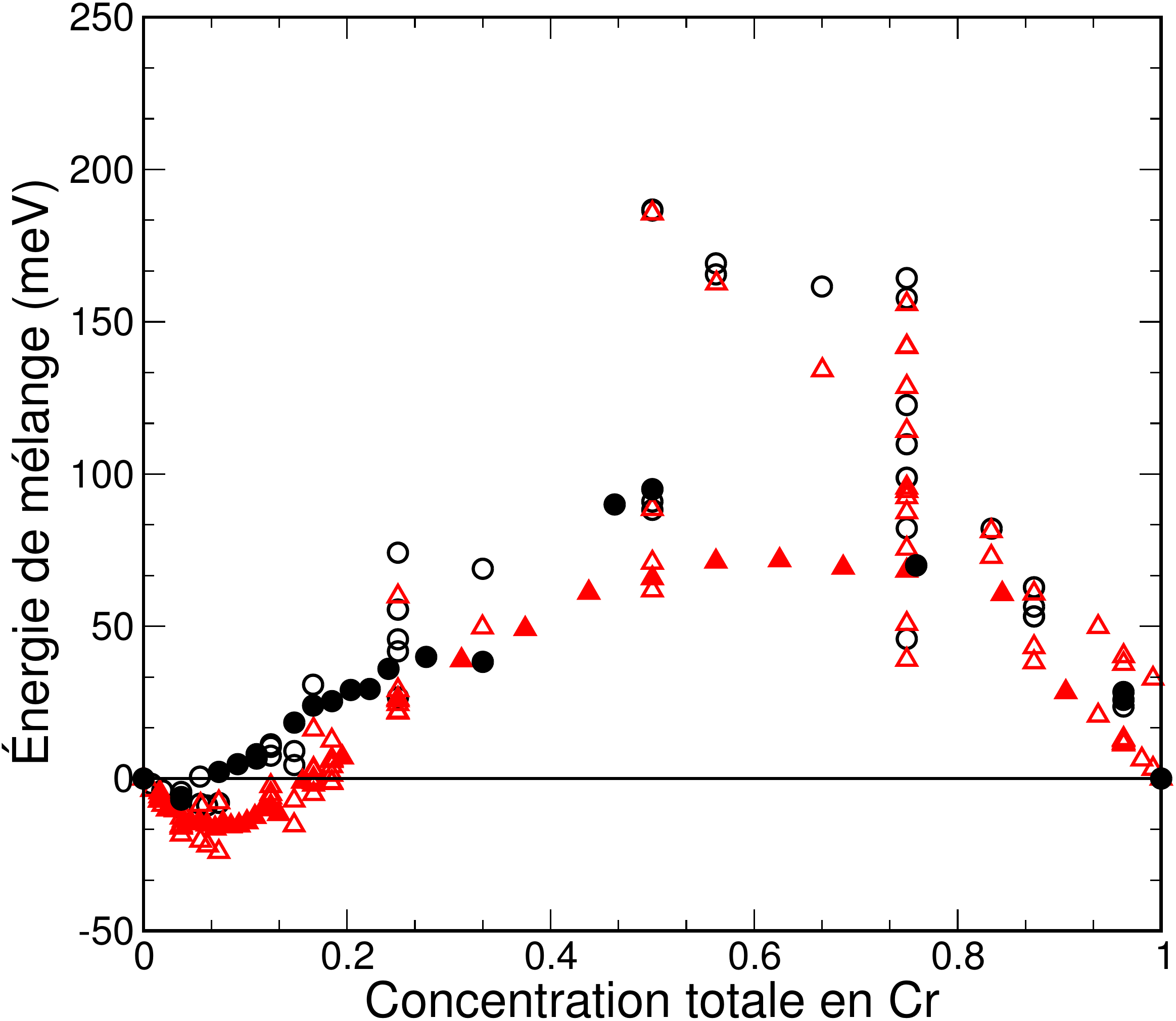}
\par\end{centering}

\caption{Énergies de mélange en fonction de la concentration de la super-cellule
en Cr des structures ordonnées (symboles vides) et SQS (symboles pleins)
calculées en PWSCF-PAW (noir) et SIESTA-NC (rouge).\label{fig:Enthalpie-de-m=0000E9lange}}
\end{figure}

Les énergies de mélange des structures ordonnées et SQS calculées
avec SIESTA et PWSCF en pseudo-potentiels à norme-conservée et PAW
sont indiquées sur la figure \ref{fig:Enthalpie-de-m=0000E9lange}.

L'énergie de mélange $\Delta E_{mix}$ change de signe aux faibles
concentrations en Cr : autour de $x_{Cr}=0.15$ en SIESTA-NC, et $x_{Cr}=0.07$
en PWSCF-PAW. Les résultats SIESTA-NC couvrent un domaine d'énergie
plus vaste que les résultats PWSCF-PAW. SIESTA-NC surestime la stabilisation
de la solution solide riche en Fe et la déstabilisation de la solution
solide équimolaire Fe$_{\mbox{0.5}}$Cr$_{\mbox{0.5}}$. Comme pour
l'impureté Cr dans la matrice Fe, les calculs SIESTA-NC surestiment
les moments magnétiques de Cr et surestiment en conséquence les interactions
(répulsives ou attractives) entre atomes de Cr.

\paragraph{Un intermétallique autour de 4\,\%\,Cr ?}

\begin{figure}[h]
\begin{centering}
\includegraphics[scale=0.4]{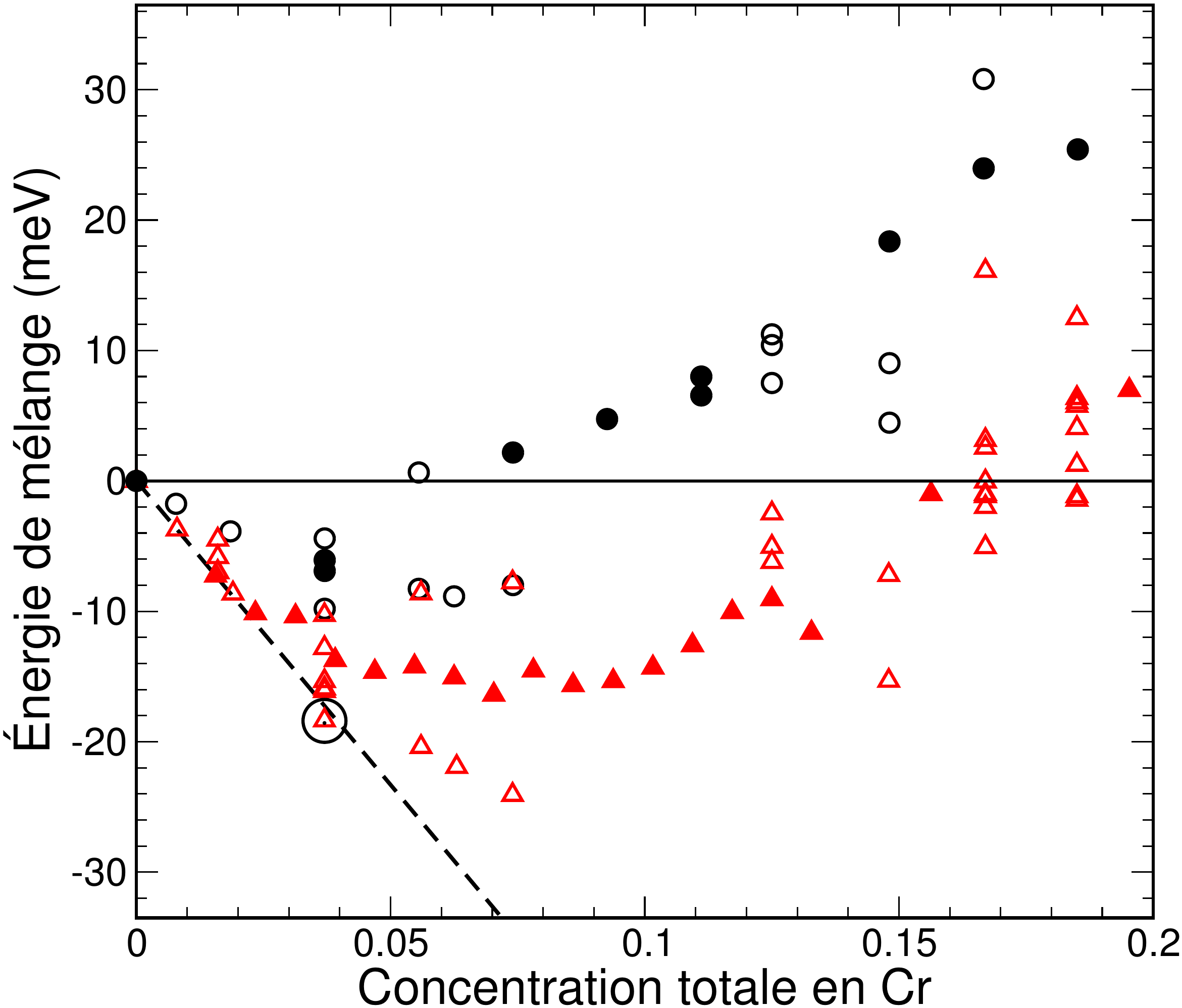}
\par\end{centering}

\caption{Énergies de mélange calculées en PWSCF-PAW (noir) et SIESTA-NC (rouge)
en fonction de la concentration globale en Cr. Les structures ordonnées
sont en symboles vides et SQS en symboles pleins. L'énergie de mise
en solution du chrome dans le fer calculée en PWSCF-PAW est représentée
en tirets pointillés. La structure Fe$_{\text{52}}$Cr$_{\text{2}}$
est entourée en noir.\label{fig:Emix_Fe52Cr2}}

\end{figure}

L'énergie de mélange des structures ordonnées et SQS de concentration
globale en chrome inférieure à 20\,\% sont représentées sur la figure
\ref{fig:Emix_Fe52Cr2}. On représente également l'énergie de mise
en solution du chrome dans le fer en tirets pointillés. Il s'agit
de l'énergie gagnée par la solution solide lorsque la concentration
en chrome augmente, dans l'hypothèse où les atomes de Cr n'interagissent
pas entre eux.

L'énergie de mélange de la structure Fe$_{\text{52}}$Cr$_{\text{2}}$
entourée en noir sur la figure \ref{fig:Emix_Fe52Cr2} est égale à
$-18.36$\,meV/atome. Elle est inférieure à l'énergie de mise en
solution. Les interactions entre atomes de Cr stabilisent donc le
système dans cette structure ordonnée, en opposition avec la tendance
générale de répulsion entre atomes de Cr. Dans cette structure à 3.7\,\%\,Cr,
le réseau est de type cubique centré $3\times3\times3$ à 2 atomes
par maille. Cette structure ordonnée, observée parallèlement par Erhart
et al. \cite{erhart_lro_2008} contient 52 atomes de Fe et 2 atomes
de Cr distants de 7.48\,Å (2.6\,$a_{Fe}$) dans la direction $\left\langle 111\right\rangle $.
Chaque atome de Cr a ses neuf premières sphères de coordination occupées
par Fe et 25\,\% de sa dizième sphère occupée par Cr. Cependant,
l'énergie de mélange de cet intermétallique est seulement $1.12$\,meV
par atome de Cr plus basse que l'énergie de mise en solution (49\,meV
par atome de Cr selon les calculs VASP-PAW de Erhart et al.). Cet
intermétallique n'existe donc certainement qu'à des températures extrêmement
basse, et ne pourraient être observée expérimentalement. Deux structures
ordonnées Fe$_{15}$Cr$_{1}$ et Fe$_{14}$Cr$_{1}$ ont également
été proposées à 0\,K par Nguyen-Manh et Dudarev, et Pareige et al.
\cite{nguyen-manh_modelStoner_2009,pareige_SRO_2009}. Selon nos calculs,
ces structures ont une énergie de mélange supérieure à l'énergie de
mise en solution. Cependant, la difference entre ces deux énergies
est ici tellement petite que la très faible variation de paramètre
de maille entre les simulations suffit certainement à l'inverser.

\paragraph{La structure B2}

La structure B2 est la structure ordonnée dont l'énergie de mélange
est la plus haute ($\Delta E_{mix}=0.186$\,eV en SIESTA-NC, 0.187\,eV
en PWSCF-PAW et 0.199\,eV en VASP-PAW selon Klaver et al. \cite{klaver_magnetism_2006}).
Cette structure B2 illustre bien la dépendance complexe de l'énergie
de liaison Cr--Cr avec l'environnement local. Si les moments magnétiques
atomiques des Cr voulaient simplement s'anti-aligner aux moments magnétiques
des Fe, alors la structure magnétique la plus stable serait celle
représentée sur la figure \ref{fig:impuret=0000E9_Cr_dans_Fe} : une
phase antiferromagnétique AF similaire à celle de Cr pur.

\begin{figure}[h]
\begin{centering}
\includegraphics[scale=0.4]{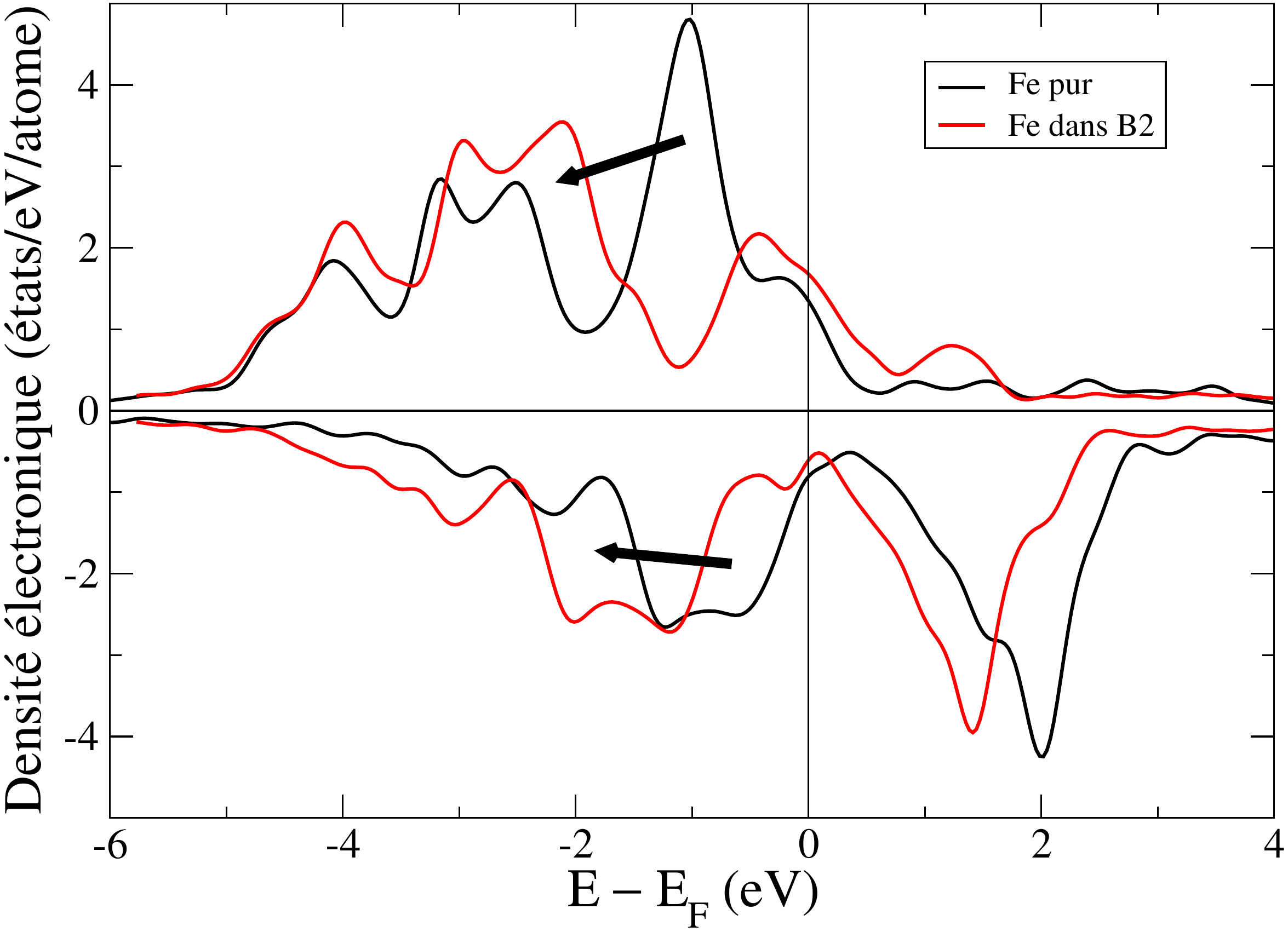}
\par\end{centering}

\begin{centering}
\includegraphics[scale=0.4]{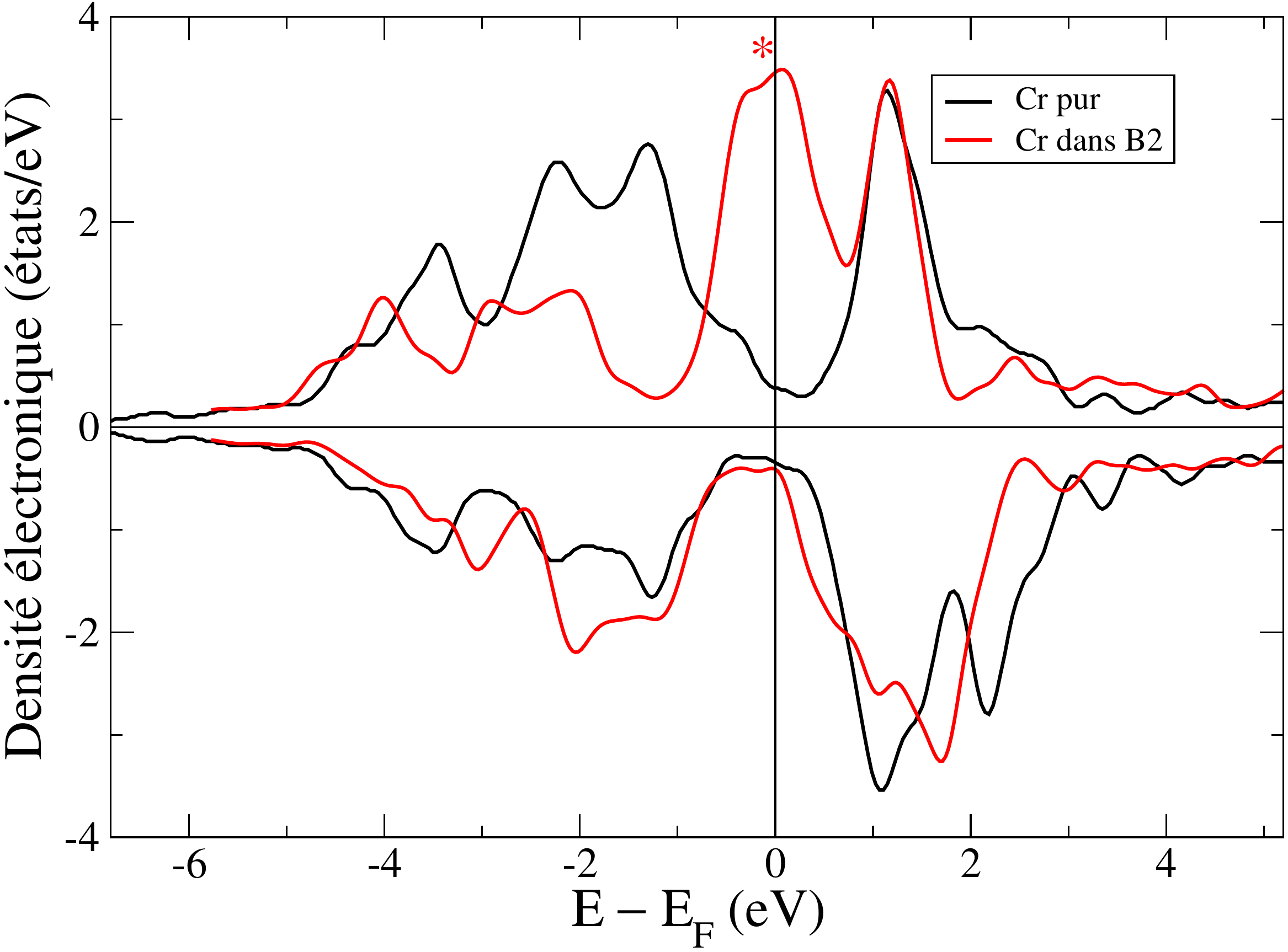}
\par\end{centering}

\caption{(en haut) PDOS du fer cubique centré ferromagnétique et de l'atome
de Fe dans la structure FeCr B2. Des transferts importants de densité
d'états sont représentés par des flêches. (en bas) PDOS du chrome
cubique centré AF et de l'atome de Cr dans la structure FeCr B2. Dans
FeCr B2, un nouveau pic de densité d'états apparaît pour l'atome de
Cr au niveau de Fermi. On l'indique par une étoile rouge. \label{fig:PDOS-B2}}
\end{figure}

Or, selon les calculs SIESTA-NC et PWSCF-PAW, Fe et Cr portent tous
deux un moment positif ($+1.70$ et $+0.25$\,$\mu_{B}$) : le système
est ferromagnétique. On représente la PDOS des éléments purs dans
la structure B2 sur la figure \ref{fig:PDOS-B2}.

La structure électronique des atomes Fe et Cr de la structure B2 est
largement modifiée par rapport à celle des éléments purs dans leurs
structures cubiques centrées magnétiques.
\begin{itemize}
\item L'atome de Fe dans la structure B2 est stabilisé par un transfert
de densité d'états du pic proche du niveau de Fermi des bandes $\uparrow$
et $\downarrow$ du fer pur. Ces transferts sont représentés par des
flèches sur la figure \ref{fig:PDOS-B2}.
\item Quant à l'atome de Cr, il est fortement déstabilisé par rapport au
chrome pur car il y a un pic densité d'états au niveau de Fermi, repéré
par une étoile rouge sur la figure \ref{fig:PDOS-B2}. La présence
d'un tel pic est très déstabilisante pour le système qui tend à se
réarranger pour minimiser la densité d'états au niveau de Fermi :
le système est instable.
\end{itemize}

Dans cette structure B2, on voit encore que c'est le chrome qui donne
toute sa complexité au problème. Ce n'est pas un élément qui s'allie
selon une vision simples des couplages ferromagnétiques / antiferromagnétiques
satisfaits ou frustrés. L'énergie totale d'un alliage à base de fer
et de chrome est ainsi nettement dépendante de l'environnement local
de chacun des atomes de Cr.

\subsubsection*{Distribution des moments magnétiques locaux}

\begin{figure}[h]
\begin{centering}
\includegraphics[scale=0.4]{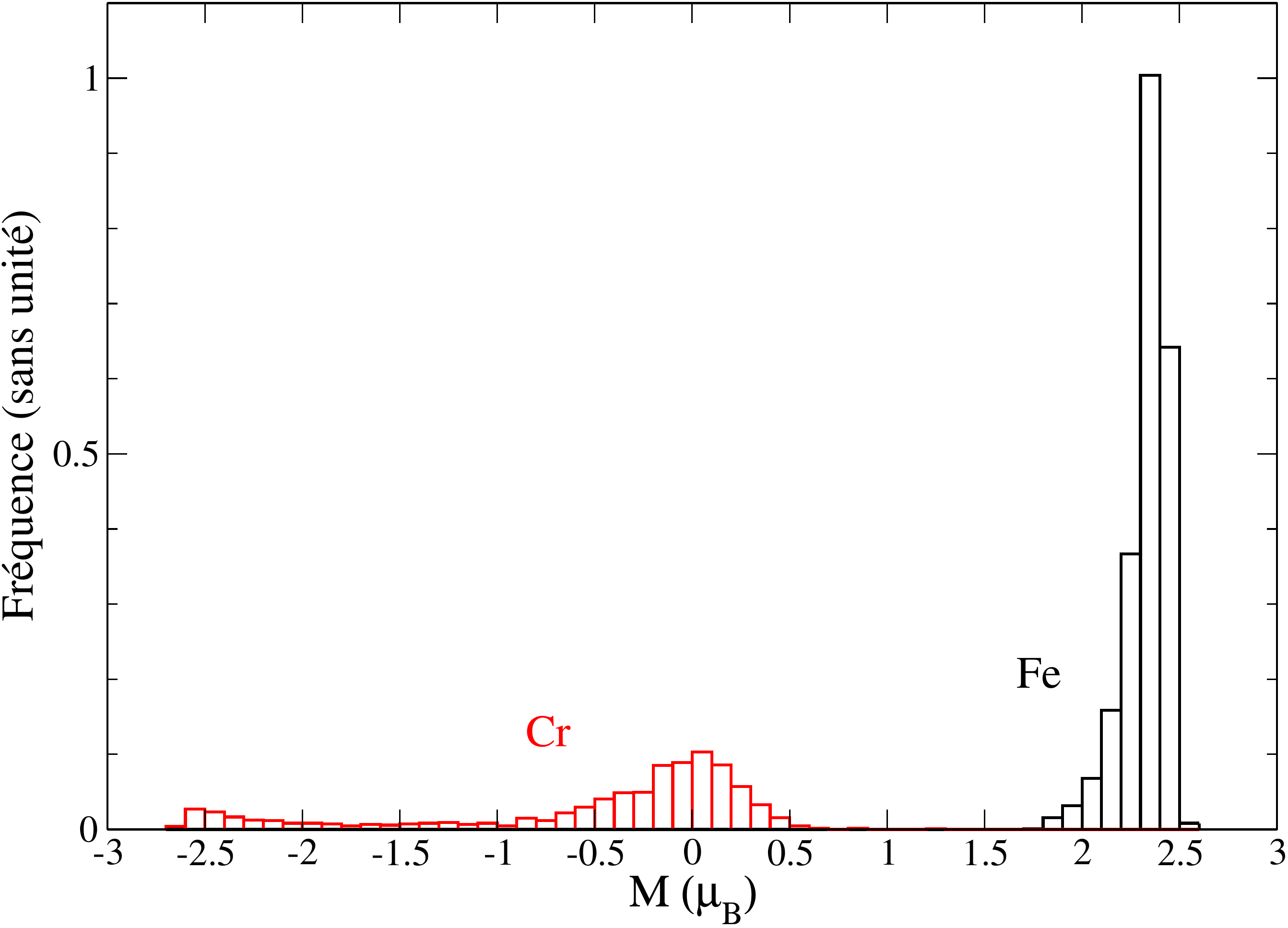}
\par\end{centering}

\caption{iagramme de distribution des moments magnétiques des atomes de Fe
(en noir) et Cr (en rouge) pour toutes les configurations SQS ou ordonnées
dont on a calculé l'énergie totale avec SIESTA-NC.\label{fig:diagramme_de_distri_M_SQS}}
\end{figure}

Nous représentons sur la figure \ref{fig:diagramme_de_distri_M_SQS}
un diagramme de distribution des moments magnétiques des atomes de
Fe et Cr pour toutes les configurations (SQS ou ordonnées) dont on
a calculé l'énergie totale avec SIESTA-NC. Ce sont ces mêmes structures
dont nous présentions les énergies de mélange à la page \pageref{fig:Enthalpie-de-m=0000E9lange}. 

Les moments magnétiques des atomes de fer sont majoritairement d'une
amplitude entre 2.2 et 2.5\,$\mu_{B}$. Le fer pur a également une
amplitude de cet ordre. Le moment magnétique est donc peu affecté
par la présence du chrome.

Au contraire, les moments magnétiques des atomes de chrome ont une
amplitude dont la distribution est beaucoup plus large. Celle-ci varie
d'environ $-2.5$\,$\mu_{B}$ à $+0.5$\,$\mu_{B}$. Le léger pic
de distribution autour de $-2.5$\,$\mu_{B}$ correspond aux structures
dans lesquelles le chrome est dilué dans la matrice de fer. Le deuxième
pic entre $-0.5$ et $+0.5$\,$\mu_{B}$ correspond à l'ensemble
des autres structures dans lesquelles les concentrations en chrome
sont supérieures à 16\,\%\,Cr. Il s'agit des structures dans lesquelles
les atomes de Cr interagissent entre eux de façon répulsive et complexe,
ce qui conduit à un affaiblissement de leurs moments magnétiques atomiques
pour diminuer l'intensité de la répulsion. Le moment magnétique de
l'atome de Cr dans l'alliage FeCr est largement dépendant du type
d'atomes de son environnement local. 

Cette observation amène un nouveau regard sur les modèles d'interaction
tenant explicitement compte du moment magnétique atomique que l'on
discutera dans le chapitre 3.

\clearpage

\section{Les surfaces libres de l'alliage : état de l'art \emph{ab initio}\label{sub:etat_de_lart_surf_DFT}}

Cette deuxième partie de l'étude \emph{ab initio} du système binaire
fer--chrome est dédiée aux surfaces libres des éléments purs et de
l'alliage. On pourra se référer à l'ouvrage très complet de Desjonquères
et Spanjaard pour une présentation très complète et didactique des
concepts de physique des surfaces \cite{desjonqueres_spanjaard_concepts_1996}.
Par calculs DFT, on étudie les surfaces à bas indices de Miller des
éléments purs, la ségrégation de Cr dans Fe, et l'effet des interactions
Cr--Cr dans ces surfaces. On a vu dans le chapitre précédent que le
magnétisme joue un rôle de premier ordre dans la physique du système
fer-chrome. Nous verrons que cet effet, amplifié en surface, a des
conséquences sur les phénomènes de ségrégation. L'effet des approximations
méthodologiques est également discuté à la lumière des résultats précédents.
Ces calculs sont essentiellement réalisés avec SIESTA car on montrera
qu'il est important que les super-cellules de calcul contiennent un
grand nombre d'atomes.

\subsection{Surfaces libres des éléments purs}

Malgré leur importance technologique, les surfaces libres de Fe et
Cr sont très peu étudiées. Ce sont en effet des surfaces extrêmement
réactives, ce qui rend les études expérimentales difficiles. La majorité
des observations sont structurales, par LEED, MEIS et AES. On y mesure
l'effet de la surface sur la distance entre les plans (les relaxations
de surface). Bien que ce soit une technique classique pour l'étude
des surfaces, les observations au microscrope à effet tunnel (STM)
des surfaces de Fe et Cr sont rares. Elles apportent cependant une
connaissance de base des propriétés électroniques. La nécessité d'un
ultra-vide et de très basses températures pour les observations expérimentales
entraîne un réel intérêt pour les calculs \emph{ab initio} afin d'en
savoir plus sur les propriétés structurales, électroniques, magnétiques
et énergétiques de ces surfaces.

\subsubsection*{Fe}

Selon Tyson et Miller \cite{tyson_surface_tensions_from_liquids_1977},
l'énergie nécessaire à la création d'une unité de surface de Fe (110)
se situerait entre 2.417 et 2.476\,J. Les propriétés structurales
des surfaces pures de Fe de bas indices de Miller ont été systématiquement
étudiées par Sokolov, Jona, Shih et Marcus \cite{legg_Fe100_1977,shih_LEEDFe110_1980,shih_LEEDFe111_1981,sokolov_Fe210_1985,sokolov_Fe_surf_trends_1984,sokolov_LEEDFe111_1986,sokolov_LEEDFe211_1984}.
Elles se comportent en accord avec les théories de Jona, Landman,
Barnett et Cleveland \cite{jona__LEED_cristallography_revue_1978,landman_surf_relax_1980,barnett_1_relaxations_highMillerindex_1983,barnett_2_surfrelax_1983,barnett_3_surfrelax_1983}
: contraction de la première ou des deux premières distances interréticulaires,
puis dilatation. S'en suivent des oscillations qui s'atténuent rapidement
en s'éloignant de la surface.

Les calculs \emph{ab initio} de Blonski et al. montrent que la surface
de Fe la plus facile à créer à $0$\,K (ayant la plus basse énergie
de surface) a l'orientation $\left(110\right)$ \cite{blonski_Fe_surfaces_2007}.
Suivent les orientations $\left(100\right)$, $\left(211\right)$
puis $\left(111\right)$. Les auteurs observent que le moment magnétique
atomique de Fe augmente en surface. En particulier, dans la surface
$\left(100\right)$, il augmente de 2.22 à 2.95\,$\mu_{B}$ ($+33$\,\%).
La surface $\left(110\right)$ est moins sensible aux variations de
moment magnétique, avec un moment magnétique local de 2.59\,$\mu_{B}$
($+17$\,\%).

\subsubsection*{Cr}

Les propriétés de surface de Cr sont mal connues. Les résultats expérimentaux
\cite{davies_STMFeCr100_1996,fawcett_Crpur_1988,klebanoff_Cr100_1984,klebanoff_Cr100bis_1985,klebanoff_Cr100ter_1985,zabel_Crmag_1999}
et théoriques \cite{bihlmayer_Cr100V100_2000,blugel_Cr100_ferro_or_antiferro_1989,eichler_Cr100_ads_2000,kolesnychenko_STM_DFT_kondo_Cr100_2005}
se limitent à l'orientation $\left(100\right)$ du fait de son intérêt
pour la magnéto-résistance géante \cite{Fert__GMR_1988} justement
due à la direction de propagation $\left[100\right]$ de l'onde de
densité de spin (voir paragraphe \ref{par:Cr_SDW}) perpendiculairement
à une interface Fe$\left(100\right)$/Cr$\left(100\right)$. Les calculs
théoriques montrent que Cr est très modifié structuralement et magnétiquement
par les surfaces libres. Par exemple, le moment magnétique passe de
0.6\,$\mu_{B}$ en volume à presque $3$\,$\mu_{B}$ ($+400$\,\%)
dans la surface (100) \cite{zabel_Crmag_1999,fawcett_Crpur_1988}.
Structuralement, la contraction des premières couches est également
notable ($>5$\,\%) \cite{eichler_Cr100_ads_2000,kolesnychenko_STM_DFT_kondo_Cr100_2005}.

\subsection{Surfaces libres de l'alliage et ségrégation à température nulle}

L'énergie de ségrégation d'un atome de chrome dans une surface de
fer est l'énergie nécessaire à l'échange d'un atome de Cr en volume
avec un atome de Fe dans la surface.

Ponomareva et al \cite{ponomareva_HsegFe100_2007} calculent en EMTO-CPA
et DFT que Cr ne devrait pas ségréger, ou de façon très limitée, dans
Fe à 0\,K. Ces auteurs montrent également que l'énergie de ségrégation
est fortement corrélée à la concentration en chrome $x_{Cr}$, comme
l'est l'énergie de mélange dont nous discutions à la section précédente.%
{} D'autres auteurs trouvent différentes énergies de ségrégation de
Cr dans Fe, de différents signes, sans toutefois donner d'explication
physique \cite{blonski_Fe_surfaces_2007,kiejna_Eseg_2008,ossowski_Cr_surfaces_2008}.
Ruban et al. \cite{ruban_surf_seg_TMalloys_1999} calculent en LMTO-LDA
une énergie de ségrégation très positive ($\unit[+0.16]{eV}$), c'est-à-dire
non-favorable. Nonas et al. \cite{nonas_impuretes_dans_Fe100_1998}
calculent en FP-KKR dans l'approximation de la densité locale (LDA)
une énergie de ségrégation très légèrement négative à $\unit[0]{K}$
($\unit[-0.03]{eV}$). Cela irait dans le sens de la ségrégation de
Cr en surface, mais reste trop faible en comparaison des phénomènes
entropiques dès quelques Kelvin. Geng \cite{geng_segGGAFLAPW_2003}
est le premier à faire des calculs en GGA tous électrons et montre
le rôle important de la fonctionnelle d'échange et corrélation. L'énergie
de ségrégation calculée est très légèrement négative ($\unit[-0.05]{eV}$),
mais le calcul n'est pas explicité dans l'article. Il montre également
qu'à 0\,K, la répulsion entre éléments de même espèce en volume comme
en surface, dans Fe comme dans Cr, induirait la ségrégation vers la
surface de Fe dans Cr $\left(100\right)$ mais pas l'inverse. Il est
cependant important de noter que tous les calculs \emph{ab initio}
de la littérature se font dans de petites super-cellules (14 à 54
atomes) et qu'une extrapolation aux systèmes plus dilués nous parait
injustifiée considérant les propriétés de l'alliage. Récemment, Ponomareva
et al. \cite{ponomareva_HsegFe100_2007} montrent par des calculs
en PAW GGA que l'énergie de ségrégation calculée est faible, positive
ou négative, et très dépendante de la taille de la super-cellule utilisée
(voir tableau \ref{tab:Hseg_impureteCr}). Kiejna et Wachowicz \cite{kiejna_Eseg_2008}
refont ces calculs en PAW-GGA mais se limitent également aux systèmes
de petites tailles (très concentrés). Les super-cellules utilisées
contiennent 1 à 2 atomes par plan en moyenne. Ainsi, ce que les auteurs
appellent énergie de ségrégation d'une impureté est en fait l'énergie
de ségrégation d'une (quasi-)monocouche de Cr du volume vers la surface.
Un exemple probant est le moment magnétique calculé par Kiejna et
Wachowicz pour Cr dans la surface pour quatre orientations différentes
\cite{kiejna_Eseg_2008}. Cr aurait son moment magnétique local maximum
dans le plan de surface dans l'orientation $\left(210\right)$, et
minimum dans l'orientation $\left(100\right)$. Or, la surface $\left(210\right)$
est la seule pour laquelle il ait utilisé une cellule plus grande
(6 atomes par plan). Il libère ainsi un degré de liberté magnétique
et l'amplitude augmente en conséquence. En fait, il observe ici un
effet de taille et non un effet de ségrégation d'impureté. C'est aussi
la raison pour laquelle il calcule une énergie de ségrégation quasi-nulle
de Cr dans Fe $\left(100\right)$.

\subsection{Les surfaces libres étudiées : $\left(100\right)$, $\left(110\right)$,
$\left(111\right)$ et $\left(211\right)$}

\begin{figure}[h]
\noindent \begin{centering}
\includegraphics[scale=0.12]{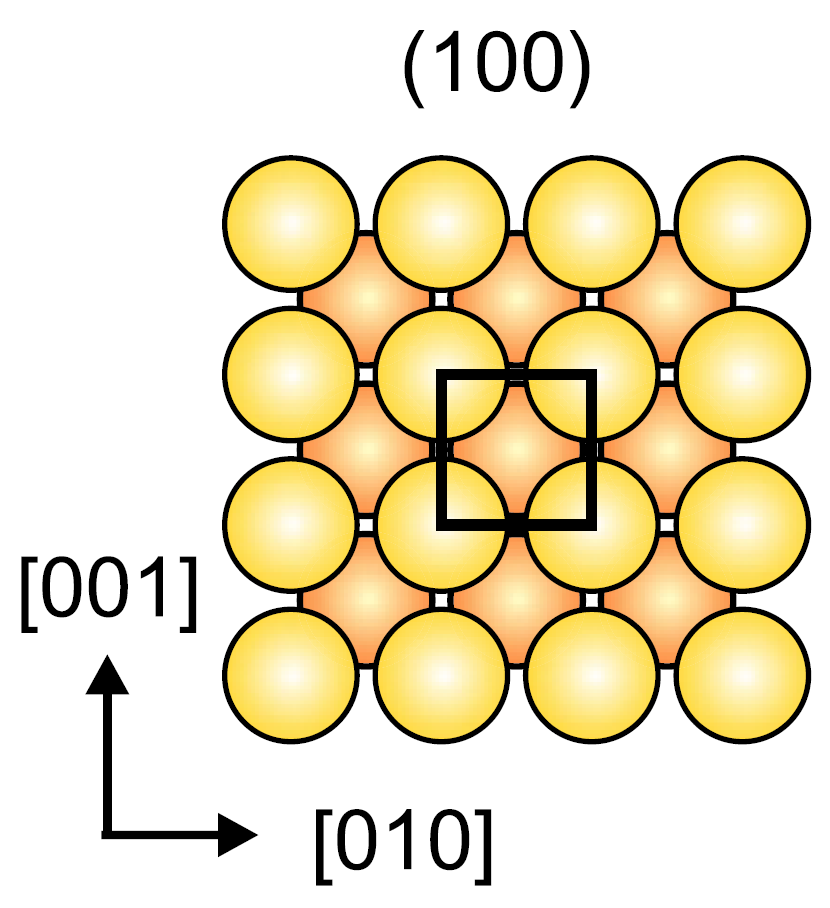}\includegraphics[scale=0.12]{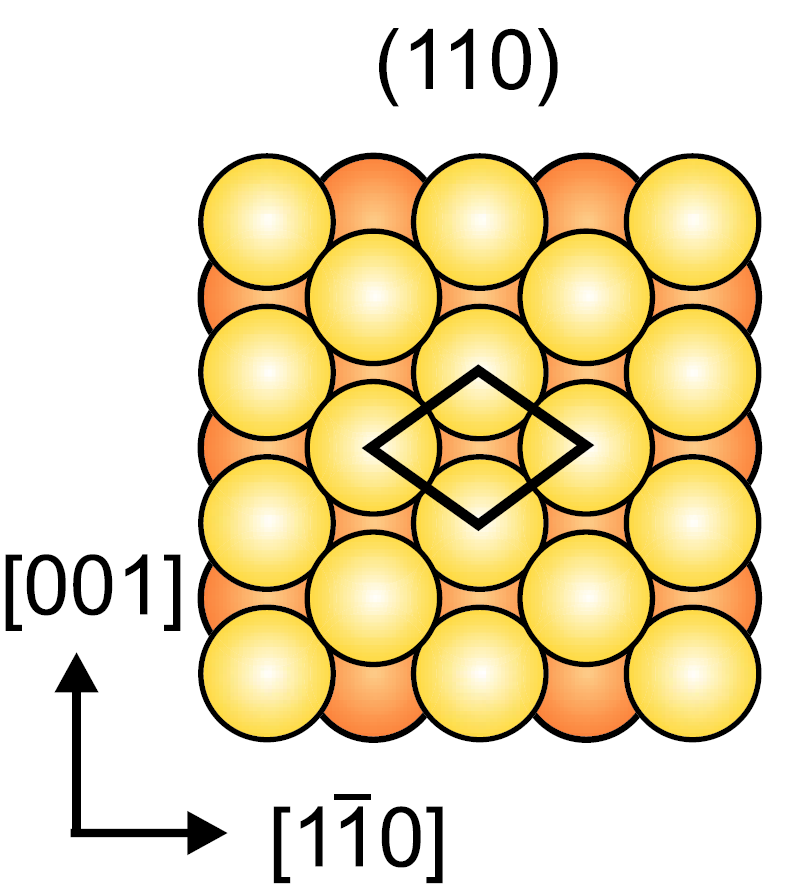}\\

\par\end{centering}

\noindent \begin{centering}
\includegraphics[scale=0.15]{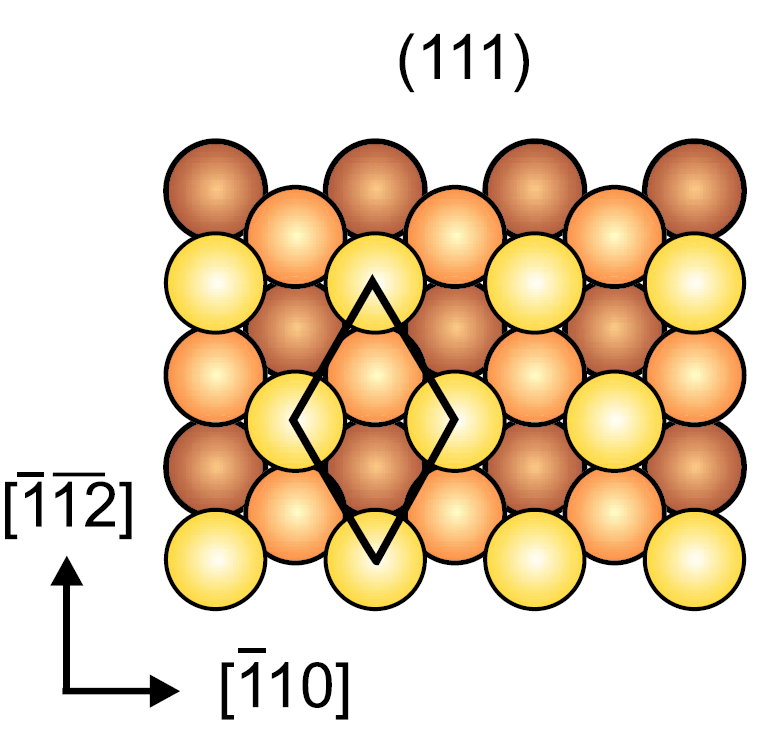}\includegraphics[scale=0.12]{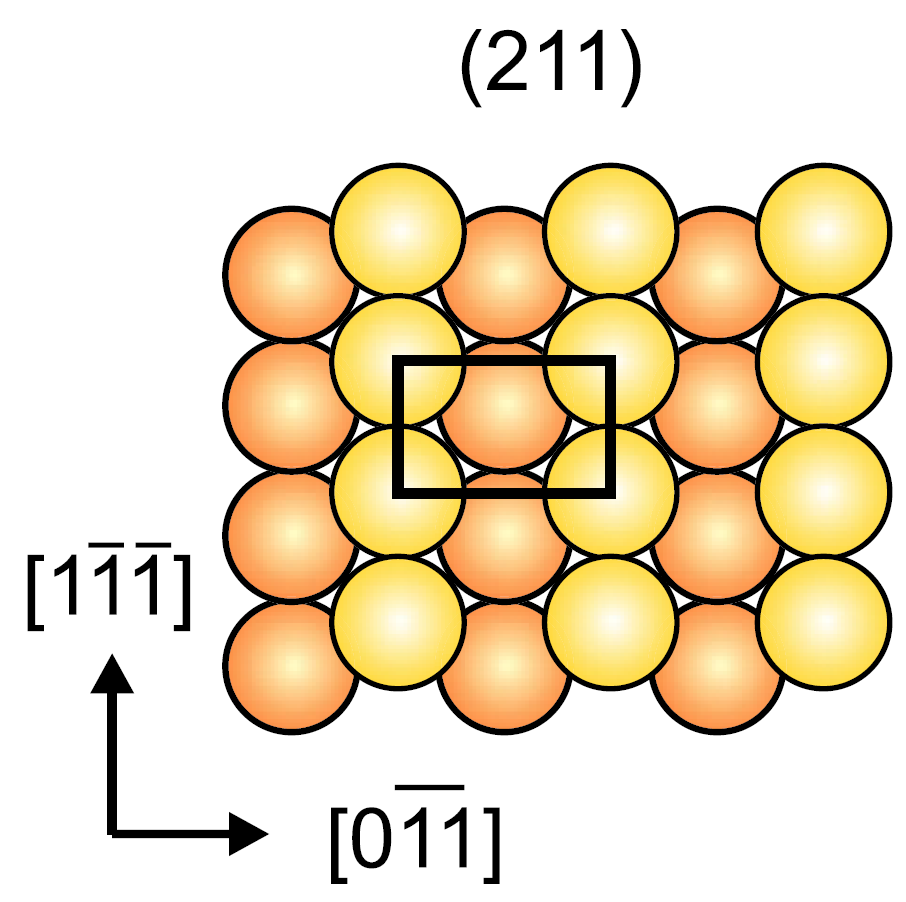}
\par\end{centering}

\caption{Vue du plan supérieur des surfaces d'orientation cristallographique
$\left(100\right)$, $\left(110\right)$, $\left(111\right)$ et $\left(211\right)$.
Figures de Blonski et Kiejna \cite{blonski_Fe_surfaces_2007}.\label{fig:cristallo-surfaces}}

\end{figure}

Dans cette étude, nous nous intéressons à quatre orientations de surface
: $\left(100\right)$, $\left(110\right)$, $\left(111\right)$ et
$\left(211\right)$. Ces surfaces libres de faibles indices de Miller
sont représentées sur la figure \ref{fig:cristallo-surfaces}.

Afin de comparer des orientations entre elles, il est commode de définir
la densité de surface $\rho$, le nombre d'atomes par unité de surface.
\begin{equation}
\rho\left(hkl\right)=\dfrac{\mbox{nombre d'atomes dans la surface d'orientation \ensuremath{\left(hkl\right)}}}{\mbox{aire de la surface}}.\label{eq:definition_densite_de_surface}
\end{equation}

Comme l'indique le tableau \ref{tab:Densit=0000E9-des-surfaces} par
ordre croissant de densité, les quatre orientations choisies sont
à la fois d'indice de Miller bas et de densités de surfaces représentatives
: de la plus faible (111) à la plus forte (110) des densités possibles
dans un cristal cubique centré.

\begin{table}[H]
\begin{centering}
\begin{tabular}{|c|c|c|c|}
\hline 
 & $\rho$ ($a^{-2}$) & $\Delta Z^{\left(1\right)}$ & $\Delta Z^{\left(2\right)}$ \\
\hline 
\hline 
$\left(111\right)$ & 0.577 & 4 & 3 \\
\hline 
$\left(211\right)$ & 0.8165 & 3 & 3 \\
\hline 
$\left(100\right)$ & 1 & 4 & 2 \\
\hline 
$\left(110\right)$ & 1.4142 & 2 & 1 \\
\hline 
\end{tabular}
\par\end{centering}

\caption{Densité de surface $\rho$ des orientations $\left(100\right)$, $\left(110\right)$,
$\left(111\right)$ et $\left(211\right)$ d'un réseau cubique centré.
L'unité d'aire est exprimée en coordonnées réduites afin de pouvoir
comparer Fe et Cr de paramètres de maille différents. $\Delta Z^{\left(1\right)}$
et $\Delta Z^{\left(2\right)}$ sont les nombres de liaisons aux premiers
et deuxièmes voisins coupées.\label{tab:Densit=0000E9-des-surfaces}}
\end{table}

Par ordre de densité de surface croissante, on a donc :
\[
\rho\left(111\right)<\rho\left(211\right)<\rho\left(100\right)<\rho\left(110\right).
\]

\subsection{Modèle d'empilement pour les calculs \emph{ab initio }\label{sub:Comment-mod=0000E9liser-des_surf_ab_initio}}

\subsubsection*{Par un empilement de couches minces séparées par du vide}

En conditions tri-périodiques, les surfaces libres $\left(hkl\right)$
sont modélisées par un empilement de $N$ plans atomiques d'orientation
$\left(hkl\right)$, et d'une couche de vide d'épaisseur $L$ selon
$\left\langle hkl\right\rangle $ \cite{levesque_uranyl_2008}.

\subsubsection*{Combien de plans empiler ?}

Le nombre de plans atomiques empilés $N$ doit être assez grand pour
que le plan central soit équivalent (structuralement, électroniquement
et magnétiquement) à un plan de volume, ce qui revient à séparer les
deux surfaces d'assez de couches atomiques pour qu'elles n'interagissent
pas entre elles. Deux stratégies sont possibles :
\begin{itemize}
\item augmenter le nombre de couches atomiques $N$ par pas jusqu'à convergence
des propriétés de la couche centrale.
\item fixer la couche centrale dans les positions de volume et augmenter
pas à pas $N$.
\end{itemize}
Cette dernière solution a l'avantage de limiter $N$, mais nous observons
avec ce dernier schéma une convergence du moment magnétique local
à des valeurs différentes de celles en volume, sauf pour $N$ aussi
grand que dans le premier schéma de convergence. C'est donc ce premier
schéma que nous utilisons dans la suite du document. 

Nous trouvons $N=14$ pour converger pour les quatre orientations.

C'est un nombre élevé, qui s'explique d'une part par les surfaces
très ouvertes que nous étudions (les relaxations structurales se propageant
plus loin vers le volume), et d'autre part par les interactions magnétiques
(il faut un certain nombre de couches pour qu'un atome de la couche
centrale ne ressente plus le très grand moment magnétique d'un atome
de surface).

\subsubsection*{Symétries selon $xy$ et concentrations surfaciques}

Nos tests montrent que $14$ couches sont nécessaires pour que les
atomes du plan central aient les propriétés d'atomes de volume. Or,
les puissances de calcul disponibles sont adaptées à l'étude de quelques
centaines d'atomes seulement. On ne peut donc considérer que quelques
atomes par plan.

Par exemple, une super-cellule $2\times2\times14$ contient déjà $56$
atomes si la maille élémentaire ne contient qu'un atome. La concentration
volumique est donc au minimum de $1$ pour $56$, soit $1.8$\,\%.
Par contre, dans le plan, la concentration minimum est déjà de $1$
pour $4$, soit $25$\,\%.

Du fait des puissances de calcul disponibles, une super-cellule de
$14$ couches atomiques ne peut contenir qu'un maximum d'une vingtaine
d'atomes par couche. Ainsi, le code SIESTA-NC permet, grâce à sa base
localisée et aux pseudo-potentiels à norme conservée, de traiter assez
aisément jusqu'à 18 atomes par couches pour l'orientation $\left(110\right)$.
Cela représente $266$\,atomes par super-cellule, ce qui est très
important pour des études systématiques nécessitant des centaines
de calculs. L'efficacité des calculs en base d'ondes localisées permet
d'envisager une étude systématique et étayée des effets de concentration,
c'est-à-dire de considérer l'ensemble des positions possibles, et
un grand nombre de concentrations. Les codes en base d'ondes planes,
moins efficaces, ne permettent pas d'envisager des études aussi systématiques.
Nous montrons dans la suite du document que cela biaise parfois l'interprétation
physique des résultats quant à la concentration surfacique d'impuretés
(parfois de 100\,\% dans certaines cellules élémentaires représentées
en noir dans la figure \ref{fig:cristallo-surfaces}).

Si la concentration surfacique est $x\left(p\right)=1$ :
\begin{enumerate}
\item Du fait des conditions périodiques, le degré de liberté de la structure
électronique en surface (magnétisme, transfert de densité de charge)
est limité.
\item Nous verrons dans la suite du document que la concentration surfacique
$x_{Cr}\left(p\right)$ est un degré de liberté aux conséquences physiques
importantes.
\end{enumerate}

\section{Notre étude \emph{ab initio} des surfaces libres des éléments purs}

\subsection{Énergies de surface\label{sub:Energies-de-surface}}

L'énergie de surface $\gamma_{i}^{\left(hkl\right)}$ est le travail
à fournir au système pour former une unité de surface d'orientation
$\left(hkl\right)$ de l'élément $i$ (Fe ou Cr dans notre étude).
C'est en conséquence une énergie par unité de surface. Connaissant
le nombre d'atomes par unité de surface ($\rho\left(hkl\right)$,
voir équation \ref{eq:definition_densite_de_surface} et tableau \ref{tab:Densit=0000E9-des-surfaces}),
c'est aussi une énergie par atome, ce qui peut être une écriture avantageuse
dans un modèle atomistique. Pour calculer l'énergie de surface $\gamma_{i}^{\left(hkl\right)}$
de l'élément $i$ dans l'orientation $\left(hkl\right)$ en DFT, nous
utilisons la relation suivante : 
\begin{equation}
\gamma_{i}^{\left(hkl\right)}=\dfrac{1}{2}\left(E_{slab,i}^{\left(hkl\right)}-NE\left(i\right)\right)\label{eq:def_Esurf}
\end{equation}
où $E_{slab,i}^{\left(hkl\right)}$ est l'énergie totale d'une super-cellule
de $N$ atomes $i$ contenant deux surfaces $\left(hkl\right)$ de
l'élément $i$, et $E\left(i\right)$ est l'énergie par atome de l'élément
$i$ en volume calculée dans son état de référence, tel que défini
dans le paragraphe \ref{sub:def_mu}. Pour Fe et Cr, il s'agit des
phases cubiques centrées magnétiques à leur paramètre de maille d'équilibre.

Dans un modèle de liaisons coupées tel que le modèle d'Ising, l'énergie
de surface est proportionnelle au nombre de liaisons coupées et à
l'énergie de cohésion. Les énergies de cohésion de Fe (4.28\,eV/atome
selon \cite{kittel_introduction_1958}) et de Cr (4.10\,eV/atome
selon \cite{kittel_introduction_1958}) diffèrent de moins de $\unit[5]{\%}$.
Ainsi, selon ce modèle, les énergies de surface de Fe seraient $\approx5$\,\%
supérieures à celles de Cr.

\begin{figure}[h]
\begin{centering}
\subfloat[]{\begin{centering}
\includegraphics[scale=0.4]{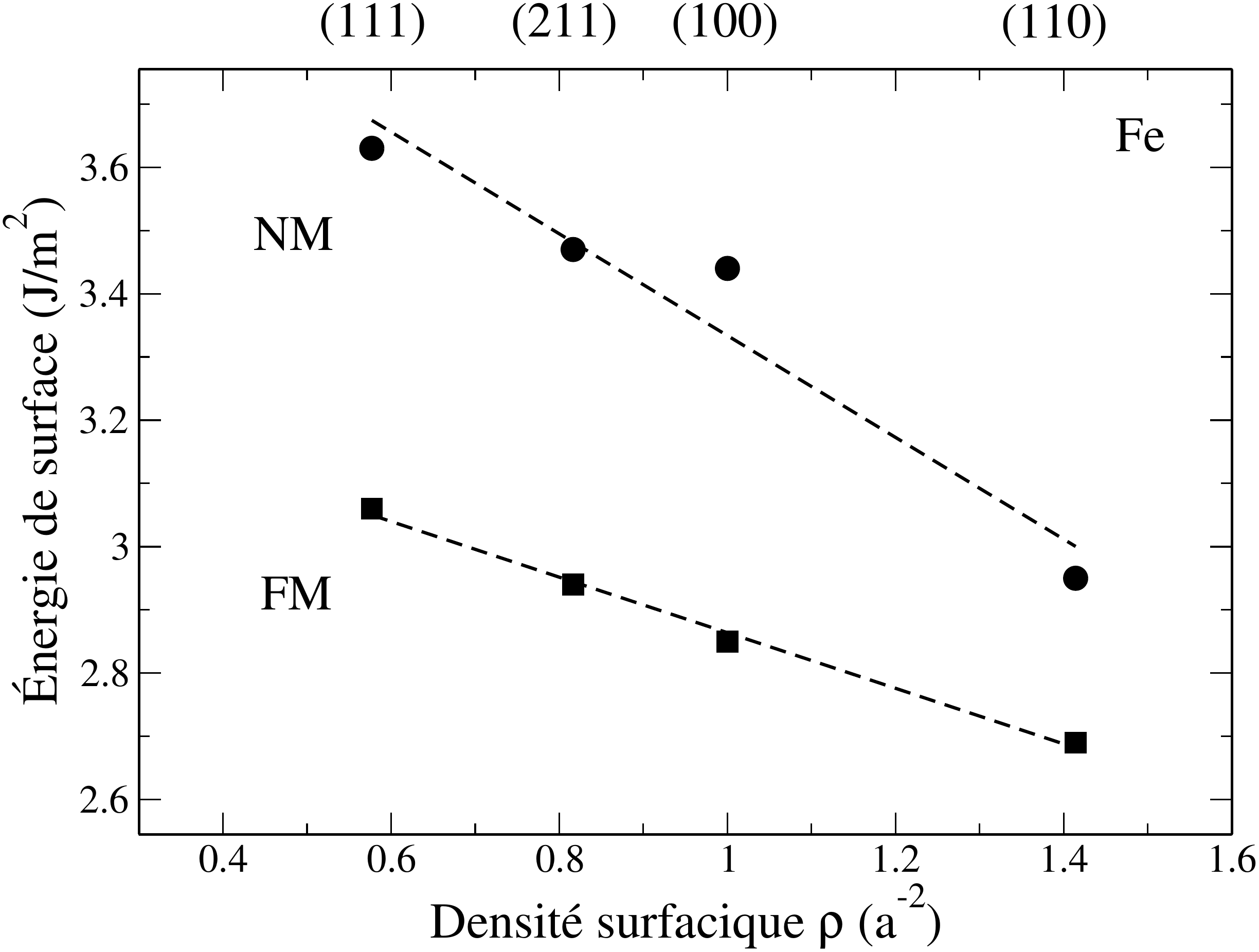}
\par\end{centering}

}
\par\end{centering}

\begin{centering}
\subfloat[]{\begin{centering}
\includegraphics[scale=0.4]{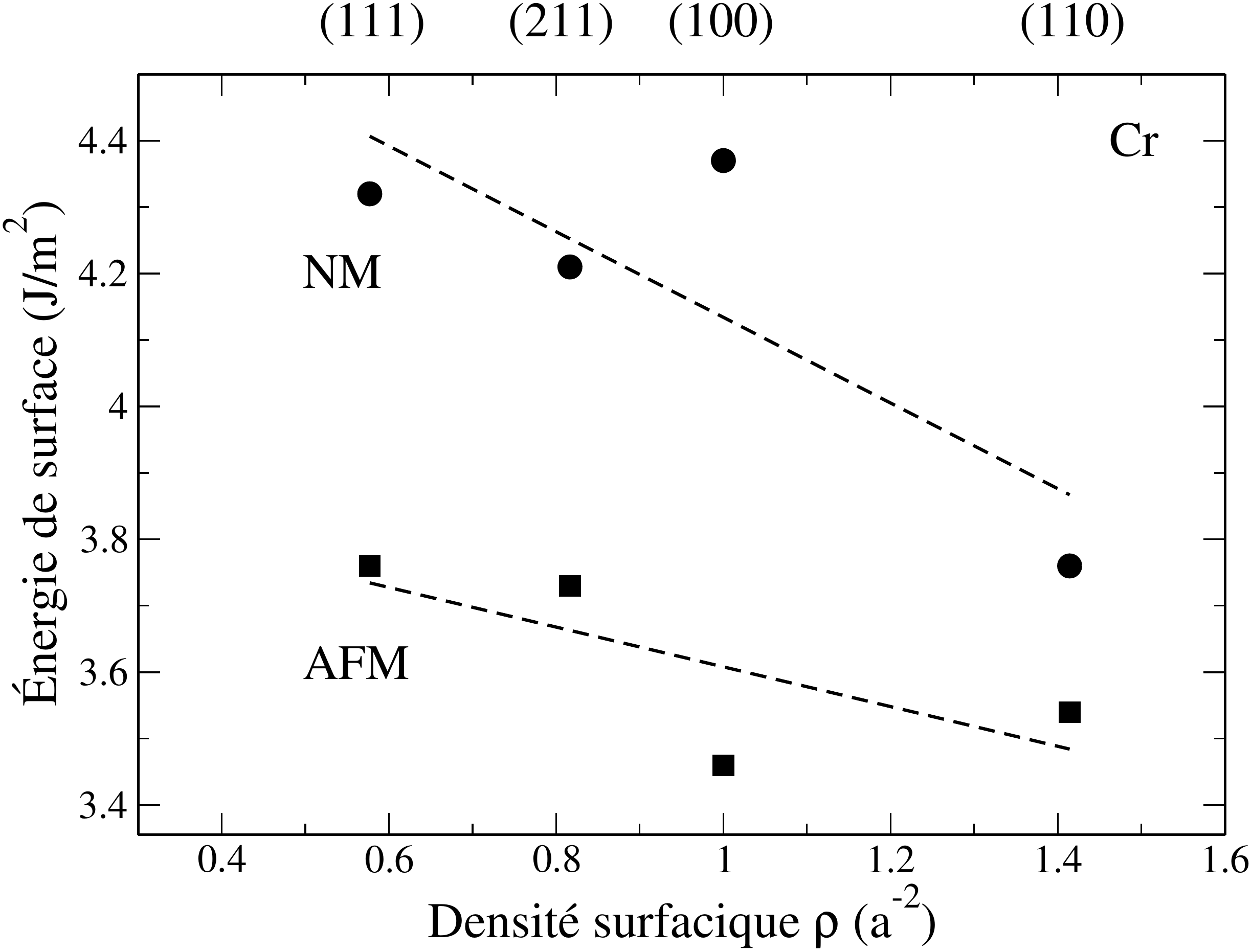}
\par\end{centering}

}
\par\end{centering}

\caption{Énergies des surfaces libres de Fe (en haut) et Cr (en bas) calculées
par SIESTA-NC avec (ronds) et sans (carrés) magnétisme pour des super-cellules
contenant $14$ plans de $9$ à $18$ atomes et un vide de $15$\,Å
en fonction de la densité de surface (se référer au tableau \ref{tab:Densit=0000E9-des-surfaces}
pour les relations entre densité surfacique et orientation de la surface
: $\rho_{111}<\rho_{211}<\rho_{100}<\rho_{110}$).\label{fig:=0000C9nergie-de-surface}}
\end{figure}

Les énergies de surface calculées dans les cas magnétiques et non-magnétiques
avec SIESTA-NC pour des super-cellules contenant 14 plans de 9 à 18
atomes par plan et une couche de vide d'épaisseur $15$\,Å sont représentées
sur la figure \ref{fig:=0000C9nergie-de-surface} en fonction de la
densité de surface (se référer au tableau \ref{tab:Densit=0000E9-des-surfaces}
pour le lien entre la densité surfacique $\rho$ de chaque orientation
de surface).

Pour Fe comme pour Cr, on observe sur la figure \ref{fig:=0000C9nergie-de-surface}
que la prise en compte du magnétisme dans les calculs abaisse l'énergie
de surface. On attend donc des atomes de surface à moment magnétique
local non-nul, en accord avec les mesures expérimentales présentées
au paragraphe \ref{sub:etat_de_lart_surf_DFT}.

On observe également sur cette figure que $\gamma_{Fe}$ est compris
entre 2.7 et 3.7\,J/m$^{\text{2}}$ quand $\gamma_{Cr}$ est compris
entre 3.5 et $4.4$\,J/m$^{\text{2}}$. Les énergies de surface de
Cr sont donc supérieures à celles de Fe d'environ 0.7\,J/m$^{\text{2}}$
:
\begin{equation}
\gamma_{Cr}>\gamma_{Fe},\forall h,k,l.\label{eq:gamma_Cr_sup_gamme_Fe}
\end{equation}

Ces calculs sont en accord qualitatif avec les premiers résultats
\emph{ab initio} systématiques sur les surfaces de métaux publiés
par Vitos et al. \cite{Vitos_surfscience}. Ces derniers calculent
en GGA des énergies de suface de fer comprises entre 2.2 et 2.8\,J/m\texttwosuperior{}
et de chrome entre 3.5 et 4.1 J/m\texttwosuperior{}. Cette observation
(eq. \ref{eq:gamma_Cr_sup_gamme_Fe}) est en contradiction avec ce
qu'on pourrait attendre dans le cadre théorique d'un modèle de liaisons
coupées. Nous verrons dans les deux chapitres suivant qu'un tel modèle
ne peut reprodruire les effets structuraux, électroniques et surtout
magnétiques des surfaces de Fe, Cr et de l'alliage FeCr.

Les énergies de surface de Fe varient quasi-linéairement avec la densité
surfacique, en accord avec les résultats très récents de Blonski et
al. \cite{blonski_Fe_surfaces_2007}. Pour Fe et Cr, l'orientation
$\left(100\right)$ a une énergie de surface en dessous de la courbe
de tendance linéaire avec la densité surfacique. Cet effet est plus
marqué pour Cr : environ $\unit[0.2]{J/m^{2}}$ en dessous de la courbe
de tendance. 

Comment expliquer cette singularité Cr$\left(100\right)$ ? Lorsqu'on
compare les calculs non-magnétiques et les magnétiques pour Fe et
Cr $\left(100\right)$ représentés sur la figure \ref{fig:=0000C9nergie-de-surface},
on observe la tendance inverse : l'énergie de surface est de nouveau
hors de la courbe de tendance, mais cette fois-ci au dessus.

Pour Cr comme pour Fe, les surfaces non-magnétiques coûtent entre
0.4 et $\unit[0.6]{J/m^{2}}$ de plus que les surfaces magnétiques.
Le magnétisme stabilise donc de façon importante les surfaces libres
des deux éléments. Cela invalide les extrapolations depuis les mesures
à haute température des tensions de surface dans la phase liquide
pour le fer et le chrome de Tyson et Miller \cite{tyson_surface_tensions_from_liquids_1977}.

Pour illustrer ce rôle important du magnétisme, notons que la différence
d'énergie entre les phases AF et NM de Cr cc est $\unit[0.028]{eV/atome}$,
alors que la différence entre les énergies de surfaces de Cr magnétiques
et NM est de $0.5$ à $\unit[1.5]{eV/atome}$, selon l'orientation.
Cependant, contrairement à Cr, le magnétisme stabilise significativement
Fe en volume de $\unit[0.555]{eV/atome}$, ce qui est du même ordre
de grandeur que la stabilisation en surface.

\subsection{Magnétisme en surface}

\begin{figure}[h]
\begin{centering}
\subfloat[Fe]{\begin{centering}
\includegraphics[scale=0.4]{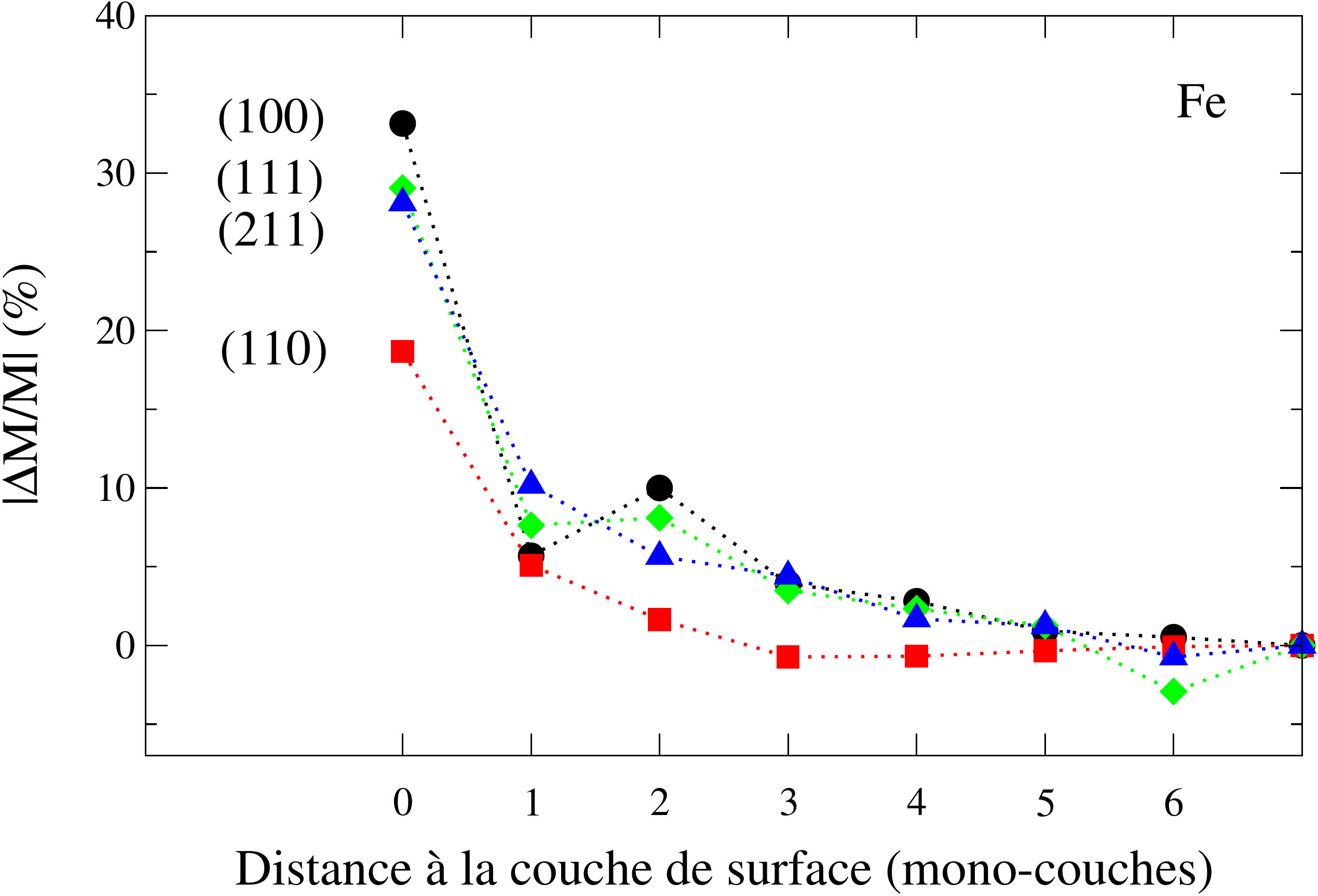}
\par\end{centering}

}
\par\end{centering}

\begin{centering}
\subfloat[Cr]{\centering{}\includegraphics[scale=0.4]{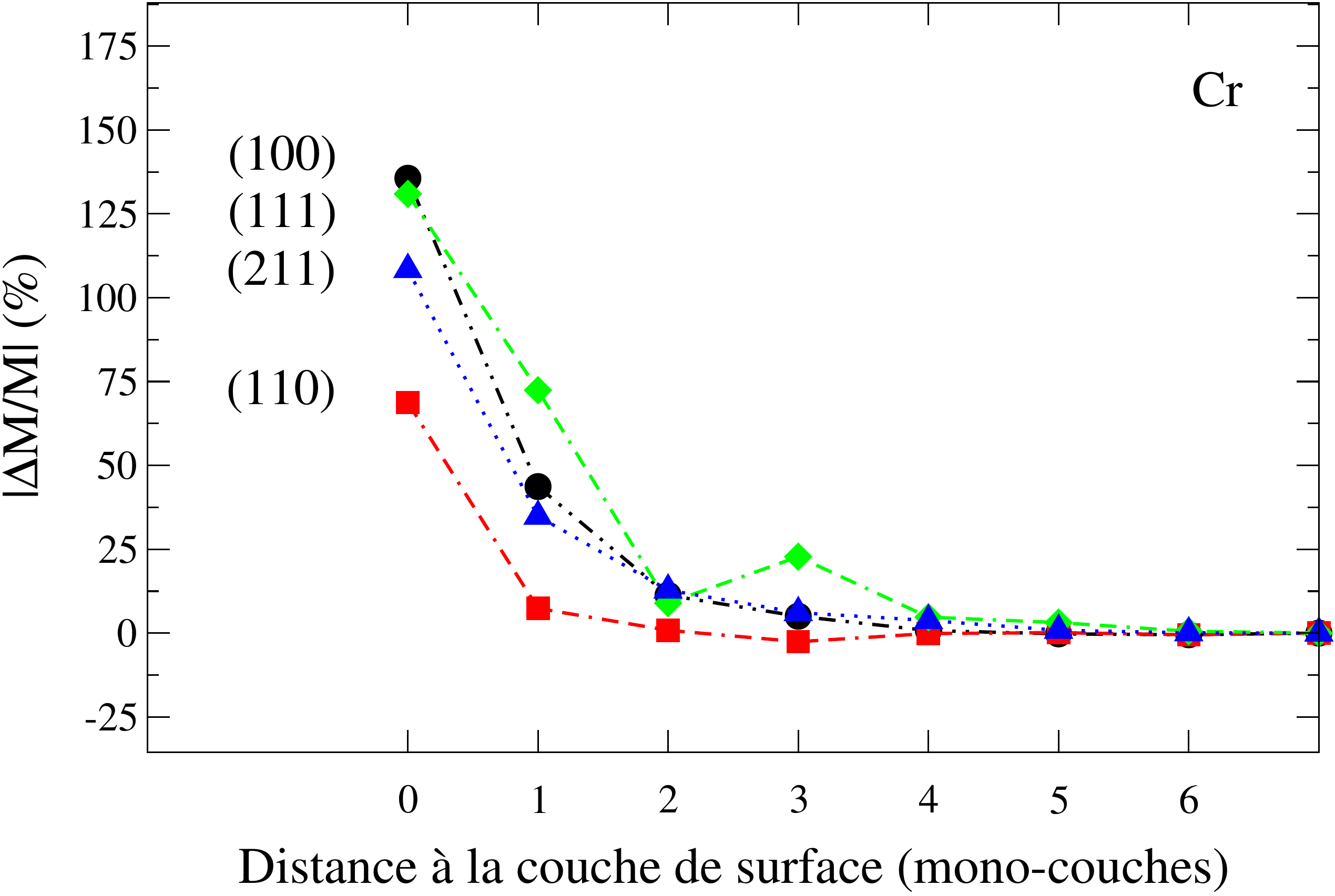}}
\par\end{centering}

\caption{Variation du moment magnétique local en fonction de la distance séparant
l'atome de la surface pour Fe et Cr. Calculs SIESTA-NC.\label{fig:Variation-du-moment_en_surface_pur}}

\end{figure}

Afin de comprendre ce fort effet du magnétisme sur la stabilisation
des surfaces, on regarde figure (\ref{fig:Variation-du-moment_en_surface_pur})
le moment magnétique porté par chaque atome de Fe ou de Cr en fonction
du plan dans lequel il se trouve. Lorsque l'atome est dans le plan
d'indice 7, c'est-à-dire au milieu de la couche mince, le moment magnétique
local est celui de l'atome de volume : l'augmentation de moment magnétique
local $\Delta M/M$ est nul. Lorsqu'on observe le moment magnétique
des atomes des plans plus proches de la surface, ceux-ci augmentent
: les atomes de surface sont moins coordinés, ce qui donne lieu au
rétrécissement de la bande $d$ et induit une augmentation du moment
magnétique local. Cette augmentation n'est cependant pas dans les
mêmes rapports pour Fe et Cr : elle est bien plus grande pour Cr.
Dans Fe, l'augmentation relative par rapport au volume $\Delta M/M$
est de 19 à 32\,\% selon l'orientation. Elle est de 71 à 136\,\%
dans Cr.

Pour les deux éléments, l'augmentation en surface est largement dépendante
de l'orientation. En dehors de l'orientation $\left(100\right)$,
plus la surface est dense, plus l'augmentation est faible. Cela est
également dû à l'augmentation de l'interaction $d$--$d$ et à l'élargissement
de la bande $d$ dans les systèmes plus denses. L'orientation $\left(100\right)$
ne vérifie pourtant pas cette affirmation. Alors que sa densité $\rho_{100}$
est intermédiaire entre $\rho_{211}$ et $\rho_{110}$, c'est dans
cette orientation que l'augmentation relative du moment magnétique
local en surface est la plus grande.

Qu'est-ce qui rend l'orientation $\left(100\right)$ si particulière
?

\subsection{Singularités de l'orientation $\left(100\right)$\label{sub:La-singuli=0000E8re-orientation100}}

Dans un réseau cubique centré, un atome dans un plan de surface $\left(100\right)$
a 4 voisins dans le plan sous-jacent (plan 1 sur la figure \ref{fig:Variation-du-moment_en_surface_pur}).
Cet atome n'a aucun premier voisin dans son plan (0 sur la figure
\ref{fig:Variation-du-moment_en_surface_pur}). Or, c'est avec les
premiers voisins que se fait le couplage antiferromagnétique.

Des calculs de liaisons fortes auto-cohérents ont prédit \cite{allan_CrsurfANTIFERRO_1978,grempel__surf_ordering_Cr_1981}
qu'un élément antiferromagnétique tel que Cr pourrait présenter, pour
cette orientation, un ferromagnétisme dans le plan, ainsi qu'une nette
augmentation du moment magnétique dans la surface. Ces prédictions
théoriques ont été confirmées expérimentalement \cite{klebanoff_Cr100_1984,klebanoff_Cr100bis_1985,klebanoff_Cr100ter_1985}
par spectroscopie ARPES (Angle-resolved photoelectron spectroscopy),
puis recalculées \emph{ab initio}. Ces résultats montrent une augmentation
de moment magnétique de l'ordre de 500 \% en surface \cite{ossowski_Cr_surfaces_2008,kolesnychenko_STM_DFT_kondo_Cr100_2005,blugel_Cr100_ferro_or_antiferro_1989}.

Comment comprendre que le coût énergétique de création d'une surface
de Cr $\left(100\right)$ soit plus faible qu'attendu, mais qu'il
soit étonnamment fort dans le cas non-magnétique ?

\begin{figure}[h]
\begin{centering}
\includegraphics[scale=0.4]{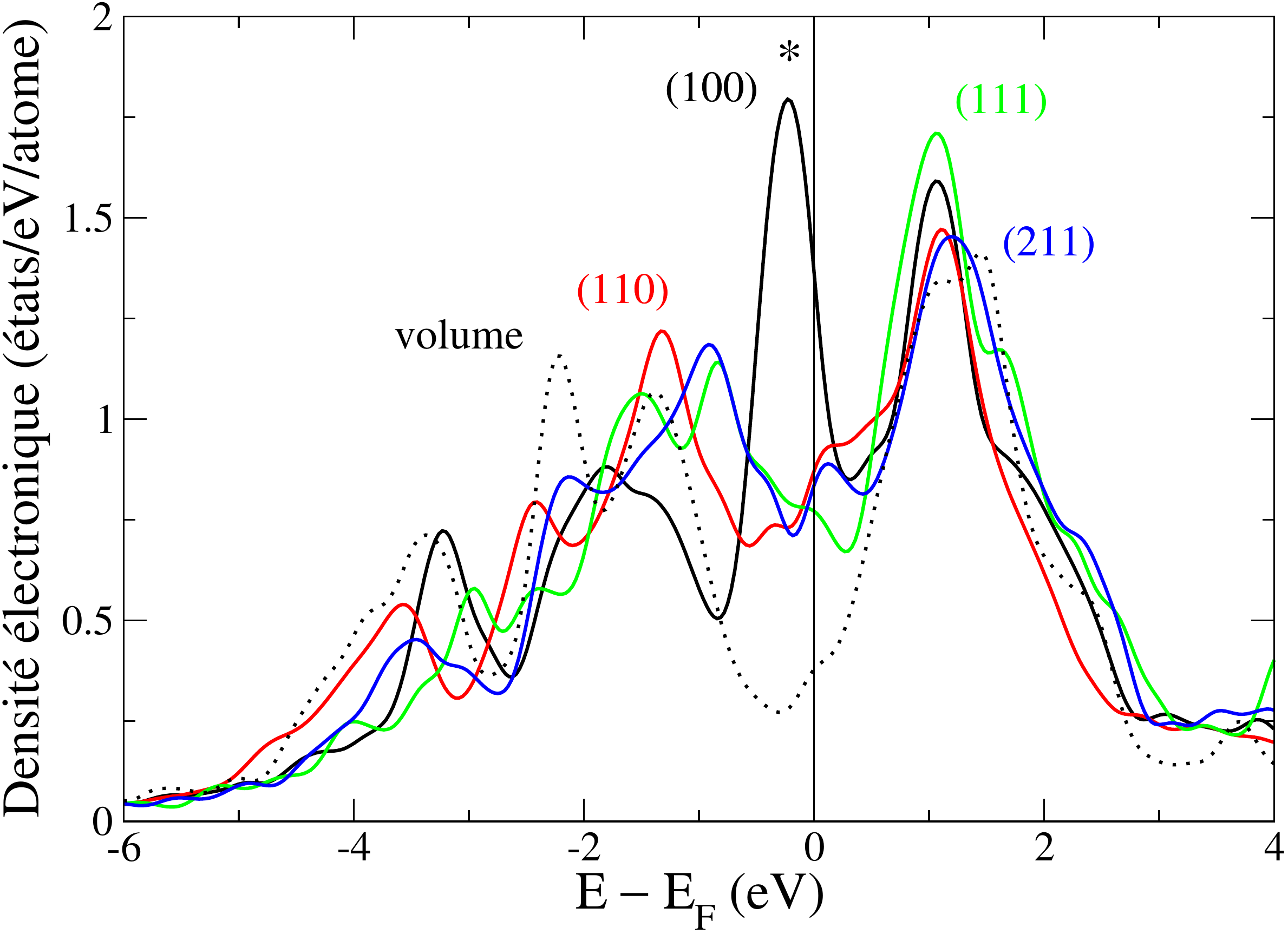}
\par\end{centering}

\caption{PDOS des atomes de volume cubique centré et du plan de surface des
orientations de Cr $\left(110\right)$, $\left(211\right)$, $\left(111\right)$
et $\left(100\right)$ non-magnétiques calculées en SIESTA-NC.\label{fig:DOS_NM_surf_Cr}}
\end{figure}

Sur la figure \ref{fig:DOS_NM_surf_Cr}, on représente la PDOS calculées
en SIESTA-NC d'un atome de Cr non-magnétique de volume cubique centré,
et d'un atome de Cr de surface dans les orientations $\left(100\right)$,
$\left(110\right)$, $\left(111\right)$ et $\left(211\right)$. Bien
que gardant l'allure générale de la PDOS des atomes de volume, les
amplitudes relatives de chacun des pics diffèrent d'une orientation
à l'autre. En particulier, un état de surface apparaît, pour l'orientation
$\left(100\right)$ uniquement, au niveau de Fermi (repéré par une
étoile sur la figure \ref{fig:DOS_NM_surf_Cr}). Cet état de surface
vient de l'apparition d'états non-liants en surface du fait de la
diminution de la coordination discutée plus haut. Il est clair qu'il
s'agit d'un nouvel état et non d'une restructuration par recombinaison
d'états déjà existants car on retrouve de part et d'autre du niveau
de Fermi les états liants et antiliants qui apparaissaient déjà dans
Cr cubique centré. La présence d'un tel état au niveau de Fermi est
très déstabilisant pour le système, qui aimerait pouvoir se réorganiser. 

Dans le molybdène et le tungsten, éléments des couches $4d$ et $5d$
de la même colonne que le chrome dans le tableau périodique, cette
réorganisation se fait structurellement. W$\left(100\right)$ reconstruit
en $\left(\sqrt{2}\times\sqrt{2}\right)R45\text{\textdegree}$ et
Mo$\left(100\right)$ en $c\left(7\sqrt{2}\times\sqrt{2}\right)R45\text{\textdegree}$
\cite{felter_Mo100_W100_1977,debe_W100_1977,daley_Mo100_1993,che_Mo100_1998}.

\begin{figure}[h]
\begin{centering}
\includegraphics[scale=0.4]{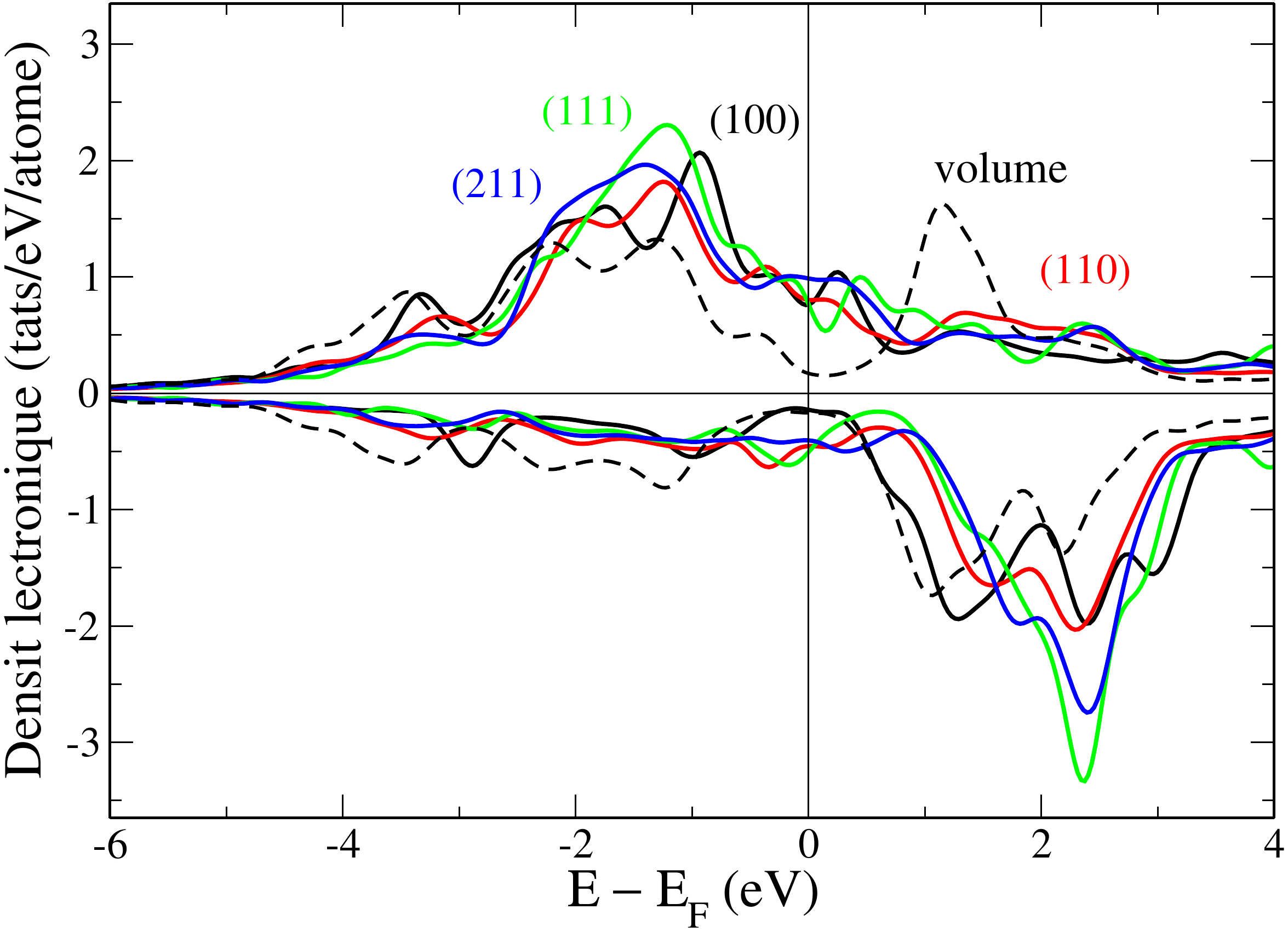}
\par\end{centering}

\caption{PDOS des atomes de volume cubique centré et du plan de surface des
orientations de Cr $\left(110\right)$, $\left(211\right)$, $\left(111\right)$
et $\left(100\right)$ magnétiques calculées en SIESTA-NC.\label{fig:DOS_M_surf_Cr}}

\end{figure}

Le Cr ne fait pas de reconstruction structurelle en surface comme
W ou Mo pour ne pas avoir d'état de surface au niveau de Fermi. Afin
d'éliminer ce pic, la bande se décompose violemment en deux bandes
de part et d'autre du niveau de Fermi, donnant lieu à un moment magnétique
local d'autant plus fort que le la densité au niveau de Fermi était
forte. Les PDOS des atomes de volume cubique centré et du plan de
surface des orientations de Cr $\left(110\right)$, $\left(211\right)$,
$\left(111\right)$ et $\left(100\right)$ calculées avec le magnétisme
en SIESTA-NC sont représentées sur la figure \ref{fig:DOS_M_surf_Cr}. 

On observe sur tous les plans de surface un net transfert de densité
électronique de la bande $\downarrow$ vers la bande $\uparrow$,
donnant lieu, dans toutes les orientations, à un moment magnétique
local augmenté par rapport au volume, comme le montre déjà la figure
\ref{fig:Variation-du-moment_en_surface_pur}. Il n'y a plus d'état
de surface dans l'orientation $\left(100\right)$. La PDOS de cette
orientation montre d'ailleurs un pseudo-gap prononcé au niveau de
Fermi, ce qui la stabilise plus que les autres surfaces.

En résumé : sans magnétisme, la surface $\left(100\right)$ est donc
moins stable que les autres surfaces libres du chrome pur à cause
d'un état de surface au niveau de Fermi. À l'opposé, le dédoublement
de la bande par le magnétisme induit un pseudo-gap au niveau de Fermi
qui stabilise particulièrement le système.

\begin{figure}[h]
\begin{centering}
\includegraphics[scale=0.55]{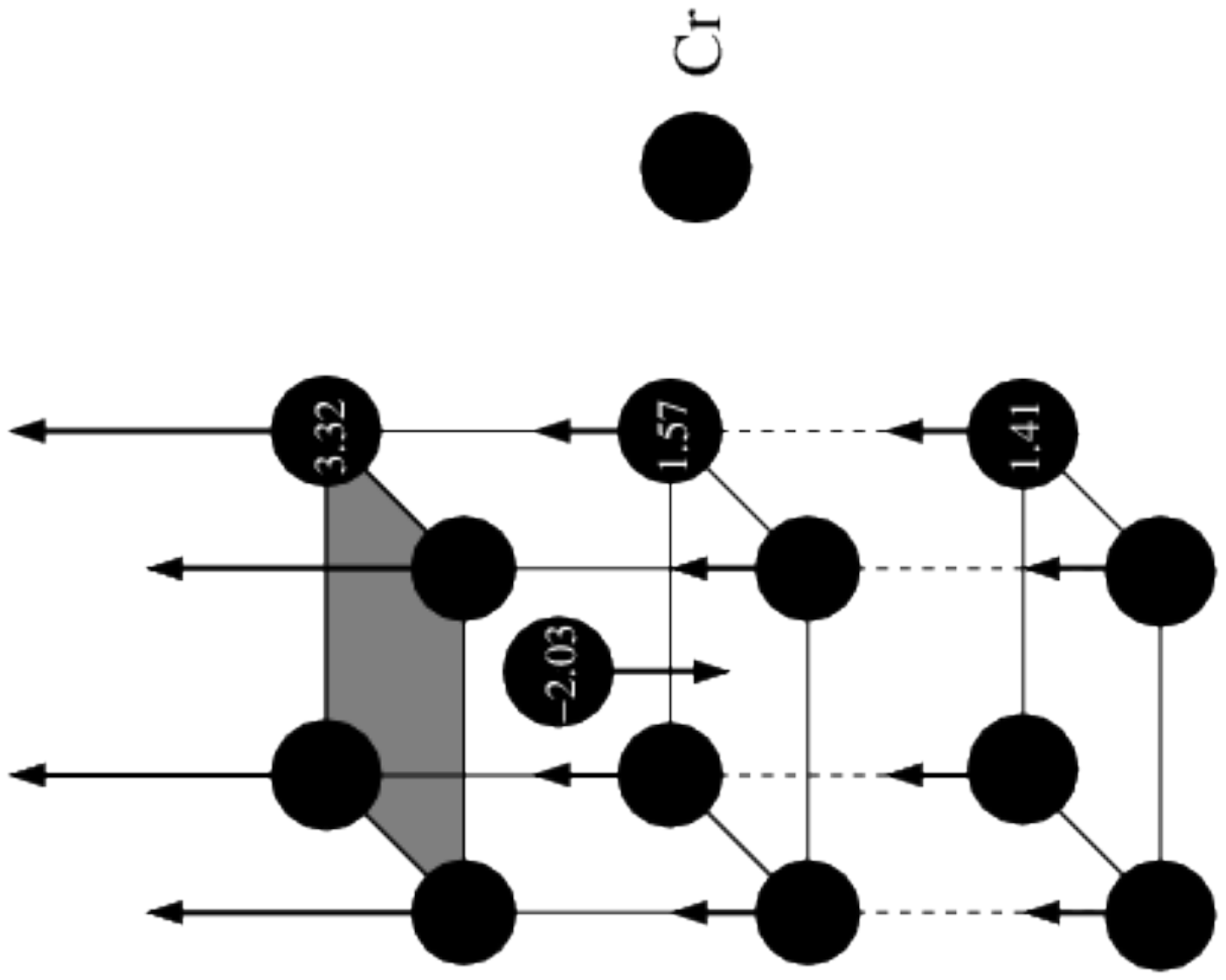}
\par\end{centering}

\caption{État magnétique de la surface de Cr $\left(100\right)$ calculée en
SIESTA-NC. Le plan grisé représente la surface. Le plan inférieur
représente le 14$\mbox{\ensuremath{^{e}}}$ plan, c'est-à-dire le
volume. Les moments magnétiques atomiques sont indiqués en blanc.
On observe un ferromagnétisme dans le plan, pour chacun des plans.\label{fig:schema_100_ferromagnetique}}

\end{figure}

Nous observons également que cette surface est ferromagnétique dans
le plan. Tous les atomes de Cr de la surface portent un moment aligné
avec les moments voisins dans le plan, comme l'indique le schéma \ref{fig:schema_100_ferromagnetique}
sur lequel nous reportons les moments magnétiques atomiques calculés
en SIESTA-NC.

\subsection{Relaxations de surface}

La présence de surfaces induit une relaxation des distances inter-réticulaires.
Une relaxation classique est du type $-+-+-\ldots$, ce qui signifie
: contraction de la distance entre les deux premiers plans, puis dilatation,
puis contraction, dilatation \ldots{} , de sorte qu'apparaissent
des oscillations dont l'amplitude diminue en s'éloignant de la surface.
Ces relaxations sont dues au lissage de l'évanescence de la densité
électronique en surface \cite{landman_surf_relax_1980,barnett_1_relaxations_highMillerindex_1983,barnett_2_surfrelax_1983,barnett_3_surfrelax_1983}.

Les relaxations relatives des trois premières distances inter-réticulaires
calculées en SIESTA-NC sont indiquées dans le tableau \ref{tab:Relaxations_surf_pures}
avec les calculs \emph{ab initio} antérieurs et les mesures expérimentales,
principalement de Sokolov, Shih, Jona et Marcus \cite{jona__LEED_cristallography_revue_1978,shih_LEEDFe110_1980,shih_LEEDFe111_1981,sokolov_LEEDFe211_1984,sokolov_LEEDFe111_1986,wang_LEEDFe100_1987}.
Des difficultés techniques dans le cycle d'auto-cohérence de la densité
électronique ont empêché la relaxation des surfaces Fe $\left(111\right)$
et $\left(211\right)$ non-magnétiques. Les surfaces de Fe et Cr $\left(100\right)$,
$\left(110\right)$ et $\left(211\right)$ ont des comportements classiques
: $d_{01}$ diminue, puis $d_{12}$ augmente, et $d_{23}$ diminue
de nouveau pour devenir négligeable. Dans Fe et Cr, l'amplitude des
relaxations dans les orientations Fe et Cr $\left(100\right)$ et
$\left(110\right)$ ainsi que Cr $\left(211\right)$ sont faibles,
inférieures à $3$\,\%. Les relaxations dans Fe $\left(211\right)$
sont d'une amplitude plus élevée ($-9.2$\,\% puis $+4.7$\,\%).
Le tableau \ref{tab:Relaxations_surf_pures} récapitule ces données,
et les compare aux rares résultats expérimentaux en LEED ou MEIS.

Les surfaces de Fe et Cr $\left(111\right)$ ont un comportement différent.
Fe$\left(111\right)$ subit une contraction de $d_{01}$ assez prononcée
($-6.2\,\%$) suivie d'une contraction violente de $d_{12}$ ($-15.2\,\%$)
et enfin d'une dilatation également violente de $d_{23}$ ($+12.9\,\%$).
Les surfaces de Cr $\left(111\right)$ subit une violente dilatation
de $d_{01}$ ($+13.0\,\%$), contrairement à toutes les autres surfaces,
suivie d'une violente contraction de $d_{12}$ ($-13.4\,\%$), puis
enfin d'une nouvelle dilatation de $d_{23}$ ($+8.2\,\%$). Ce type
de comportement est lié à l'hybridation d'orbitales électroniques
dans certaines orientations cristallographiques%
.

Dans Fe, l'amplitude des relaxations sur les trois premières couches
est corrélée à la densité surfacique (se référer au tableau \ref{tab:Densit=0000E9-des-surfaces}).
Plus celle-ci est ouverte, c'est-à-dire plus sa densité surfacique
est faible, et plus la relaxation est importante : $\Delta_{111}>\Delta_{211}>\Delta_{100}>\Delta_{110}$.
Ce n'est plus vrai pour Cr pour lequel nous avons montré que les effets
magnétiques sont très forts. Ils supplantent les effets élastiques.

Les surfaces non-magnétiques montrent des relaxations supérieures
de presque un ordre de grandeur aux surfaces magnétiques. Le seul
degré de liberté de ces plans atomiques pour s'accomoder de la présence
de la surface est structural : ils relaxent. Dans les surfaces magnétiques,
le degré de liberté magnétique permet de limiter les relaxations structurales
qui coûtent en énergie élastique. 

\begin{table}[h]
\begin{centering}
\begin{tabular}{|cc|c|c|c|c|}
\hline 
\multicolumn{2}{|c|}{Élément} & Orientation & $d_{01}$ (\%) & $d_{12}$(\%) & $d_{23}$(\%) \\
\hline 
\hline 
 &  & 111 & $-14.31$ & $-21.20$ & $13.92$ \\
\cline{3-6} 
\multicolumn{2}{|c|}{Cr} & 211 & $-14.51$ & $0.14$ & $2.38$ \\
\cline{3-6} 
\multicolumn{2}{|c|}{non-magnétique} & 100 & $10.58$ & $2.43$ & $-1.88$ \\
\cline{3-6} 
 &  & 110 & $-3.53$ & $0.78$ & $-0.10$ \\
\hline 
\hline 
 &  & 111 & $13.05$ & $-13.44$ & $8.23$ \\
\cline{3-6} 
\multicolumn{2}{|c|}{Cr} & 211 & $-1.42$ & $-0.21$ & $1.43$ \\
\cline{3-6} 
\multicolumn{2}{|c|}{magnétique} & 100 & $-1.25$ & $2.05$ & $-0.39$ \\
\cline{3-6} 
 &  & 110 & $-0.86$ & $0.44$ & $-0.38$ \\
\hline 
\hline 
 &  & 111 & {*} & {*} & {*} \\
\cline{3-6} 
\multicolumn{2}{|c|}{Fe} & 211 & {*} & {*} & {*} \\
\cline{3-6} 
\multicolumn{2}{|c|}{non-magnétique} & 100 & $15.95$ & $9.03$ & $-0.75$ \\
\cline{3-6} 
 &  & 110 & $-1.39$ & $0.53$ & $0.41$ \\
\hline 
\hline 
 &  & 111 & $-6.24$ & $-15.19$ & $12.89$ \\
\cline{3-6} 
 &  & 111 LEED \cite{sokolov_LEEDFe111_1986} & $-16.9\pm3$ & $-9.8\pm3$ & $4.2\pm3.6$ \\
\cline{3-6} 
 &  & 111 MEIS \cite{xu_MEISFe111_1990} & $-29.0\pm7$ & $6\pm5$ &  \\
\cline{3-6} 
 &  & 211 & $-9.16$ & $4.74$ & $-0.16$ \\
\cline{3-6} 
\multicolumn{2}{|c|}{Fe} & 211 LEED \cite{sokolov_LEEDFe211_1984} & $-10.4\pm2.6$ & $5.4\pm2.6$ & $-1.3\pm3.4$ \\
\cline{3-6} 
\multicolumn{2}{|c|}{magnétique} & 100 & $-3.00$ & $2.90$ & $1.15$ \\
\cline{3-6} 
 &  & 100 LEED \cite{wang_LEEDFe100_1987} & $-5\pm2$ & $5\pm2$ &  \\
\cline{3-6} 
 &  & 110 & $-0.20$ & $0.27$ & $-0.40$ \\
\cline{3-6} 
 &  & 110 LEED \cite{shih_LEEDFe110_1980} & $0.5\pm2$ &  &  \\
\cline{3-6} 
 &  & 110 MEIS \cite{xu_Fe_relaxations_1991} & $1\pm2$ & $0.5\pm2$ &  \\
\hline 
\end{tabular}
\par\end{centering}

\caption{Relaxations relatives des trois premières distances inter-réticulaires
$d_{01}$, $d_{12}$ et $d_{23}$ calculées avec SIESTA-NC pour les
surfaces de Fe et Cr $\left(110\right)$, $\left(211\right)$, $\left(111\right)$
et $\left(100\right)$ dans les cas magnétiques et non-magnétiques.
Les orientations sont classées par ordre de densité surfacique croissant
: $\rho_{111}<\rho_{211}<\rho_{100}<\rho_{110}$.\protect \\
{*} les calculs de Fe $\left(111\right)$ et $\left(211\right)$ non-magnétiques
ne convergent pas (voir texte).\label{tab:Relaxations_surf_pures}}
\end{table}

Ossowski, Blonski et Kiejna ont également calculé les relaxations
de surfaces de Fe et Cr \cite{blonski_Fe_surfaces_2007,ossowski_Cr_surfaces_2008}.
Ils trouvent les mêmes tendances avec une atténuation systématique
de quelques pourcents. Cette faible différence est certainement due
à l'influence du magnétisme plus faible dans les calculs de type VASP-PAW.

\subsection{Effet des relaxations sur les énergies de surface}

\begin{figure}[h]
\begin{centering}
\includegraphics[scale=0.4]{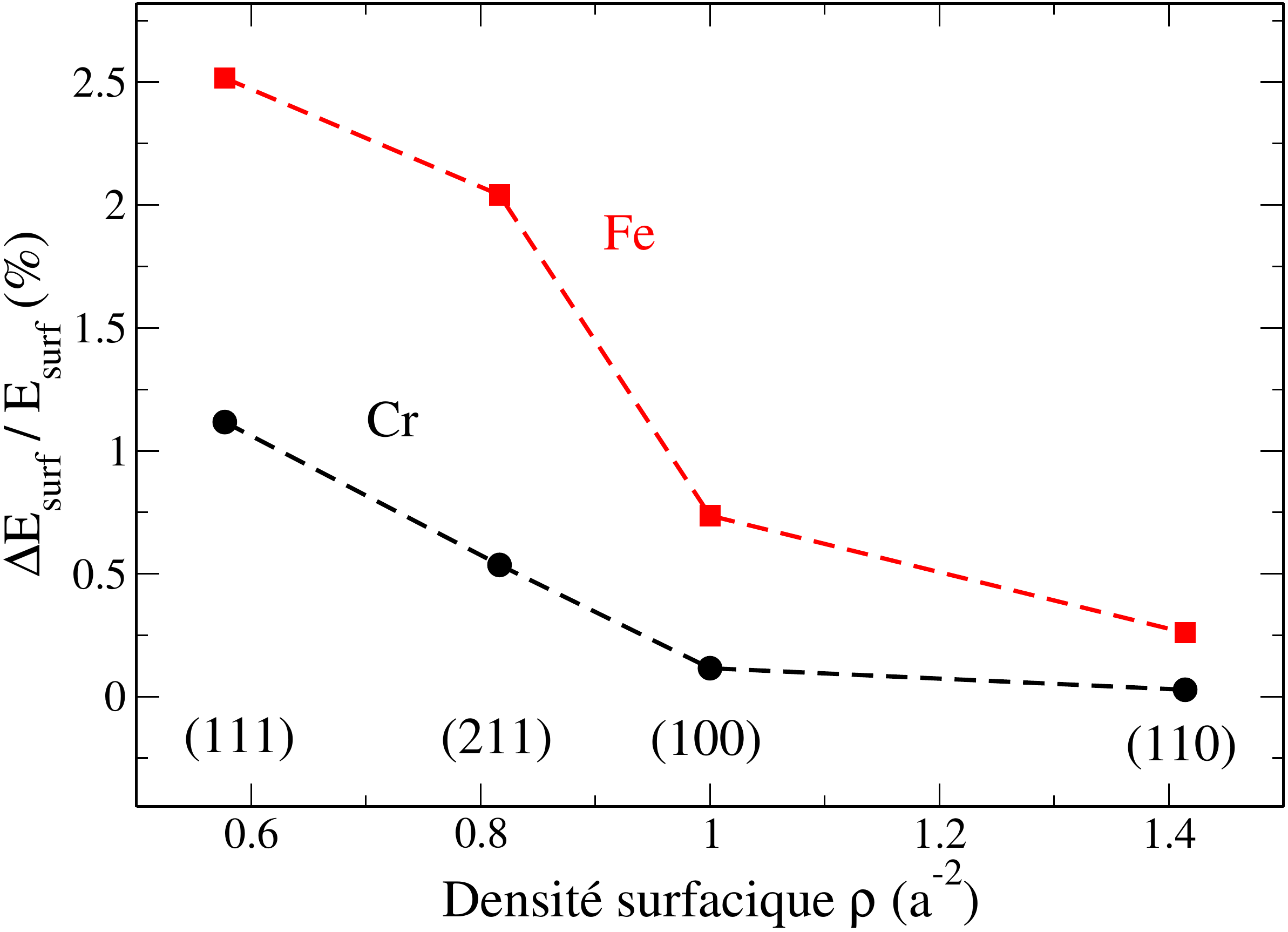}
\par\end{centering}

\caption{Différence relative entre les énergies des surfaces libres relaxées
et non relaxées de fer et de chrome. On pourra se référer au tableau
\ref{tab:Densit=0000E9-des-surfaces} pour les relations entre densité
surfacique et orientation de la surface ($\rho_{111}<\rho_{211}<\rho_{100}<\rho_{110}$).\label{fig:Esurf_relax}}
\end{figure}

L'amplitude des relaxations discutées ci-dessus est parfois non-négligeable.
Afin de vérifier l'influence des relaxations sur les énergies de surface,
on a calculé ces dernières avec et sans relaxations des positions
atomiques. On représente sur la figure \ref{fig:Esurf_relax} la différence
relative entre les énergie des surfaces relaxées et non-relaxées.

L'effet des relaxations sur l'énergie de surface est faible. Elles
stabilisent les surfaces de moins de 3\,\%. En accord avec les résultats
précédents, l'effet des relaxations diminue avec la densité surfacique.
Les surfaces de fer sont plus sensibles, énergétiquement, à l'effet
des relaxations. Cet effet reste cependant minime, de l'ordre de 0.5
à 1\,\% plus présent que dans le chrome.

\subsection{Transferts de charge en surface}

Dans Fe, des transferts de charge de $0.04$ ont lieu de la couche
de surface ($0$) vers les couches adjacentes ($1$ et $2$). C'est
également le cas dans Cr $\left(111\right)$. Ce sont 0.06\,électron
par atome qui sont transférés pour Cr $\left(110\right)$ et $\left(211\right)$.
La surface perd alors 0.10\,électron par atome. Cr $\left(100\right)$
présente le comportement inverse : chaque atome du plan de surface
gagne $0.13$ électrons qui viennent de façon quasi-égale des quatre
couches suivantes. Les plans sous-jacents à la surface alimentent
l'amplification du moment magnétique des atomes du plan de surface.

\section{Ségrégation du chrome dans les surfaces libres de fer\label{sec:seg_DFT}}

Christensen et al. \cite{christensen_phasediagramsforSURFACEalloys_1997}
font une étude \emph{ab initio} systématique en LMTO-CPA pour 24 métaux
de transition. Cela représente 552 combinaisons d'impuretés--hôte.
Chelikowsky \cite{Chelikowsky_collectionSURFseg_1997} compile 40
expériences de ségrégation (dont Leygraph et al. pour Fe--Cr \cite{leygraf_surface_1974}).
Les calculs de Christensen et al. sont en accord avec 38 des 40 expériences.
Seuls la ségrégation de Cu dans Pt et de Cr dans Fe induisent des
désaccords qualitatifs entre calculs et expérience. Alors que Leygraph
et al. semblent observer un enrichissement des surfaces libres (100)
et (110) en chrome, Christensen et al. calculent des énergies de ségrégation
de $+0.36$\,eV, signe d'un appauvrissement des surfaces en chrome.
Les observations AES de Leygraph et al. à température ambiante sont
cependant à composition nominale de chrome assez élevée (16\,\%\,Cr)
après des traitements thermiques à températures supérieures à 800\,K.
Une manière de rappeler la singularité du système fer--chrome.

\subsection{Définition de l'énergie de ségrégation}

\begin{figure}[h]
\begin{centering}
\includegraphics[scale=0.4]{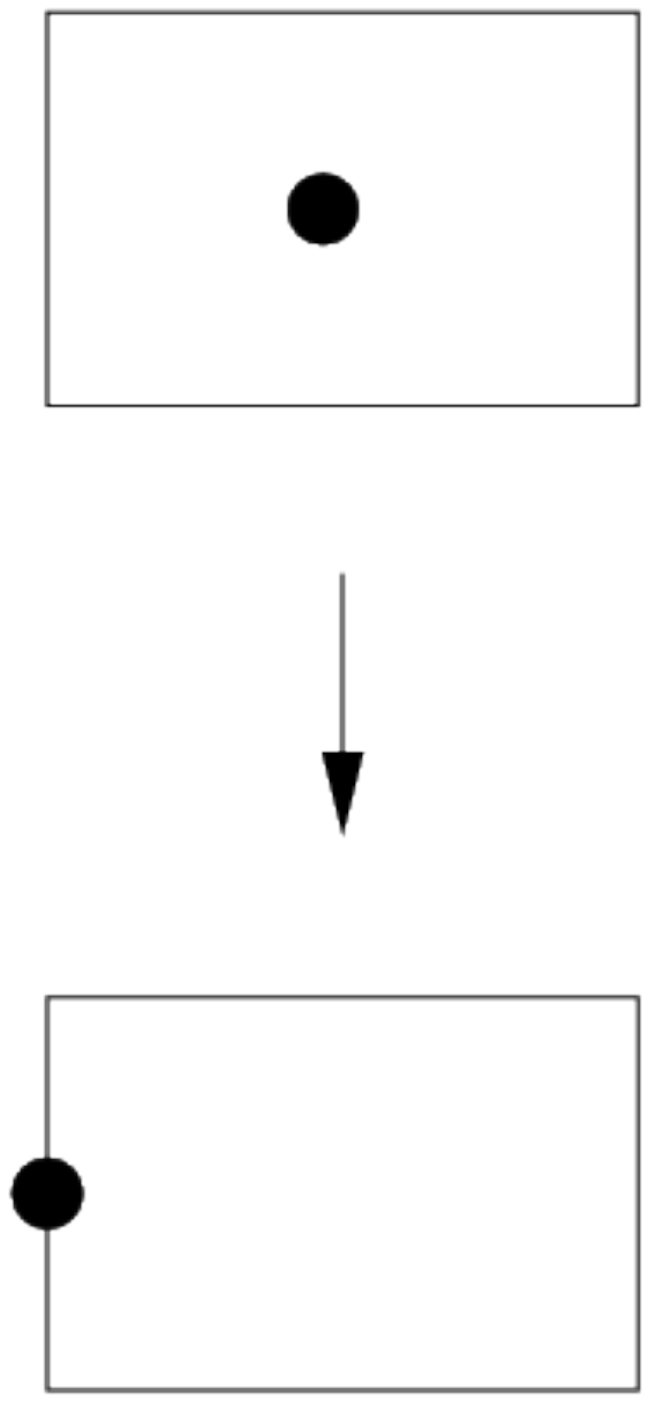}
\par\end{centering}

\caption{Schéma du bilan énergétique pour le calcul de l'énergie de ségrégation.
Dans l'état final, l'atome noir est dans le plan de surface ($p=0$
dans la relation \ref{eq:def_Eseg}).\label{fig:Sch=0000E9ma-du-bilan_Hseg}}
\end{figure}

L'énergie de ségrégation est la force motrice de la ségrégation de
surface. Elle représente la variation d'énergie qui résulte de l'échange
entre un atome de type A de volume et un atome de type B en surface.
La surface n'est pas nécessairement le plan de surface. Il peut s'agir
d'un plan sous-jacent. On calcule l'énergie de ségrégation $\Delta E_{seg}^{0}\left(p\right)$
d'un atome de Cr vers le plan de surface $p$ dans l'orientation $\left(hkl\right)$
: 
\begin{equation}
\Delta E_{seg}^{0}\left(p\right)=E_{slab}\left(p\right)-E_{slab}\left(v\right)\label{eq:def_Eseg}
\end{equation}
avec
\begin{equation}
E_{slab}\left(v\right)\equiv E_{slab}\left(7\right)
\end{equation}
où $E_{slab}\left(p\right)$ est l'énergie totale d'une super-cellule
contenant une couche mince de Fe avec une impureté Cr dans le plan
$p$, et $E_{slab}\left(v\right)$ est l'énergie de la même super-cellule
avec Cr dans le plan du milieu de la couche mince (le plan de volume
qui est le plan $p=7$ dans nos supercellules). La figure \ref{fig:Sch=0000E9ma-du-bilan_Hseg}
illustre le bilan énergétique \ref{eq:def_Eseg}.

\subsection{Ségrégation dans le plan (100)}

Nous avons vu au paragraphe \ref{par:dissolutionCrdansFe} que la
dissolution d'une impureté Cr dans une matrice Fe est exothermique.
L'énergie de mise en solution correspondante calculée \emph{ab initio}
va de $\unit[-0.47]{eV}$ en SIESTA-NC à $\unit[-0.12]{eV}$ en PWSCF-PAW
comme nous le montrions à la section \ref{sub:def_mu}. Nous avons
également vu que les pseudo-potentiels ont une grande influence sur
les propriétés magnétiques calculées. En particulier, l'approximation
des pseudo-potentiels à norme conservée surestime le moment magnétique
local de Cr. Qu'en est-il en surface ?

Le calcul \emph{ab initio} de l'énergie de ségrégation de Cr dans
des surfaces de Fe a fait l'objet d'un regain d'intérêt ces dernières
années \cite{ponomareva_HsegFe100_2007,kiejna_Eseg_2008}. Le tableau
\ref{tab:Hseg_impureteCr} récapitule l'ensemble des énergies de ségrégation
calculées par le passé à notre connaissance, ainsi que la méthode
et la taille des super-cellules de calcul utilisées (pour les calculs
périodiques).

\begin{table}[h]
\begin{centering}
\begin{tabular}{|c|c|c|c|c|}
\hline 
Orientation & Référence & Méthode & $\Delta E_{seg}^{0}\left(0\right)$ (eV) & Super-cellule \\
\hline 
\hline 
 & Ce travail & SIESTA-NC-GGA & $-0.295$ & $3\times3\times14$$^{\text{a}}$ \\
\cline{2-5} 
 & Ce travail & SIESTA-NC-GGA & $-0.305$ & $6\times6\times14$ \\
\cline{2-5} 
 & Ce travail & PWSCF-US-GGA & $0.183$ & $3\times3\times14$ \\
\cline{2-5} 
 & Ce travail & PWSCF-PAW-GGA & $0.141$ & $3\times3\times14$ \\
\cline{2-5} 
 & \cite{ponomareva_HsegFe100_2007} & VASP-PAW-GGA & $-0.043$ & $2\times2\times7$ \\
\cline{2-5} 
 & \cite{ponomareva_HsegFe100_2007} & VASP-PAW-GGA & $-0.02$ & $2\times4\times7$ \\
\cline{2-5} 
 & \cite{ponomareva_HsegFe100_2007} & VASP-PAW-GGA & $-0.009$ & $3\times3\times7$ \\
\cline{2-5} 
 & \cite{ponomareva_HsegFe100_2007} & VASP-PAW-GGA & $0.080$ & $2\times2\times9$ \\
\cline{2-5} 
$\left(100\right)$ & \cite{ponomareva_HsegFe100_2007} & VASP-PAW-GGA & $0.090$ & $3\times3\times9$ \\
\cline{2-5} 
 & \cite{ponomareva_HsegFe100_2007} & VASP-PAW-GGA & $0.069$ & $2\times2\times11$ \\
\cline{2-5} 
 & \cite{ponomareva_HsegFe100_2007} & VASP-PAW-GGA & $0.064$ & $3\times3\times11$ \\
\cline{2-5} 
 & \cite{kiejna_Eseg_2008} & VASP-PAW-GGA & $0.076$ & $2\times2\times11$ \\
\cline{2-5} 
 & \cite{geng_segGGAFLAPW_2003} & FLAPW-GGA & $0.31$ & $1\times1\times14$ \\
\cline{2-5} 
 & \cite{geng_segGGAFLAPW_2003} & FLAPW-GGA & $0.04$ & $c\left(2\times2\times14\right)$ \\
\cline{2-5} 
 & \cite{geng_segGGAFLAPW_2003} & FLAPW-GGA & $-0.11$ & $2\times2\times14$ \\
\cline{2-5} 
 & \cite{geng_segGGAFLAPW_2003} & FLAPW-GGA & $-0.15$ & $c\left(4\times4\times14\right)$ \\
\cline{2-5} 
 & \cite{nonas_impuretes_dans_Fe100_1998} & FP-KKR-LDA & $-0.05$ &  \\
\cline{2-5} 
 & \cite{ruban_surf_seg_TMalloys_1999} & LMTO-ASA+M-LDA & $0.2$ &  \\
\cline{2-5} 
 & \cite{ponomareva_HsegFe100_2007} & EMTO-CPA-LDA & $0.180$ & $7$\,couches \\
\cline{2-5} 
 & \cite{ponomareva_HsegFe100_2007} & EMTO-CPA-LDA & $0.130$ & $13$\,couches \\
\hline 
$\left(110\right)$ & \cite{kiejna_Eseg_2008} & VASP-PAW-GGA & $-0.001$ & $2\times2\times11$ \\
\cline{2-5} 
 & Ce travail & SIESTA-NC-GGA & $-0.260$ & $3\times3\times14$$^{\text{b}}$ \\
\hline 
$\left(111\right)$ & \cite{kiejna_Eseg_2008} & VASP-PAW-GGA & $0.014$ & $2\times2\times15$ \\
\hline 
$\left(210\right)$ & \cite{kiejna_Eseg_2008} & VASP-PAW-GGA & $0.281$ & $1\times2\times20$ \\
\hline 
\end{tabular}
\par\end{centering}

\caption{Énergies de ségrégation calculées \emph{ab initio} de Cr dans Fe dans
les orientations $\left(100\right)$, $\left(110\right)$, $\left(111\right)$
et $\left(211\right)$. La méthode de calcul employée et la taille
de la super-cellule utilisées sont indiquées. Les configurations $c\left(x\times x\times z\right)$
sont des surfaces reconstruites. \label{tab:Hseg_impureteCr}\protect \\
$^{\text{a}}$ 9 atomes par plan\protect \\
$^{\text{b}}$ 18 atomes par plan}
\end{table}

Nous observons pour l'orientation $\left(100\right)$ une très forte
dispersion des énergies de ségrégation calculées. Même en restant
dans des approximations de calcul tout-à-fait comparables, par exemple
VASP-PAW-GGA, l'énergie de ségrégation de Cr dans Fe$\left(100\right)$
varie de $-0.04$ à $+0.18$\,eV. Cette observation a déjà été faite
par Ponomareva et al. mais n'a, à notre connaissance, pas été expliquée
\cite{ponomareva_HsegFe100_2007}.

Afin de donner une explication à ce phénomène, nous indiquons également
dans le tableau \ref{tab:Hseg_impureteCr} la taille des super-cellules
utilisées pour les calculs. Cela nous permet de rappeler l'importance
des effets de taille dans les calculs \emph{ab initio} de surfaces.
Ces études sont en effet limitées à quelques dizaines d'atomes. Comme
nous l'indiquions au paragraphe \ref{sub:Comment-mod=0000E9liser-des_surf_ab_initio},
il faut empiler un certain nombre de couches atomiques perpendiculairement
à la surface. Un atome par plan revient en fait à une monocouche !
Ce nombre de couches, que nous avons pris égal à $14$, limite grandement
l'aire de la surface. Si on met quatre atomes dans le plan de surface,
cela fait déjà $4\times4\times14=224$ atomes dans la super-cellule,
ce qui est grand pour un calcul \emph{ab initio}. Nous verrons dans
la suite du document qu'une étude de la taille des super-cellule des
calculs PAW, comme l'indique le tableau \ref{tab:Hseg_impureteCr},
permet de mieux appréhender ce problème. En effet, nous montrons qu'une
super-cellule à un atome par plan revient à calculer l'énergie de
ségrégation d'une monocouche entière de Cr, ce qui est différent d'une
énergie de ségrégation d'impureté.

Rappelons ici l'avantage des calculs SIESTA en pseudo-potentiels à
norme conservée qui permettent de considérer des cellules aussi grandes
plus sereinement qu'avec des bases d'ondes planes. 

\begin{figure}[h]
\begin{centering}
\includegraphics[scale=0.4]{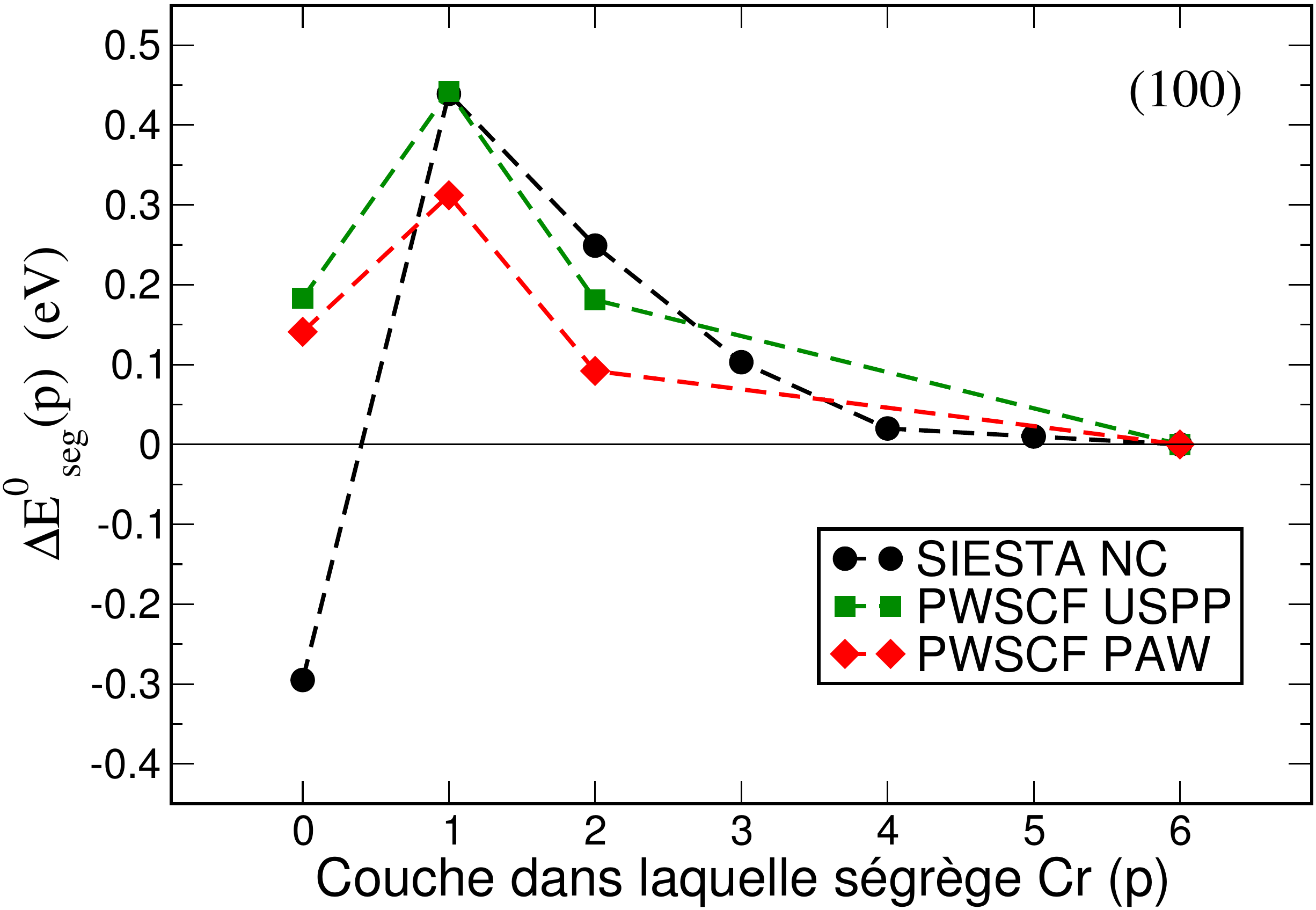}
\par\end{centering}

\caption{Énergie de ségrégation d'une impureté Cr dans Fe$(100)$ magnétique
dans les 6 premières couches de surface ($p=0$ à 5, $p=6$ est le
plan de volume), calculée avec les méthodes suivantes : (noir) SIESTA
NC, (rouge) PWSCF PAW, et (vert) PWSCF US, dans une super-cellule
$3\times3\times14$.\label{fig:Hseg_SIESTA_PAW_100_110}}
\end{figure}

Comme pour le volume, nous avons réalisé des calculs de surface avec
SIESTA-NC, mais également avec PWSCF-PAW et PWSCF-US. L'énergie de
ségrégation calculée dans les 6 plans de surface de Fe $\left(100\right)$
est représentée sur la figure \ref{fig:Hseg_SIESTA_PAW_100_110}.

On observe une différence significative de comportement entre les
résultats NC d'une part, et US et PAW d'autre part. Ces calculs sont
réalisés pour les mêmes tailles de cellule : $3\times3\times14$.
Alors que les calculs US et PAW montrent une ségrégation d'impureté
dans le plan de surface endothermique (respectivement +0.18 et +0.14
eV), les calculs NC avec SIESTA trouvent $\Delta E_{seg\left(100\right)}^{0}$
négatif, c'est-à-dire exothermique ($-0.29$ eV). C'est une différence
qualitative importante que l'on peut expliquer par la surestimation
des effets de stabilisation liés au magnétisme par les pseudopotentiels
NC. Ceux-ci surestiment le moment magnétique du chrome en surface
et en conséquence sa stabilisation. Il serait intéressant ici de pouvoir
controler le moment magnétique de l'atome de chrome, à moment magnétique
du fer constant, et de quantifier ainsi son influence sur l'énergie
de ségrégation.

On observe également, indépendemment du type de pseudopotentiel, un
phénomène jamais décrit à notre connaissance : quelle que soit l'approximation
utilisée, l'énergie de ségrégation dans le plan sous-jacent la surface
($p=1$) est très élevée ($>0.3$ eV). Nous reviendrons sur ce point
plus tard mais notons qu'il est particulièrement lié aux interactions
magnétiques à l'échelle atomique.

\subsection{Comparaison (100) / (110)}

\begin{figure}[h]
\begin{centering}
\includegraphics[scale=0.4]{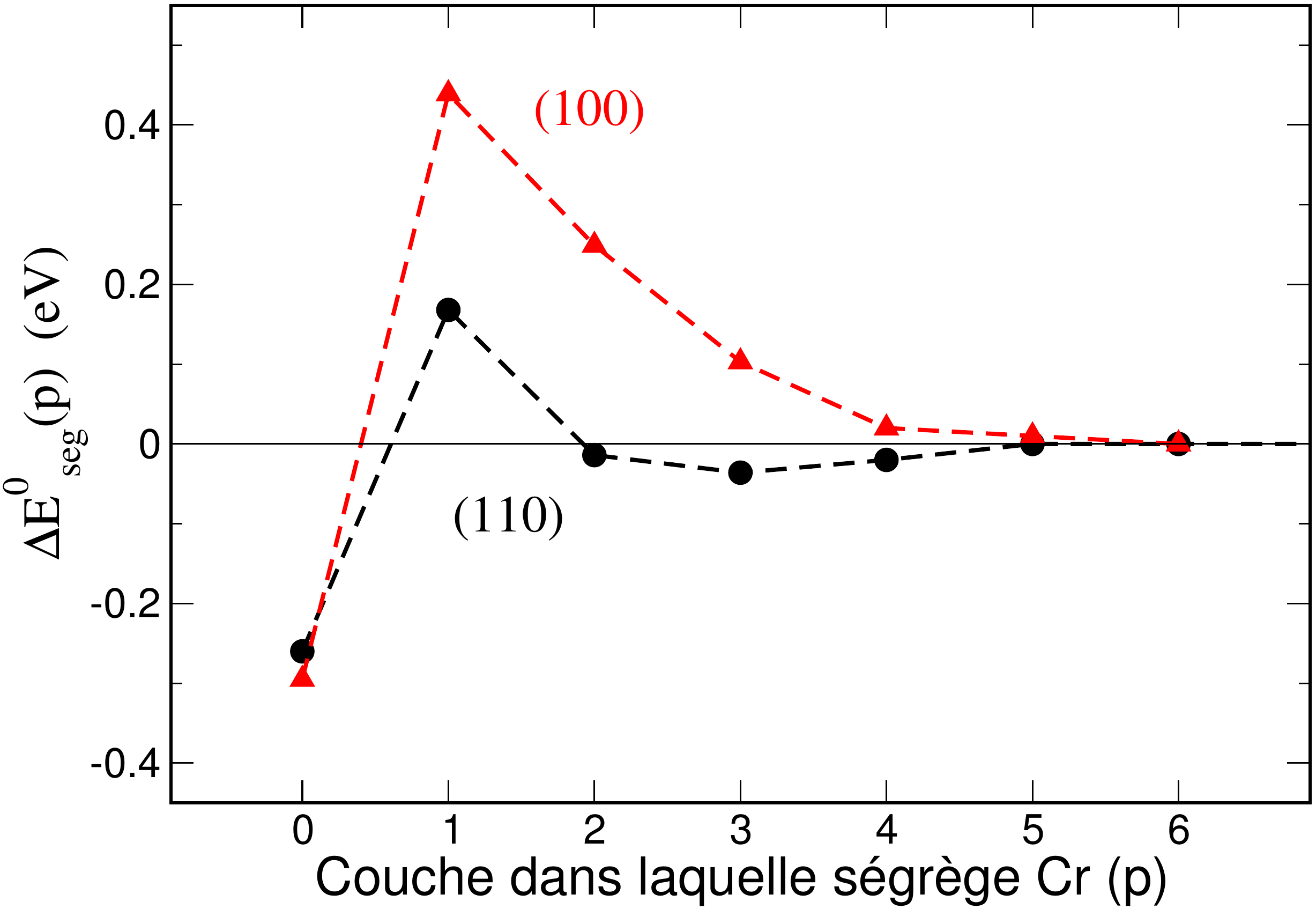}
\par\end{centering}

\caption{Énergie de ségrégation d'impureté du chrome dans le fer (110) en noir
et (100) en rouge. Les cellules sont de types $3\times3\times14$
contenant 9 atomes par plan (100) et 18 atomes par plan (110). Les
calculs sont réalisés avec SIESTA-NC.\label{fig:seg100110}}
\end{figure}

Nous avons également calculé l'énergie de ségrégation d'impureté du
chrome dans le fer (110) qui est représentée sur la figure \ref{fig:seg100110}
avec l'énergie de ségrégation d'impureté dans Fe (100).

La ségrégation d'impureté dans Fe (110) a la même tendance que dans
Fe (100) avec une plus faible ampleur. L'énergie de ségrégation dans
le plan sous-jacent de la surface ($p=1$) est exothermique, égale
à 0.17\,eV. La ségrégation dans le plan de surface est également
endothermique en SIESTA-NC de $-0.26$\,eV.

La diminution de l'amplitude des mêmes effets s'explique par la densité
surfacique plus élevée de l'orientation (110). Les effets du magnétisme
sont moins forts dans la (110) que dans la (100).

\subsection{Effet du magnétisme sur la ségrégation d'impureté}

\begin{figure}[h]
\begin{centering}
\includegraphics[scale=0.4]{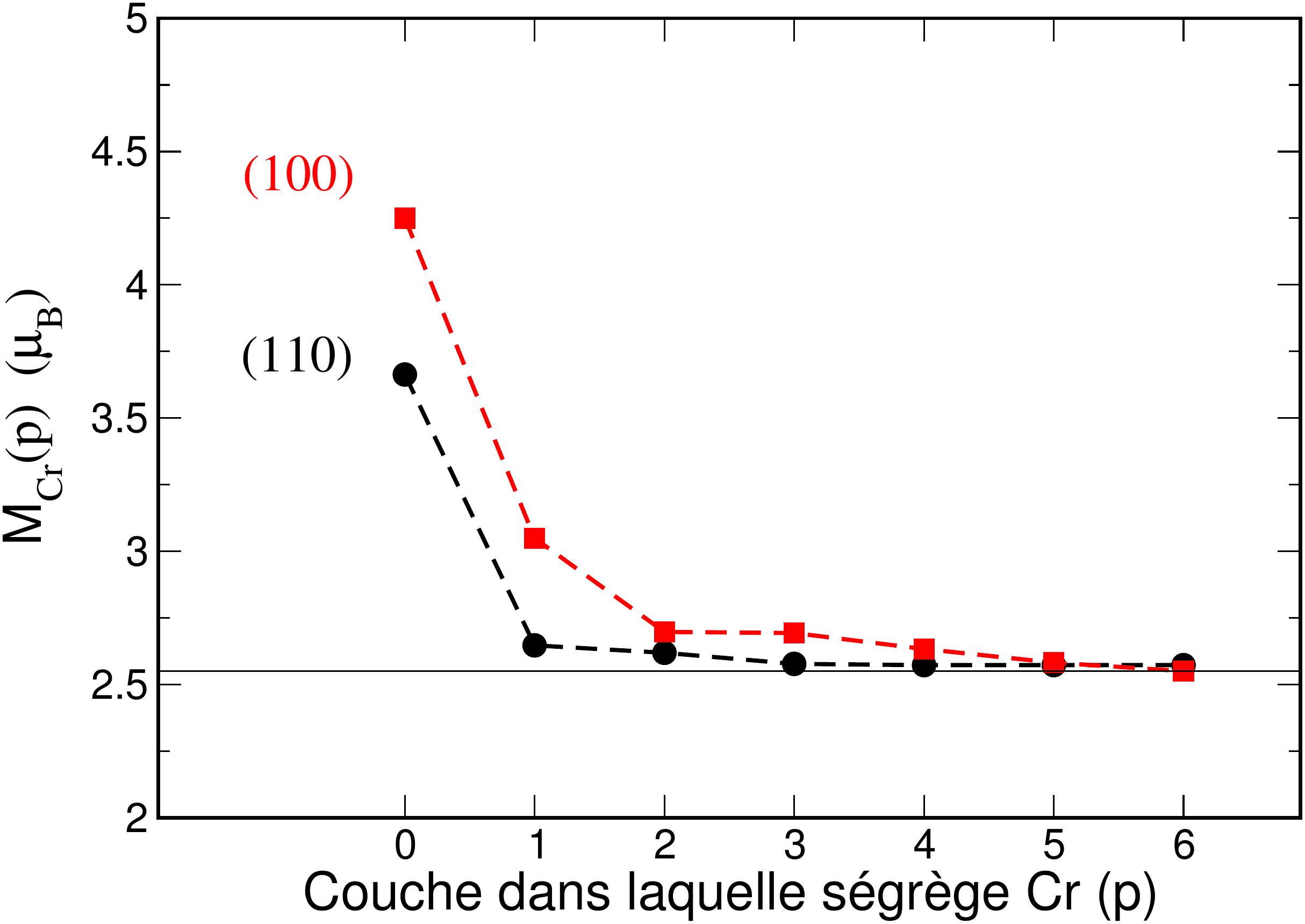}
\par\end{centering}

\caption{Moment magnétique de l'atome d'impureté Cr dans les différents plans
de volume ($p=6$) jusqu'à la surface ($p=0$) dans Fe (100) en rouge
et (110) en noir. Ces résultats sont issus de calculs SIESTA-NC.\label{fig:mu_seg}}
\end{figure}

Nous avons décrit la stabilisation des surfaces des éléments purs
par l'augmentation du moment magnétique atomique en surface. Nous
avions décrit que c'est l'élément chrome qui est le plus sensible
à une augmentation de son moment magnétique atomique en surface. On
trace l'évolution du moment magnétique atomique de l'impureté Cr dans
Fe (100) sur la figure \ref{fig:mu_seg}.

C'est dans l'orientation (100) que le moment magnétique du chrome
$M_{Cr}$ est le plus augmenté en présence de la surface car c'est
l'orientation la moins dense. Dans le plan de surface, le moment magnétique
atomique du chrome est égal à 4.25\,$\mu_{B}$. Cette augmentation
a lieu pour les mêmes raisons que dans les surfaces libres. Dans Fe
(100), le moment magnétique augmente à partir du plan $p=1$, c'est-à-dire
du plan sous-jacent la surface.

Dans l'orientation (110), le moment magnétique n'augmente notablement
que dans le plan de surface ($p=0$). Le chrome a alors un moment
magnétique égal à 3.66\,$\mu_{B}$.

\begin{figure}[h]
\begin{centering}
\includegraphics[scale=0.4]{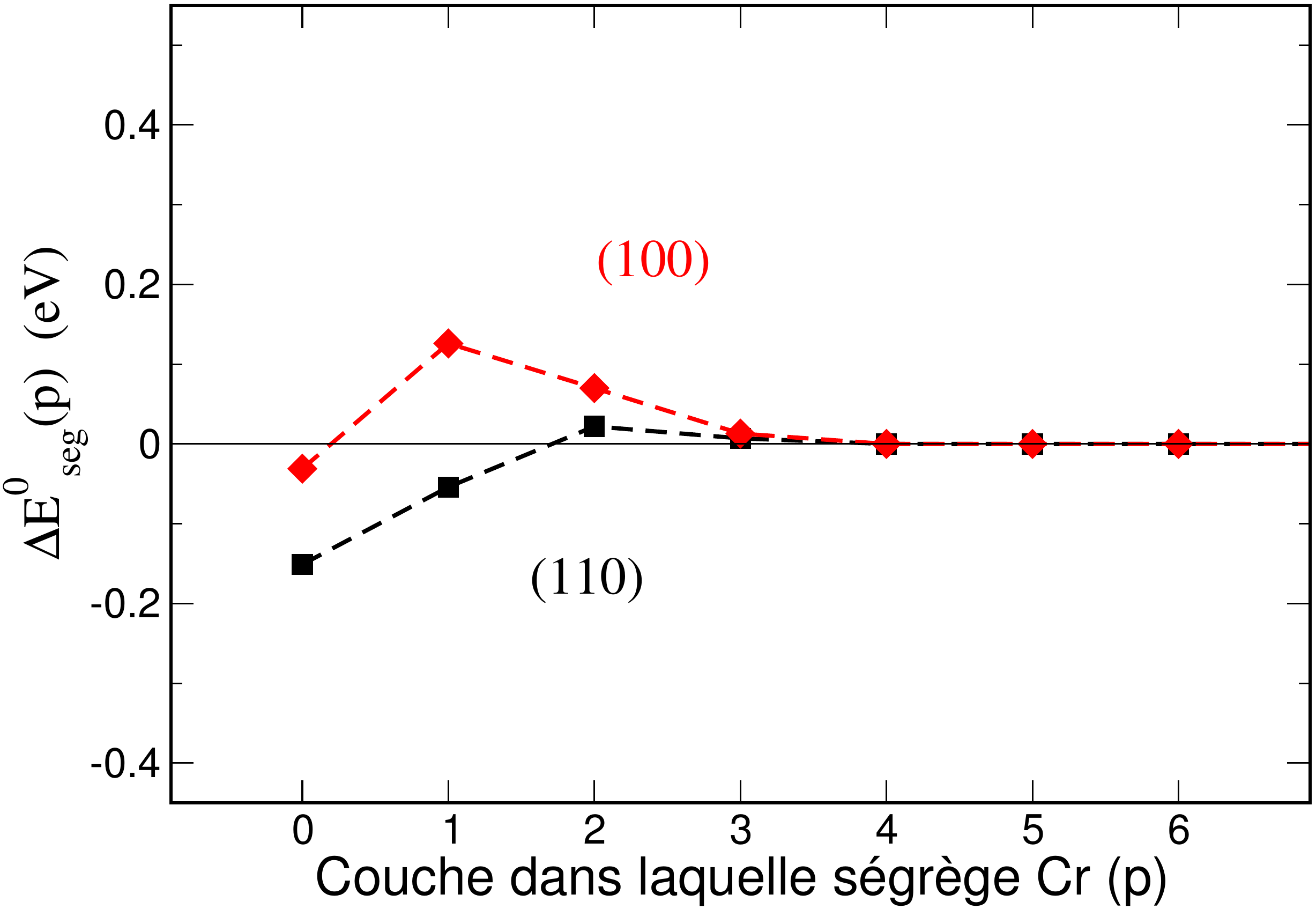}
\par\end{centering}

\caption{Énergie de ségrégation d'impureté du chrome non-magnétique dans le
fer non-magnétique (110) en noir et (100) en rouge. Les cellules sont
de types $3\times3\times14$ contenant 9 atomes par plan (100) et
18 atomes par plan (110). Les calculs sont réalisés avec SIESTA-NC.
Le système entier est non-magnétique.\label{fig:esegNM}}
\end{figure}

Afin de comprendre la relation entre la nette augmentation du moment
magnétique des impuretés et la ségrégation d'impureté dans Fe, on
calcule les énergies de ségrégation du chrome non-magnétique dans
le fer non-magnétique. Ces énergies de ségrégation sont représentées
sur la figure \ref{fig:esegNM}.

En comparant les figures \ref{fig:seg100110} et \ref{fig:esegNM},
il semble clair que les énergies de ségrégation du chrome dans le
plan de surface du fer sont fortement corrélées au magnétisme. Cependant,
on observe toujours une augmentation de l'énergie de ségrégation dans
le plan sous-jacent de la surface (100), puis une légère stabilisation
dans la surface. Dans le plan (110), il n'y a plus de déstabilisation
dans le plan $p=1$, mais la stabilisation reste importante dans le
plan de surface. Le profil de ségrégation semble ainsi lié de façon
très fine à la structure électronique de la surface et dépend de l'orientation.
Le magnétisme explique partiellement ce phénomène atypique. Des études
complémentaires sont nécessaires, par exemple en imposant un moment
magnétique nul à seulement l'un ou l'autre des éléments.

On notera enfin que tous ces calculs relaxés ont été comparés à leur
équivalent non-relaxé. On observe que l'effet des relaxations est
négligeable. On continue cependant de réaliser des calculs relaxés.

\section{Interactions entre impuretés de chrome dans les surfaces de fer}

\subsection{Interaction Cr--Cr en surface}

Afin d'évaluer l'effet de la taille des super-cellules, nous avons
augmenté progressivement le nombre d'atomes de Cr présents dans notre
grande super-cellule de calcul. La première étape est donc l'étude
de l'énergie d'interaction de deux Cr en surface.

\begin{figure}[h]
\begin{centering}
\includegraphics[scale=0.4]{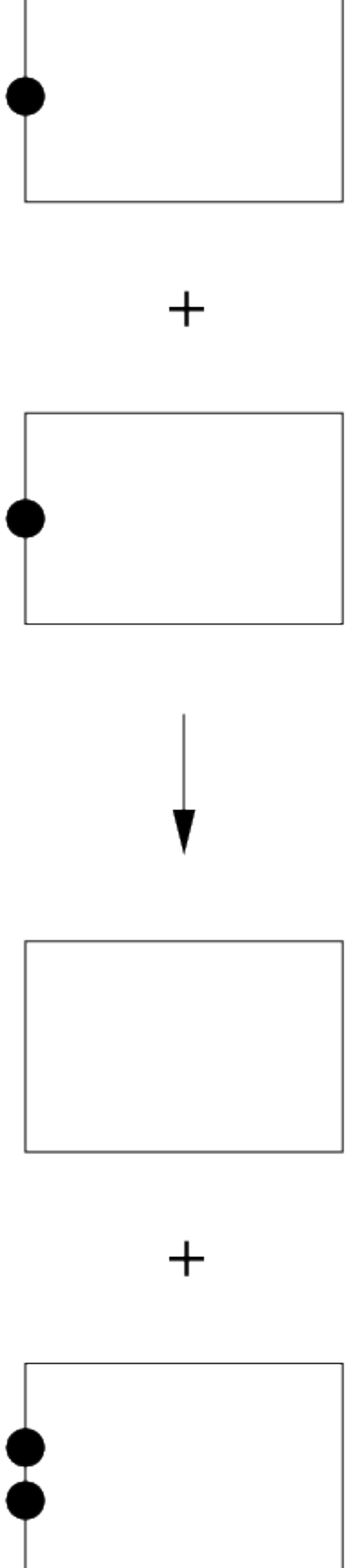}
\par\end{centering}

\caption{Schéma du bilan énergétique défini dans la relation \ref{eq:def_Ebind_CrCrsurf}
pour le calcul de l'énergie de liaison Cr--Cr en surface. \label{fig:bilan_energetique_pour_le_calcul_de_ECrCr_en_surface}}
\end{figure}

Nous avons calculé l'énergie de liaison $E_{liaison}^{(hkl)p}$ entre
Cr $i$$^{\text{e}}$ voisins à la surface de Fe (100) et (110). $E_{liaison}^{(hkl)i}$
se calcule \emph{ab initio} suivant la relation \ref{eq:def_Ebind_CrCrsurf}:
\begin{equation}
E_{liaison}^{(hkl)i}=E_{tot}^{(hkl)i}\left(Fe_{N-2}Cr_{2}\right)+NE\left(Fe\right)-2E_{tot}^{(hkl)}\left(Fe_{N-1}Cr_{1}\right)\label{eq:def_Ebind_CrCrsurf}
\end{equation}
où $E_{tot}^{(hkl)}\left(Fe_{N-2}Cr_{2}\right)$ est l'énergie totale
d'une super-cellule de Fe $\left(hkl\right)$ contenant 2 Cr $i$$^{\mbox{e}}$
voisins en surface et $E_{tot}^{(hkl)}\left(Fe_{N-1}Cr_{1}\right)$
est l'énergie totale d'une super-cellule contenant un seul Cr en surface.
Ce bilan est représenté sur la figure \ref{fig:bilan_energetique_pour_le_calcul_de_ECrCr_en_surface}.

Bien que les résultats SIESTA-NC et PWSCF-PAW soient en désaccord
sur l'énergie de ségrégation d'impureté, les énergies d'interaction
Cr--Cr calculées par SIESTA-NC sont fiables. En effet, en volume,
on rappelle que l'énergie de liaison Cr--Cr premiers voisins calculée
est $+0.37$ et $+0.32$\,eV respectivement en SIESTA-NC et PWSCF-PAW.
Considérant la taille des supercellules de calcul, ces calculs n'ont
été réalisé qu'en SIESTA-NC. On fera l'hypothèse d'une décroissance
de l'interaction de surface similaire entre PAW et NC, comme ça l'est
en volume.

\begin{figure}[h]
\begin{centering}
\includegraphics[scale=0.4]{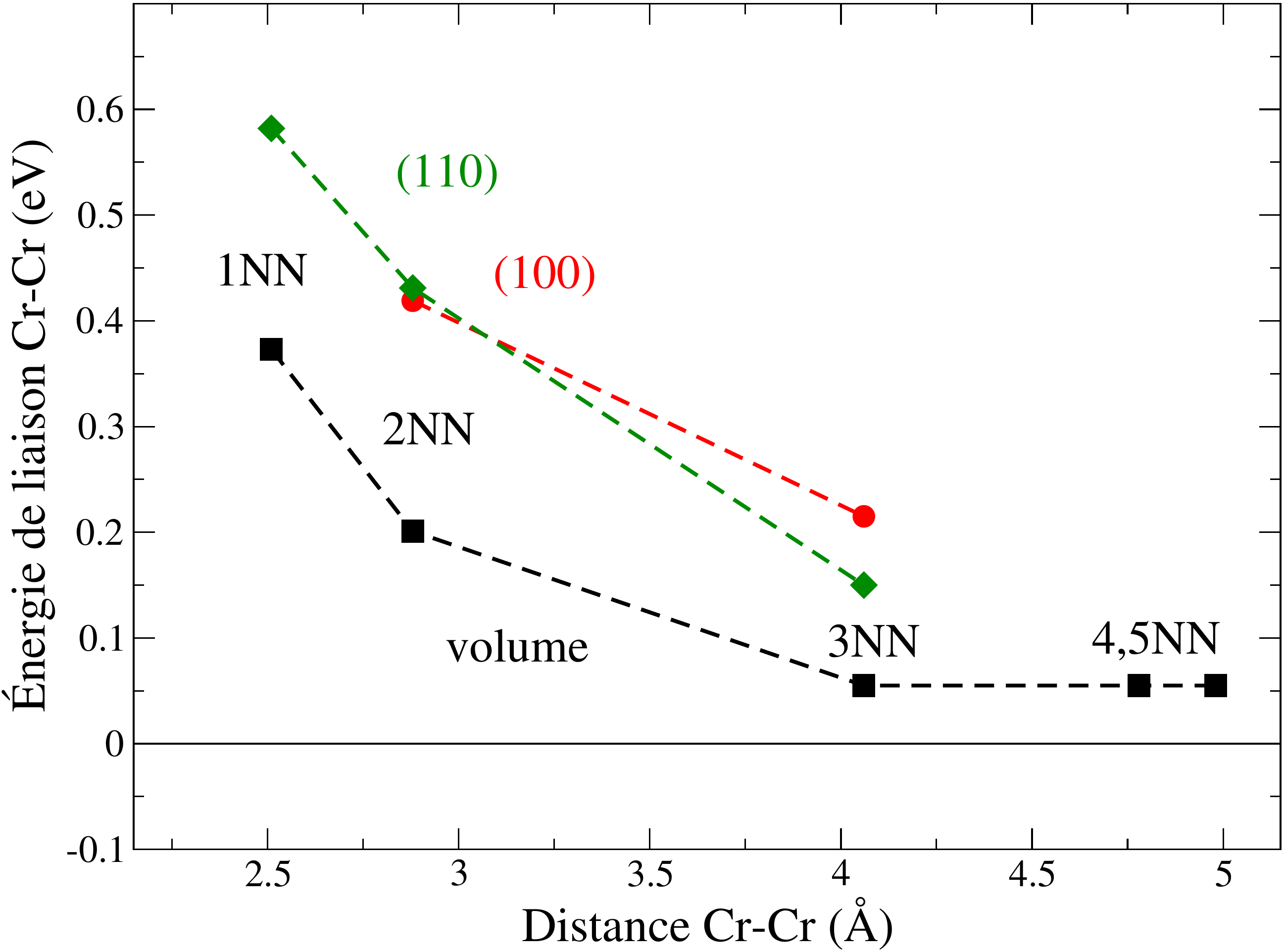}
\par\end{centering}

\caption{Énergie d'interaction Cr--Cr (noir) en volume , (rouge) dans Fe $\left(100\right)$,
(vert) dans Fe $\left(110\right)$. Il n'y a pas de premiers voisins
dans une surface (100).\label{fig:Erepulsion_CrCr_surf}}
\end{figure}

Les énergies de liaison Cr--Cr calculées selon le bilan \ref{eq:def_Ebind_CrCrsurf}
dans les orientations $\left(100\right)$ et $\left(110\right)$ sont
représentées sur la figure \ref{fig:Erepulsion_CrCr_surf}.

Comme nous l'indiquions au paragraphe \ref{sub:La-singuli=0000E8re-orientation100},
il n'y a pas de premiers voisins dans le plan dans le cas d'une surface
cubique centrée $\left(100\right)$. C'est la raison pour laquelle
$E_{liaison}^{(hkl)1}$ n'apparaît pas sur la figure \ref{fig:Erepulsion_CrCr_surf}.

Alors que les énergies de liaison Cr--Cr dans le volume de Fe sont
déjà largement répulsives (voir figure \ref{fig:EbindCrCr}), on observe
qu'elles sont encore supérieures d'environ 0.2 eV par liaison dans
les surfaces.

\subsection{Trois impuretés Cr dans Fe(100)}

\begin{figure}[h]
\begin{centering}
\includegraphics[scale=0.3]{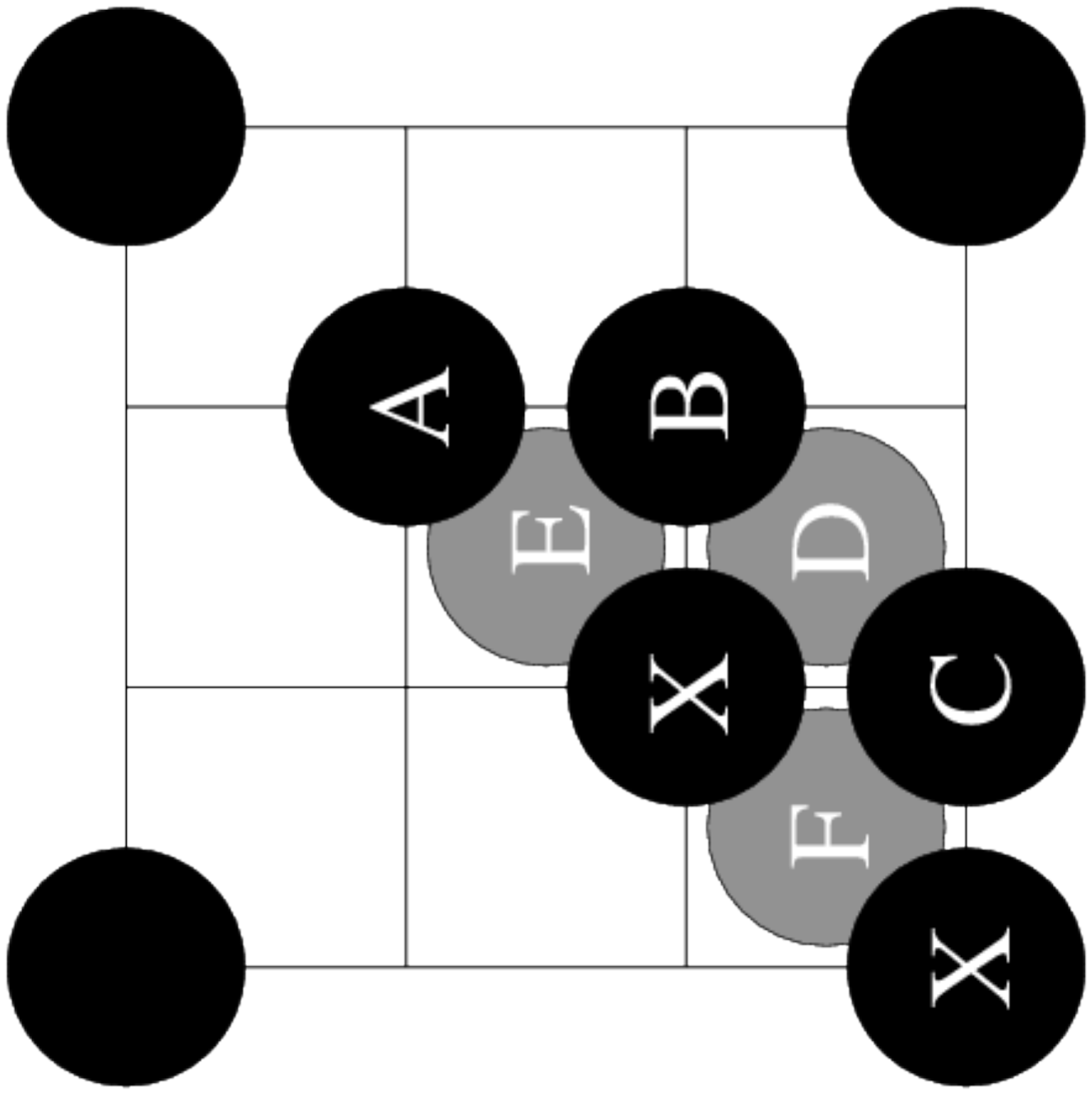}
\par\end{centering}

\caption{Ensemble des configurations possibles pour 3 atomes de Cr dans la
surface Fe $\left(100\right)$. Les atomes de Fe ne sont pas représentés.
Les atomes de Cr noirs sont dans le plan de surface ($i=0$). Deux
des atomes de Cr, notés X, sont communs à toutes les configurations.
Les atomes de Cr gris sont dans le plan sous-jacent ($i=1$). Les
6 configurations correspondent aux configurations XXA à XXF.\label{fig:positions_3Cr_dans_Fe100}}
\end{figure}

Pour trois atomes de Cr dans le plan de surface Fe $\left(100\right)$,
c'est-à-dire pour une concentration volumique de 2.4\,\%\,Cr, le
nombre de configurations que peuvent prendre ces impuretés augmente
largement. Ainsi, pour nos calculs utilisant des super-cellules à
9 atomes par plan pour l'orientation $\left(100\right)$, 7 configurations
sont possibles. Elles sont représentées sur la figure \ref{fig:positions_3Cr_dans_Fe100}.
Elles correspondent, dans les cas notés A, B et C, à 3 atomes de Cr
directement dans le plan de surface (3 atomes noirs de la figure \ref{fig:positions_3Cr_dans_Fe100},
dont l'un est A, B ou C), c'est-à-dire à une concentration surfacique
en Cr $c_{0}=0.33$ et une concentration du plan sous-jacent $c_{1}=0$.
Si au lieu de choisir le troisième atome parmi A, B ou C nous choisissons
D, E ou F, alors il y a deux atomes dans le plan de surface et un
atome dans le plan sous-jacent, pour une concentration $c_{0}=0.22$
et $c_{1}=0.11$.

\begin{figure}[h]
\begin{centering}
\includegraphics[scale=0.4]{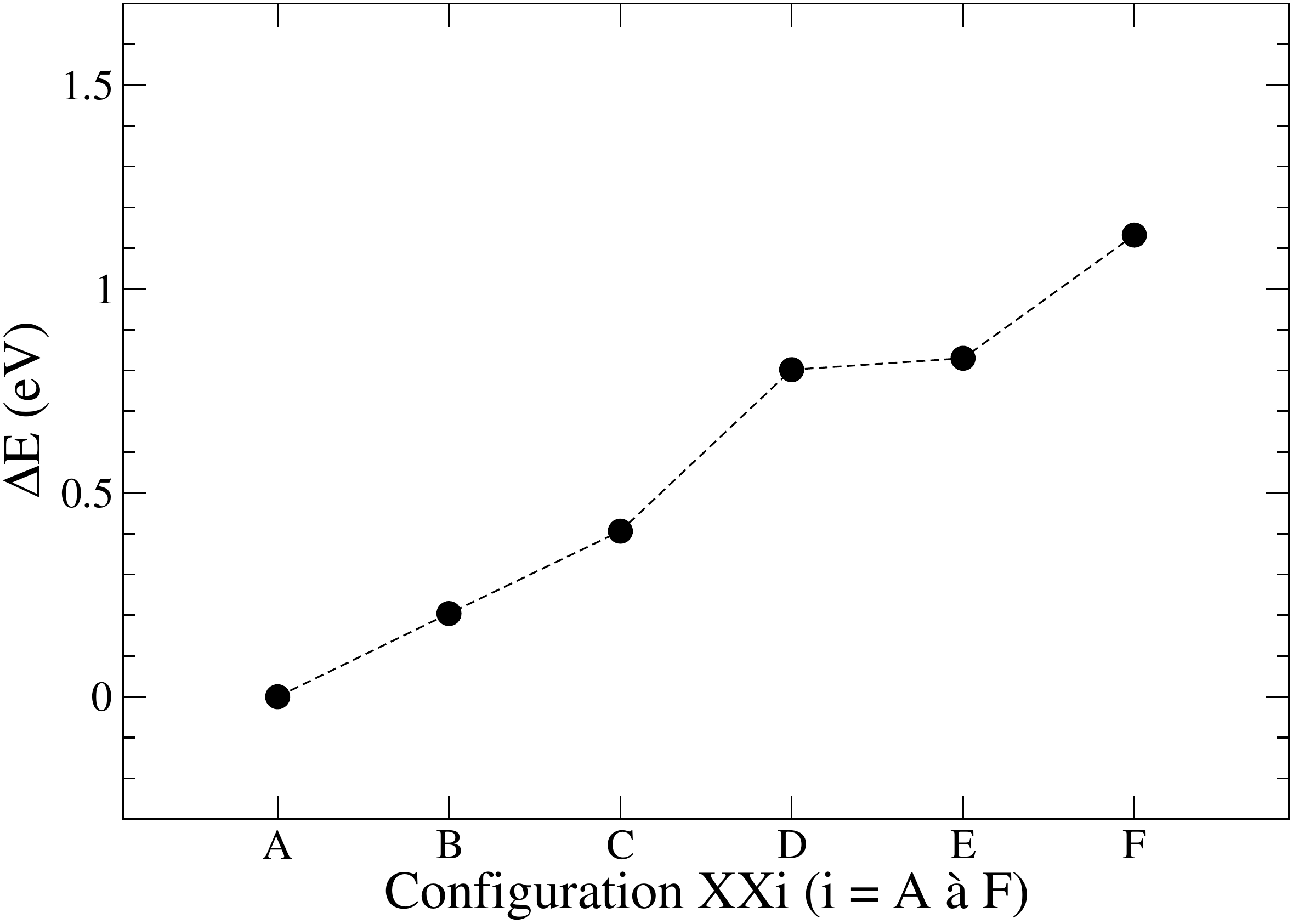}
\par\end{centering}

\caption{Différence d'énergie entre 3 Cr dans une surface de Fe(100) dans les
positions décrites dans la figure \ref{fig:positions_3Cr_dans_Fe100}.
L'énergie totale de la configuration A fait référence en énergie.\label{fig:Energie_de_3_Cr_dans_Fe100}}
\end{figure}

La différence d'énergie $\Delta E$ entre ces sept configurations
est représentée sur la figure \ref{fig:Energie_de_3_Cr_dans_Fe100}.
La référence en énergie est la configuration A qui est la configuration
de plus basse énergie totale.

La configuration XXA est la position dans laquelle les atomes de Cr
sont les plus éloignés les uns des autres : les liaisons Cr--Cr ne
sont qu'entre 3$^{\text{e}}$ voisins. Dans le cas où le troisième
atome serait en position D, E ou F, il serait dans le plan sous-jacent
à la surface, créant des frustrations entre Cr premiers voisins. Plus
les Cr sont éloignés les uns des autres, plus la structure est stable.

Toutes les configurations dans lesquelles existe une liaison Cr--Cr
aux premiers voisins, c'est-à-dire les configurations D, E et F, sont
nettement plus hautes en énergies (de $\approx0.5$ à presque 1 eV).
Ces configurations sont donc très largement défavorables.

\subsection{Monocouche de Cr sur Fe $\left(100\right)$ et $\left(110\right)$}

Malgré les possibilités offertes par SIESTA-NC, il n'est pas possible,
avec les puissances de calcul disponibles actuellement, d'envisager
l'étude systématique du remplissage de la surface par Cr en augmentant
progressivement le nombre d'atomes de Cr. En effet, au dessus de 3
impuretés, le nombre de configurations devient trop grand. Nous pouvons
par contre étudier la monocouche de Cr sur Fe $\left(100\right)$
et $\left(110\right)$, à l'autre extrêmité du domaine de concentration
en surface. Les concentrations surfaciques sont donc $c_{0}=1$, pour
7\,\%\,Cr en volume, pour les deux orientations.

Rappelons que les surfaces de Fe $\left(100\right)$ et $\left(110\right)$
sont modélisées par des empilements de 14 plans avec respectivement
9 et 18 atomes par plan.

\begin{figure}[H]
\begin{centering}
\includegraphics[scale=0.4]{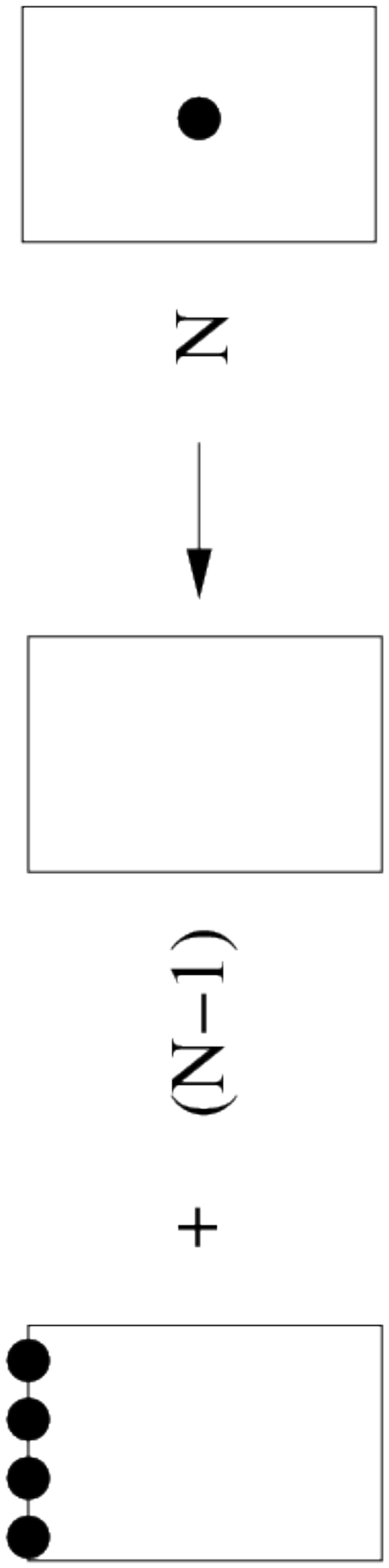}
\par\end{centering}

\caption{Bilan pour la formation d'une monocouche de Cr (en noir) sur une surface
de Fe $\left(100\right)$ ou $\left(110\right)$. Les atomes de Fe
ne sont pas représentés.\label{fig:Bilan-pour-la_monocouche}}
\end{figure}

Le calcul de l'énergie de formation d'une telle monocouche n'est pas
unique : il dépend du choix de la référence en énergie. Selon le bilan
représenté sur la figure \ref{fig:Bilan-pour-la_monocouche}, la formation
de la monocouche entière à partir d'impuretés de Cr infiniment diluées
dans la matrice nécessite environ 1 eV pour les deux orientations.

Il est moins coûteux en énergie de former une monocouche atomique
de Cr sur Fe $\left(100\right)$ que sur Fe $\left(110\right)$. Ceci
est dû à l'absence d'interactions Cr--Cr premiers voisins très coûteuses
en énergie dans un plan $\left(100\right)$. Ainsi, les répulsions
entre Cr restent limitées entre deuxièmes voisins dans Fe $\left(100\right)$,
contrairement à Fe $\left(110\right)$.

La discussion est cependant ici seulement qualitative. Il est important
de garder à l'esprit que créer une mono-couche de Cr revient à remplacer
une des deux surfaces de Fe $\left(hkl\right)$ par :
\begin{itemize}
\item une surface de Cr $\left(hkl\right)$,
\item une interface Fe$\left(hkl\right)$/Cr$\left(hkl\right)$.
\end{itemize}
Or, nous avons vu au chapitre paragraphe \ref{sub:Energies-de-surface}
que l'énergie de surface de Cr$\left(hkl\right)$, $\gamma_{Cr}^{100}$,
est environ $0.3$\,J/m$^{2}$ supérieure à $\gamma_{Fe}^{100}$.
Il s'agit donc ici de faire une comparaison entre d'une part la ségrégation
de Cr en surface (relativement favorable du côté très dilué en Cr)
et d'autre part la création de liaisons Cr--Cr en surface (très défavorable)
et d'une interface Fe/Cr.

\begin{figure}[h]
\begin{centering}
\includegraphics[scale=0.55]{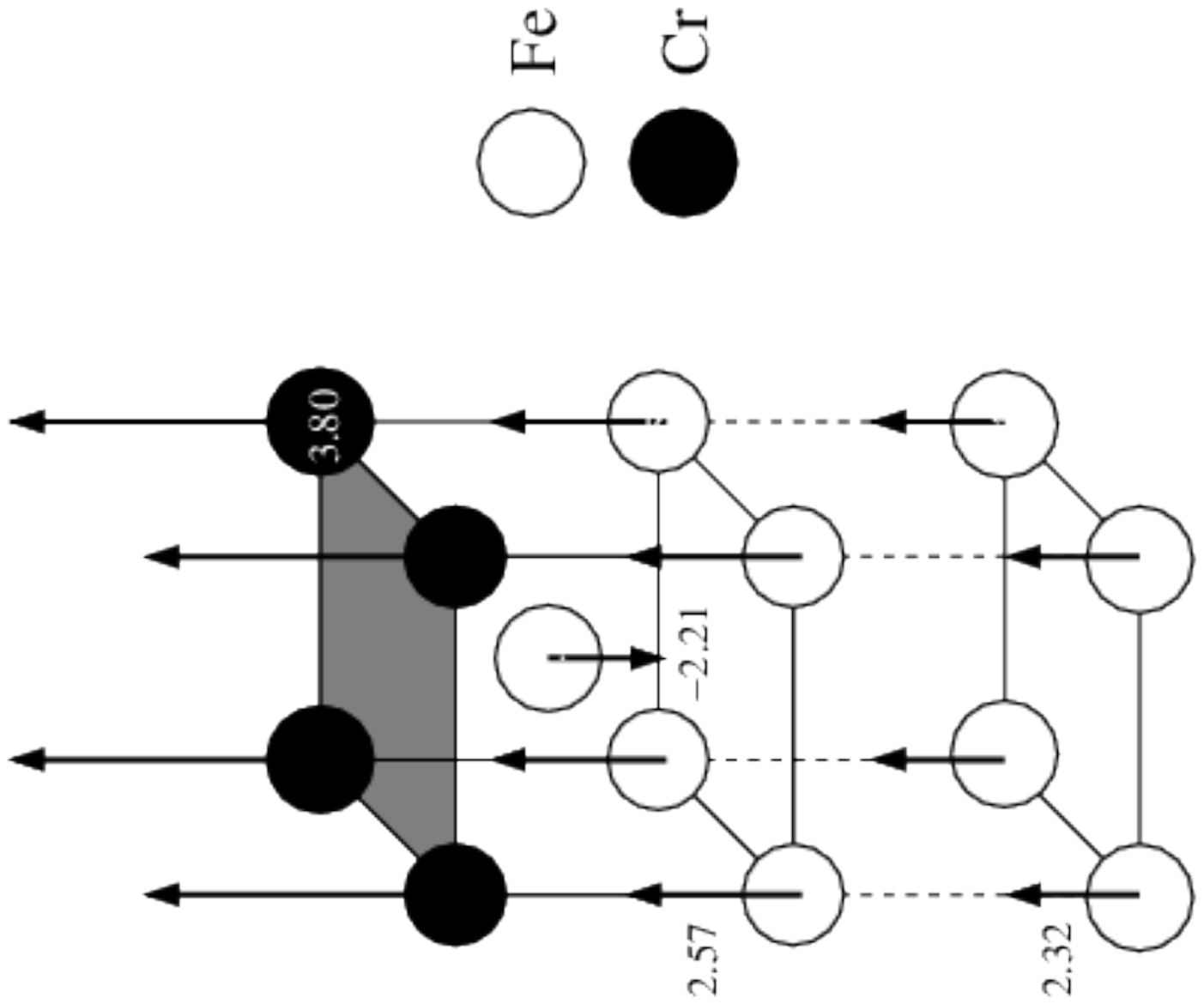}
\par\end{centering}

\caption{Configuration magnétique la plus stable calculée avec SIESTA-NC d'une
monocouche de Cr sur Fe $(100)$. Les moments magnétiques atomiques
sont indiqués en magnétons de Bohr.\label{fig:Monocouche-de-Cr_sur_Fe_100}}
\end{figure}

Dans l'orientation $\left(100\right)$, les atomes de Cr sont ferromagnétiques
dans le plan, portant un moment anti-parallèle à ceux des atomes de
Fe du plan sous-jacent. Cette configuration magnétique est représentée
sur la figure \ref{fig:Monocouche-de-Cr_sur_Fe_100}. Les Cr ont tous
un moment magnétique de $-3.80$\,$\mu_{B}$. Les Fe du plan sous-jacent
($i=1$) ont un moment magnétique local de $2.21$\,$\mu_{B}$ alors
que ceux de la troisième couche ($i=2$) ont un moment de $2.57$\,$\mu_{B}$.
Les moments des couches suivantes décroissent quasi-linéairement jusqu'à
la sixième couche où ils retrouvent un moment magnétique de volume
classique de $2.22$\,$\mu_{B}$.

Le cas de Cr sur Fe $\left(110\right)$ est plus complexe. Deux configurations
magnétiques sont possibles à l'état fondamental : antiferromagnétisme
ou ferromagnétisme dans le plan. Nous trouvons la configuration antiferromagnétique
dans le plan plus stable de $0.147$\,eV par atome de Cr. Les Cr
portent des moments de $+2.64$ ou $-2.79$\,$\mu_{B}$ selon qu'ils
sont alignés ou anti-alignés aux atomes de Fe sous-jacents.

\subsection{Effet de la concentration en chrome}

\begin{figure}[h]
\begin{centering}
\includegraphics[scale=0.4]{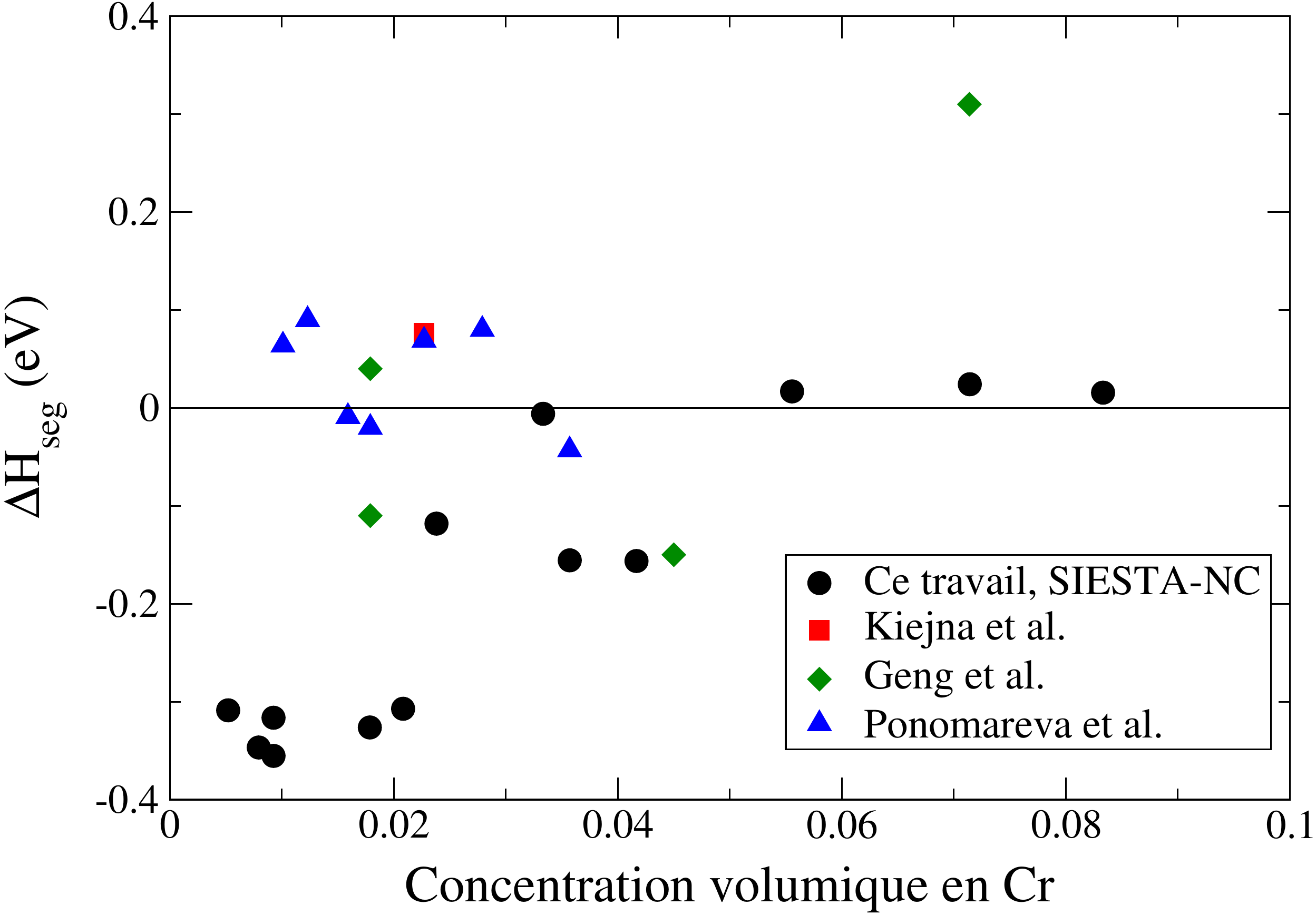}
\par\end{centering}

\caption{Énergies de ségrégation de Cr dans Fe $\left(100\right)$ pour différentes
concentrations volumiques déduites des super-cellules utilisées. (1)
Ce travail (2) selon Kiejna et al. \cite{kiejna_Eseg_2008} (3) selon
Geng et al. \cite{geng_segGGAFLAPW_2003} (4) selon Ponomareva et
al. \cite{ponomareva_HsegFe100_2007}.\label{fig:Hseg_c}}
\end{figure}

Nous présentions ci-dessus une étude \emph{ab initio} de l'énergie
de ségrégation du chrome dans le fer à 0\,K. Ces calculs sont toujours
réalisés à concentration volumique faible. Nous étudions cependant
le domaine entier des concentrations surfaciques. Ces calculs, ainsi
que ceux issus de la littérature sont résumés dans le tableau \ref{tab:Hseg_impureteCr}
pour l'orientation $\left(100\right)$. Sur la figure \ref{fig:Hseg_c},
nous représentons les énergies de ségrégation calculées en fonction
de la concentration volumique déduite de la taille de la super-cellule
de calcul. 

Les calculs SIESTA-NC montrent un changement de signe de l'énergie
de ségrégation avec la concentration volumique. En-dessous de 3 à
5\,\%\,Cr, l'énergie de ségrégation est négative. Le chrome a tendance
à ségréger en surface. Au-dessus, l'énergie de ségrégation devient
très légèrement positive. L'impureté reste donc en volume. Il faut
cependant rester prudant du fait de la surestimation de l'énergie
de ségrégation par SIESTA-NC.

Il s'agit de la première étude à si faibles concentrations surfaciques
et volumiques. Les super-cellules de calcul de Kiejna, Geng, Ponomareva
et de leurs collaborateurs sont tellement concentrées en surface qu'il
ne s'agit plus d'études de la ségrégation d'impureté \cite{kiejna_Eseg_2008,geng_segGGAFLAPW_2003,ponomareva_HsegFe100_2007}.
Cependant, il convient de considérer l'effet des pseudo-potentiels
à norme conservée qui surestiment l'effet du magnétisme et ses conséquences
sur la ségrégation.

La conclusion est donc qualitative. D'une part, il est nécessaire
de relire de façon critique la littérature du fait de l'utilisation
de super-cellules de calcul trop petites latéralement. D'autre part,
l'énergie de ségrégation d'une impureté de chrome dans le fer dépend
de concentration globale en chrome. Les données \emph{ab initio }ne
peuvent dans ce cas précis pas éclairer la situation car les tendances
ne sont pas nettes et l'influence des approximations de calcul n'est
pas négligeable.

\section{Conclusions}

Dans ce chapitre, nous avons décrit une étude \emph{ab initio} des
éléments fer et chrome, de leurs alliages et de leurs surfaces libres.

\subsection*{Générales}

L'un des objectifs de ce chapitre était de quantifier au mieux l'effet
des approximations des calculs ab initio, en particulier des pseudopotentiels.
Le système fer-chrome est particulièrement complexe. En particulier,
ses propriétés magnétiques le rendent sensible aux pseudopotentiels.
Une lecture critique des résultats \emph{ab initio} est donc particulièrement
nécessaire, dans ce manuscrit comme dans la littérature. Sous cet
éclairage, nous pouvons faire les conclusions suivantes.

Une fonctionnelle d'échange et corrélation de type GGA est nécessaire
pour reproduire l'état fondamental des éléments fer et chrome cubiques
centrés. Fe est ferromagnétique et Cr antiferromatique commensurable,
ce qui constitut une approximation de l'onde de spin incommensurable
observée dans le chrome pur. L'approximation des pseudo-potentiels
a un effet faible sur le volume d'équilibre des éléments. Le moment
magnétique des atomes de Cr est très sensible au volume atomique,
ce qui le rend très variable selon l'approximation des pseudo-potentiel
mise en œuvre dans le calcul.

\subsection*{En volume}

Nous confirmons que l'énergie de mélange de l'alliage change de signe
avec la concentration en chrome. La concentration pour laquelle ce
changement de signe se produit dépend légèrement des approximations
de calcul. Au-dessous d'environ 10\,\%\,Cr, l'addition de Cr à la
solution solide $\alpha$ est favorable. Ceci est dû à la mise en
solution favorable de Cr dans Fe du fait d'interactions magnétiques
antiferromagnétiques du chrome dilué dans le fer. L'énergie de mise
en solution dépend de l'amplitude des moments magnétiques, c'est-à-dire
de l'approximation des pseudo-potentiels. Au-delà de cette concentration
seuil, les interactions répulsives à longue portée entre atomes de
Cr déstabilisent le mélange. L'énergie de mélange devient alors positive.

La mise en solution du fer dans la solution solide $\alpha'$ riche
en chrome est défavorable. Le moment magnétique de l'impureté de fer
est alors presque nul, ce qui est très coûteux en énergie.

Les éléments Fe et Cr ont des paramètres de maille très proches. Les
relaxations sont en conséquence très faibles. Cela nous permet d'envisager
un modèle sur réseau rigide dans la suite de ce travail.

Le moment magnétique de l'atome de chrome est très sensible à son
environnement chimique local, ce qui n'est pas le cas du fer.

\subsection*{En surface}

Quelle que soit l'orientation, il est plus facile de créer une unité
de surface de fer que de chrome.

Les surfaces renforcent les phénomènes magnétiques par rétrecissement
de la bande $d$ qui est la principale responsable du magnétisme dans
le solide. Le magnétisme est alors particulièrement important pour
les surfaces. On calcule des énergies de surfaces plus faibles que
ne l'attendaient les expérimentateurs après observations à haute température
où il n'y a pas d'effet fort du magnétisme. Le moment magnétique du
chrome est plus sensible à la surface que le fer. En particulier,
l'orientation $\left(100\right)$ du chrome, et dans une moindre mesure
du fer, est particulièrement stabilisée par un ferromagnétisme dans
le plan.

Les phénomènes de ségrégation du chrome dans le fer sont également
liés au magnétisme à l'échelle atomique. La ségrégation de Cr dans
les surfaces de Fe est quasiment négligeable. Selon les approximations,
la tendance peut pourtant changer. Il y a tendance à la ségrégation
en SIESTA-NC alors que les calculs PWSCF-PAW plus robustes montrent
que la ségrégation est défavorable. Ces différences sont liées à la
surestimation du magnétisme par les pseudo-potentiels à norme conservée.
Le magnétisme étant un facteur stabilisant du chrome dilué en surface,
les pseudo-potentiels NC surestiment la stabilisation de Cr dilué
ce plan.

La répulsion entre atomes de chrome est plus forte en surface qu'en
volume car le moment magnétique y est plus fort. Il est très coûteux
en énergie de ségréger plusieurs atomes de chrome en surface. La mono-couche
de Cr sur Fe est ainsi très peu probable en température.

Du fait de cette répulsion très forte entre atomes dans le plan de
surface, il est important de tenir compte de la taille latérale de
la super-cellule, c'est-à-dire de sa concentration surfacique. Les
énergies de ségrégation présentées dans la littérature sont souvent
obtenues pour des concentrations surfaciques de 25 ou 50\,\%\,Cr,
ce qui représente déjà de très fortes interactions répulsives en surface.\\

\chapter{Propriétés thermodynamiques de l'alliage\label{cha:thermo_volume}}

\malettrine{C}{}e troisième chapitre est voué à la construction d'un
hamiltonien pour modéliser les propriétés thermodynamique de volume
de la solution solide fer--chrome cubique centrée. Cet hamiltonien
doit permettre de calculer des limites de solubilité dans tout le
domaine de température et de servir de base pour le calculs d'isothermes
de ségrégation superficielle (chapitre 4). À plus long terme, notre
but est de l'utiliser pour des études cinétiques. Il doit donc reproduire
l'essentiel des résultats \emph{ab initio} à 0\,K et les limites
de solubilité à haute température mesurées expérimentalement, mais
également rester efficace numériquement.

Nous commençons ce chapitre par une revue des modèles thermodynamiques
proposés pour la solution solide fer--chrome cubique centrée, en commençant
par les modèles classiques, puis en montrant leurs limites quant à
la prise en compte du magnétisme à l'échelle atomique. Les modèles
récemment proposés pour pallier à ces limites sont discutés.

Nous proposons ensuite une approche phénoménologique simple, basée
sur l'hamiltonien bien connu d'Ising \cite{Ising_1925}. Par son efficacité
numérique, celle-ci permet le calcul des limites de solubilité dans
tout le domaine de concentration et température et d'envisager les
calculs cinétiques. Nous donnons aux constantes de couplage une dépendance
en concentration locale ajustée sur les résultats \emph{ab initio}
présentés au chapitre précédent ainsi qu'une dépendance en température
ajustée sur les mesures expérimentales de limites de solubilité à
haute température.

Le diagramme de phases du système fer--chrome est d'abord calculé
dans le cadre d'une approximation de champ moyen de point (approximation
de Bragg-Williams) qui sera aussi utilisée pour le calcul des isothermes
de ségrégation dans le chapitre suivant.

Ensuite, nous réalisons des simulations Monte Carlo pour valider et
étudier les effets qui ne sont pas pris en compte dans l'approximation
de champ moyen de point : effet de la portée des interactions et de
l'étendue de la zone sur laquelle on définit la concentration locale.\vspace{1cm}

\section{Les modèles existants}

Depuis la découverte de la fragilisation à basse température, de nombreux
modèles énergétiques ont été proposés pour le calcul du diagramme
de phases de l'alliage fer--chrome.

Les premiers modèles sont empiriques et macroscopiques. Certains tiennent
compte de l'effet du magnétisme aux températures proches des températures
de Curie et Néel \cite{williams_miscibility_1974,andersson_thermodynamic_1987,hertzman_thermodynamic_1983,chuang_thermodynamic_1987,lin_thermodynamic_1987},
mais les calculs \emph{ab initio} ont mis en évidence l'effet du magnétisme
à l'échelle atomique (chapitre 2, \cite{olsson_ab_2003,klaver_magnetism_2006}).
D'abord, ces calculs montrent que l'énergie de mélange de l'alliage
est contrôlée par des couplages complexes entre interactions chimiques
et magnétiques. Il a ainsi été observé le rôle des frustrations entre
moments magnétiques des atomes de Cr, voulant à la fois être anti-parallèles
entre eux et anti-parallèles aux moments magnétiques des atomes de
fer de la matrice. Ils ont également mis en évidence l'importance
de la dépendance en composition chimique locale du moment magnétique
du chrome. C'est cette dépendance, difficile à prendre en compte dans
les modèles simples, que différents auteurs ont tenté d'implémenter
dans des hamiltoniens originaux. Nous les décrivons ci-dessous.

\subsection{Les modèles macroscopiques empiriques}

\subsubsection*{Le modèle de solution régulière perturbé par une composante magnétique}

Comme nous l'indiquons au chapitre bibliographique de ce manuscrit,
les mesures expérimentales se sont limitées, jusque dans les années
1970, aux températures hautes. Les limites de solubilité alors mesurées
sont en accord avec le modèle de la solution régulière de Gibbs :
l'enthalpie de mélange est quadratique \cite{masumoto_chaleurspe_1953,backhurst__1958,becket__1938,dench_adiabatic_1963,kubaschewski_calc_miscibily_1965,kubaschewski_thermodynamics_1960,martens_heat_1956,williams__1957,williams_further_1958}.

Dès 1974, Williams fait l'hypothèse d'un écart au modèle de solution
régulière dû au mélange ferromagnétique--antiferromagnétique \cite{williams_miscibility_1974}.
Il propose de décomposer l'enthalpie de mélange $\Delta H$ comme
la somme de l'enthalpie de mélange de la solution régulière $\Delta H_{mix}^{r\acute{e}g}$
et d'une perturbation liée aux phénomènes magnétiques (seulement supputés
à l'époque) $\Delta H_{mix}^{mag}$ :
\begin{equation}
\Delta H_{mix}=\Delta H_{mix}^{r\acute{e}g}+\Delta H_{mix}^{mag}
\end{equation}
où l'enthalpie de mélange magnétique $\Delta H_{mix}^{mag}$ est proportionnelle
à la température de Curie $T_{C}$ de l'alliage, elle-même proportionnelle
à la teneur en Cr de l'alliage :

\begin{equation}
\Delta H_{mix}^{mag}\propto T_{C}\propto-x_{Cr}.
\end{equation}

\begin{figure}[h]
\begin{centering}
\includegraphics[scale=0.4]{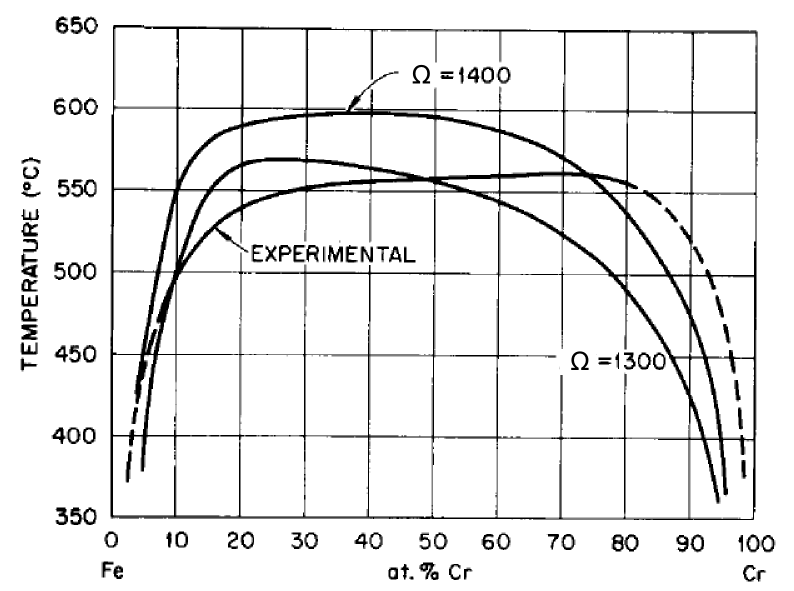}
\par\end{centering}

\caption{Comparaison entre la limite de solubilité expérimentale \cite{williams__1957,williams_further_1958}
(trait plein, traits pointillés pour les extrapolations depuis les
hautes températures) et déduites du modèle de Williams \cite{williams_miscibility_1974}
pour une énergie d'ordre $\Omega$ de 1300 et 1400\,$\unit{cal/g}$.
Dans ce modèle, l'enthalpie de mélange est décrite comme la somme
d'une contribution chimique régulière et d'une contribution magnétique
dépendant de la température de Curie de l'alliage. Figure de Williams
\cite{williams_miscibility_1974}.\label{fig:limite_sol_williams_1974-1}}
\end{figure}

Le rapport entre l'enthalpie de mélange régulière et la perturbation
magnétique $\left(\Delta H_{mix}^{reg}/\Delta H_{mix}^{mag}\right)$
est choisi empiriquement pour obtenir une température critique proche
de la mesure de 830\,K à 50\,\%\,Cr \cite{williams__1957,williams_further_1958}.
La lacune de miscibilité calculée est représentée sur la figure \ref{fig:limite_sol_williams_1974-1}
pour différentes énergies d'ordre $\Omega$ définies par :
\begin{equation}
\Delta H_{mix}^{r\acute{e}g}=-x_{Cr}x_{Fe}\Omega
\end{equation}

La contribution magnétique applatit le haut de la lacune de miscibilité
de la solution régulière et stabilise la solution solide $\alpha$'
riche en Cr dans tout le domaine de température. On note également
que les solubilités réciproques tendent vers zéro à 0\,K. Or, les
calculs \emph{ab initio} récents montrent que l'enthalpie de mise
en solution de Cr dans Fe est favorable à 0\,K, au contraire de Fe
dans Cr. Ils semblent donc indiquer que ce serait la solution solide
$\alpha$ qui serait stabilisée à basse température par le magnétisme.

Ce modèle de solution régulière perturbée par une contribution magnétique
est à notre connaissance le premier à souligner un effet du magnétisme
sur la lacune de miscibilité.

\subsubsection*{Calphad}

La méthode numérique CALPHAD\nomenclature{CALPHAD}{CALculation of PHAse Diagrams -- Calcul de diagrammes de phases}
(Calculation of Phase Diagrams) est une méthode de minimisation de
l'enthalpie libre pour le calcul de diagrammes de phases. L'enthalpie
libre est ajustée sur des observations expérimentales (limites de
solubilité, mesures calorimétriques, magnétisme ...) \cite{schmid-fetzer_CALPHADassessment_2007,saunders_CALPHAD_1998}
ou des calculs \emph{ab initio}\cite{dupin_calphadabinitio_2001}.

Comme Williams, Sundman, Hertzman et Andersson \cite{hertzman_thermodynamic_1983,andersson_thermodynamic_1987},
Lin et al. \cite{chuang_thermodynamic_1987,lin_thermodynamic_1987}
introduisent une composante magnétique à l'enthalpie libre de mélange
de l'alliage. Celle-ci s'exprime :

\begin{equation}
\Delta G_{mix}=\Delta H_{mix}-T\Delta S_{mix}+\Delta G_{mo}\label{eq:sundman1}
\end{equation}
 où $\Delta H_{mix}$ et $\Delta S_{mix}$ sont les termes d'enthalpie
et d'entropie de configuration classiques. $\Delta G_{mo}$ est une
contribution liée à la mise en ordre magnétique. Cette contribution
s'exprime ainsi :
\begin{equation}
\Delta G_{mo}=RT\ln\left(M+1\right)f\left(\tau\right)\label{eq:sundman2}
\end{equation}
où $\tau$ est le rapport $\frac{T}{T_{C}}$ entre la température
et la température de Curie de l'alliage en Kelvin, et $M$ est le
moment magnétique moyen par atome en magnétons de Bohr ($\mu_{B}$).
$f$ est un développement empirique en puissances de $\tau$ proposé
par Hillert et Jarl \cite{hillert_model_1978}. La fonction $\tau$
est différente de part et d'autre de la température de Curie de l'alliage
$T_{C}$. Cette dernière et le moment magnétique moyen par atome $M$
sont exprimés empiriquement par :
\begin{equation}
M=x_{Fe}M\left(Fe\right)+x_{Cr}M\left(Cr\right)+x_{Fe}x_{Cr}M^{0}
\end{equation}

\begin{equation}
T_{C}=x_{Fe}T_{C}\left(Fe\right)+x_{Cr}T_{N}\left(Cr\right)+x_{Fe}x_{Cr}\left[T_{C}^{0}+\left(x_{Cr}-x_{Fe}\right)T_{C}^{1}\right]
\end{equation}
où $T_{C}\left(Fe\right)$ et $T_{N}\left(Cr\right)$ sont les températures
de Curie et de Néel du fer et du chrome purs. $M\left(Fe\right)$
et $M\left(Cr\right)$ sont les moments magnétiques atomiques du fer
pur ferromagnétique et du chrome pur antiferromagnétique, et $T_{C}^{0}$,
$T_{C}^{1}$ et $M^{0}$ sont ajustés sur les limites de solubilité
et la température de Curie expérimentales de l'alliage. 

\begin{figure}[h]
\begin{centering}
\includegraphics[scale=0.5]{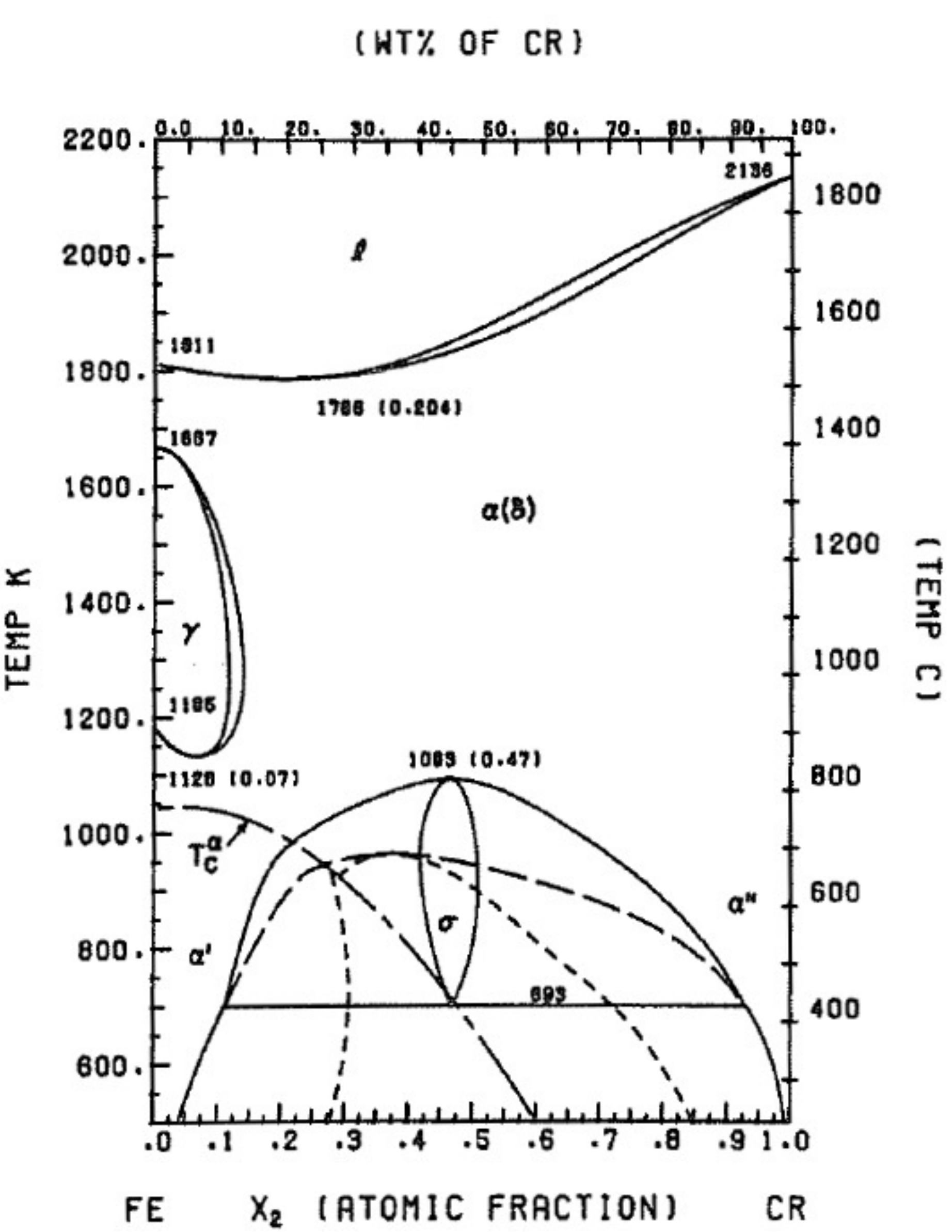}
\par\end{centering}

\caption{Diagramme de phases calculé par Chuang, Lin et al. \cite{chuang_thermodynamic_1987,lin_thermodynamic_1987}
avec une paramétrisation CALPHAD tenant compte d'une enthalpie libre
de mélange magnétique. Les limites de solubilité métastable sont en
traits pointillés longs. Les limites spinodales sont en traits pointillés
serrés. \label{fig:ddp_lin_1987-1}}
\end{figure}

Nous représentons sur la figure \ref{fig:ddp_lin_1987-1} le diagramme
de phases complet Fe--Cr déduit par l'ajustement CALPHAD de Chuang,
Lin et al. \cite{chuang_thermodynamic_1987,lin_thermodynamic_1987}.
La solubilité de Cr dans Fe y est supérieure à celle de Fe dans Cr
à haute température. Ils trouvent une température critique de 965\,K
à 40\,\%\,Cr ainsi qu'une large asymétrie du domaine biphasé $\alpha+\alpha'$,
comme des limites de décomposition spinodale.

Les diagrammes de phases produits par les modèles CALPHAD font actuellement
référence auprès des technologues. Cependant, les observations expérimentales
sont rares en-dessous de 600\,K. Toutes les fonctions constitutives
de l'enthalpie libre de l'alliage ajustées sur ces mesures à haute
température sont extrapolées aux basses températures. Elles ne reproduisent
pas le changement de signe de l'énergie de mélange calculée \emph{ab
initio} à 0\,K. Le diagramme de phases calculé est donc asujetti
à la validité de l'extrapolation.

Très récemment, après la soutenance de ce travail de thèse a été publié
par Bonny et al. un nouvel ajustement CALPHAD qui montre une solubilité
du chrome dans le fer supérieure à basse température par la prise
en compte d'une enthalpie de mélange qui change de signe avec la concentration
en chrome \cite{bonny_calphad_2010}.

Enfin, notons que les modèles CALPHAD ne sont pas atomistiques.

\subsection{Les potentiels inter-atomiques\label{sub:Les-potentiels-inter-atomiques}}

Les potentiels empiriques permettent de faire des simulations atomistiques
sur réseau relaxé. On peut ainsi calculer un diagramme de phases avec
entropie de vibration ou réaliser des simulations de dynamique moléculaire,
par exemple pour l'étude du mouvement des dislocations \cite{patinet_these_2009}.
Pour les métaux, ces potentiels sont généralement dans le formalisme
de l'atome immergé (EAM, ou \og Embedded Atom Model \fg{} en anglais
\nomenclature{EAM}{Embedded Atom Model -- Modèle de l'atome immergé})
\cite{daw_EAM_1983,daw_EAM_1984}. Ce formalisme tient compte de la
nature à N corps de la cohésion métallique et reproduit par exemple
le renforcement des liaisons restantes quand un atome perd une liaison.
De plus, elle reste assez simple pour être mise en œuvre dans des
simulations comprenant plusieurs millions d'atomes. La forme générale
de ces potentiels s'écrit :
\begin{equation}
E=\dfrac{1}{2}\sum_{i\ne j;r_{ij}<R_{c}}\Phi_{t_{i}t_{j}}\left(r_{ij}\right)+\sum_{i}F_{t_{i}}\left(\bar{\rho}_{i}\right)\;;\;\bar{\rho}_{i}=\sum_{i\ne j;r_{ij}<R_{c}}\rho_{t_{i}}\left(r_{ij}\right)
\end{equation}
 où $t_{i}$ et $t_{j}$ désignent la nature des atomes sur les sites
$i$ et $j$ distants de $r_{ij}$. $R_{c}$ est le rayon de coupure
au-delà duquel l'interaction est considérée nulle. Le premier terme
est une énergie de répulsion de paires. Le deuxième terme est le terme
d'immersion à N corps qui dépend de la densité électronique locale
$\bar{\rho}_{i}$ égale à la somme des densités électroniques atomiques
$\rho_{t_{i}}$.

Dans un alliage binaire, trois potentiels d'interaction sont mis en
jeu, deux homo-atomiques et un hétéro-atomique. Pour l'étude du système
Fe--Cr, le potentiel Fe--Fe (Cr--Cr) est proposé par Ackland et al.
\cite{ackland_mendelev_FeFe_2004} (Wallenius et al. \cite{wallenius_EAM1_2004,wallenius_EAM2_2004}).
Jusqu'à récemment, les potentiels EAM publiés dans la littérature
ne reproduisaient pas le changement de signe de l'énergie de mélange
\cite{farkas_embedded_1996,yifang_eam_1996,konishi_potentielfinnis_1999,chakarova_developpement_2006}
car ils étaient ajustés sur les mesures d'enthalpie de mélange de
l'alliage désordonné à haute température de Dench \cite{dench_adiabatic_1963}.
En 2004, Wallenius et al. \cite{wallenius_EAM1_2004,wallenius_EAM2_2004}
développent un jeu de potentiels EAM  qui reproduit les calculs \emph{ab
initio} d'Olsson et al. \cite{olsson_ab_2003}. Pour cela, il est
nécessaire de faire appel à un potentiel Fe--Cr différent pour chaque
concentration. Finalement, deux nouvelles approches plus efficaces
des potentiels EAM ont été développées : les potentiels à deux bandes
(2BM) et les potentiels dépendants de la concentration (CDM).

\subsubsection*{Les potentiels EAM à deux bandes électroniques (2BM)}

\begin{figure}[H]
\begin{centering}
\includegraphics[scale=0.4]{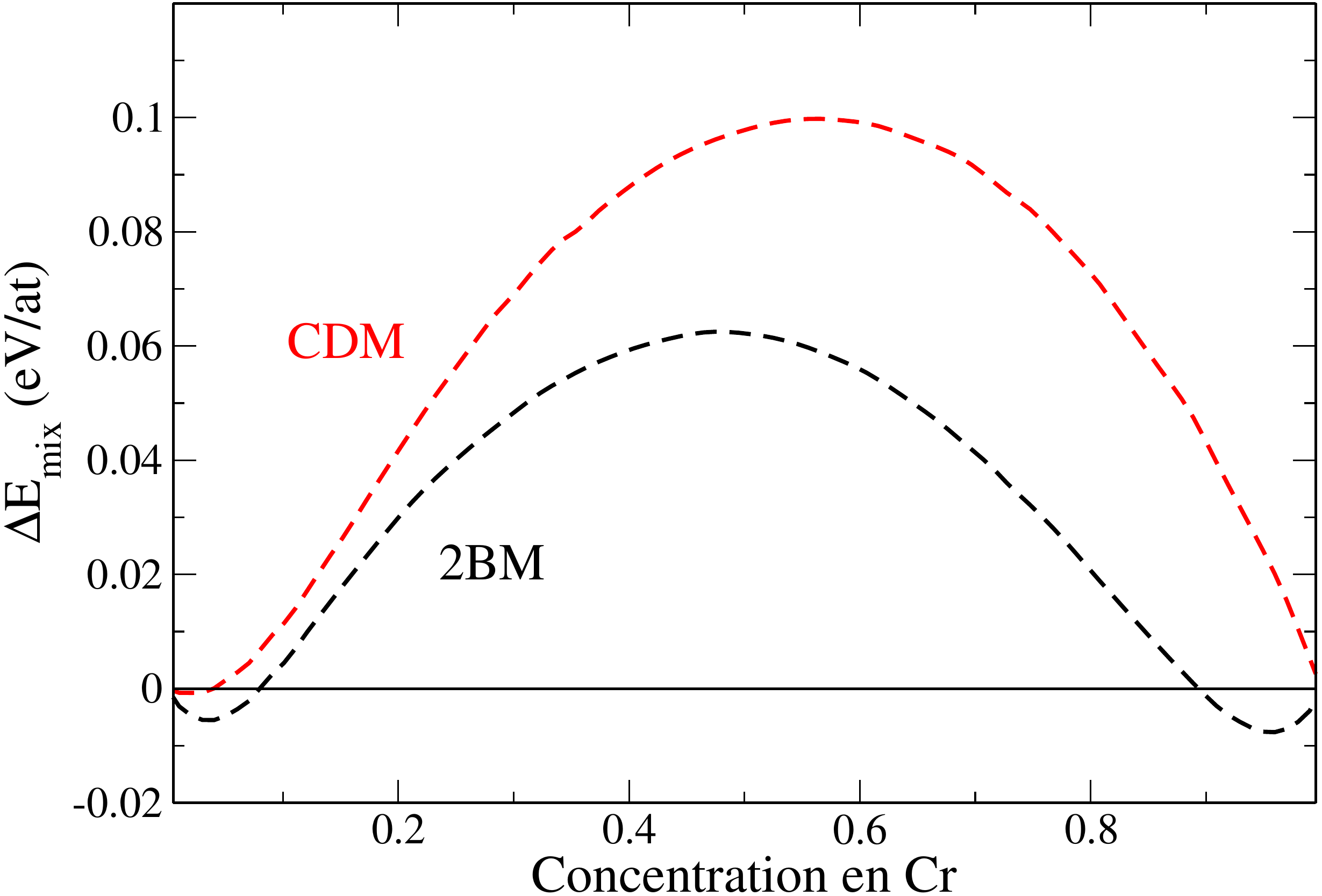}
\par\end{centering}

\caption{Énergies de mélange déduites des potentiel inter-atomiques (noir)
2BM de Olsson et al. \cite{olsson_two-band_2005} et (rouge) CDM de
Caro et al. \cite{Caro_CDM_2005} en fonction de la concentration
en Cr.\label{fig:Emix_2BM_CDM}}

\end{figure}

Dans ce deuxième type de potentiel développé par Ackland et Reed \cite{ackland_two-band_2003,ackland_two-band_2006}
puis appliqué au potentiel hétéro-atomique Fe--Cr par Olsson et al.
\cite{olsson_two-band_2005,olsson_erratum:_2006}, l'énergie de site
du potentiel EAM tient compte de la bande $s$ des éléments Fe et
Cr en plus de la bande $d$ utilisée classiquement par les potentiels
EAM. Celui-ci a alors deux fonctions dépendantes de l'environnement
local par atome : une première pour la bande $s$ et une seconde pour
la bande $d$. Le nouveau degré de liberté introduit par la paramétrisation
de la bande $s$ permet de reproduire le changement de signe de l'énergie
de mélange aux faibles concentrations en Cr. L'énergie de mélange
déduite de ce potentiel est reproduite sur la figure \ref{fig:Emix_2BM_CDM}.

On observe sur la figure \ref{fig:Emix_2BM_CDM} que cette paramétrisation
est très satisfaisante dans le domaine riche en Fe mais moins dans
le domaine riche en Cr. La mise en solution de Fe dans Cr serait exothermique
à 0\,K, en désaccord avec les résultats \emph{ab initio}. Le fer
serait ainsi plus soluble dans le chrome que le chrome dans le fer.
On s'attend donc à un diagramme de phases favorisant, au moins à basse
température, le domaine $\alpha'$ riche en chrome, en contradiction
avec les résultats expérimentaux. Les bons résultats de ce potentiel
du coté riche en fer d'intérêt technologique a permis de nombreuses
études numériques du vieillissement de l'alliage à environ 10\,\%\,Cr
: précipitation, mouvement des dislocations, vieillissement et cascades
de déplacements par simulations Monte Carlo cinétiques et de dynamique
moléculaire \cite{kineticOrMD_A,kineticOrMD_B,kineticOrMD_C,kineticOrMD_D}.
Cependant, les précipités de la phase $\alpha$' ont une concentration
en fer surestimée, et leur contribution à l'énergie totale du système
peut alors être biaisée par ce potentiel donnant lieu à une énergie
de mélange négative à ces concentrations. Ce biais augmente avec le
nombre d'interfaces entre phases riches en fer et en chrome (c'est-à-dire
avec le nombre de précipités) \cite{caro_thermodynamicsFeCu_2005,lopasso_phase_2003}.
De plus, une mauvaise description de la phase $\alpha$' peut induire
une mauvaise description des cinétiques de décomposition $\alpha-\alpha$'.

Très récemment, Bonny, Pasianot et Malerba ont montré qu'il est théoriquement
possible de reproduire une énergie de mélange asymétrique avec des
potentiels EAM dans le formalisme 2BM \cite{bonny_fitting_c_dep_pot_2009}.
Ils ont depuis développé un dernier jeu de potentiel Fe--Cr et Cr--Cr
qui semble prometteur \cite{terentyev_newpot_2010}.

\subsubsection*{Les potentiels EAM dépendant de la concentration (CDM)}

Un autre type de potentiel empirique a été développé parallèlement
par Caro et al. \cite{Caro_CDM_2005,caro_implications_2006}. Le terme
de paire du potentiel EAM est multiplié par un élément dépendant de
la concentration locale, ce qui permet de reproduire le changement
de signe de l'énergie de mélange calculée \emph{ab initio} et de ne
pas être symétrique des deux côtés du diagramme de phases.

L'énergie de mélange calculée en champ moyen de point par Malerba
et al. \cite{Malerba_revue_2008} à partir de ce potentiel inter-atomique
est présentée sur la figure \ref{fig:Emix_2BM_CDM}. Ce potentiel
reproduit bien l'asymétrie de l'énergie de mélange à 0\,K : la solubilité
de Cr dans Fe est supérieure à celle de Fe dans Cr.

\begin{figure}[h]
\begin{centering}
\includegraphics[scale=0.4]{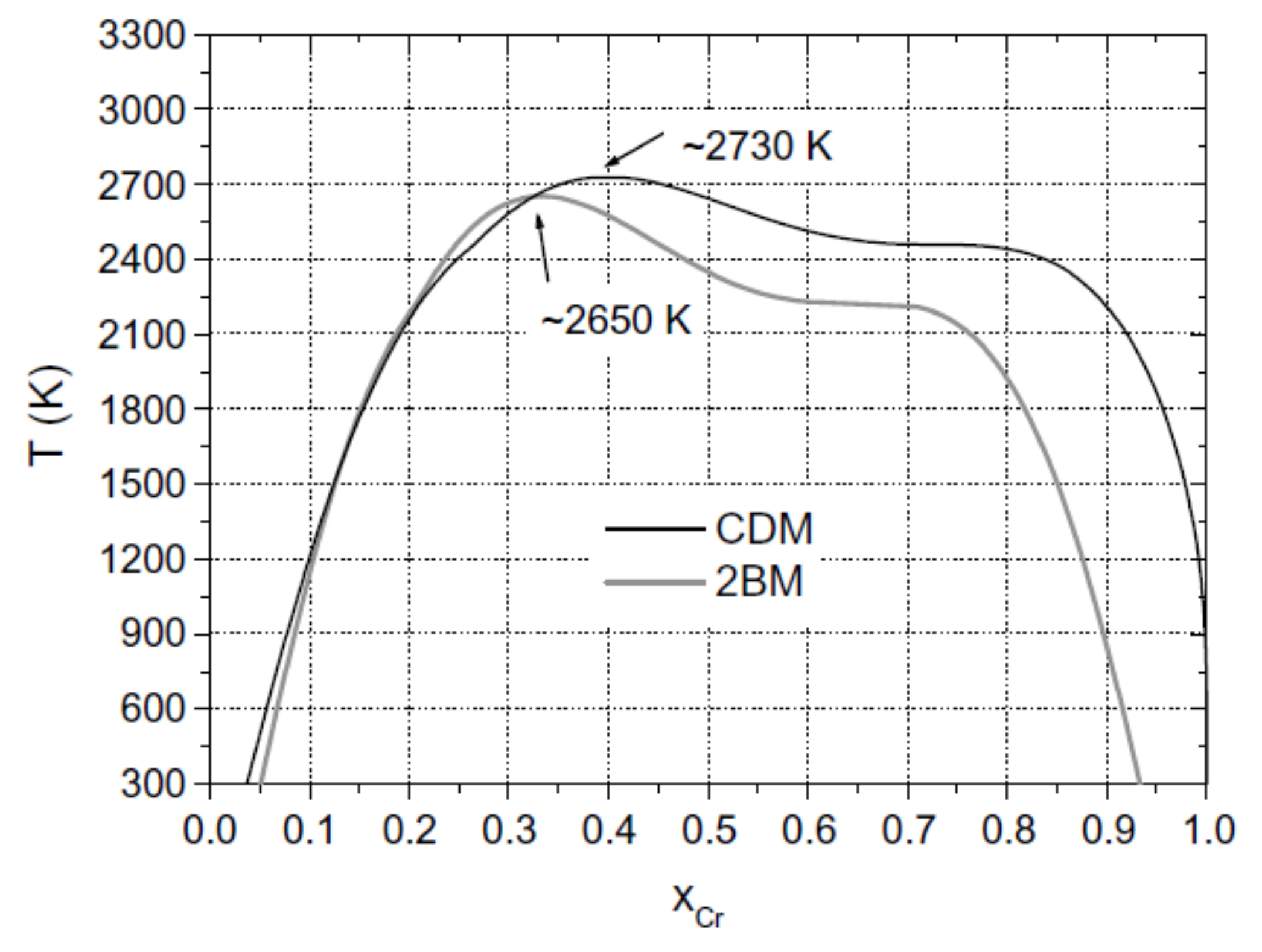}
\par\end{centering}

\caption{Diagrammes de phases calculés en champ moyen de point par Malerba
et al. \cite{Malerba_revue_2008} en utilisant les potentiels 2BM
\cite{Caro_CDM_2005} et CDM \cite{olsson_two-band_2005}. Figure
issue de Malerba et al. \cite{Malerba_revue_2008}. \label{fig:ddp_2BM_CDM_selonMalerba}}
\end{figure}

On reproduit sur la figure \ref{fig:ddp_2BM_CDM_selonMalerba} les
limites de solubilité calculées par Malerba et al. en utilisant les
potentiels CDM et 2BM dans l'approximation statistique de champ moyen
de point \cite{Malerba_revue_2008}.

La solubilité de Fe dans Cr calculée en champ moyen avec le potentiel
2BM semble trop importante, mais les résultats expérimentaux sont
rares dans ce domaine de concentration. La limite de solubilité dans
le domaine riche en Cr déduite du potentiel 2BM est moins pentue qu'avec
le potentiel CDM. Cela changerait la composition en fer des précipités
$\alpha'$ que l'on calculerait avec l'un ou l'autre potentiel inter-atomique
d'un ordre de grandeur pour $T>300$\,K. Du côté riche en Fe, les
limites de solubilité sont similaires jusqu'à 30\,\%\,Cr à environ
2650\,K.

Les températures critiques calculées en champ moyen sur site pour
les deux potentiels EAM sont de l'ordre de 2700\,K, soit environ
trois fois supérieures à la température critique de référence ($\approx$1000\,K).
Les auteurs mettent en cause l'absence d'entropie vibrationnelle dans
les calculs. Seule l'entropie de configuration est en effet considérée
ici. Lorsque l'entropie vibrationelle est prise en compte, la température
critiques est d'environ 740\,K pour le potentiel 2BM mais reste très
élevée (plus de 2000\,K) pour le potentiel CDM. Rappelons que ces
modèles sont développés en ayant certaines études en objectifs. Le
potentiel 2BM a par exemple été construit pour reproduire un grand
nombre de propriétés liées aux défauts cristallins pour simuler des
propriétés liées aux simulations, aux cinétiques ... L'entropie vibrationelle
ne fait pas partie des propriétés sur lesquelles il a été ajusté.
C'est la raison pour laquelle leur \og réponse \fg{} à l'entropie
vibrationnelle peut être très différente d'un potentiel à l'autre.

\subsection{Les modèles tenant compte du moment magnétique atomique\label{sub:modele_d_ackland}}

\subsubsection*{Le modèle d'Ackland}

Dans l'hamiltonien d'Ising classique, l'énergie totale du système
est la somme des énergies de paires sur réseau rigide \cite{Ising_1925}.
Les énergies de paires dépendent d'une variable associée à chaque
nœud $i$ du réseau et d'une constante de couplage unique.

Différents auteurs proposent dans les années 70 d'étendre l'hamiltonien
d'Ising à l'étude des alliages magnétiques en ajoutant un degré de
liberté \cite{cadeville_magnetism_1987}. Ackland propose d'appliquer
ce modèle aux alliages fer--chrome \cite{ackland_magnetically_2006}.
Il y a deux variables pour chaque nœud du réseau : 
\begin{itemize}
\item une variable d'espèce chimique $S_{i}=\pm1$ (Cr ou Fe),
\item et un moment magnétique atomique $\sigma_{i}=\pm1$ ($\uparrow$ ou
$\downarrow$). L'amplitude des moments magnétiques est constante
($\left|\sigma_{i}\right|=1$). 
\end{itemize}
Ackland ne peut pas ajuster parfaitement les constantes de couplage
sur les calculs \emph{ab initio} car l'amplitude du moment magnétique
$\sigma_{i}$ ne dépend pas de l'environnement chimique local. Ne
voulant que mettre en évidence les grandes tendances de ce modèle,
il choisit d'imposer égales toutes les constantes de couplage aux
premiers et aux deuxièmes voisins. Il n'a donc pas d'échelle de température
réelle à comparer avec l'expérience, bien qu'il suffirait pour cela
de donner une valeur aux constantes de couplages. Le degré de liberté
supplémentaire permet de prendre en compte une première approximation
de l'entropie magnétique. Le système a la possibilité de faire varier
les moments magnétiques, même si l'amplitude est unique. Il peut donc
explorer un plus grand nombre de configurations.

Ackland construit un hamiltonien qui favorise le couplage entre moments
magnétiques de même signe ($\sigma_{i}\sigma_{j}=+1$) entre atomes
de fer, et les couplages entre moments magnétiques de signes opposés
($\sigma_{i}\sigma_{j}=-1$) entre Fe et Cr ainsi qu'entre atomes
de Cr. Dans le cas où toutes les interactions chimiques et magnétiques
sont limitées aux 1$^{\text{ers}}$ et 2$^{\text{e}}$ voisins et
sont de même force (ce qui est équivalent à une constante de couplage
égale à $\frac{1}{T}$ dans le modèle d'Ising classique), l'hamiltonien
du système s'écrit :
\begin{equation}
H=\sum_{i,j>i}\dfrac{1}{2}\left(S_{i}+S_{j}\right)\sigma_{i}\sigma_{j}+\dfrac{1}{2}\left(1-S_{i}S_{j}\right)\sigma_{i}\sigma_{j}\label{eq:Ackland_Ising}
\end{equation}
où le premier terme décrit les interactions entre atomes de même espèces
($\frac{1}{2}\left(S_{i}+S_{j}\right)=\pm1$ et $\frac{1}{2}\left(1-S_{i}S_{j}\right)=0$)
et le deuxième terme décrit les interactions entre atomes d'espèces
différentes ($\frac{1}{2}\left(S_{i}+S_{j}\right)=0$ et $\frac{1}{2}\left(1-S_{i}S_{j}\right)=1$).

Le diagramme de phases correspondant à cet hamiltonien déduit de simulations
Monte Carlo est en accord qualitatif avec les principales propriétés
du diagramme de phases expérimental :
\begin{itemize}
\item la lacune de miscibilité $\alpha$--$\alpha$' asymétrique,
\item une large solubilité de Cr dans Fe (jusque 23\,\%\,Cr) mais pas
l'inverse,
\item la diminution de la température de Curie lors de l'ajout de Cr,
\item la température de Curie du fer est très supérieure à la température
de Néel du chrome.
\end{itemize}
Dans un deuxième article dédié à la phase $\sigma$ et à la ségrégation
de surface libre (100), Ackland \cite{ackland_orderedsigma_2009}
propose de ne tenir compte des interactions de paires magnétiques
jusqu'aux cinquièmes voisins. Cela permet de reproduire la longue
portée des interactions Cr--Cr mise en évidence par les calculs \emph{ab
initio }\cite{klaver_magnetism_2006,olsson_ab_2003} et les développements
en amas de Lavrentiev et al. \cite{lavrentiev_magnetic_CE_2009,lavrentiev_CEmag2_2010}.
L'énergie totale de $N$ voisins en interaction est la somme des énergies
de paires magnétiques dont la constante de couplage dépend des atomes
formant la paire :
\begin{equation}
H_{i}\left(N\right)=\sum_{j=1}^{N}H_{ij},\label{eq:acklandsurf}
\end{equation}
où $H_{ij}$ dépend de l'espèce sur les sites $i$ et $j$ :
\[
H_{ij}=A_{N}\sigma_{i}\sigma_{j}\mbox{ lorsque }S_{i}+S_{j}=2,\mbox{ (interaction Cr-Cr)},
\]
\[
H_{ij}=\sigma_{i}\sigma_{j}\mbox{ lorsque }S_{i}+S_{j}=0\mbox{ (interaction Cr-Fe)},
\]
\[
H_{ij}=-\sigma_{i}\sigma_{j}\mbox{ lorsque }S_{i}+S_{j}=-2\mbox{ (interaction Fe-Fe)},
\]
où le paramètre $A_{N}$ est ajusté pour obtenir un rapport supérieur
à 1 entre les températures de Curie $T_{C}$ et de Néel $T_{N}$ du
fer et du chrome. Les interactions Cr--Fe et Fe--Fe sont choisies
égales à 1. Il n'y a donc pas d'échelle de température réelle à comparer
avec l'expérience.

\begin{figure}[H]
\begin{centering}
\includegraphics[scale=0.35]{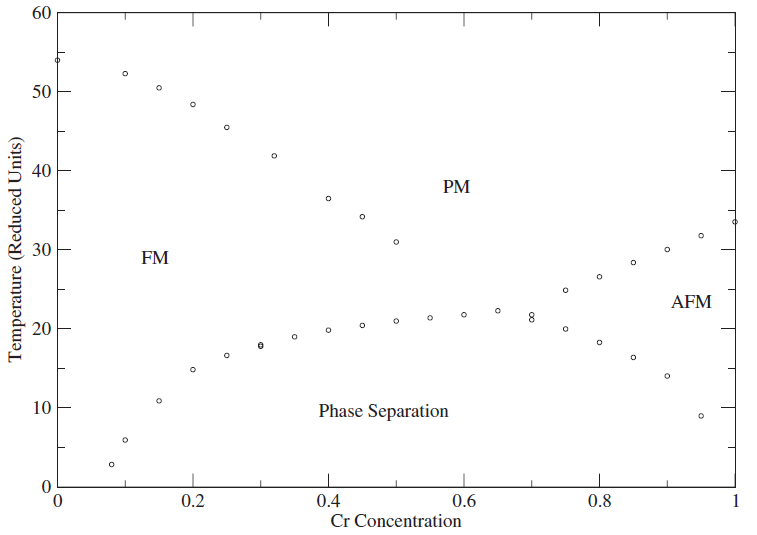}
\par\end{centering}

\caption{Diagramme de phases déduit par Ackland avec un modèle de paires magnétiques
aux cinquièmes voisins décrit par l'hamiltonien \ref{eq:acklandsurf}.
Les trois phases ferromagnétiques (FM), paramagnétiques (PM) et antiferromagnétiques
(AF) sont dans un réseau cubique centré. Figure de Ackland \cite{ackland_orderedsigma_2009}.\label{fig:Ackland_ddp_5NN}}
\end{figure}

Ackland calcule en Monte Carlo le diagramme de phases représenté sur
la figure \ref{fig:Ackland_ddp_5NN}.

La limite principale de ce modèle est qu'il ne reproduit ni l'inversion
du signe de l'énergie de mélange, ni la dépendance en composition
chimique locale de l'amplitude du moment magnétique des atomes de
Cr. 

Cependant, la relative simplicité et les résultats de ce modèle montrent
qu'il capture une part importante de la physique de l'alliage en introduisant
un degré de liberté supplémentaire : le moment magnétique atomique.

\subsubsection*{La CVM de tétraèdres chimiques et magnétiques d'Inden et Schön}

Inden et Schön proposent un premier modèle classique dans lequel l'énergie
du système est décomposée en énergies de tétraèdres \cite{inden_thermodynamicFeCr_2008}:
\begin{equation}
U=6N\sum_{i,j,k,l}\epsilon_{i,j,k,l}^{\alpha,\beta,\gamma,\delta}\rho_{i,j,k,l}^{\alpha,\beta,\gamma,\delta}\label{eq:CVM}
\end{equation}
où $N$ est le nombre d'atomes du système cubique centré considéré,
et $\rho_{i,j,k,l}^{\alpha,\beta,\gamma,\delta}$ est la probabilité
de trouver un amas $\left(\alpha,\beta,\gamma,\delta\right)$ occupé
par les espèces $\left(i,j,k,l\right)$. $\epsilon_{i,j,k,l}^{\alpha,\beta,\gamma,\delta}$
est l'énergie associée à cet amas que Inden et Schön ajustent sur
des mesures calorimétriques à haute température \cite{dench_adiabatic_1963}.
Ils traitent ensuite leur hamiltonien avec la méthode de champ moyen
CVM (méthode des variations d'amas). Ce modèle, comme le modèle d'Ising
classique, ne tient pas compte du moment magnétique atomique. Le diagramme
de phases calculé est très éloigné du diagramme de phases expérimental.

Dans un deuxième temps, Inden et Schön ajoutent aux énergies de tétraèdres
classiques $\epsilon_{i,j,k,l}^{\alpha,\beta,\gamma,\delta}$ une
énergie $\epsilon_{M_{i},M_{j},M_{k},M_{l}}^{\alpha,\beta,\gamma,\delta}$
tenant compte des moments magnétiques $\left(M_{i},M_{j},M_{k},M_{l}\right)$
portés par les atomes $\left(i,j,k,l\right)$ du tétraèdre $\left(\alpha,\beta,\gamma,\delta\right)$.
L'énergie du système s'écrit alors :
\begin{equation}
U=6N\sum_{i,j,k,l}\left(\epsilon_{i,j,k,l}^{\alpha,\beta,\gamma,\delta}+\epsilon_{M_{i},M_{j},M_{k},M_{l}}^{\alpha,\beta,\gamma,\delta}\right)\rho_{i,j,k,l}^{\alpha,\beta,\gamma,\delta}\label{eq:CVM_mag}
\end{equation}
Contrairement au modèle d'Ackland dans lequel les atomes ne peuvent
avoir que des moments magnétiques $+1$ ou $-1$, trois valeurs sont
autorisées pour Fe et deux pour Cr. La part magnétique de l'énergie
de tétraèdre est réduite à la somme des interactions de paires magnétiques
aux premiers et deuxièmes voisins. Ces énergies de paires sont ajustées
sur les températures de Curie et Néel expérimentales de l'alliage.

Ce nouvel hamiltonien leur permet de construire un diagramme de phases
fer--chrome avec une lacune de miscibilité reproduisant les observations
expérimentales à haute température et la variation des températures
de Curie et Néel de l'alliage. La solubilité du fer dans le chrome
est cependant trop forte, et Inden n'indique pas les limites à 0\,K
que l'on imagine nulles par extrapolation. Les interactions magnétiques
de l'alliage sont elles aussi ajustées sur les mesures à haute température.

Il serait intéressant de reprendre l'ajustement des énergies d'amas
sur les calculs \emph{ab initio} à 0\,K plutôt que sur les propriétés
à haute température. Les calculs \emph{ab initio} montrent qu'à basse
température, c'est le moment magnétique du chrome qui est le plus
sensible à l'environnement chimique local : une grande amplitude et
une discrétisation suffisante sont nécessaire.

L'espace des configurations magnétiques possibles est plus grand que
celui du modèle d'Ackland puisque le nombre des amplitudes autorisées
est supérieur. L'entropie magnétique est donc mieux prise en compte.
La température critique déduite de cet hamiltonien est d'environ 800\,K.
Cette température beaucoup plus basse que les températures critiques
déduites en champ moyen des potentiels EAM ($\approx2700$\,K) confirme
le rôle important de l'entropie magnétique.

\subsubsection*{Les développements en amas chimiques et magnétiques de Lavrentiev
et al. \label{sub:cluster_exp_classique}}

Les développements en amas permettent de représenter l'énergie totale
d'un alliage sur réseau rigide comme une somme d'énergies de configurations
à plusieurs sites (singulets, paires, triplets, tétraèdres \cite{inden_thermodynamicFeCr_2008}
\ldots{})\cite{lavrentiev_classicalCE_FeCr_2007,nguyen-manh_classicalCE2_FeCr_2008}.
C'est une façon classique d'étudier les propriétés énergétiques d'une
solution solide dans un large domaine de concentration et de température.
À chaque site est associée une variable d'occupation de site $S_{i}$
prenant les valeurs $\pm1$ (Fe ou Cr). L'énergie totale s'exprime
: 
\begin{equation}
E\left(\vec{S}_{i}\right)=J^{\left(0\right)}+\sum_{\gamma}D_{\gamma}J_{\gamma}\prod_{i\in\gamma}S_{i}\label{eq:devamas}
\end{equation}

où $\vec{S_{i}}=\left\{ S_{i}\right\} _{i=1,\ldots,N}$ est la configuration
du système d'énergie totale $E\left(\vec{S}_{i}\right)$. $J^{\left(0\right)}$
est une constante. Chaque type d'amas $\gamma$ a une énergie effective
$J_{\gamma}$ ajustée sur les énergies de mélange des structures ordonnées
calculées \emph{ab initio}. $D_{\gamma}$ est le nombre total d'amas
de type $\gamma$ dans le système. Inden et Schön n'utilisent par
exemple que des amas de type tétraèdres \cite{inden_thermodynamicFeCr_2008}.

Dans une première étude, Lavrentiev et al. \cite{lavrentiev_classicalCE_FeCr_2007,nguyen-manh_classicalCE2_FeCr_2008}
montrent qu'il est possible de reproduire les résultats \emph{ab initio}
avec l'hamiltonien classique \ref{eq:devamas} sans interactions magnétiques
au prix d'un développement en de nombreux amas. 12 amas de 2 à 5 atomes
et une portée des interactions jusqu'aux 6$^{\mbox{e}}$ voisins sont
nécessaires pour reproduire le changement de signe des énergies de
mélange calculées \emph{ab initio}, alors que seuls quelques amas
de deux et trois atomes suffisent à reproduire des énergies de mélange
dans des systèmes plus simples sans interactions magnétiques \cite{drautz_obtainingCE_2006}.
Le calcul de propriétés d'équilibre avec un tel développement serait
alors très lourd numériquement. Le calcul du diagramme de phases,
des isothermes de ségrégation ou des propriétés de diffusion serait
difficile. En accord avec les résultats du modèle d'Ising ou de la
CVM de tétraèdres classiques, cela démontre que la physique de l'alliage
n'est pas captée par une unique variable d'espèce.

Dans une deuxième série d'études, Lavrentiev et al. ajoutent donc
un nouveau degré de liberté au système, comme l'ont fait Ackland ou
Inden et Schön, mais en allant un pas plus loin. Les moments magnétiques
$\vec{M}_{i}$ de chaque atome $i$ sont pris en compte, mais contrairement
à Ackland ou Inden, ce sont à la fois leur direction et leur amplitude
qui sont variables. Ce moment magnétique dépend de la configuration
chimique locale. 

Avec ce nouveau degré de liberté, seules les interactions non-magnétiques
classiques de sites et paires aux premiers et deuxièmes voisins associées
à des interactions magnétiques de paires jusqu'aux cinquièmes voisins
sont nécessaires pour reproduire les énergies de mélanges calculées
\emph{ab initio}.

L'énergie totale s'exprime :
\begin{eqnarray}
E\left(\vec{S}_{i}\right) & = & J^{\left(0\right)}+J^{\left(1\right)}\sum_{i}S_{i}+J^{\left(2\right)}\sum_{i,j\ne i}S_{i}S_{j}\\
 &  & -\sum_{i,j\ne i}\left(J_{ij}^{\left(0\right)}+J_{ij}^{\left(1\right)}\left(S_{i}+S_{j}\right)+J_{ij}^{\left(2\right)}S_{i}S_{j}\right)\vec{M}_{i}\vec{M}_{j}\nonumber \\
 &  & +\sum_{i}\left(A^{\left(0\right)}+A^{\left(1\right)}+\sum_{j\ne i}A_{ij}^{\left(2\right)}S_{i}S_{j}\right)\vec{M}_{i}^{2}\nonumber \\
 &  & +\sum_{i}\left(B^{\left(0\right)}+B^{\left(1\right)}+\sum_{j\ne i}B_{ij}^{\left(2\right)}S_{i}S_{j}\right)\vec{M}_{i}^{4}\nonumber 
\end{eqnarray}
où $J^{\left(n\right)}$ sont les constantes de couplages non-magnétiques
à courte distance entre $n$$^{\mbox{e}}$ voisins. $A^{\left(n\right)}$
et $B^{\left(n\right)}$ sont les termes de site qui déterminent l'amplitude
du moment magnétique. Les $J_{ij}^{\left(n\right)}$ sont les paramètres
d'interactions magnétiques.

En plus du terme d'interaction de paires magnétique $\vec{M}_{i}\vec{M}_{j}$,
des développements de type Ginzburg-Landau en $\vec{M}_{i}^{2}$ et
$\vec{M}_{i}^{4}$ sont ajoutés à l'hamiltonien de l'équation \ref{eq:devamas}.
Ces termes déterminent l'amplitude des moments magnétiques et empêchent
leur divergence.

Ces auteurs montrent que les moments magnétiques des configurations
les plus stables sont le plus souvent colinéaires. C'est en particulier
le cas des structures à forte concentration en l'un ou l'autre des
éléments.

La direction et l'amplitude des moments magnétiques étant libres,
ce modèle permet d'explorer l'ensemble des configurations magnétiques
possibles et donc de prendre complètement en compte l'entropie magnétique.
Les entropies magnétiques dans les modèles d'Ackland ou d'Inden sont
sous-estimées car seule une partie de l'espace des configurations
magnétiques peut être échantillonnée. Cette approximation est justifiée
si l'échantillonnage correspond à l'ensemble des configurations les
plus favorables énergétiquement.

La contre-partie de ce modèle est qu'il nécessite un traitement statistique
numériquement coûteux. Pour chaque configuration chimique, un grand
nombre de pas Monte Carlo doivent être réalisés pour échantillonner
l'immense espace des configurations magnétiques. Jusqu'à aujourd'hui,
le diagramme de phases complet n'a pas été calculé. Seul un diagramme
de phases partiel ne concernant que l'équilibre $\alpha-\gamma$ du
domaine riche en fer a été publié \cite{lavrentiev_dudi_magnetic_bccfcc_2010}.

\section{Développement d'un modèle d'interactions de paires dépendant de la
concentration locale et de la température\label{sec:Notre-mod=0000E8le-thermodynamique}}

L'ensemble des modèles présentés ci-dessus montre qu'il serait souhaitable
de prendre en compte explicitement le moment magnétique atomique pour
décrire les phénomènes physiques à l'échelle atomique dans l'alliage
fer--chrome. Ce qui est singulier à cet alliage est que l'amplitude
du moment magnétique porté par les atomes de Fe et surtout de Cr dépend
fortement de l'environnement chimique local \cite{dudarev_magnetic_2005,lavrentiev_magnetic_CE_2009,nguyen-manh_classicalCE2_FeCr_2008,lavrentiev_classicalCE_FeCr_2007,ackland_magnetically_2006,lavrentiev_CEmag2_2010,inden_thermodynamicFeCr_2008}.

Cependant, l'introduction de ce degré de liberté supplémentaire alourdit
considérablement les calculs. Ceci est particulièrement vrai dans
le modèle de Lavrentiev et al. dans lequel à la fois la direction
et l'amplitude des moments magnétiques sont libres \cite{dudarev_magnetic_2005,lavrentiev_magnetic_CE_2009,nguyen-manh_classicalCE2_FeCr_2008,lavrentiev_classicalCE_FeCr_2007,lavrentiev_CEmag2_2010}.
Cette lourdeur numérique est le principal frein à l'utilisation des
modèles de Lavrentiev, Ackland ou Inden et al. \cite{dudarev_magnetic_2005,lavrentiev_magnetic_CE_2009,nguyen-manh_classicalCE2_FeCr_2008,lavrentiev_classicalCE_FeCr_2007,ackland_magnetically_2006,ackland_two-band_2003,ackland_two-band_2006,lavrentiev_CEmag2_2010}
pour réaliser des calculs d'équilibre (diagramme de phases, isothermes
de ségrégation) et, à l'avenir, des cinétiques de ségrégation et de
décomposition.

Nous cherchons un modèle atomistique léger qui reproduise à la fois
les résultats \emph{ab initio} décrits au chapitre \ref{cha:DFT}
et le diagramme de phase expérimental à haute température.

\subsection{Interactions de paires sur réseau rigide \label{sec:L'hamiltonien-du-syst=0000E8me}}

Nous choisissons comme point de départ le modèle le plus simple pour
reproduire une lacune de miscibilité : des interactions de paires
sur réseau rigide.

Dans un alliage binaire sans défauts A$_{x_{A}}$B$_{x_{B}}$, les
énergies de paires sont définies comme $\epsilon_{AA}^{\left(i\right)}$,
$\epsilon_{BB}^{\left(i\right)}$ et $\epsilon_{AB}^{\left(i\right)}$,
où $i$ est la portée de l'interaction en sphères de coordination.
Le moteur au mélange ou à la démixtion est alors l'énergie d'ordre
$\Omega$ :
\begin{eqnarray}
\Omega & = & \sum_{\left(i\right)}Z^{\left(i\right)}V^{\left(i\right)}\label{eq:omega_paires}
\end{eqnarray}
où $Z^{\left(i\right)}$ est le nombre total de $i$$^{\text{e}}$
voisins de chaque atome en volume, et où
\begin{equation}
V^{\left(i\right)}=\frac{1}{2}\left(\epsilon_{AA}^{\left(i\right)}+\epsilon_{BB}^{\left(i\right)}-2\epsilon_{AB}^{\left(i\right)}\right).
\end{equation}
Les propriétés thermodynamiques dépendant uniquement des $V^{\left(i\right)}$
et pas de la répartition entre les $\epsilon_{AA}^{\left(i\right)}$,
$\epsilon_{BB}^{\left(i\right)}$ et $\epsilon_{AB}^{\left(i\right)}$
\cite{ducastelle_order_1991}. Selon le signe de l'énergie d'ordre
$V^{\left(i\right)}$, le système aura tendance à maximiser soit le
nombre de paires homo-atomiques ($V^{\left(i\right)}<0$), soit le
nombre de paires hétéro-atomiques ($V^{\left(i\right)}>0$). Ce modèle
de paires et les différents diagrammes de phases qui peuvent en résulter
selon le traitement statistique mis en œuvre (Monte Carlo, champs
moyens) sont discutés par de Fontaine ou Ducastelle \cite{de_fontaine_1979,ducastelle_order_1991}
:
\begin{itemize}
\item si $\Omega<0$, le système a tendance à la démixtion,
\item si $\Omega>0$ et qu'on se limite à des interactions entre premiers
voisins alors le système a tendance à l'ordre.
\end{itemize}
On illustre cela par un traitement de champ moyen de point de ce modèle
très simple. C'est ce qu'on appelle l'approximation de Bragg-Williams
du modèle d'Ising.

\subsection{L'approximation de Bragg-Williams}

L'approximation de champ moyen de point (Bragg-Williams) permet une
étude à la fois analytique et numérique très efficace. Dans cette
approximation, la concentration est homogène. Le modèle d'interactions
de paires constantes sur réseau rigide décrit au paragraphe \ref{sec:L'hamiltonien-du-syst=0000E8me}
est alors équivalent au modèle de la solution régulière \cite{porter_easterling_1992}.

Dans l'approximation de Bragg-Williams, les concentrations sont homogènes.
Le nombre de paires entre un élément $m$ et un élément $n$ est donc
égal à $x_{m}x_{n}$, ce qui revient à négliger complètement l'ordre
à courte distance (SRO) dans la solution solide. L'énergie de mélange
$\Delta E_{mix}$ d'un alliage modèle A$_{x_{A}}$B$_{x_{B}}$ sans
défauts s'exprime alors \cite{porter_easterling_1992} : 
\begin{equation}
\Delta E_{mix}=-x_{A}x_{B}\Omega\label{eq:Hmix_solreg}
\end{equation}
où $\Omega$ est l'énergie d'ordre du système décrite dans l'équation
\ref{eq:omega_paires}. Dans l'approximation de Bragg-Williams, dans
le cas d'un système à tendance à la démixtion, le diagramme de phases
dépend uniquement de l'énergie d'ordre $\Omega$ et non de la répartition
de l'énergie sur les différentes sphères de coordination.

Dans un modèle sur réseau rigide, l'entropie de vibration est négligée.
L'entropie de configuration $\Delta S_{mix}$ s'écrit : 
\begin{equation}
\Delta S_{mix}=-k_{B}\left(x_{A}\ln x_{A}+x_{B}\ln x_{B}\right)
\end{equation}
où $k_{B}$ est la constante de Boltzman. L'energie libre $\Delta F_{mix}$
de l'alliage s'exprime donc :
\begin{eqnarray}
\Delta F_{mix} & = & \Delta E_{mix}-T\Delta S_{mix}\\
 & = & -x_{A}x_{B}\Omega+k_{B}T\left(x_{A}\ln x_{A}+x_{B}\ln x_{B}\right).\label{eq:elibreBW}
\end{eqnarray}

Le potentiel chimique d'alliage $\Delta\mu$ s'écrit :

\begin{equation}
\Delta\mu=-\left(x_{A}-x_{B}\right)\Omega+k_{B}T\ln\left(\dfrac{x_{B}}{x_{A}}\right).
\end{equation}

À l'équilibre, dans un système à tendance à la démixtion, le potentiel
chimique d'alliage est uniforme dans les deux phases, ce qui permet
de déduire une expression analytique de la limite de solubilité en
fonction de la concentration en chrome $T\left(x_{B}\right)$ :
\begin{equation}
T\left(x_{B}\right)=\dfrac{\Omega\left(x_{A}-x_{B}\right)}{k_{B}\ln\left(\dfrac{x_{B}}{x_{A}}\right)},\forall x_{B}\ne x_{A}\label{eq:Tc_analytique_solreg}
\end{equation}

La température critique $T_{c}$ correspondant au haut de la lacune
de miscibilité à laquelle le système redevient monophasé s'écrit :

\begin{equation}
T_{c}=\lim\limits _{x_{B}\rightarrow0.5}T\left(x_{B}\right)=-\dfrac{\Omega}{2k_{B}}\label{eq:relation_T_Omega_solreg}
\end{equation}
qui ne dépend que de l'énergie d'ordre $\Omega$. On peut également
déduire la limite spinodale $T_{spinodal}\left(x_{B}\right)$ à laquelle
le potentiel chimique d'alliage devient invariant avec la concentration
:

\begin{equation}
T_{spinodal}=-\dfrac{2\Omega}{k_{B}}x_{A}x_{B}\label{eq:spinodale_analytique_solreg}
\end{equation}

La température critique expérimentale $T_{c}^{exp}$ de la solution
solide fer--chrome est d'environ 960\,K. On peut déduire de l'équation
\ref{eq:Tc_analytique_solreg} l'énergie d'ordre $\Omega_{r\acute{e}g}$
du système dans ce modèle de solution solide :

\begin{equation}
\Omega_{r\acute{e}g}=-0.166\,\textrm{eV}
\end{equation}

\begin{figure}[h]
\begin{centering}
\includegraphics[scale=0.4]{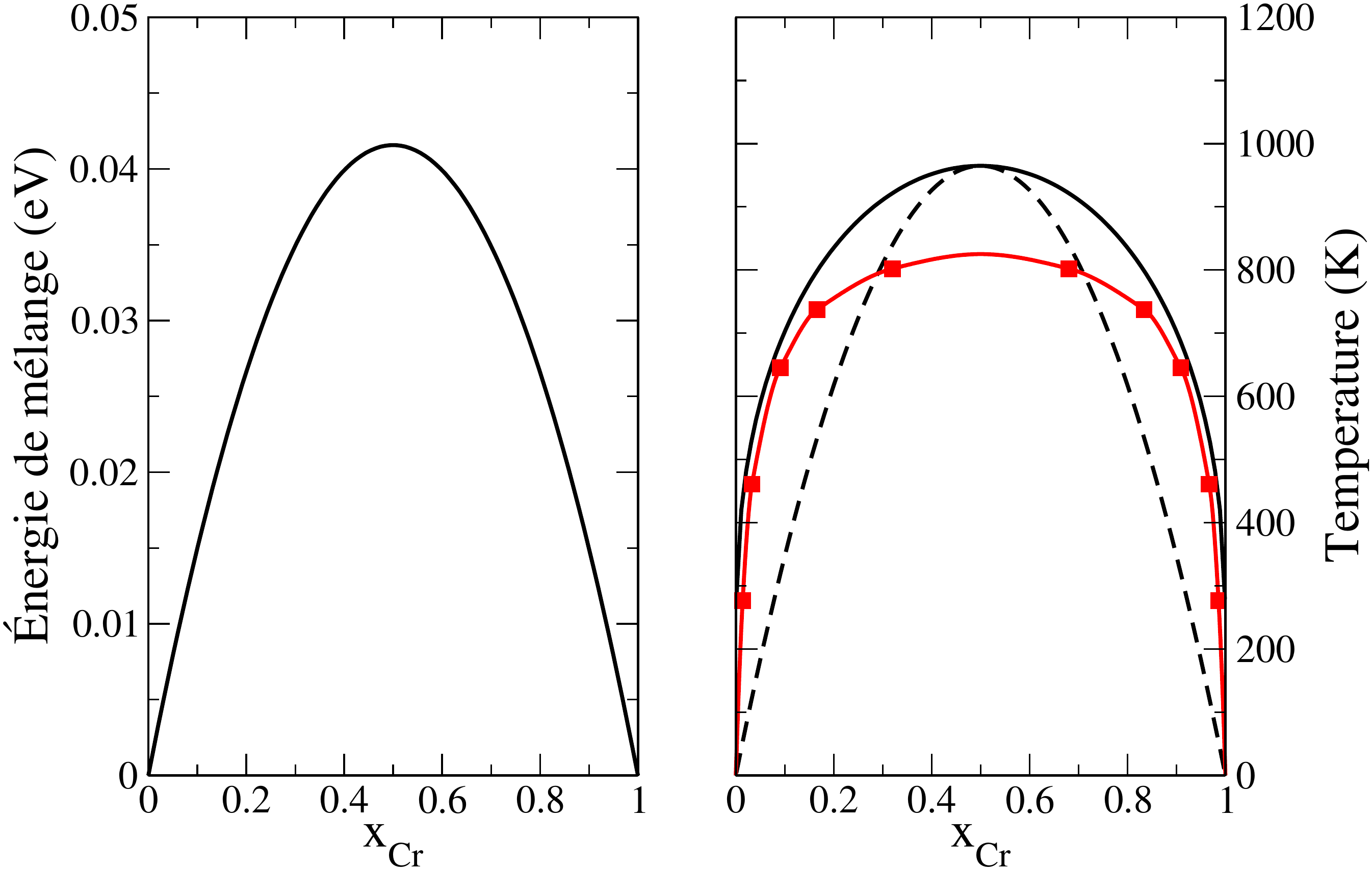}
\par\end{centering}

\caption{(À gauche) énergie de mélange dans l'approximation de Bragg-Williams
pour l'énergie d'ordre $\Omega_{r\acute{e}g}=-0.166$\,eV ajustée
sur la température critique expérimentale de 960\,K. (À droite en
noir) limites de solubilité (trait plein) et limites spinodales (pointillés)
déduites de l'énergie d'ordre $\Omega_{r\acute{e}g}$ dans le modèle
de la solution régulière. (À droite en rouge) diagramme de phases
exact déduit des simulations Monte Carlo pour l'énergie d'ordre $\Omega_{r\acute{e}g}$.\label{fig:solution_reg_Hmix_et_ddp}}
\end{figure}

Grâce aux expressions analytiques \ref{eq:Hmix_solreg}, \ref{eq:Tc_analytique_solreg}
et \ref{eq:spinodale_analytique_solreg}, on trace l'énergie de mélange
de l'alliage, la lacune de miscibilité et les limites spinodales sur
la figure \ref{fig:solution_reg_Hmix_et_ddp}. 

On reproduit également sur la figure \ref{fig:solution_reg_Hmix_et_ddp}
le diagramme de phases exact de ce modèle d'interaction de paires
constantes déduit de simulations Monte Carlo semi-grand-canoniques
\footnote{Ici comme dans la suite du document, les interpolations numériques
sont faites par splines Akima \cite{akima_splines_1970}.%
} dont le détail est donné en annexes \ref{sec:Principe-des-simulations_monte_carlo}
et \ref{sec:Ensembles_canoniques_semi-grand-canonique}. 

Du fait de la forme \ref{eq:omega_paires} de l'énergie d'ordre $\Omega$,
il est aisé de réaliser dans les expressions \ref{eq:elibreBW} de
l'énergie libre de l'alliage, \ref{eq:Tc_analytique_solreg} des limites
de solubilité et \ref{eq:spinodale_analytique_solreg} des limites
spinodales ainsi que sur la figure \ref{fig:solution_reg_Hmix_et_ddp}
que toutes ces propriétés sont symétriques par rapport à la concentration
$x_{A}=x_{B}=0.5$. Il ne s'agit pas d'un effet du traitement statistique
mais du modèle même d'énergies de paires constantes. Ces symétries
sont en accord avec les mesures calorimétriques de Dench et al. \cite{dench_adiabatic_1963}
pour qui la solution solide est régulière à haute température, mais
en désaccord avec les résultats \emph{ab initio} présentés au paragraphe
\ref{sub:Energies-de-m=0000E9lange_DFT}. Ces derniers montrent que
la variation de l'énergie de mélange avec la concentration en chrome
n'est pas symétrique à basse température. Le problème principal de
ce modèle est donc que l'énergie d'ordre ne dépend pas de la concentration
locale.

\subsection{Dépendance en concentration locale de l'énergie de mélange\label{sub:param_hmix}}

Les limites de solubilité sont difficiles à mesurer à basse température
à cause des cinétiques lentes. Du fait des calculs \emph{ab initio}
à 0\,K, on peut cependant s'attendre à ce qu'elles soient asymétriques
de part et d'autre du diagramme de phases. Des observations sous irradiation
qui en première approximation accélère le vieillissement ont été compilées
par Bonny et al. \cite{bonny_onthe_aa_demixtion_2008}. Ils confirment
l'asymétrie de la lacune de miscibilité à basse température. Un consensus
général n'est cependant pas trouvé \cite{xiong_grrrrbonny_2010,Malerba_revue_2008}.

Afin de reproduire les énergies de mélange \emph{ab initio} dans les
approximations PWSCF-PAW et SIESTA-NC qui changent de signe avec la
concentration, nous choisissons de donner à l'énergie d'ordre $\Omega$
une dépendance en concentration locale portée exclusivement par les
paires hétéro-atomiques. Les énergies de paires homo-atomiques sont
fixées par les propriétés des éléments purs.%
\footnote{Il n'est pas nécessaire de décrire ici la paramétrisation des énergies
de paires homo-atomiques car les propriétés thermodynamiques de volume
sont entièrement déterminées par les $V_{i}=1/2\left(\epsilon_{FeFe}^{\left(i\right)}+\epsilon_{CrCr}^{\left(i\right)}-2\epsilon_{FeCr}^{\left(i\right)}\right)$.
Nous reviendrons sur leur paramétrisation sur les énergies cohésion
des éléments purs dans le chapitre 3.%
} Par soucis de simplicité, on choisit les rapports $\eta^{\left(i\right)}=\epsilon_{mn}^{\left(i\right)}/\epsilon_{mn}^{\left(1\right)}$
constants. En effet, si les interactions aux différentes portées évoluaient
différemment avec la composition locale, on pourrait dans certains
cas stabiliser des phases ordonnées à basse température qui n'ont
à ce jour pas été observées expérimentalement.

Les énergies de mélange de la solution solide sont ajustées sur les
calculs \emph{ab initio} SQS PWSCF-PAW et SIESTA-NC par un polynôme
de Redlich-Kister \cite{redlich_kister_1948}. Ce formalisme, largement
répandu dans la communauté CALPHAD, permet une expression analytique
simple, continue et dérivable plusieurs fois de l'énergie de mélange.
On ne pose pas ici la problématique de la définition de la concentration
locale puisqu'en champ moyen sur site la composition locale est assimilée
à la composition nominale. Ce travail peut être comparé à celui de
Caro et al. dans lequel les énergies d'interaction sont également
dépendantes de la concentration locale \cite{Caro_CDM_2005,stukowski_efficient_2009}.

Nous considérons un alliage binaire Fe$_{x_{Fe}}$Cr$_{x_{Cr}}$ sans
défaut. Afin de simplifier les écritures, on notera dorénavant $x_{Cr}=1-x_{Fe}\equiv x$.

L'énergie de mélange s'écrit :
\begin{equation}
\Delta E_{mix}=-x\left(1-x\right)\sum_{p=0}^{n}L^{\left(p\right)}\left(1-2x\right)^{p}
\end{equation}
où $n$ est l'ordre de la paramétrisation, et $L^{\left(p\right)}$
est le paramètre d'interaction de l'ordre $p$ qui a la forme :
\begin{equation}
L^{\left(p\right)}=a^{\left(p\right)}+b^{\left(p\right)}T
\end{equation}
et qui dépend de la température. Les calculs \emph{ab initio} étant
à $0$\,K, $b^{\left(p\right)}=0$, $\forall p$. La meilleure paramétrisation
de $\sum_{p=0}^{n}L^{\left(p\right)}\left(1-2x\right)^{p}$ est telle
que :
\[
\Delta E_{mix}=-x\left(1-x\right)\Omega\left(x\right)
\]
avec
\begin{equation}
\Omega\left(x\right)=\left(x-\alpha\right)\left(\beta x^{2}+\gamma x+\delta\right)\label{eq:omega_PAWc}
\end{equation}
où les valeurs des paramètres $\alpha$, $\beta$, $\gamma$ et $\delta$
pour les paramétrisations PWSCF-PAW et SIESTA-NC sont indiquées dans
le tableau \ref{tab:parametres_Omega} pour une expression de l'énergie
d'ordre $\Omega$ en eV.

\begin{table}[h]
\begin{centering}
\begin{tabular}{|c|c|c|}
\hline 
 & PWSCF-PAW & SIESTA-NC \\
\hline 
$\alpha$ & 0.07 & 0.16 \\
\hline 
$\beta$ & $-2.2883$ & $-2.3481$ \\
\hline 
$\gamma$ & 4.43903 & 4.38102 \\
\hline 
$\delta$ & $-2.48044$ & $-2.48041$ \\
\hline 
\end{tabular}
\par\end{centering}

\caption{Paramètres de dépendance en concentration de l'énergie d'ordre $\Omega\left(x\right)=\left(x-\alpha\right)\left(\beta x^{2}+\gamma x+\delta\right)$
ajustés sur les calculs \emph{ab initio} d'énergie de mélange dans
les approximations PWSCF-PAW et SIESTA-NC.\label{tab:parametres_Omega}}
\end{table}

\begin{figure}[h]
\begin{centering}
\includegraphics[scale=0.4]{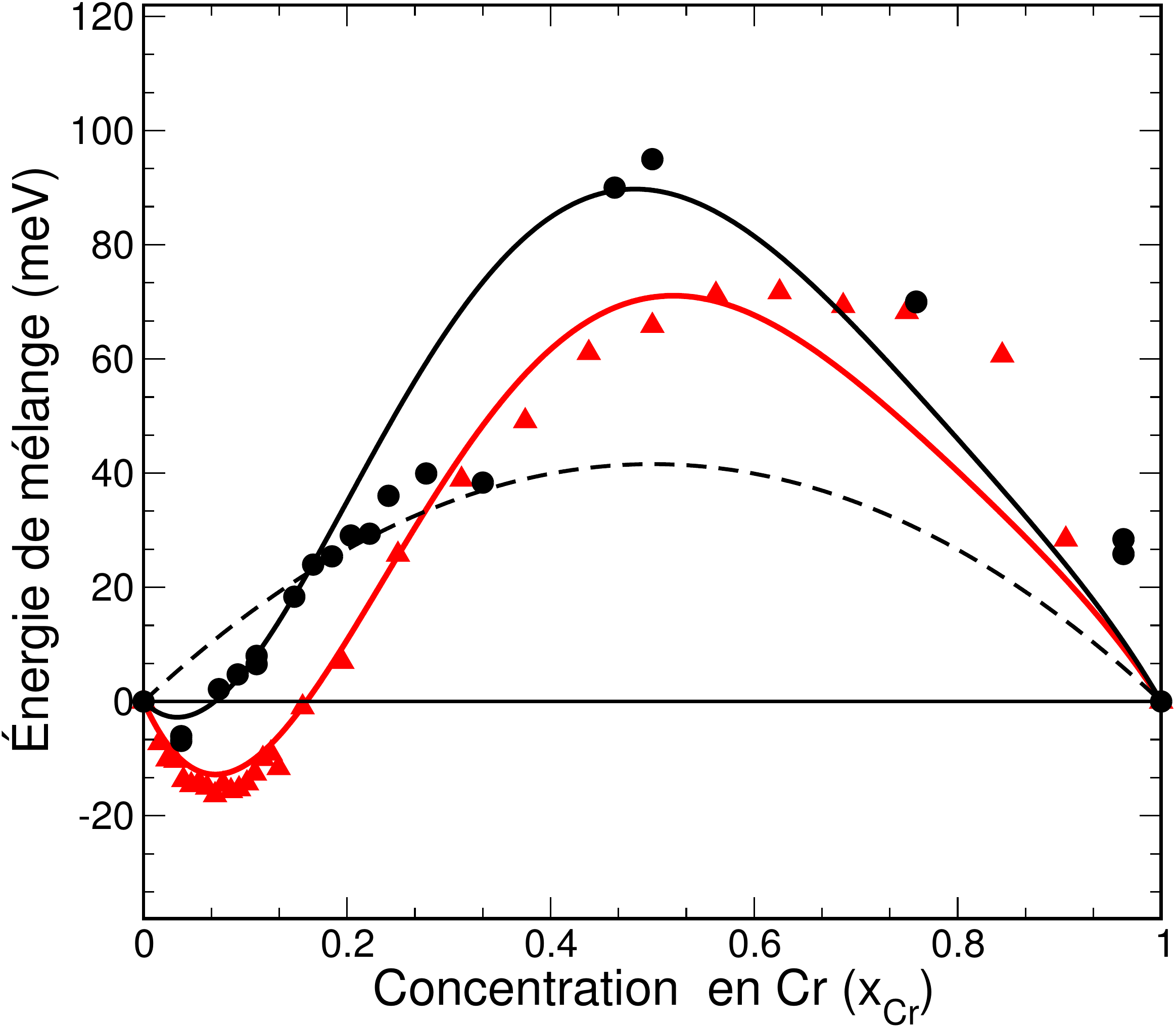}
\par\end{centering}

\caption{Les symboles noirs (rouges) représentent les énergies de mélange de
solutions solides quasi-aléatoires (SQS) calculées dans les approximations
PWSCF-PAW (SIESTA-NC). L'ajustement de Redlich-Kister de ces énergies
de mélange sont en traits pleins. La solution régulière décrite au
paragraphe \ref{sub:FeCr_solreg} est en trait pointillé.\label{fig:hmix}}
\end{figure}

Les énergies de mélange à 0\,K correspondant aux paramétrisations
PWSCF-PAW et SIESTA-NC sont représentées sur la figure \ref{fig:hmix}
et comparées à l'énergie de mélange de la solution régulière déjà
représentée sur la figure \ref{fig:solution_reg_Hmix_et_ddp}.

La dépendance en concentration permet d'obtenir une énergie de mélange
dont la variation est non-symétrique. Celle-ci est négative pour $x<0.07$
(PWSCF-PAW) ou 0.16 (SIESTA-NC), puis positive. Le maximum de $\Delta E_{mix}$
est dans le cas PWSCF-PAW à $x=0.48$ et 0.089\,eV, et dans le cas
SIESTA-NC pour $x=0.52$ et 0.071\,eV.

Les énergies de paires homo-atomiques sont déduites des propriétés
des éléments purs et ne sont pas fonction de la concentration. L'équation
\ref{eq:omega_paires} permet alors de déduire les énergies de paires
hétéro-atomiques de l'énergie d'ordre $\Omega$ et des énergies de
paires homo-atomiques : 

\begin{equation}
\epsilon_{FeCr}^{\left(1\right)}=-\dfrac{\Omega\left(x\right)}{\sum_{i}Z^{\left(i\right)}\eta^{\left(i\right)}}+\dfrac{\epsilon_{FeFe}^{\left(1\right)}+\epsilon_{CrCr}^{\left(1\right)}}{2}\label{eq:EFeCr}
\end{equation}
où 
\begin{equation}
\eta^{\left(i\right)}=\dfrac{\epsilon_{mn}^{\left(i\right)}}{\epsilon_{mn}^{\left(1\right)}}\label{eq:etaeta}
\end{equation}

\begin{figure}[H]
\begin{centering}
\includegraphics[scale=0.4]{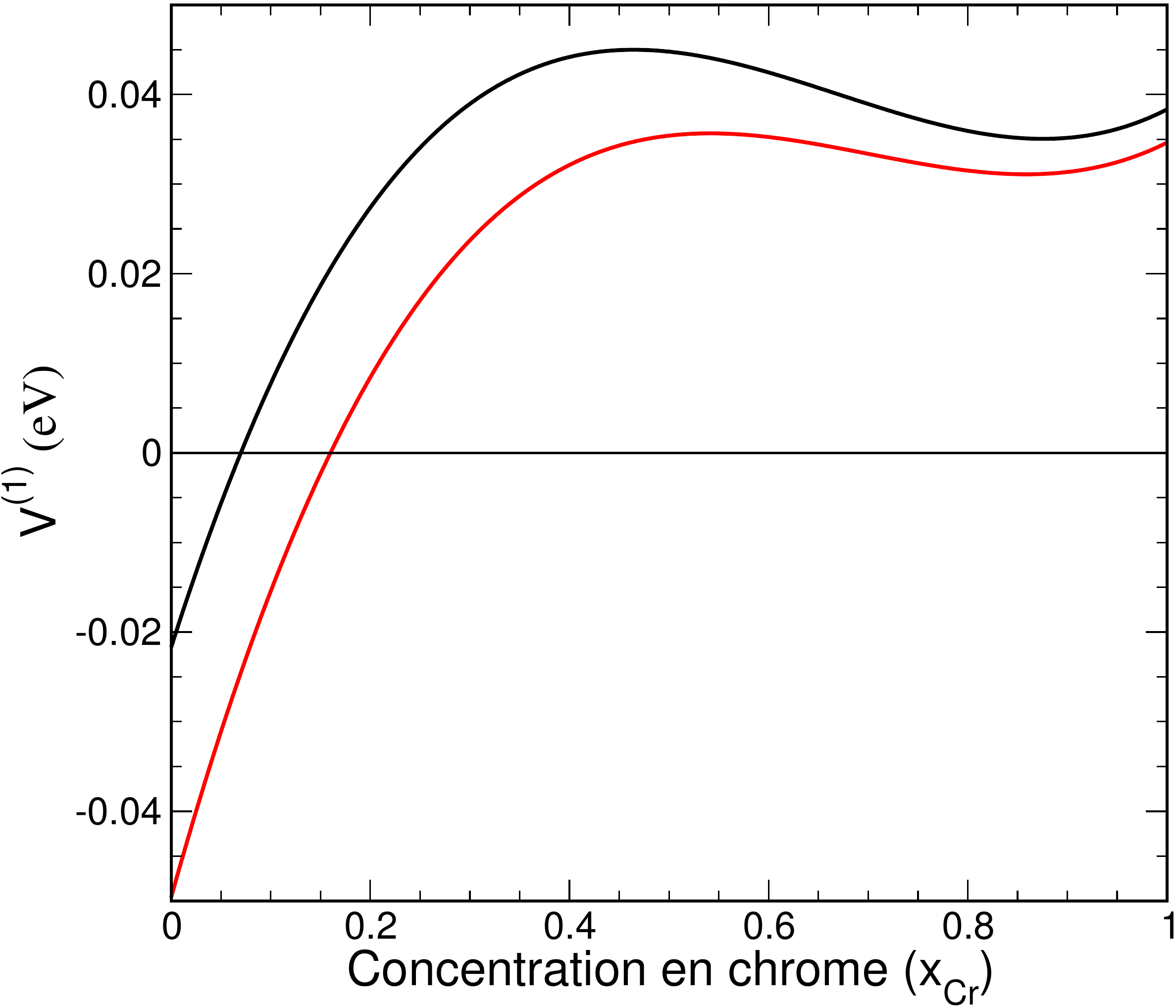}
\par\end{centering}

\caption{$V^{\left(1\right)}=1/2\left(\epsilon_{FeFe}^{\left(1\right)}+\epsilon_{CrCr}^{\left(1\right)}-2\epsilon_{FeCr}^{\left(1\right)}\right)$
en fonction de la concentration locale en chrome $x$ déduite de l'énergie
d'ordre $\Omega\left(x\right)$ ajustée sur les calculs \emph{ab initio}
PWSCF-PAW (en noir) et SIESTA-NC (en rouge) dans le cas d'un modèle
aux premiers voisins seulement.\label{fig:VAB_c}}

\end{figure}

Pour exemple, dans le cas d'un modèle aux premiers voisins seulement,
on trace $V^{\left(1\right)}$ en fonction de la concentration en
chrome $x$ sur la figure \ref{fig:VAB_c} pour les ajustements PWSCF-PAW
et SIESTA-NC de l'énergie d'ordre.

On observe sur la figure \ref{fig:VAB_c} qu'il est favorable de créer
des paires hétéro-atomiques par rapport à des paires homo-atomiques
pour des concentrations $x$ inférieures à 0.07 pour la paramétrisation
PWSCF-PAW et $0.16$ pour la paramétrisation SIESTA-NC, en lien avec
le changement de signe de l'énergie de mélange calculé \emph{ab initio}
à 0\,K. On observe également un changement de signe de la dérivée
seconde de $\epsilon_{FeCr}^{\left(1\right)}$ autour de $x=0.9$
dont les conséquences seront discutées dans la suite du manuscrit.
Notons cependant que les énergies de mélange calculées \emph{ab initio}
dans le domaine de concentration riche en chrome sont moins nombreuses
qu'à l'autre extrémité du diagramme de phases. L'ajustement de Redlich-Kister
y est en conséquence moins précis.

\subsection{Compositions locales ou globales en champ moyen de point}

Dans l'approximation de champ moyen de point, la probabilité d'occupation
de chaque site du réseau est égale à la composition globale (nominale)
du système. On perd donc dans cette approche toute information locale.
Elle ne permet pas de décrire l'ordre à courte distance d'une solution
solide puisque les probabilités de trouver un élément sur chacun des
sites ne dépend pas de son environnement mais seulement de la concentration
globale.

Dans l'ensemble canonique, associer composition locale et composition
nominale est incompatible avec la notion de démixtion où localement
les atomes d'un même type précipitent pour que la concentration locale
soit égale à la limite de solubilité. Dans cet ensemble, le diagramme
de phases sera donc différent de celui obtenu dans l'ensemble grand-canonique
et ne vérifiera pas la règle du bras de levier. C'est une des raisons,
en plus des difficultés numériques décrites en annexe \ref{sec:Ensembles_canoniques_semi-grand-canonique},
pour lesquelles nous avons réalisés les calculs de diagrammes des
phases dans l'ensemble grand-canonique.

Grâce aux simulations Monte Carlo que l'on présente dans la suite
du chapitre, on vérifie l'effet de la dépendance en concentration
globale de l'énergie d'ordre sur le diagramme de phases.

\subsection{Conséquences sur le diagramme de phases de la dépendance en concentration
locale de l'hamiltonien\label{sub:FeCr_solreg}}

\begin{figure}[h]
\begin{centering}
\includegraphics[scale=0.4]{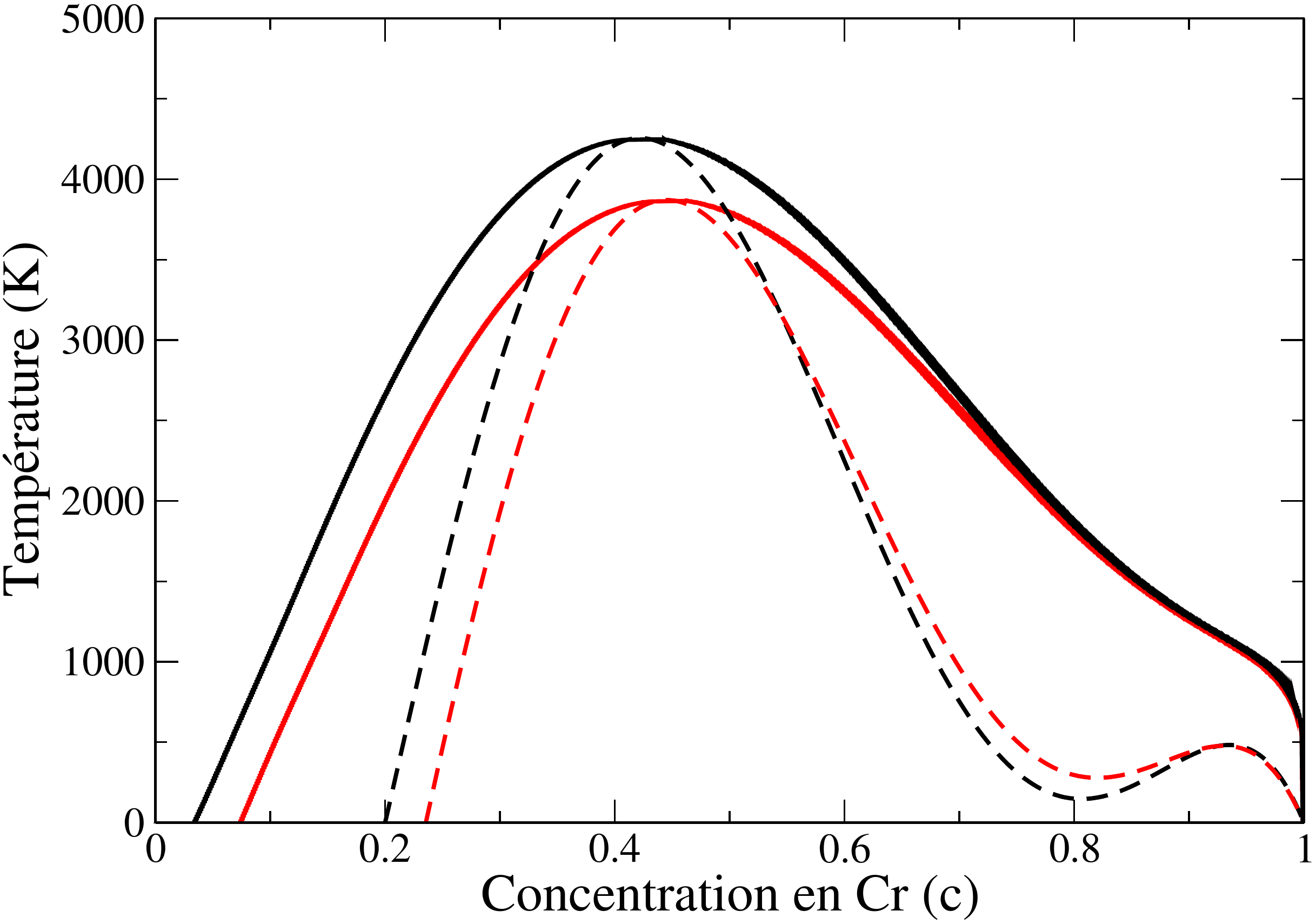}
\par\end{centering}

\caption{Diagrammes de phases en champ moyen de point déduits des modèles d'interactions
de paires sur réseau rigide dépendant de la concentration ajustés
sur les calculs \emph{ab initio} PWSCF-PAW en noir et SIESTA-NC en
rouge. Les limites spinodales sont en tirets pointillés.\label{fig:Diagramme-de-phases}}
\end{figure}

Pour une énergie d'ordre $\Omega$ donnée par l'équation \ref{eq:omega_PAWc},
les limites de solubilité déduites de l'expression \ref{eq:Tc_analytique_solreg}
sont :
\begin{equation}
T\left(x\right)=\dfrac{1}{k_{B}\ln\left(\dfrac{x}{1-x}\right)}\left(\left(1-2x\right)\Omega\left(x\right)+x\left(1-x\right)\dfrac{d\Omega\left(x\right)}{dx}\right)
\end{equation}
où
\begin{equation}
\dfrac{d\Omega\left(x\right)}{dx}=3\beta x^{2}+2\left(\gamma-\alpha\beta\right)x+\delta-\alpha\gamma
\end{equation}
Ces expressions permettent de déduire, par la construction géométrique
de la tangente commune, les limites de solubilité $\alpha$--$\alpha$'
présentées dans la figure \ref{fig:Diagramme-de-phases}.

Les limites de solubilité ne sont plus symétriques. On observe une
large solubilité d'environ 10\,\%\,Cr dans Fe à $0$\,K. À l'autre
extrêmité du diagramme de phases, Fe est insoluble dans Cr.

On observe également un épaulement de la limite de solubilité, que
ce soit pour les paramétrisation PWSCF-PAW ou SIESTA-NC, entre 70
et 90\,\%\,Cr. Autour de $x=0.5$, la prise en compte du changement
de signe de la dérivée seconde de l'énergie de d'ordre induit l'épaulement
que les expérimentateurs associaient jusqu'ici à la dissymétrie de
la phase $\sigma$ \cite{williams_further_1958,andersson_thermodynamic_1987}. 

La limite de solubilité est dorénavant non-nulle à 0\,K du côté riche
en Fe, en contradiction avec le diagramme de phases de référence CALPHAD
de Andersson et Sundman \cite{andersson_thermodynamic_1987}, mais
en accord avec les calculs thermodynamiques de Bonny et al. \cite{bonny_onthe_aa_demixtion_2008}
à partir des potentiels 2BM et CDM et de l'hamiltonien CE.

On calcule également analytiquement les limites spinodales :
\begin{equation}
T_{spinodal}\left(x\right)=-\dfrac{x\left(1-x\right)}{k_{B}}\left(2\Omega\left(x\right)-2\left(1-2x\right)\dfrac{d\Omega}{dx}-x\left(1-x\right)\dfrac{d^{2}\Omega}{dx^{2}}\right)
\end{equation}
où
\begin{equation}
\dfrac{d^{2}\Omega}{dx^{2}}=6\beta x+2\gamma-2\alpha\beta.
\end{equation}
Celles-ci sont représentées en traits pointillés longs sur la figure
\ref{fig:Diagramme-de-phases}.

Du côté riche en Cr, la limite spinodale montre un épaulement singulier
entre 70 et 90\,\%\,Cr. Celui-ci peut être relié au changement de
signe de la dérivée seconde de l'énergie de paires hétéro-atomique.
Ce phénomène a lieu à des températures inférieures à 500\,\textdegree{}C,
difficilement accessibles à l'expérience du fait des cinétiques lentes.
C'est à notre connaissance la première fois qu'un tel comportement
est proposé.

Les températures critiques calculées dans l'approximation de champ
moyen de point sont de 4200\,K (3800\,K) à partir des paramétrisations
PWSCF-PAW (SIESTA-NC). Ces températures sont bien au-dessus des températures
critiques expérimentales mesurées à 1000\,K environ. Les potentiels
CDM et 2BM traités en champ moyen montrent le même problème de température
critique trop élevée \cite{Malerba_revue_2008,bonny_onthe_aa_demixtion_2008}. 

Il n'est pas étonnant que les températures critiques calculées avec
notre modèle d'interactions de paires sur réseau rigide soient trop
élevées. En effet, il ne tient pas compte des entropies non-configurationnelles.

\subsection{Stratégies pour tenir compte de l'entropie non-configurationnelle}

Les diagrammes de phases que nous avons calculés jusqu'ici ainsi que
ceux proposés par Bonny et al. dans l'article \cite{bonny_onthe_aa_demixtion_2008}
ne tiennent pas compte des propriétés thermodynamiques suivantes (au
moins) :
\begin{itemize}
\item l'amplitude des moments magnétiques atomiques diminue quand la température
augmente. Cela diminue les énergies de paires quand $T$ augmente
aux concentrations intermédiaires \cite{lavrentiev_magnetic_CE_2009}.
\item l'entropie magnétique \cite{dudarev_magnetic_2005,lavrentiev_magnetic_CE_2009,lavrentiev_CEmag2_2010}.
\item l'entropie vibrationnelle \cite{desplat_bley_Svib_1996,fultz_phonon_Fe_Cr_1995}.
\end{itemize}
C'est pour ces raisons que les diagrammes de phases que nous rapportons
dans le paragraphe \ref{sub:Les-potentiels-inter-atomiques} ont des
températures critiques trop élevées. Pour abaisser ces dernières,
nous pourrions choisir d'introduire, par exemple, à chaque température
et concentration, un calcul de contribution entropique vibrationnelle
dans le cadre théorique de l'approximation harmonique. Malgré tout,
il n'est pas possible à notre connaissance de calculer rigoureusement
toutes les dépendances en température. Obligé d'introduire une paramètre
ajustable relié à la température, nous le choisissons toujours dans
le même esprit : le plus simple possible. Cela nous permet de toujours
contrôler au mieux le modèle, de pouvoir y revenir facilement (contrairement
à un potentiel EAM par exemple). À défaut de pouvoir être rigoureux,
nous choisissons donc de compenser toutes les entropies non-configurationnelles
en donnant à l'énergie d'ordre une dépendance linéaire en température.
Cela nous permet de plus de ne pas impacter l'efficacité numérique
du modèle.

\subsection{Dépendance en température\label{sub:Dependance-en-temperature}}

Pour conserver la simplicité du modèle de Redlich-Kister, nous donnons
une simple dépendance linéaire en température à l'énergie d'ordre
$\Omega$. Il s'agit du plus bas développement de Landau réalisable.
L'énergie d'ordre s'exprime dorénavant :
\begin{equation}
\Omega\left(c,T\right)=\Omega\left(c\right)\left(1-\dfrac{T}{\Theta}\right)\label{eq:Omega_PAW_T}
\end{equation}
où $\Theta$ est une température en Kelvin ajustée sur la température
critique expérimentale d'environ $960$\,K. Cela revient à donner
une dépendance en température aux énergies de paires. On trouve :
\begin{eqnarray}
\Theta_{PWSCF-PAW} & = & 1480\,\textrm{K}\\
\Theta_{SIESTA-NC} & = & 1520\,\mbox{K}
\end{eqnarray}

\begin{figure}[h]
\begin{centering}
\includegraphics[scale=0.4]{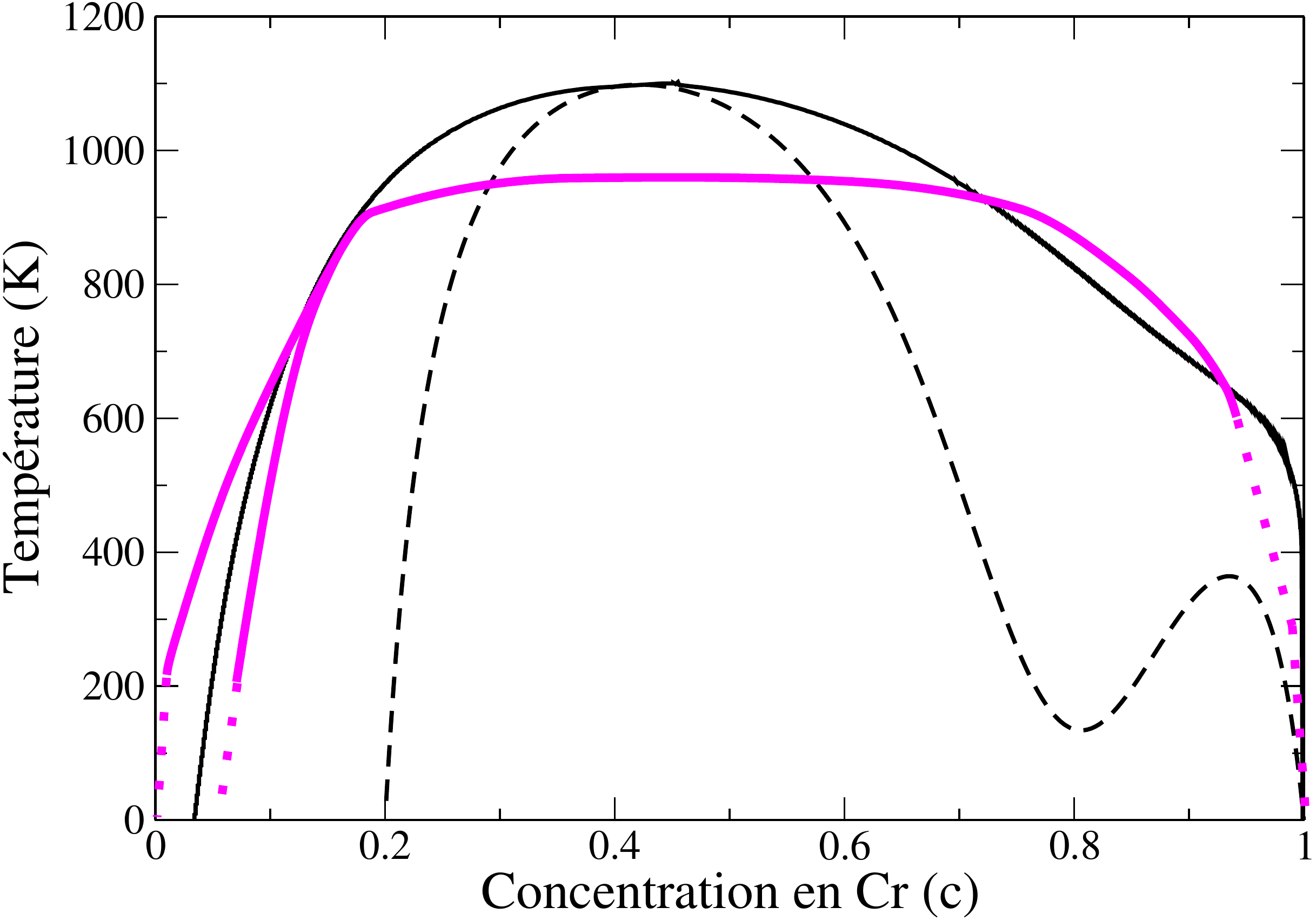}
\par\end{centering}

\caption{En trait plein noir, le diagramme de phases correspondant à la paramétrisation
PWSCF-PAW de l'énergie de mélange, calculé en champ moyen de point,
tenant compte de la dépendance en concentration de l'énergie d'ordre
et de la correction en température. Les limites spinodales sont représentées
en tirets pointillés noirs. En traits pleins magenta, on représente
les limites de solubilité inférieures et supérieures proposées par
Xiong et al. sur la base d'une revue des données expérimentales \cite{xiong_grrrrbonny_2010}.
Les extrapolations des limites de Xiong et al. à basse température
sont en traits pointillés magenta. \label{fig:Diagramme-de-phases_Tdep}}
\end{figure}

Le diagramme de phases correspondant est représenté sur la figure
\ref{fig:Diagramme-de-phases_Tdep} pour la paramétrisation PWSCF-PAW
de l'énergie d'ordre. Notons qu'il ne s'agit pas d'une simple homotéthie
du diagramme de phases de la figure \ref{fig:Diagramme-de-phases}
correspondant à l'hamiltonien indépendant de la température car l'entropie
de mélange $\Delta S_{mix}$ reste indépendante de la température.

Le diagramme de phases de la figure \ref{fig:Diagramme-de-phases_Tdep}
est semblable à celui décrit au paragraphe \ref{sub:param_hmix} ci-dessus.
La limite de solubilité à 0\,K du côté riche en Fe est 0.04. Fe n'est
pas soluble dans Cr à 0\,K, et la limite de solubilité monte de façon
très raide avec la température. La température critique calculée en
champ moyen est d'environ 1100\,K. Ces résultats sont en bon accord
avec les compilations expérimentales de Xiong et al. \cite{xiong_grrrrbonny_2010}
et de Bonny et al. \cite{bonny_onthe_aa_demixtion_2008}.

On observe toujours un épaulement des limites spinodales entre 70
et 90\,\%\,Cr lié au changement de signe de la dérivée seconde de
l'énergie de paire hétéro-atomique dans ce domaine de concentration.
Aucune mesure expérimentale n'a jamais mis en évidence un tel comportement
dans le système fer-chrome car s'ils apparaissaient, ce serait à des
températures telles que l'équilibre serait atteint en plusieurs années
ou décennies \ldots{}

\begin{figure}[h]
\begin{centering}
\includegraphics[scale=0.4]{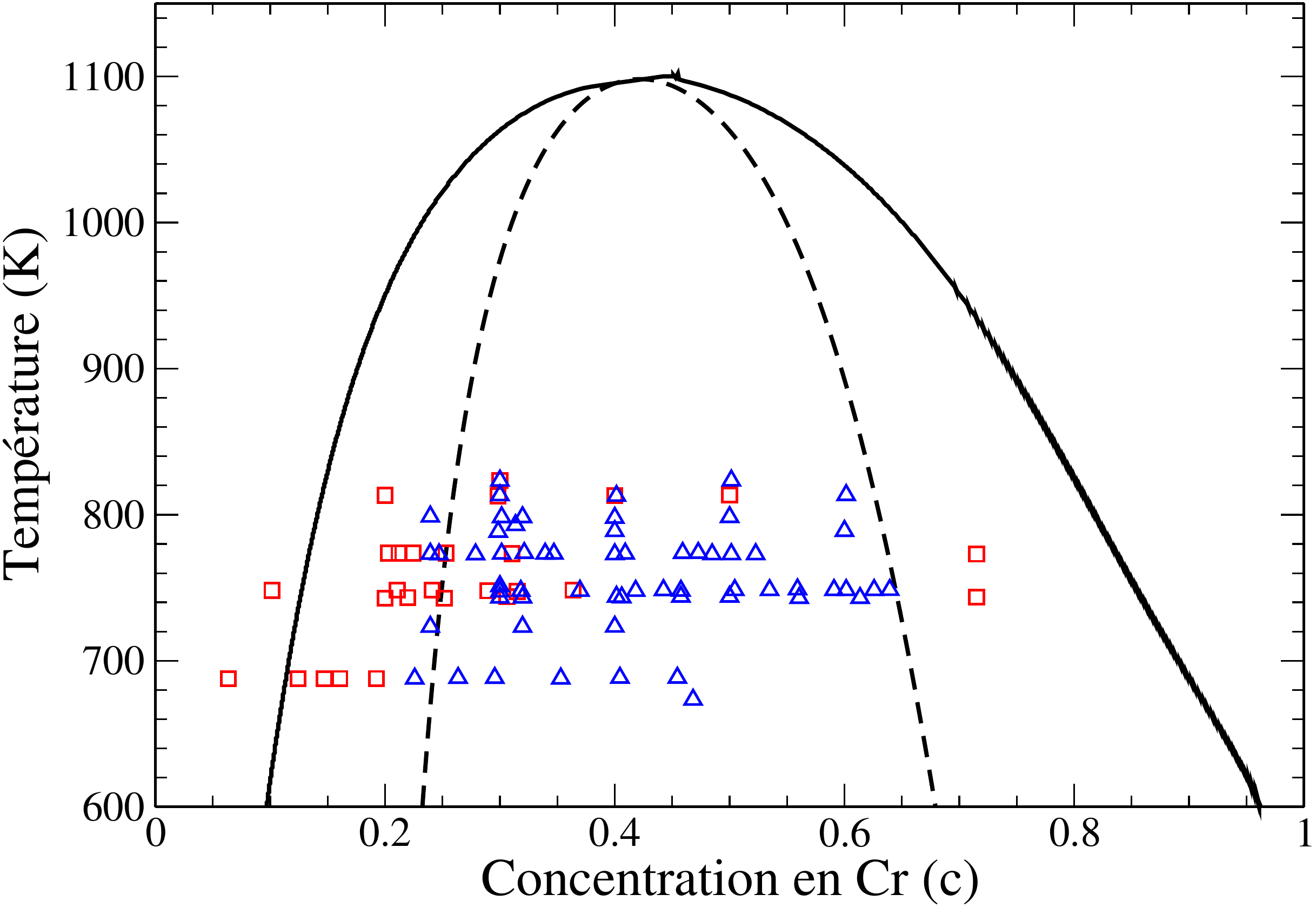}
\par\end{centering}

\caption{Limites de solubilité $\alpha-\alpha'$ (trait noir) et limites spinodales
(pointillés noirs) calculées en champ moyen à partir de notre modèle.
Les mesures expérimentales montrant un régime de démixtion par (en
rouge) germination--croissance--coalescence et (en bleu) décomposition
spinodale. Les données expérimentales sont compilées par Xiong et
al. \cite{xiong_grrrrbonny_2010}. Se référer à cet article pour le
détail expérimental.\label{fig:limites_spinodales_champ_moyen}}
\end{figure}

De nombreuses mesures expérimentales ont cependant été réalisées entre
650 et 800\,K afin de déterminer les limites de décomposition spinodales.
Ces mesures ont été très récemment compilées par Xiong et al. \cite{xiong_grrrrbonny_2010}
et nous les comparons à notre diagramme de phases dans la figure \ref{fig:limites_spinodales_champ_moyen}.

On ne peut définir de limite nette entre les régimes de germination--croissance--coalescence
et de décomposition spinodale. La transition entre ces deux régimes
est continue, de part et d'autre de la limite spinodale représentée
en pointillés dans la figure \ref{fig:limites_spinodales_champ_moyen}.
En conséquence, nos limites spinodales sont très satisfaisantes dans
ce domaine de température.

\section{Comparaison avec les simulations Monte Carlo}

Les simulations Monte Carlo permettent d'échantillonner de manière
exacte l'espace des phases des systèmes de taille finie, aux incertitudes
numériques près. On peut donc en déduire le diagramme de phases le
plus exact numériquement possible du système décrit par l'hamiltonien
présenté au paragraphe \ref{sec:Notre-mod=0000E8le-thermodynamique}.
On vérifie ainsi les résultats de champ moyen présentés ci-dessus. 

En champ moyen de point, la concentration est homogène et le diagramme
de phases d'un système à tendance à la démixtion ne dépend que de
l'énergie d'ordre. Dans les simulations Monte Carlo, deux portées
sont à distinguer : celle pour le calcul de la concentration locale
et celle des interactions.

\subsection{Définition de la concentration locale}

\begin{figure}[H]
\begin{centering}
\includegraphics[scale=0.4]{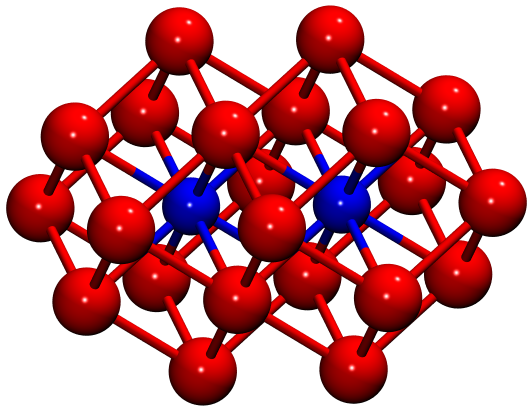}
\par\end{centering}

\caption{Dans le cas illustré ici, l'interaction de la paire AB (sphères bleues)
dépend de la composition moyenne mesurée sur l'ensemble des sites
1$^{\text{ers}}$ et 2$^{\text{e}}$ voisins (sphères rouges). \label{fig:Illustration_clocal}}
\end{figure}

Nous définissons dans l'annexe \ref{cha:def_clocal_methodo_annexe}
l'algorithme que nous proposons pour définir la concentration locale.
La zone à considérer s'étale sur les $i$ premières sphères de coordinations
des atomes dont nous calculons l'énergie de paire. Nous montrons dans
l'annexe \ref{cha:def_clocal_methodo_annexe} que le choix de la taille
de cette zone a des conséquences critiques sur l'efficacité numérique
des calculs. Pour exemple, nous illustrons sur la figure \ref{fig:Illustration_clocal}
l'ensemble des sites (en rouge) à considérer pour le calcul de la
concentration locale autour des sites bleus lorsque le calcul de la
concentration locale se fait sur les deux premières sphères de coordination
($i=2$) pour deux sites bleus deuxièmes voisins.

Dans la suite du manuscrit, on notera la concentration locale simplement
$x$. Ce n'est pas en contradiction avec la notation utilisée pour
les calculs de champ moyen de point puisque la concentration locale
y était égale à la concentration globale. On définit en fait $x$
comme la concentration locale, que ce soit en champ moyen ou en Monte
Carlo.

Quand un système tend à la démixtion, le diagramme de phases ne dépend
que de l'énergie d'ordre totale $\Omega$, somme des $V^{\left(i\right)}$
sur l'ensemble des voisins $i$. Il faut cependant décider de la répartition
de l'énergie sur chacune de ces sphères de coordination, c'est-à-dire
définir les valeurs des $\eta^{\left(i\right)}$ définis au paragraphe
\ref{eq:etaeta}. C'est également en annexe \ref{sub:Mod=0000E8le-d'int=0000E9raction}
que nous discutons ce point. En conclusion de cette annexe, nous choisissons
$\eta^{\left(i\right)}$ égal au rapport entre d'une part la distance
entre $i$$^{\text{e}}$ voisins et d'autre part la distance entre
premiers voisins.

Le principe des simulations Monte Carlo telles que nous les implémentons
est rappelé succinctement en annexe \ref{sec:Principe-des-simulations_monte_carlo}.
Ces simulations se font dans l'ensemble semi-grand-canonique, ce qui
permet une estimation précise des limites de solubilité dont nous
détaillons la méthodologie en annexe  \ref{sec:Ensembles_canoniques_semi-grand-canonique}.

\subsection{Paramètres des simulations Monte Carlo}

Tous les calculs Monte Carlo sont réalisés en conditions périodiques.
Les super-cellules contiennent typiquement $10^{5}$\,atomes. Les
effets de taille de la super-cellule de calcul sont toujours vérifiés. 

On observe qu'environ $10^{4}$ pas Monte Carlo (échanges ou permutations
selon l'ensemble thermodynamique) acceptés par atome permettent d'atteindre
la solution stationnaire (voir annexe \ref{sec:Principe-des-simulations_monte_carlo}).

\subsection{Influence de la portée des interactions\label{sub:Influencede-la-port=0000E9e_des_interactions}}

\begin{figure}[h]
\begin{centering}
\includegraphics[scale=0.4]{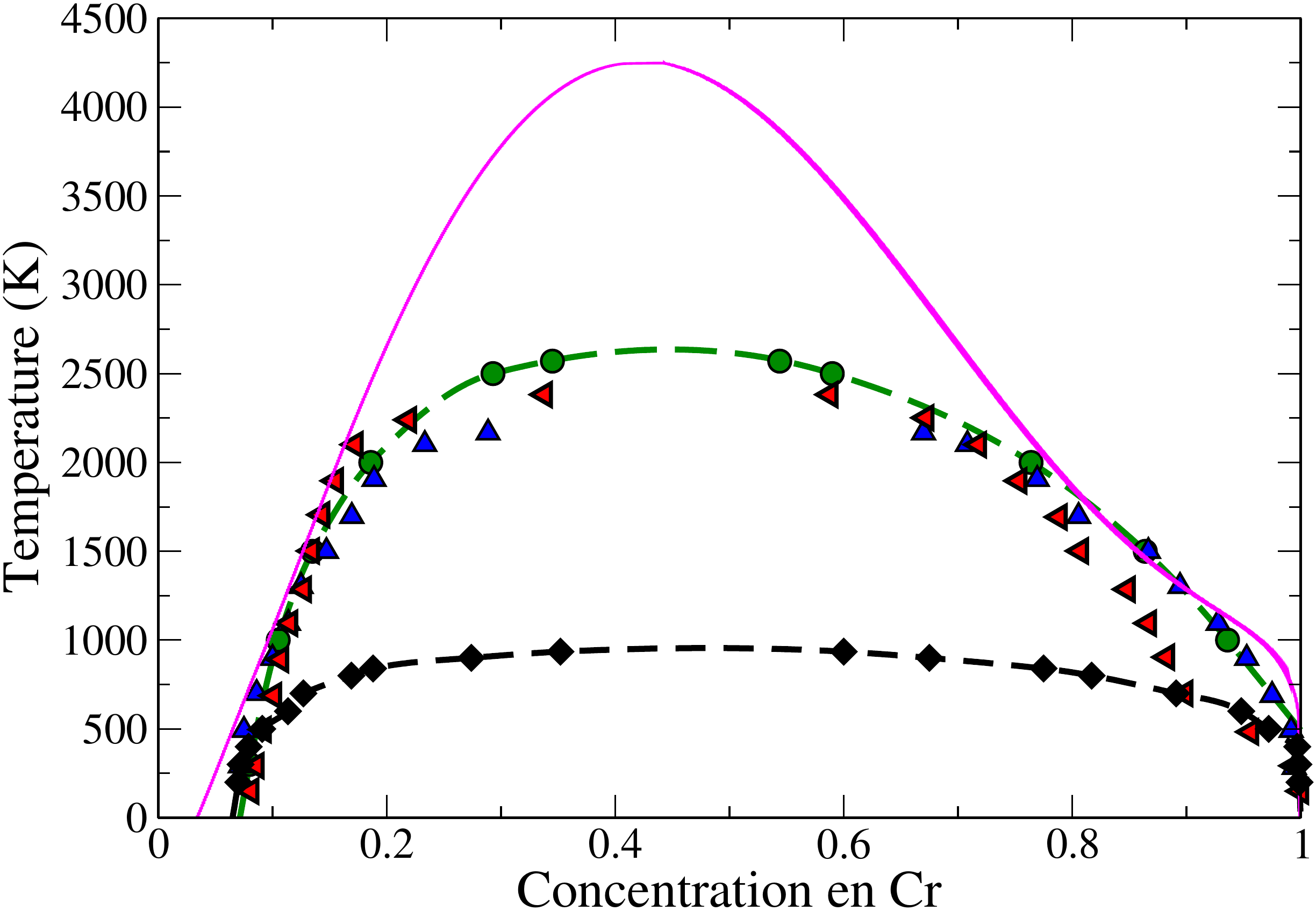}
\par\end{centering}

\caption{Limites de solubilité déduites des simulations Monte Carlo pour une
énergie d'ordre $\Omega$ ajustée sur les résultats \emph{ab initio}
PWSCF-PAW avec différentes portées d'interaction. (bleu) aux deuxièmes
voisins soit $i=2$, (rouge) aux cinquièmes voisins, soit $i=5$,
et (vert) pour une composition locale assimilée à la composition globale.
Les points correspondent aux résultats des simulations Monte Carlo.
Le diagramme de phases tenant compte de la correction en température
aux deuxièmes voisins est représenté en noir. Celui issu des calculs
de champ moyen de point sans correction en température est en trait
continu magenta.\label{fig:MC_portee_interactions}}
\end{figure}

Nous discutons ici et évaluons l'influence de la portée des interactions
sur les diagrammes de phases déduits des simulations Monte Carlo.
Les limites de solubilité calculées à partir des énergies d'ordre
$\Omega$ ajustées sur les calculs \emph{ab initio} SQS PWSCF-PAW
sont représentées sur la figure \ref{fig:MC_portee_interactions}
pour des portées aux deuxièmes et cinquièmes voisins ainsi que pour
une portée infinie (la concentration locale est alors égale à la concentration
globale du système). L'hamiltonien utilisé ne dépend pas de la température
car les différences sont trop faibles entre les trois portées présentées
pour être perceptibles une fois la correction en température prise
en compte. On présente également le diagramme de phases pour une portée
aux deuxièmes voisins avec l'hamiltonien dépendant de la température.

La portée des interactions n'a pas d'influence très forte sur les
limites de solubilité calculées. La température critique augmente
légèrement avec la portée des interactions. Les résultats aux troisièmes
et quatrièmes voisins ($i=3$ et 4), qui ne sont pas représentés sur
la figure \ref{fig:MC_portee_interactions} pour plus de lisibilité,
suivent également cette tendance. Théoriquement, le champ moyen de
point devient exact quand la portée des interactions tend vers l'infini.
Il est donc rassurant qu'en augmentant la portée des interactions
les résultats Monte Carlo se rapprochent des résultats en champ moyen.
Cependant, la température critique calculée en Monte Carlo en assimilant
compositions locales et globales ($\approx2600$\,K) est inférieure
d'environ 38\,\% à celle calculée en champ moyen (en magenta) sans
correction en température ($\approx4200$\,K). C'est une différence
supérieure à la différence classique entre champ moyen de point et
Monte Carlo.

Du côté riche en Fe, la figure \ref{fig:MC_portee_interactions} montre
des limites de solubilité calculées à 0\,K de l'ordre de $7$\,\%\,Cr.
Les limites de solubilité ajustées sur les résultats SIESTA-NC sont
d'environ $16$\,\%\,Cr. À très basse température ($T<500$\,K),
la solubilité calculée de Fe dans Cr est nulle.

Les températures critiques calculées avec l'hamiltonien indépendant
de la température varient d'environ $2300$\,K aux deuxièmes voisins
à $2600$\,K quand la composition locale est assimilée à la composition
globale. En désaccord avec les mesures expérimentales, les température
critiques présentées sur la figure \ref{fig:Diagramme-de-phases}
sont trop élevées d'un facteur deux à trois, selon la portée des interactions.
Ce problème de température critique est similaire à celui exposé par
Malerba, Bonny et al. \cite{Malerba_revue_2008,bonny_onthe_aa_demixtion_2008}
quant aux diagrammes de phases calculés avec les potentiels CDM et
2BM. 

Afin de baisser la température critique de notre modèle énergétique,
nous donnons la même dépendance en température à l'énergie d'ordre
$\Omega$ qu'en champ moyen de point. Le diagramme de phases déduit
des simulations Monte Carlo avec la paramétrisation PWSCF-PAW corrigé
en température est reproduit sur la figure \ref{fig:MC_portee_interactions}.
La température critique est dorénavant de 950\,K, en accord avec
les résultats expérimentaux.

\subsection{Comparaison champ moyen -- Monte Carlo pour les modèles }

\begin{figure}[h]
\begin{centering}
\includegraphics[scale=0.4]{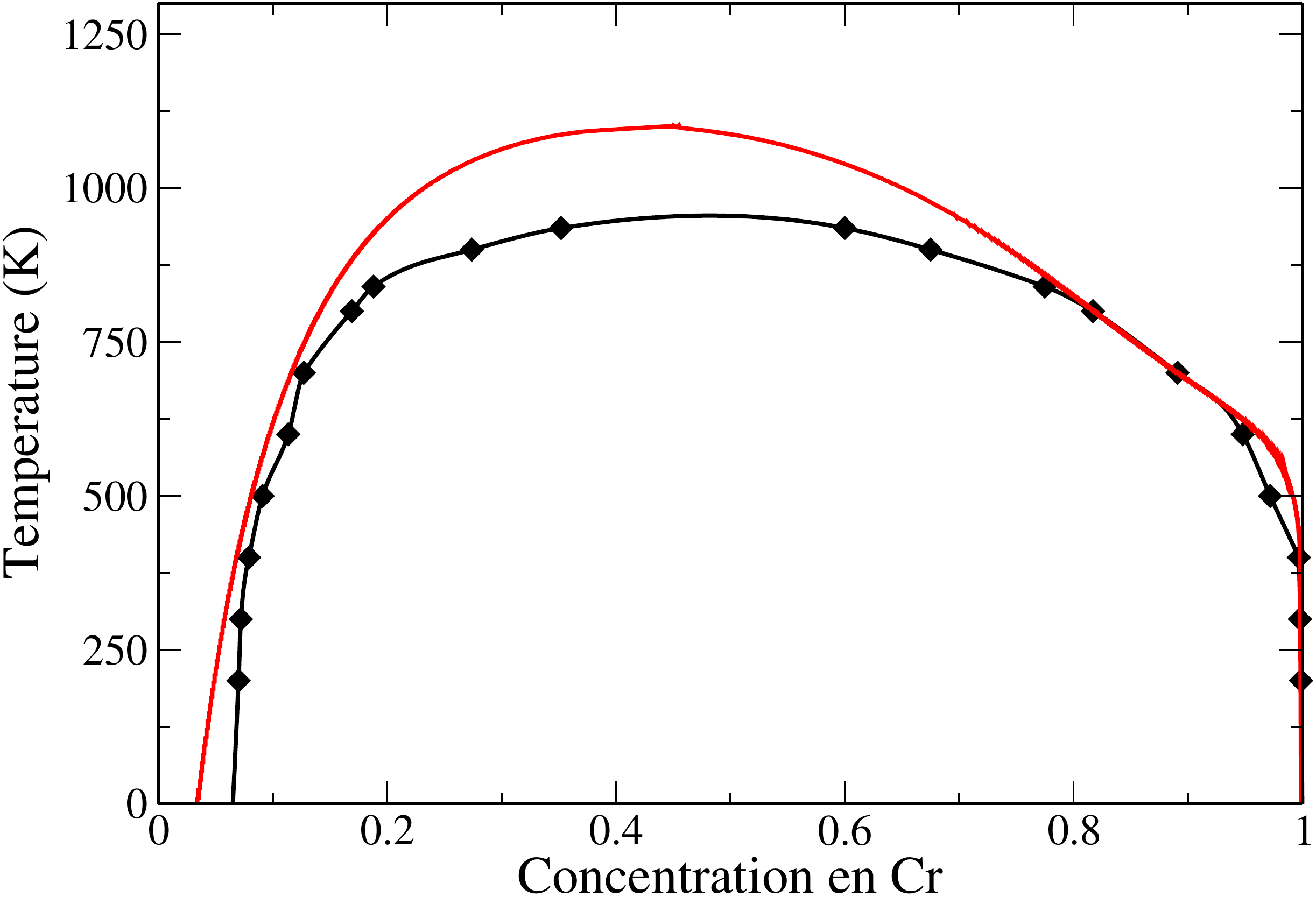}
\par\end{centering}

\caption{Limites de solubilité calculées en champ moyen de point (en rouge)
et en Monte Carlo aux deuxièmes voisins (en noir) avec l'hamiltonien
dépendant à la fois de la concentration et de la température. La paramétrisation
utilisée est PWSCF-PAW. Les losanges correspondent aux résultats des
simulations Monte Carlo.\label{fig:comparaison_MC_MF}}
\end{figure}

On représente sur la figure \ref{fig:comparaison_MC_MF} les limites
de solubilité calculées en champ moyen de point et en Monte Carlo,
avec des interactions aux deuxièmes voisins et l'hamiltonien dépendant
à la fois de la concentration et de la température.

Nous décrivions ci-dessus que sans la correction en température, le
décalage entre Monte Carlo et champ moyen est d'environ 38\,\%, ce
qui est très supérieur au décalage habituel \cite{de_fontaine_1979}.
Avec la correction en température, la température critique calculée
en champ moyen est de 1100\,K. Elle est d'environ 950\,K en Monte
Carlo. La différence est de 14\,\%, en accord avec les résultats
théoriques de Fontaine \cite{de_fontaine_1979}.

\section{Comparaison avec l'expérience et les autres modèles énergétiques\label{sub:comparaison_modeles_thermo}}

\subsection{Comparaison avec les limites de solubilité proposées par Xiong et
al. et Bonny et al.}

\begin{figure}[H]
\begin{centering}
\includegraphics[scale=0.4]{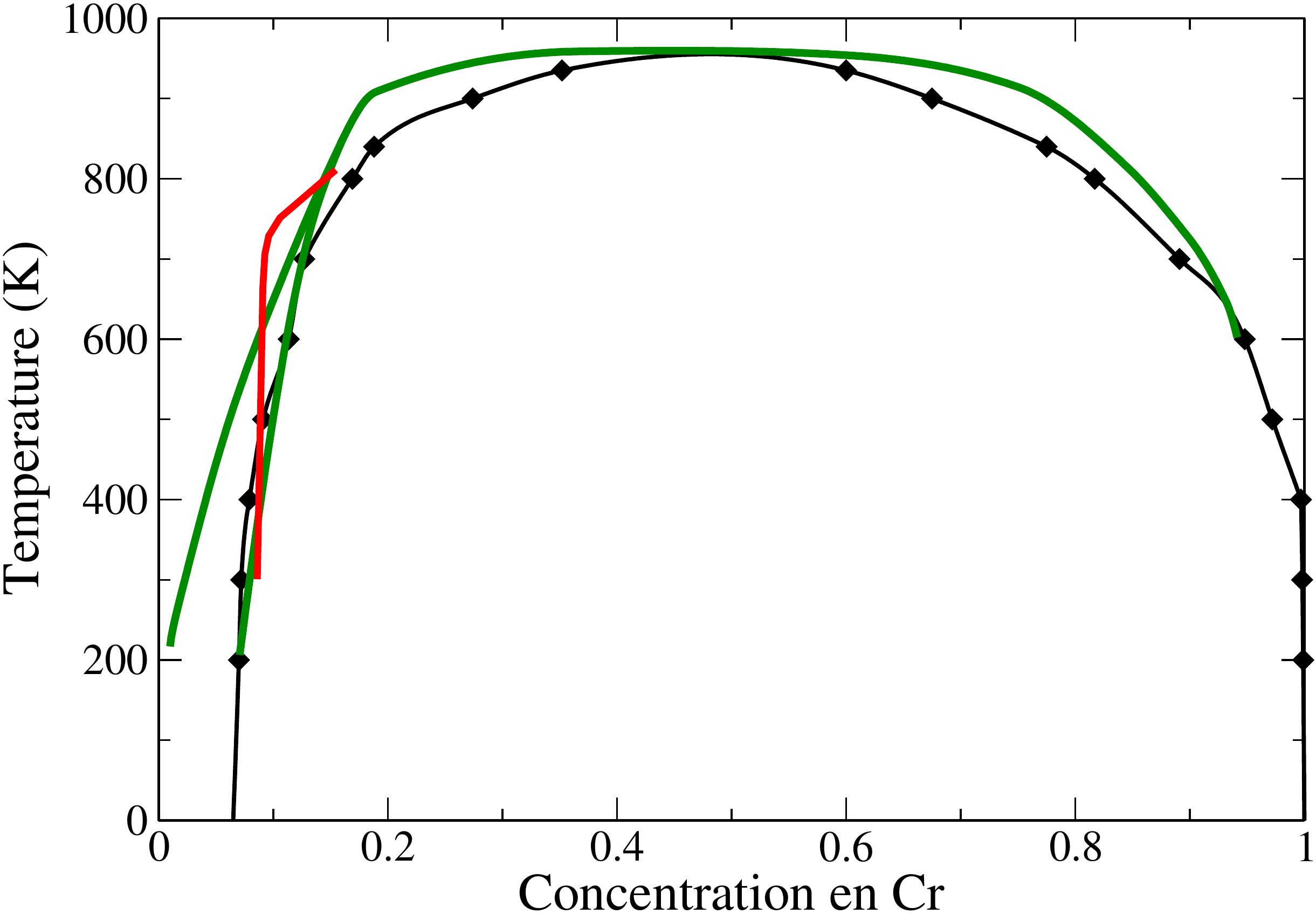}
\par\end{centering}

\caption{Limites de solubilité $\alpha-\alpha$' déduites des simulations Monte
Carlo (en noir). Les bornes inférieures et supérieures proposées par
Xiong et al. \cite{xiong_grrrrbonny_2010} sont en vert. La limite
proposée par Bonny et al. \cite{bonny_onthe_aa_demixtion_2008} tenant
compte des expériences sous irradiation est en rouge.\label{fig:nous_exp}}
\end{figure}

Sur la figure \ref{fig:nous_exp}, on compare le diagramme de phases
calculé en champ moyen ainsi qu'en Monte Carlo avec les limites de
solubilité proposées par Xiong et al. \cite{xiong_grrrrbonny_2010}
et Bonny et al. \cite{bonny_onthe_aa_demixtion_2008} dans leurs revues
critiques des résultats expérimentaux.

Notre modèle est en accord qualitatif avec ces deux études. La solubilité
du chrome dans le fer est large. À 250\,K, Xiong et al. considèrent
qu'elle devrait être comprise entre 1 et 7\,\%\,Cr. Les limites
déduites de notre hamiltonien montrent 5 (7)\,\%\,Cr en champ moyen
(Monte Carlo). La température critique est légèrement surévaluée en
champ moyen d'environ 15\,\%, ce qui est classique avec ce type de
traitement statistique. La solubilité du fer dans le chrome est inférieure
à 1\,\% jusqu'aux températures supérieures à 600\,K. À température
nulle, la solubilité de chrome dans Fe calculée est de 3\,\% en champ
moyen et 6\,\% en Monte Carlo.

\subsection{Comparaison avec les autres modèles énergétiques}

\begin{figure}[H]
\begin{centering}
\includegraphics[scale=0.4]{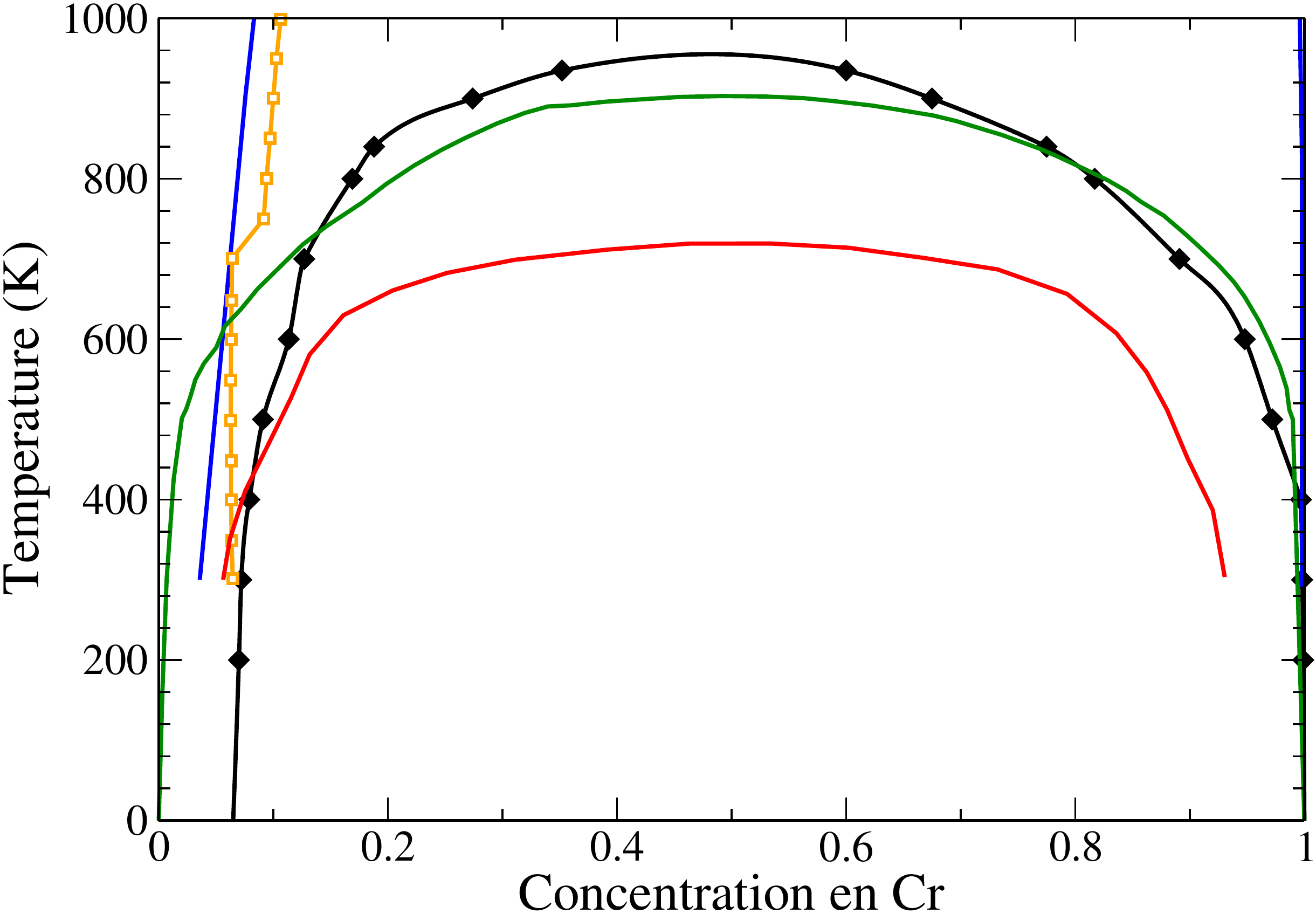}
\par\end{centering}

\caption{Limites de solubilité $\alpha$--$\alpha$' (points et traits noirs)
de nos simulations Monte Carlo avec le modèle aux deuxièmes voisins
dépendant de la concentration locale et de la température. (orange)
développement en amas magnétiques de Lavrentiev, Dudarev et al. \cite{lavrentiev_classicalCE_FeCr_2007,bonny_onthe_aa_demixtion_2008}.
(rouge) potentiel 2BM d'Olsson et al. \cite{olsson_electronic_2006,olsson_erratum:_2006,Malerba_revue_2008}
tenant compte de l'entropie de vibration\cite{Malerba_communication_perso}.
(bleu) potentiel CDM de Caro et al. \cite{Caro_CDM_2005,Malerba_revue_2008}
tenant compte de l'entropie de vibration \cite{Malerba_communication_perso}.
(vert) paramétrisation CALPHAD d'Andersson et Sundman \cite{andersson_thermodynamic_1987}.\label{fig:limites_MC_2nn_Tdep-1}}
\end{figure}

Les limites de solubilité $\alpha$--$\alpha'$ dans l'ensemble du
domaine de concentration calculées avec les modèles CDM \cite{Caro_CDM_2005},
2BM \cite{olsson_two-band_2005,olsson_erratum:_2006}, développements
en amas \cite{lavrentiev_classicalCE_FeCr_2007}, CALPHAD d'après
Andersson et Sundman \cite{andersson_thermodynamic_1987} sont représentées
dans la figure \ref{fig:limites_MC_2nn_Tdep-1}. Elles sont comparées
avec les limites de solubilité que nous calculons par simulations
Monte Carlo avec le modèle aux deuxièmes voisins dépendant de la concentration
locale et de la température. Notre modèle reproduit au mieux à la
fois la température critique d'environ 950\,K, une solubilité non
nulle du chrome dans le fer $\alpha'$ ainsi qu'une solubilité nulle
du fer dans le chrome $\alpha$.

\begin{figure}[h]
\begin{centering}
\includegraphics[scale=0.4]{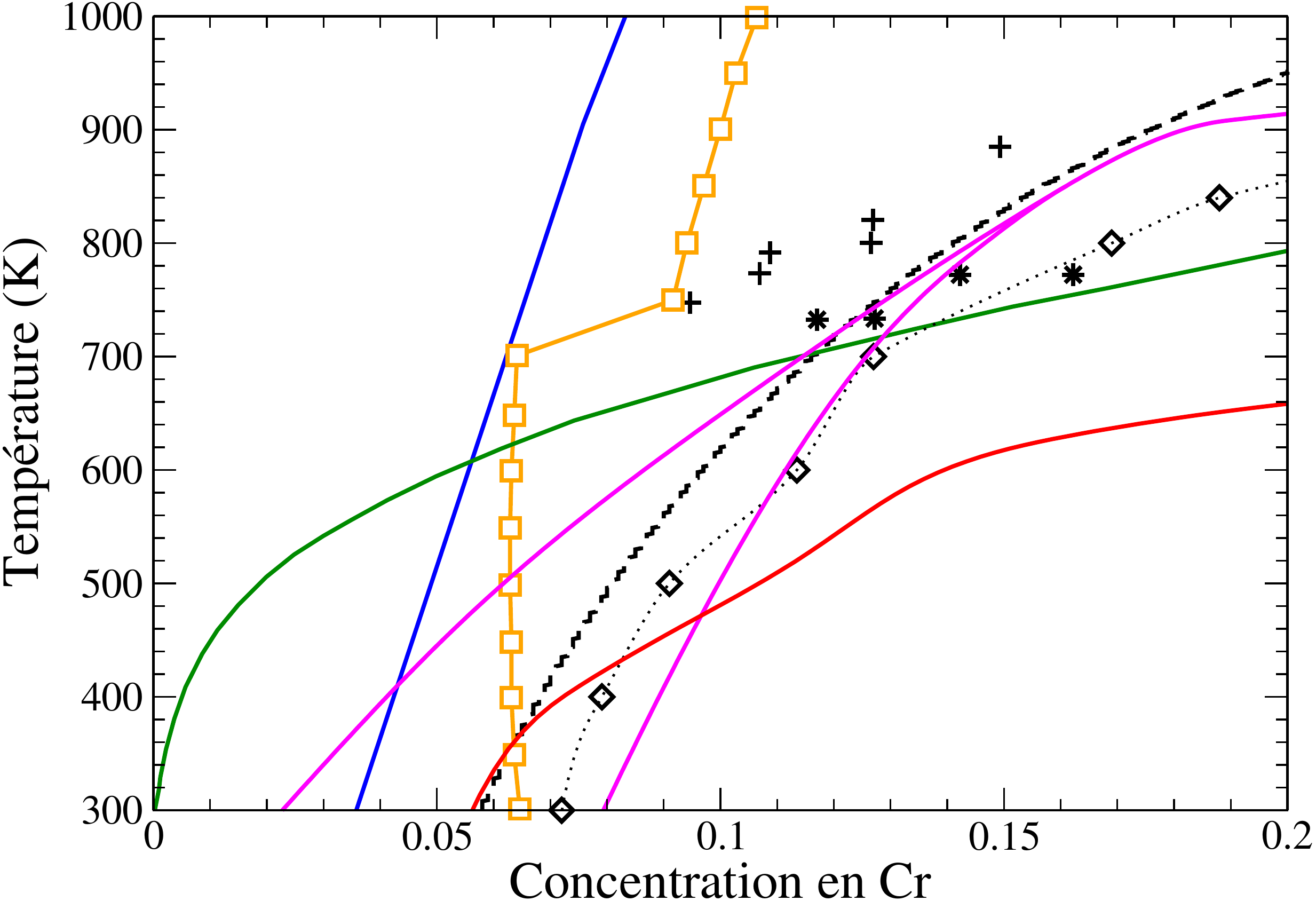}
\par\end{centering}

\caption{Limites de solubilité calculées à partir des modèles \protect \\
(orange) développements en amas \cite{lavrentiev_classicalCE_FeCr_2007,bonny_onthe_aa_demixtion_2008}
d'après Bonny et al. \cite{bonny_onthe_aa_demixtion_2008}, \protect \\
(rouge) potentiels 2BM \cite{olsson_electronic_2006,olsson_erratum:_2006,Malerba_revue_2008,Malerba_communication_perso},\protect \\
(bleu) potentiels CDM \cite{Caro_CDM_2005,Malerba_revue_2008,Malerba_communication_perso},\protect \\
(vert) CALPHAD \cite{andersson_thermodynamic_1987},\protect \\
(traits pointillés noirs et losanges noirs) notre modèle dans l'approximation
de champ moyen de point, (trait pleins et carrés noirs) notre modèle
par simulations Monte Carlo.\protect \\
On représente également les limites de solubilité expérimentales d'équilibre
de Kuwano \cite{kuwano_mssbauer_1988} (croix noires) et de Dubiel
et Inden \cite{dubiel_mossbauer_longtermannealed_1987}(étoiles noires).\protect \\
(traits magenta) limites de solubilité proposées par Xiong et al.
après lecture critique des données expérimentales de la littérature
\cite{xiong_grrrrbonny_2010}.\label{fig:ddp_Ferich_comparaison_modeles}}
\end{figure}

Un focus sur le côté riche en fer du diagramme de phases d'après les
mêmes auteurs mais aussi dans l'approximation de la solution régulière
et à partir de notre modèle sont comparées dans la figure \ref{fig:ddp_Ferich_comparaison_modeles}.
Les limites de solubilité expérimentales mesurées par spéctroscopie
Mössbauer après des recuits longs de Kuwano \cite{kuwano_mssbauer_1988}
et de Dubiel et Inden \cite{dubiel_mossbauer_longtermannealed_1987}
sont également indiquées. Nous choisissons de ne pas représenter les
mesures expérimentales sous irradiation car l'influence de l'irradiation
sur les limites de solubilité fait débat \cite{bonny_onthe_aa_demixtion_2008,xiong_grrrrbonny_2010}
(on peut penser en particulier à des effets de désordre balistiques
induits par les collisions nucléaires, où à des phénomènes de ségrégation
induite par l'irradiation, sous l'effet de couplage entre flux de
défauts et flux d'espèces chimiques).

Le domaine biphasé calculé par Andersson et Sundman \cite{andersson_thermodynamic_1987}
dans l'approche CALPHAD est très proche de celui d'une solution régulière.
Ces deux modèles sont incompatibles avec les récentes mesures expérimentales
et les calculs \emph{ab initio} du côté riche en fer. Le domaine biphasé
calculé par Bonny et al. \cite{bonny_onthe_aa_demixtion_2008} à partir
des potentiels 2BM \cite{olsson_electronic_2006,olsson_erratum:_2006}
et CDM \cite{Caro_CDM_2005} est trop stable en température : il y
a un facteur deux entre la température critique expérimentale et la
température critique déduite des modèles. Le potentiel 2BM produit
une énergie de mélange symétrique (négative également du côté riche
en chrome), ce qui induit une trop grande solubilité de Fe dans Cr.
Le développement en amas classique de Lavrentiev et al. \cite{lavrentiev_classicalCE_FeCr_2007}
semble également mener à une température critique trop élevée. 

Comme on peut l'observer sur la figure \ref{fig:ddp_Ferich_comparaison_modeles}
notre modèle est en accord quantitatif avec les mesures expérimentales.
Il est en accord avec les revues expérimentales de Xiong et al. \cite{xiong_grrrrbonny_2010}
et Bonny et al. \cite{bonny_onthe_aa_demixtion_2008}. Il semble meilleur
que les modèles proposés précédemment pour reproduire des limites
de solubilité $\alpha-\alpha'$.

Les autres modèles conservent tout de même d'autres avantages : les
potentiels EAM permettent des calculs tenant compte des lacunes et
interstitiels, par exemple pour les simulations incluant des dislocations.
Il n'y a cependant pas de frein théorique à l'intégration des lacunes
et interstitiels dans notre modèle. Les modèles CALPHAD permettent
quant à eux de modéliser la lacune de miscibilité, mais aussi d'intégrer
les phases $\sigma$, $\gamma$ ou liquides, c'est à dire de calculer
le diagramme de phases complet du système. L'approche CALPHAD n'est
cependant pas atomistique et ne permet de plus que l'étude des propriétés
à l'équilibre. Il s'agit d'utiliser le modèle adéquate au système
que l'on veut étudier.

\section{Conclusions}

Dans ce chapitre, nous avons comparé l'ensemble des modèles énergétiques
proposés dans la littérature. Afin de reproduire les résultats \emph{ab
initio}, les modèles les plus récents introduisent un nouveau degré
de liberté : la variation du moment magnétique atomique en fonction
de l'environnement chimique local.  Ce modèle semble reproduire les
propriétés essentielles de l'alliage à 0\,K mais est numériquement
coûteux pour calculer le diagramme de phases, les isothermes de ségrégation
ou des cinétiques.

Nous développons un hamiltonien léger, ajusté sur les calculs \emph{ab
initio} présentés dans le chapitre \ref{cha:DFT} et sur les limites
de solubilité expérimentales à haute températures. Il s'agit d'un
modèle d'interactions de paires aux deuxièmes voisins sur réseau rigide.
Nous choisissons de donner aux interactions Fe--Fe et Cr--Cr des énergies
constantes, alors que nous donnons à l'interaction Fe--Cr une dépendance
en concentration que nous ajustons par un polynôme de Redlich-Kister
sur les énergies de mélange calculées\emph{ ab initio }à 0\,K. 

L'entropie de configuration est correctement prise en considération,
mais notre modèle est sur réseau rigide et il n'y a pas de degré de
liberté lié au moment magnétique. Les termes non-configurationnels
d'entropies vibrationnelle et magnétique sont cependant pris en compte
indirectement par une dépendance en température de l'énergie d'ordre
ajustée sur la température critique expérimentale.

Nous commençons par traiter statistiquement cet hamiltonien en champ
moyen de point. Cela permet d'extraire le maximum d'information analytique
du modèle. Les limites de solubilité sont en excellent accord avec
les compilations les plus récentes des mesures expérimentales. Les
limites spinodales sont également en très bon accord avec les résultats
expérimentaux.

Nous avons également déduit de cet hamiltonien les limites de solubilité
par simulations Monte Carlo semi-grand-canoniques. Nous montrons que
la portée des interactions n'a pas de conséquences importantes sur
le diagramme de phases. De même, le volume sur lequel est calculée
la concentration locale a des conséquences critiques sur l'efficacité
des calculs mais a peu d'influence sur les limites de solubilité.
Ce volume doit être assez étendu pour contenir un nombre d'atomes
suffisant pour reproduire le changement de signe de l'énergie de mélange.
Le diagramme de phases calculé est en accord très satisfaisant avec
l'expérience. La température critique est de 940\,K. La solubilité
du chrome dans le fer est de l'ordre de 7\,\% à 0\,K. À cette même
température, la solubilité du fer dans le chrome est nulle. La limite
de solubilité $\alpha'$ de Fe dans Cr reste très faible sur un large
domaine de température, alors que la limite $\alpha$ dépend plus
nettement de la température.

Par une comparaison avec les simulations Monte Carlo, nous montrons
qu'un traitement statistique de champ moyen de point de cet hamiltonien
est raisonnable. La température critique est supérieure de 14\,\%
à la température critique déduite des simulations Monte Carlo, ce
qui est en accord avec le décalage théorique dû à la mauvaise prise
en compte en champ moyen de l'entropie de configuration à haute température.
Sans la correction en température, ce décalage est cependant bien
plus important (de l'ordre de 38\,\%).

Ces résultats de champ moyen de point comme de Monte Carlo sont en
accord avec les limites de solubilité proposées récemment par Bonny
et al. \cite{bonny_onthe_aa_demixtion_2008} ainsi que Xiong et al.
\cite{xiong_grrrrbonny_2010} à la suite d'une compilation des données
expérimentales de la littérature. On peut donc envisager d'utiliser
ce modèle pour le calcul d'isothermes de ségrégation et ainsi prévoir
les profils de concentration des surfaces libres de l'alliage.

\chapter{Ségrégation de surfaces\label{cha:S=0000E9gr=0000E9gation}}

\malettrine{L}{}es propriétés d'une surface sont fondamentalement
corrélées à la composition chimique de la surface à l'échelle nanométrique.
Celle-ci peut être très différente de la composition volumique de
l'alliage. En catalyse hétérogène, par exemple, ce sont ces propriétés
chimiques uniques des surfaces qui sont utilisées pour accélérer des
réactions chimiques \cite{johnson_book_segregation_1979}. Les interfaces
sont aussi les premières barrières face à la corrosion. Ce sont elles
qui réagissent avec le milieu. Dans le contexte nucléaire, les surfaces
libres et joints de grains sont le lieu de la ségrégation induite
par irradiation. Les défauts créés en volume par l'irradiation neutronique
vont s'annihiler en surface, créant un flux qui se couple à la diffusion
d'autre élements. L'objectif de ce chapitre est d'utiliser le modèle
énergétique proposé au chapitre précédent et les calculs \emph{ab
initio} du chapitre 2 afin de calculer des isothermes de ségrégation.
Il s'agit donc de prédire la composition des surfaces en fonction
de la composition volumique.

Dans cette étude, nous ne tenons pas compte de l'environnement (oxygène,
hydrogène ...), du potentiel de surface, des défauts en surface, des
éléments d'addition \ldots{} Il ne s'agit donc pas d'expliquer l'oxydation.

Nous abordons ce chapitre par une présentation des modèles d'Ackland
et de Ropo et al. qu'ils utilisent pour calculer des isothermes de
ségrégation. Nous présentons ensuite un modèle général aux alliages
binaires qui permet de différentier les différents moteurs à la ségrégation.

Nous adaptons ensuite notre modèle énergétique présenté au chapitre
3 pour qu'il tienne compte de la présence des surfaces et qu'ils reproduisent
les propriétés calculées \emph{ab initio} : énergie de ségrégation
d'impureté et énergies de surface des éléments purs.

Enfin, nous décomposons dans l'approximation de champ moyen les effets
de chacun des moteurs à la ségrégation de surface de l'alliage fer--chrome.
Nous discutons également de l'effet de la dépendance en concentration
locale de l'énergie d'ordre. Nous calculons enfin les isothermes de
ségrégation des alliages fer--chrome.\vspace{1cm}

\section{Bibliographie expérimentale}

Nous discutions au chapitre 2 de la difficulté des observations expérimentales
sur les surfaces libres du fer et du chrome et de leurs alliages à
cause de leur réactivité. Toutes ces études sont, à notre connaissance,
dans l'orientation (100).

À haute température (600\,K), le chrome déposé en quantité inférieure
à une monocouche sur Fe (100) diffuse de 75 à 90\,\% vers le volume,
dans les trois premiers plans de surface. Au contraire, si deux à
trois mono-couches sont déposées sur Fe (100), les atomes de Cr ne
diffusent plus \cite{davies_STMFeCr100_1996}.

Ces résultats sont confirmés par Pfandzelter et al. \cite{pfandzelter_intermixing_surf_1996}
qui montrent que des atomes de Cr déposés sur Fe ne diffusent que
dans la première couche sous la surface (labellée $i=1$ dans les
calculs \emph{ab initio} sur la ségrégation dans le chapitre 2). À
570\,K, lorsqu'ils déposent une monocouche de Cr sur Fe (100), la
concentration du plan de surface $c_{0}$ est de $45\pm3$\,\%\,Cr,
et celle du deuxième plan $c_{1}$ est de 55$\pm5$\,\%\,Cr. Le
troisième plan a une concentration $c_{2}$ nulle. Lorsqu'ils déposent
deux monocouches atomiques de Cr, $c_{0}=0.70\pm0.05$, $c_{1}=1.00\pm0.15$
et $c_{2}=0.30\pm0.25$. Toutes ces observations par microscopie à
effet tunnel sont vérifiées par spectroscopie Auger \cite{venus_ARAES_CrsegFe100_1996,unguris_magnetism_1992,Pierce_muCr_sur_Fe100}.

\section{Les modèles énergétiques de ségrégation dans la littérature}

\subsection*{Le modèle d'Ackland}

Nous avons déjà décrit le modèle d'Ackland \cite{ackland_magnetically_2006}
dans le paragraphe \ref{sub:modele_d_ackland}. Il s'agit d'un hamiltonien
d'Ising avec deux degrés de liberté par site $i$ du réseau rigide
: une variable d'espèce ($S_{i}=\pm1$ pour Fe ou Cr) et un moment
magnétique atomique d'amplitude constante ($\sigma_{i}\pm1$). 

\begin{figure}[H]
\begin{centering}
\includegraphics[scale=0.4]{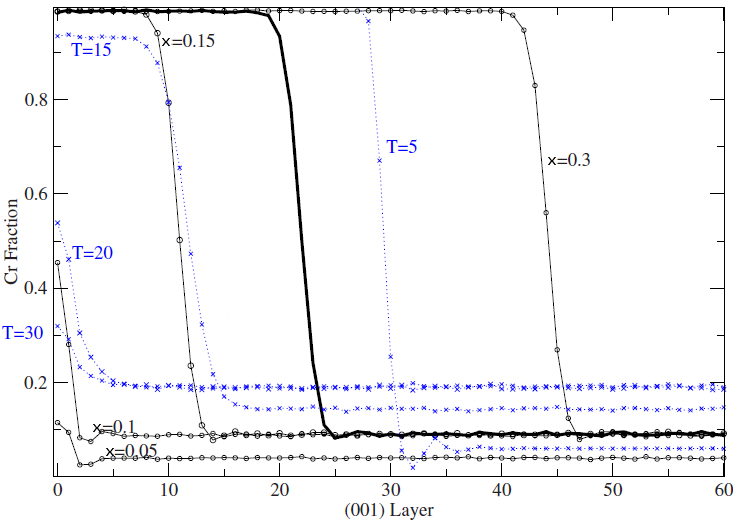}
\par\end{centering}

\caption{Profils de concentration à différentes concentrations volumiques $x$
et température $T$. La température est en unités réduites correspondant
au diagramme de phases de la figure \ref{fig:Ackland_ddp_5NN}. La
température critique est environ $T_{c}\approx22$. La concentration
volumique (température) de référence est $x=0.2$ ($T=10$). Le profil
de concentration en trait noir épais correspond à ces références.
Les profils de concentration de couleur bleue correspondent à l'évolution
du profil de concentration volumique $x=0.2$ à différentes températures.
Les traits noirs correspondent à l'évolution des profils de concentration
à température fixe $T=10$ pour différentes concentrations volumiques.
Figure de Ackland \cite{ackland_orderedsigma_2009}.\label{fig:Ackland_isothermes}}
\end{figure}

Ackland utilise cet hamiltonien pour calculer par simulations Monte
Carlo les profils de ségrégation dans l'orientation $\left(100\right)$.
On les reproduit sur la figure \ref{fig:Ackland_isothermes}.

Dans le domaine monophasé $\alpha$ riche en fer, la surface est enrichie
en chrome ($x_{0}\left(x_{v}=0.05\right)\approx0.11$ et $x_{0}\left(x_{v}=0.1\right)\approx0.46$).
Dans le domaine biphasé, la précipitation de la phase $\alpha$' riche
en chrome a lieu préférentiellement en surface, quelles que soient
la concentration et la température.

Le choix du paramètre $A_{N}$ pour que $T_{C}$ soit supérieure à
$T_{N}$ induit une énergie de cohésion du chrome inférieure à celle
du fer. Dans un modèle d'interactions de paires constantes, cela signifie
que l'énergie de surface du chrome est inférieure à celle du fer ($T_{C}/T_{N}>1\Rightarrow A_{N}\Rightarrow E_{coh}^{Cr}<E_{coh}^{Fe}\Rightarrow\gamma_{Cr}<\gamma_{Fe}$).
Or, les calculs \emph{ab initio} (voir le chapitre 2) montrent qu'à
basse température au moins, les énergies de surface du chrome sont
supérieures à celles du fer. Ainsi, du fait de la paramétrisation
de l'hamiltonien \ref{eq:acklandsurf} sur le rapport $T_{C}/T_{N}$,
il est favorable énergétiquement de créer une surface de chrome plutôt
qu'une surface de fer.

Dans le domaine monophasé $\alpha$, c'est la répulsion entre atomes
de chrome qui empêche la surface de se recouvrir de cette élément.
Dans le domaine biphasé, la phase $\alpha$' riche en chrome précipite
préférentiellement en surface pour y diminuer le nombre d'atomes de
fer. Les calculs \emph{ab initio} montrent au contraire une énergie
de surface de Cr supérieure à celle de Fe. Ainsi, si le modèle d'Ackland
reproduisait les résultats \emph{ab initio}, la phase $\alpha'$ précipiterait
peut-être en volume.

La principale approximation de l'hamiltonien d'Ackland est l'amplitude
constante du moment magnétique atomique. Or, les résultats \emph{ab
initio} montrent qu'en plus de dépendre de l'environnement chimique
local, le moment magnétique des atomes augmente en surface. L'intensité
de cette augmentation dépend de l'espèce chimique et de l'orientation
mais a toujours lieu. Les calculs \emph{ab initio }montrent que cela
induit par exemple une répulsion plus intense entre atomes de chrome
dans les surfaces.

\subsection*{Le modèle de Ropo et al.\label{sub:modele_ropo}}

Ropo et al. \cite{ropo_Crseg_2007} font des calculs de structure
électronique d'énergie de ségrégation. Ils modélisent la surface libre
$\left(100\right)$ de l'alliage Fe--Cr par : d'une part une mono-couche
atomique de concentration homogène (la surface) et d'autre part un
empilement de huit mono-couches de concentration fixe (le volume).
Le système est donc entièrement défini par deux concentrations : celle
du plan de surface et celle du volume. En égalisant les potentiels
chimiques en surface et en volume, ils déduisent la concentration
de surface correspondant à chaque concentration de volume. 

Ces calculs sont réalisés dans l'hypothèse d'un désordre substitutionnel
parfait. La structure électronique de chaque sous-système (plan de
surface / volume) est ainsi traitée dans l'approximation du potentiel
cohérent (CPA). Or, dans le domaine de concentration où l'alliage
tend à démixer, la concentration n'est plus homogène. Cela n'a donc
pas de sens physique d'y calculer des isothermes de ségrégation dans
l'approximation CPA.

Ropo et al. calculent une enthalpie de mélange qui change de signe
à environ 7\,\%\,Cr et dont on peut attendre une limite de démixtion
à 0\,K de l'ordre de 5\,\%\,Cr. L'ensemble des isothermes que les
auteurs proposent au-delà de cette concentration volumique n'ont donc
pas de sens.

\begin{figure}[H]
\begin{centering}
\includegraphics[scale=0.3]{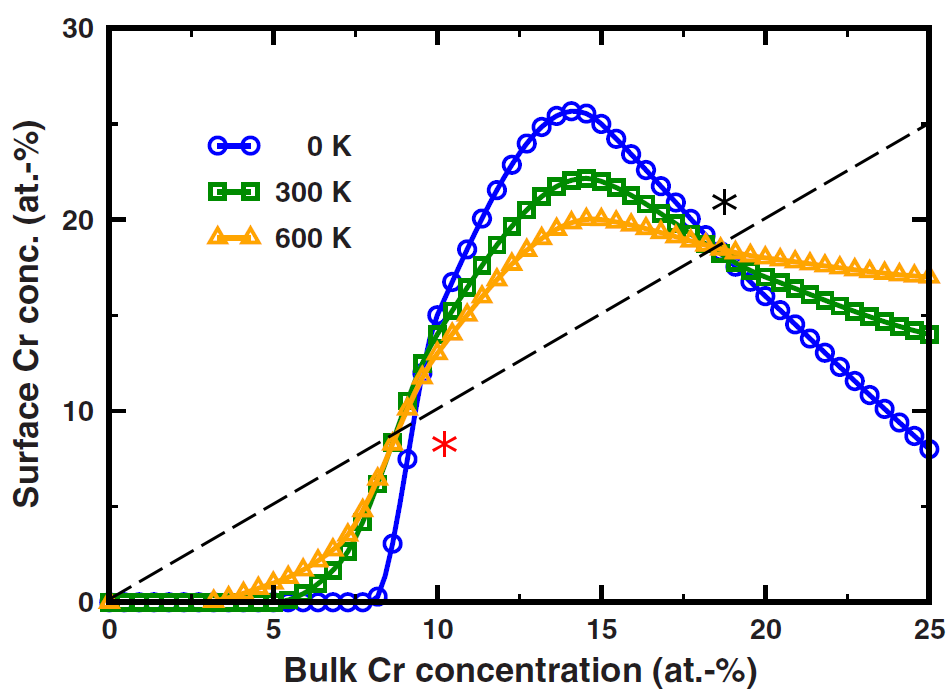}
\par\end{centering}

\caption{Évolution de la concentration surfacique en Cr en fonction de la concentration
volumique à 0, 300 et 600\,K. Les calculs sont de type DFT-GGA-EMTO-CPA.
La ligne en traits pointillés correspond au cas où surface et volume
ont la même concentration. Ses intersections avec les isothermes de
ségrégation sont représentées par des étoiles rouges et noires. Figure
de Ropo et al. \cite{ropo_Crseg_2007}.\label{fig:ropo}}
\end{figure}

On reproduit cependant sur la figure \ref{fig:ropo} ces isothermes
calculées par Ropo et al. en EMTO-CPA \cite{ropo_Crseg_2007} à 0,
300 et 600\,K. L'entropie est prise en compte par une contribution
configurationnelle uniquement.

Dans une isotherme de ségrégation, il ne peut y avoir d'effet de la
température où la concentration surfacique est égale à la concentration
volumique dans le cas où seule l'entropie de configuration est prise
en compte. C'est bien ce qu'on observe sur la figure \ref{fig:ropo}
à la concentration indiquée par l'étoile noire. Ce n'est par contre
pas le cas à l'intersection indiquée par l'étoile rouge, ce qui ne
nous semble pas justifié.

Entre 0 et 5\,\%\,Cr, où ces résultats nous semblent valides, les
isothermes montrent une surface très appauvrie en chrome. La concentration
surfacique ne dépasse pas 1\,\%\,Cr à 0\,K. Ces calculs EMTO-CPA
permettent d'obtenir les énergies de surface des éléments purs à 0\,K.
Ils ne les indiquent pas mais celles du chrome sont supérieures à
celles du fer. C'est la raison pour laquelle la surface est appauvrie
en chrome.

\subsection*{Alliages binaires standards -- le modèle classique avec interactions
de paires\label{sub:modele_bernard}}

Le modèle le plus classique de ségrégation de surface est le modèle
de Tréglia et Legrand \cite{treglia_PtNi_1987}. La base de ce modèle
pour un alliage A$_{1-x}$B$_{x}$ est le hamiltonien d'Ising que
nous décrivons au chapitre 3 : 

\begin{equation}
E=\frac{1}{2}\sum_{i,j,n,m}\epsilon_{ij}^{nm}p_{i}^{n}p_{j}^{m}\label{eq:4_1}
\end{equation}
 où $\epsilon_{ij}^{nm}$ est l'énergie d'interaction entre un atome
$i$ sur le site $n$ et un atome $j$ sur le site $m$ ($i,j=A,B$).
$p_{i}^{n}$ est la probabilité d'occupation de site égale à 1 (0)
si le site est (n'est pas) occupé par un atome de type $i$.

Pour un alliage binaire sans défaut, $p_{A}^{n}+p_{B}^{n}=1$. On
peut réécrire l'équation \ref{eq:4_1} avec une seule variable $p_{n}=p_{B}^{n}$
:
\begin{equation}
E=E_{0}+\sum_{n}p_{n}\sum_{m\ne n}\left(\tau_{nm}-V_{nm}\right)+\sum_{n,m\ne n}p_{n}p_{m}V_{nm}\label{eq:4_2}
\end{equation}
 où
\begin{equation}
E_{0}=\frac{1}{2}\sum_{n,m\ne n}\epsilon_{AA}^{nm},\,\tau_{nm}=\frac{1}{2}\left(\epsilon_{BB}^{nm}-\epsilon_{AA}^{nm}\right),\label{eq:tau}
\end{equation}
 et
\begin{equation}
V_{nm}=\frac{1}{2}\left(\epsilon_{AA}^{nm}+\epsilon_{BB}^{nm}-2\epsilon_{AB}^{nm}\right).
\end{equation}

La quantité $\tau_{nm}$ peut être reliée à la différence entre les
énergies de surface des éléments A et B. $V_{nm}$ est la différence
entre les énergies des interactions de paires homo- et hétéro-atomiques
qui caractérisent la tendance du système à s'ordonner ou à démixer.

Pour cette présentation du modèle, on se limite aux interactions entre
premiers voisins. Si l'on fait l'hypothèse que les interactions ne
sont pas modifiées en présence de la surface, alors on peut réécrire
$\tau_{nm}=\tau$ et $V_{nm}=V$, ce qui permet de réécrire l'équation
\ref{eq:4_2} en :
\begin{equation}
E=E_{0}+\left(\tau-V\right)\sum_{n}p_{n}Z^{n}+V\sum_{n,m\ne n}p_{n}p_{m}\label{eq:4_3}
\end{equation}
où $Z^{n}$ est le nombre de coordination du site $n$. 

Selon notre convention de signe, $V>0$ ($<0$) indique une tendance
à l'ordre (démixtion), et $\tau<0$ ($>0$) indique que l'élément
A (B) est celui de plus basse énergie de surface.

On pourrait rajouter un terme à l'équation \ref{eq:4_3} pour tenir
compte des différences de taille entre les éléments A et B \cite{friedel_electronic_1954,ducastelle_elastic_1970}
mais cette différence est inférieure à 1\,\% entre les atomes de
Fe et Cr. 

En champ moyen de point, les fonctions de corrélation à deux sites
$\left\langle p_{n}p_{m}\right\rangle $ peuvent se factoriser en
fonction des corrélations sur site $\left\langle p_{n}\right\rangle \left\langle p_{m}\right\rangle $.
On peut donc écrire $\left\langle p_{n}\right\rangle =x_{n}$ et $\left\langle p_{n}\right\rangle \left\langle p_{m}\right\rangle =x_{n}x_{m}$.
On choisit un modèle dans lequel $x_{n}$ ne dépend pas de $n$. Il
y a une concentration homogène dans le plan. La concentration peut
être différente d'un plan parallèle à la surface à un autre. On définit
donc des concentrations par plan $x_{p}$ ($x_{n}=x_{p},\forall n\in p$).
$p=0$ indique le plan de surface, $p=1$ le premier plan sous-jacent
à la surface et $p=v$ un plan de volume qui ne voit plus l'effet
de la surface. 

L'énergie interne du système est la somme des énergies internes dues
à chaque plan :
\begin{eqnarray}
E & = & E_{0}+N\left(\tau-V\right)\sum_{p}x_{p}\left(Z_{pp}+\sum_{{k=-b,b\atop b\in\mathbb{N}^{*}}}Z_{pp+k}\right)\\
 &  & +NV\sum_{p}x_{p}\left(Z_{pp}x_{p}+\sum_{{k=-b,b\atop b\in\mathbb{N}^{*}}}Z_{pp+k}x_{pp+k}\right)\nonumber 
\end{eqnarray}
où $N$ est le nombre d'atomes par plan. $Z_{pp+k}$ est le nombre
de liaisons entre les plans $p$ et $p+k$. La portée maximale des
interactions et l'orientation de la surface déterminent la valeur
maximum de $k$. Dans le cas d'une surface $\left(100\right)$ d'un
cristal cubique centré avec des interactions aux premiers voisins
seulement, il n'y a de liaisons qu'entre plans adjacents. Ainsi, dans
cet exemple, la valeur maximum de $k$ est 1.

L'entropie de configuration du système est également la somme des
entropies de configuration par plan et s'écrit :
\begin{equation}
S=-Nk_{B}\sum_{p}\left[x_{p}\ln x_{p}+\left(1-x_{p}\right)\ln\left(1-x_{p}\right)\right]
\end{equation}
Ce modèle étant sur réseau rigide, il néglige l'entropie de vibration.
Le grand potentiel $G$ du système s'écrit donc :
\begin{eqnarray}
G & = & E-TS-N\sum_{p}x_{p}\Delta\mu
\end{eqnarray}
où $\Delta\mu$ est le potentiel chimique de l'alliage. Le profil
de concentration d'équilibre est celui qui minimise le grand potentiel
($\frac{\partial G}{\partial x_{p}}=0,\forall p$). Cela conduit au
système d'équations couplées suivant :
\begin{eqnarray}
\forall p,\frac{x_{p}}{1-x_{p}} & = & \exp\left\{ -\beta\left[\left(\tau-V\right)\left(Z_{pp}+\sum_{{k=-b,b\atop b\in\mathbb{N}^{*}}}Z_{pp+k}\right)\right.\right.\label{eq:segregation_eq}\\
 &  & \left.\left.+V\left(Z_{pp}x_{p}+\sum_{{k=-b,b\atop b\in\mathbb{N}^{*}}}Z_{pp+k}x_{pp+k}\right)\right]\right\} \nonumber 
\end{eqnarray}

Dans la pratique, on choisit un plan $v$ à partir duquel tous les
plans ont la concentration du volume. Seuls les plans de 0 à $v-1$
ont donc leur concentration libre. Cela permet d'écrire la relation
suivante entre les plans de surface et de volume :
\begin{equation}
\frac{x_{p}}{1-x_{p}}=\frac{x_{v}}{1-x_{v}}\exp\left\{ -\beta\Delta E_{seg}\left(p\right)\right\} \label{eq:couple1}
\end{equation}
où $\Delta E_{seg}\left(p\right)$ est l'énergie de ségrégation pour
le plan $p$. Comme nous le discutons plus avant dans la suite du
chapitre, nous choisissons un raccord au 20$^{\text{e}}$ plan avec
des interactions jusqu'aux deuxièmes voisins. Afin de présenter une
version simple des équations, on se place pour les équations \ref{eq:40}
à \ref{eq:43} dans le cas d'un alliage cubique centré orienté $\left(100\right)$
avec des interactions entre premiers voisins seulement et un raccord
entre volume et surface au troisième plan ($x_{3}=x_{4}=\ldots=x_{v}$).
Dans ce cas très simple on peut écrire $\Delta E_{seg}$ :
\begin{eqnarray}
\Delta E_{seg}\left(0\right) & = & -Z_{pp+1}\left(\tau-V\right)\label{eq:40}\\
 &  & +2V\left[Z_{pp}\left(x_{0}-x_{v}\right)+Z_{pp+1}\left(x_{1}-2x_{v}\right)\right]\nonumber \\
\Delta E_{seg}\left(1\right) & = & 2V\left[Z_{pp}\left(x_{1}-x_{v}\right)+Z_{pp+1}\left(x_{0}+x_{2}-2x_{v}\right)\right]\\
\Delta E_{seg}\left(2\right) & = & 2V\left[Z_{pp}\left(x_{2}-x_{v}\right)+Z_{pp+1}\left(x_{1}-x_{v}\right)\right]\\
\Delta E_{seg}\left(3\right) & = & \Delta E_{seg}\left(4\right)=\ldots=\Delta E_{seg}\left(v\right)=0.\label{eq:43}
\end{eqnarray}
où les énergies de ségrégation pour la limite infiniment diluée ($\lim_{x_{k}\rightarrow0,\forall k}\Delta E_{seg}\left(p\right)$)
peuvent être calculées \emph{ab initio}.

Les énergies de ségrégation définies dans les équations \ref{eq:40}
à \ref{eq:43} sont la somme de deux contributions.

La première ne dépend pas de la concentration. Il s'agit du terme
proportionnel au nombre de liaisons perdues par le plan en l'absence
de plans supérieurs. Dans cet exemple, seul le plan 0 perd des liaisons.
Ce terme est également proportionnel à la différence d'énergie de
surface des deux éléments. Cet effet d'énergie de surface tend à faire
ségréger l'élément de plus basse énergie de surface.

La seconde dépend de la concentration des différents plans et est
proportionnel à $V$. Dans un alliage à tendance à la démixtion ($V<0$),
le système favorise les paires homo-atomiques. Cet effet d'alliage
tend à faire ségréger l'élément minoritaire. Dans un alliage à tendance
à l'ordre ($V>0$), le système favorise les paires hétéro-atomiques.
L'effet d'alliage tend donc à faire ségréger l'élément majoritaire
en surface, puis à alterner les plans riches en A et en B.

Dans ce modèle, la ségrégation est donc le résultat de la compétition
ou de la synergie entre l'effet d'énergie de surface et l'effet d'alliage.

\section{Modèle d'interactions de paires dépendant de la concentration locale
en présence de surfaces libres\label{sec:modele_surfaces}}

Nous utilisons pour les surfaces le modèle énergétique développé au
chapitre 3. Il s'agit donc d'un modèle d'interactions de paires dépendantes
de la concentration locale et de la température sur réseau rigide.

Dans le chapitre 3, nous avons discuté de l'effet de l'énergie d'ordre
sur le diagramme de phases de l'alliage. Nous avons montré qu'il n'était
pas nécessaire de décrire individuellement chaque énergie de paire
$\epsilon_{FeFe}^{\left(i\right)}$, $\epsilon_{CrCr}^{\left(i\right)}$
et $\epsilon_{FeCr}^{\left(i\right)}$ pour décrire la tendance à
la démixtion de l'alliage. Seul le signe de l'énergie d'ordre $\Omega$%
\footnote{On rappelle que l'énergie d'ordre s'exprime : $\Omega=\sum_{i}V^{\left(i\right)}=\sum_{i}1/2\left(\epsilon_{FeFe}^{\left(i\right)}+\epsilon_{CrCr}^{\left(i\right)}-2\epsilon_{FeCr}^{\left(i\right)}\right)$.%
} est déterminant.

L'équation \ref{eq:segregation_eq} indique que dans un modèle d'interactions
de paires la ségrégation de surface est également due au terme $\tau$
de l'équation \ref{eq:tau}. Celui-ci est proportionnel à la différence
d'énergies de surfaces des éléments purs. Il est donc important, en
présence de surfaces (ou plus exactement de liaisons coupées), de
discuter précisément de la distribution de l'énergie sur les paires
homo- et hétéro-atomiques.

Nous ajustons ce modèle afin qu'il reproduise les résultats \emph{ab
initio} sur les surfaces. Notre modèle est donc une déclinaison dépendante
en concentration et température du modèle de Tréglia et Legrand présenté
au paragraphe \ref{sub:modele_bernard}.

Tréglia et Legrand montrent que la ségrégation est la conséquence
de deux moteurs. On s'attachera donc à reproduire ces deux effets
:
\begin{enumerate}
\item il faut que les énergies de surfaces des éléments purs soient reproduites.
On montrait précédemment que les énergies de surfaces évaluées à partir
de mesures de tension de surface dans la phase liquide à haute température
sont erronées car on ne prend pas en compte des phénomènes magnétiques
dans la phase solide cubique centrée.
\item il faut que l'effet d'alliage particulier en volume avec le changement
de signe de l'énergie de mélange soit reproduit.
\end{enumerate}

\subsection{Ajustement des énergies de paires homo-atomiques sur les énergies
de surface \emph{ab initio}}

On ajuste les énergies de paires homo-atomiques sur les énergies de
surface des éléments purs calculées \emph{ab initio}. L'effet d'énergie
de surface décrit dans les équations \ref{eq:40} à \ref{eq:43} reproduit
exactement les résultats \emph{ab initio}.

Dans un modèle de paires, l'énergie de surface $\gamma_{A}$ de l'élément
A est proportionnelle au nombre de liaisons coupées pour créer une
unité de surface :
\begin{equation}
\gamma_{A}=-\dfrac{1}{2}\sum_{i}\sum_{q}Z_{p,p+q}^{\left(i\right)}\epsilon_{AA}^{\left(i\right)},\forall i,q\in\mathbb{N}^{*}\label{eq:Esurf}
\end{equation}
 où $Z_{p,p+q}^{\left(i\right)}$ est le nombre de $i$$^{\text{e}}$
voisins entre les plans $p$ et $p+q,\forall q\in\mathbb{N}$. Le
nombre total de voisins d'un atome $Z=\sum_{i}Z^{\left(i\right)}$
s'écrit :
\begin{equation}
\sum_{i}\sum_{q}2Z_{p,p+q}^{\left(i\right)}+Z_{p,p}^{\left(i\right)},\forall i,q\in\mathbb{N}^{*}\label{eq:Ztot}
\end{equation}

On indique dans le tableau \ref{tab:voisins_en_surface} le nombre
de liaisons coupées entre deux plans dans les orientations $\left(100\right)$,
$\left(110\right)$, $\left(111\right)$ et $\left(211\right)$ d'un
réseau cubique centré.

\begin{center}
\begin{table}[h]
\begin{centering}
\begin{tabular}{|c|c|c|c|c|}
\hline 
 & $\left(100\right)$ & $\left(110\right)$ & $\left(111\right)$ & $\left(211\right)$ \\
\hline 
\hline 
$Z_{p,p}^{\left(1\right)}$ & 0 & 4 & 0 & 2 \\
\hline 
$Z_{p,p+1}^{\left(1\right)}$ & 4 & 2 & 3 & 2 \\
\hline 
$Z_{p,p+2}^{\left(1\right)}$ & 0 & 0 & 0 & 1 \\
\hline 
$Z_{p,p+3}^{\left(1\right)}$ & 0 & 0 & 1 & 0 \\
\hline 
$Z_{p,p+q>3}^{\left(1\right)}$ & 0 & 0 & 0 & 0 \\
\hline 
\hline 
$Z_{p,p}^{\left(2\right)}$ & 4 & 2 & 0 & 0 \\
\hline 
$Z_{p,p+1}^{\left(2\right)}$ & 0 & 2 & 0 & 2 \\
\hline 
$Z_{p,p+2}^{\left(2\right)}$ & 1 & 0 & 3 & 1 \\
\hline 
$Z_{p,p+3}^{\left(2\right)}$ & 0 & 0 & 0 & 0 \\
\hline 
$Z_{p,p+q>3}^{\left(2\right)}$ & 0 & 0 & 0 & 0 \\
\hline 
\end{tabular}
\par\end{centering}

\caption{$Z_{p,p+q}^{\left(i\right)}$ est le nombre de $i$$^{\text{e}}$
voisins entre les plans $p$ et $p+q,\forall q\in\mathbb{N}$ dans
les orientations $\left(100\right)$, $\left(110\right)$, $\left(111\right)$
et $\left(211\right)$ d'un réseau cubique centré.\label{tab:voisins_en_surface}}
\end{table}

\par\end{center}

Pour les quatre orientations (100), (110), (111) et (211), la portée
maximale des interactions est entre un plan $p$ et un plan $p+3$.
Dans les plans (100) et (111), il n'y a pas de premiers voisins dans
le plan ($Z_{p,p}^{\left(1\right)}=0$). Le plan (100) a, comme nous
en discutions au paragraphe \ref{sub:La-singuli=0000E8re-orientation100},
tous ses premiers voisins dans les plans adjacents ($Z_{p,p+q}^{\left(1\right)}=0,\forall q\neq1$).
L'orientation (111) est également particulière car elle n'a ni premiers
ni deuxièmes voisins dans le plan ($Z_{p,p}^{\left(1\right)}=Z_{p,p}^{\left(2\right)}=0$).
Elle est donc la plus proche de l'adatome isolé.

En volume, les énergies de paires homo-atomiques sont ajustées sur
les énergies de cohésion, alors que dans les plans ayant des liaisons
coupées du fait de la surface, on a des énergies de paires dépendant
de l'orientation et ajustées sur les énergies de surface calculées
\emph{ab initio}.

Des énergies de surface calculées \emph{ab initio} en SIESTA-NC, on
déduit grâce à la relation \ref{eq:Esurf} et au tableau \ref{tab:voisins_en_surface}
les énergies de paires homo-atomiques aux 1$^{\text{ers}}$ et 2$^{\text{e}}$
voisins. On les résume dans le tableau \ref{tab:Eint_differentes_orientations}.

\begin{table}[H]
\begin{centering}
\begin{tabular}{|c|c|c|c|c|c|}
\hline 
(eV) & volume & $\left(100\right)$ & $\left(110\right)$ & $\left(111\right)$ & $\left(211\right)$ \\
\hline 
\hline 
$\epsilon_{FeFe}^{\left(1\right)}$ & 0.311 & 0.608 & 0.563 & 0.852 & 0.682 \\
\hline 
$\epsilon_{FeFe}^{\left(2\right)}$ & 0.270 & 0.527 & 0.487 & 0.738 & 0.591 \\
\hline 
$\epsilon_{CrCr}^{\left(1\right)}$ & 0.325 & 0.740 & 0.697 & 1.040 & 0.854 \\
\hline 
$\epsilon_{CrCr}^{\left(2\right)}$ & 0.281 & 0.641 & 0.603 & 0.900 & 0.739 \\
\hline 
\end{tabular}
\par\end{centering}

\caption{Énergies d'interactions homo-atomiques des éléments purs Fe et Cr
aux 1$^{\text{ers}}$ et 2$^{\text{e}}$ voisins en eV. En volume,
elles sont ajustées sur les énergies de cohésion expérimentales. En
surface, elles sont ajustées sur les énergies de surface calculées
\emph{ab initio} des orientations $\left(100\right)$, $\left(110\right)$,
$\left(111\right)$ et $\left(211\right)$. Le rapport entre énergies
de paire premiers et deuxièmes voisins $\eta^{\left(2\right)}$ est
égal à 0.866 (voir relation \ref{eq:eta}).\label{tab:Eint_differentes_orientations}}
\end{table}

Pour pouvoir reproduire chacune des énergies de surface, les énergies
de paires sont dépendantes de l'orientation de la surface. Cet ajustement
des énergies de paires sur les calculs \emph{ab initio} permet de
tenir compte, de façon indirecte, de l'effet des surfaces sur les
propriétés énergétiques (énergies de paires, stabilisation des surfaces
par le magnétisme).

\subsection{Ajustement des énergies de ségrégation à dilution infinie sur les
résultats \emph{ab initio}}

On peut déduire une expression analytique \ref{eq:40} à \ref{eq:43}
ou plus généralement \ref{eq:segregation_eq} des énergies de ségrégation
dans le plan de surface à dilution infinie $\Delta E_{seg}^{0}\left(p\right)$
: 
\begin{eqnarray}
\Delta E_{seg}^{0}\left(p\right) & = & \lim_{x_{v}\rightarrow0}\Delta E_{seg}\left(p\right).\label{eq:analytique_Eseg0}
\end{eqnarray}
Or, nous avons calculé ces dernières \emph{ab initio} à la section
\ref{sec:seg_DFT}. Si l'on se base sur les calculs PWSCF-PAW, elle
est de $+0.180$ eV pour l'orientation (100). Afin de reproduire cette
énergie de ségrégation à dilution infinie dans le plan de surface
avec notre modèle énergétique, on ajoute des énergies de site sur
les trois premiers plans de surface. Ces énergies de site $E_{site}\left(p\right)$
compensent les résultats analytiques décrits dans l'équation \ref{eq:analytique_Eseg0}.
On a donc :
\begin{eqnarray}
\Delta E_{seg}^{0}\left(p\right) & = & \Delta E_{seg}^{DFT}\left(p\right)\\
 & = & \lim_{x_{v}\rightarrow0}\Delta E_{seg}\left(p\right)+E_{site}\left(p\right),
\end{eqnarray}
où $\Delta E_{seg}^{DFT}\left(p\right)$ est l'énergie de ségrégation
dans le plan de surface à dilution infinie calculée \emph{ab initio}.
On ajuste ici les énergies de site uniquement sur les énergies de
ségrégation dans le plan de surface. Des développements supplémentaires
sont nécessaires pour reproduire le profil de ségrégation complet
sur les 2 à 3 premiers plan de surface que les calculs \emph{ab initio}
ont montré atypiques.

\begin{figure}[H]
\begin{centering}
\includegraphics[scale=0.35]{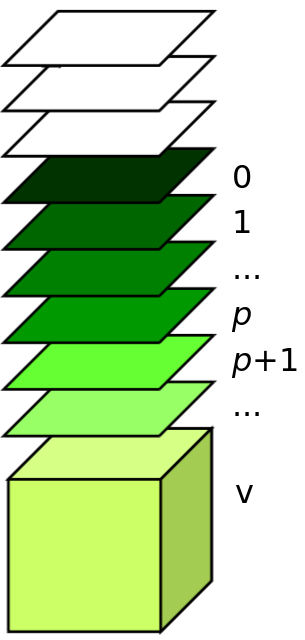}
\par\end{centering}

\caption{Schéma du modèle d'empilement. Les plans blancs sont des plans des
lacunes. La concentration dans chaque plan $p$ de 0 à $\infty$ est
homogène. Pour $p\rightarrow\infty$ la concentration est égale à
la concentration volumique.\label{fig:illu_modele}}
\end{figure}

Afin de donner un sens plus physique de l'énergie de site $E_{site}\left(p\right)$,
on peut montrer qu'il s'agit d'une expression équivalente à une énergie
d'interaction avec des lacunes en surface. Notre modèle correspond
donc à un empilement de plans d'atomes de Fe et Cr puis de plans purs
en lacunes. On schématise ce modèle sur la figure \ref{fig:illu_modele}.

\subsection{Les dépendances en concentration locale et en température}

La dépendance en concentration des énergies de paires hétéro-atomiques
est présentée dans l'équation \ref{eq:omega_PAWc} du chapitre précédent.
L'ensemble des paramètres de cette dépendance en concentration est
indiqué dans le tableau \ref{tab:parametres_Omega}. Cette dépendance
en concentration de l'énergie de paire hétéro-atomique, et en conséquence
de l'énergie d'ordre, induit une modification des développements de
Tréglia et Legrand à partir de l'équation \ref{sub:modele_bernard}
dans laquelle $\tau_{nm}$ et $V_{nm}$ sont indépendants des sites
$n$ et $m$. La correction en température est également prise en
compte selon la définition \ref{eq:Omega_PAW_T}.

\subsubsection*{Définition de la concentration locale}

Dans le cadre de cette approche phénomènologique, il n'apparait pas
de critère évident pour choisir l'étendue de la région sur laquelle
on définit une concentration locale. Nous montrons cependant au chapitre
3 que ce choix n'a pas de conséquences sur les limites de solubilité
calculées en volume. Dans le cadre du modèle de surface dans l'approximation
de champ moyen, nous définissons la concentration locale $x$, dont
dépend l'énergie d'interaction entre un atome A sur un plan $p$ et
un atome B sur un plan $q$, comme la moyenne des concentrations des
plans $p$ et $q$.
\begin{equation}
x=\dfrac{x_{p}+x_{q}}{2}.\label{eq:def_cloc_mf_surf}
\end{equation}

On prendra soin de ne pas confondre la concentration locale $x$ prise
en considération pour le calcul des énergies de paires hétéro-atomiques
et la concentration $x_{p}$ de chaque plan $p$.

\subsubsection*{Effet sur le modèle de Legrand et Tréglia}

Avec un modèle aux premiers voisins seulement, on conserve $V_{nm}^{\left(1\right)}=V_{nm}$.
Ce dernier est fonction des énergies de paires hétéro-atomiques et
devient alors dépendant de la concentration locale $x$ avec $\epsilon_{FeCr}$.
En réécrivant l'énergie totale de l'alliage $E$ :
\begin{equation}
E=E_{0}+\sum_{n,m\neq n}p_{n}\left(\tau-V_{nm}\right)+\sum_{n,m\neq n}p_{n}p_{m}V_{nm}
\end{equation}
qui peut s'exprimer sous la forme :
\begin{equation}
\begin{array}{ccc}
E & = & E_{0}+\sum_{p}Z_{p,p}x_{p}\left(\tau-V_{pp}\right)+\sum_{p}Z_{p,p+1}\left(x_{p}+x_{p+1}\right)\left(\tau-V_{pp+1}\right)\\
 &  & +\sum_{p}x_{p}^{2}V_{pp}Z_{p,p}+\sum_{p}2x_{p}x_{p+1}V_{pp+1}Z_{p,p+1}
\end{array}
\end{equation}
et en appliquant la condition d'équilibre : 
\begin{equation}
\dfrac{\partial F}{\partial x_{p}}=0,
\end{equation}
on déduit des énergies de ségrégation en champ moyen plus complexes
que dans le modèle de Tréglia et Legrand :

\begin{eqnarray}
\Delta E_{seg}^{p} & = & Z_{p,p}\left[\left(\tau-V_{pp}\right)-\left(\tau-V_{vv}\right)\right]+Z_{p,p}\left(c_{p}\dfrac{\partial\left(\tau-V_{pp}\right)}{\partial x_{p}}-c_{v}\dfrac{\partial\left(\tau-V_{vv}\right)}{\partial x_{v}}\right)\label{eq:44}\\
 &  & +Z_{p,p+1}\left[\left(\tau-V_{pp-1}\right)+\left(\tau-V_{pp+1}\right)-2\left(\tau-V_{vv}\right)\right]\nonumber \\
 &  & +Z_{p,p+1}\left[\left(x_{p}+x_{p-1}\right)\dfrac{\partial\left(\tau-V_{pp-1}\right)}{\partial x_{p}}+\left(x_{p}+x_{p+1}\right)\dfrac{\partial\left(\tau-V_{pp+1}\right)}{\partial x_{p}}-4c_{b}\dfrac{\partial\left(\tau-V_{vv}\right)}{\partial x_{v}}\right]\nonumber \\
 &  & +2Z_{p,p}\left(x_{p}V_{pp}-x_{v}V_{vv}\right)+Z_{p,p}\left(x_{p}^{2}\dfrac{\partial V_{pp}}{\partial x_{p}}-x_{v}^{2}\dfrac{\partial V_{vv}}{\partial x_{v}}\right)\nonumber \\
 &  & +2Z_{p,p+1}\left(x_{p-1}V_{pp-1}+x_{p+1}V_{pp+1}-2x_{v}V_{vv}\right)\nonumber \\
 &  & +2Z_{p,p+1}\left(x_{p}x_{p-1}\dfrac{\partial V_{pp-1}}{\partial x_{p}}+x_{p}x_{p+1}\dfrac{\partial V_{pp+1}}{\partial x_{p}}-2x_{v}^{2}\dfrac{\partial V_{vv}}{\partial x_{v}}\right)\nonumber 
\end{eqnarray}
L'énergie de ségrégation dans le plan de surface s'écrit alors :

\begin{eqnarray}
\Delta E_{seg}^{0} & = & Z_{p,p}\left[\left(\tau-V_{00}\right)-\left(\tau-V_{vv}\right)\right]\label{eq:45}\\
 &  & +Z_{p,p}\left(x_{0}\dfrac{\partial\left(\tau-V_{00}\right)}{\partial x_{0}}-x_{v}\dfrac{\partial\left(\tau-V_{vv}\right)}{\partial x_{v}}\right)\nonumber \\
 &  & +Z_{p,p+1}\left[\left(\tau-V_{01}\right)-2\left(\tau-V_{vv}\right)\right]\nonumber \\
 &  & +Z_{p,p+1}\left[\left(x_{0}+x_{1}\right)\dfrac{\partial\left(\tau-V_{01}\right)}{\partial x_{0}}-4x_{v}\dfrac{\partial\left(\tau-V_{vv}\right)}{\partial x_{v}}\right]\nonumber \\
 &  & +2Z_{p,p}\left(x_{0}V_{00}-x_{v}V_{vv}\right)+Z_{p,p}\left(x_{0}^{2}\dfrac{\partial V_{00}}{\partial x_{0}}-x_{v}^{2}\dfrac{\partial V_{vv}}{\partial x_{v}}\right)\nonumber \\
 &  & +2Z_{p,p+1}\left(x_{1}V_{01}-2x_{v}V_{vv}\right)+2Z_{p,p+1}\left(x_{0}x_{1}\dfrac{\partial V_{01}}{\partial x_{0}}-2x_{v}^{2}\dfrac{\partial V_{vv}}{\partial x_{v}}\right)\nonumber 
\end{eqnarray}

Dans le cas où l'interaction de paire hétéro-atomique $\epsilon_{AB}$
ne dépend plus de la concentration locale $x$, on retrouve les équations
\ref{eq:40} à \ref{eq:43} de champ moyen non-dépendant de la concentration
locale.

On observe dans les équations \ref{eq:44} et \ref{eq:45} que les
énergies de ségrégation conservent une partie liée à la différence
d'énergie de surface des éléments purs (première ligne des équations
\ref{eq:44} et \ref{eq:45}) et une partie liée au signe de l'énergie
d'ordre (deuxième ligne des équations). Un troisième terme apparaît,
lié aux variations des énergies d'interactions de paires lorsqu'on
se rapproche de la surface.

Les équations de champ moyen déduites ici sont limitées aux interactions
aux premiers voisins pour des raisons de lisibilité. Dans ce dernier
chapitre, comme dans le chapitre 3 dédié aux propriétés thermodynamique
de volume, la portée des interactions est limitée aux deuxièmes voisins.
Le rapport $\eta^{\left(i\right)}$ entre énergies de paires aux premiers
et deuxièmes voisins est celui utilisé en conclusion du chapitre 2,
c'est-à-dire 0.86.

\subsection{Bilan sur le modèle}

On a présenté dans cette section \ref{sec:modele_surfaces} un modèle
énergétique associé à un traitement statistique de champ moyen dans
le plan. Celui-ci reproduit exactement :
\begin{itemize}
\item les énergies de surface des éléments purs calculées \emph{ab initio},
\item l'énergie de ségrégation de Cr dans le plan de surface du fer à dilution
infinie calculées \emph{ab initio},
\item une énergie d'ordre telle que l'énergie de mélange change de signe
à 7\,\%\,Cr tel que cela a été calculé \emph{ab initio}.
\end{itemize}
Il ne reproduit pas l'énergie de ségrégation sur le plan sous-jacent
de la surface.

Le modèle énergétique et la paramétrisation de l'énergie d'ordre étant
les mêmes que dans le chapitre 3, le diagramme de phases volumique
reste inchangé. On pourra se référer à la figure \ref{fig:comparaison_MC_MF}.

Les énergies de paires homo-atomiques ne dépendent pas de la température.
Les énergies de surface ne dépendent donc pas de la température. Quand
la température s'approche de la température critique, on ne retrouve
donc pas les énergies de surface qu'on peut déduire de mesures de
tension de surface dans la phase liquide. On surestime donc l'effet
de l'énergie de surface à très haute température.

\section{Isothermes de ségrégation}

Dans cette sous-section, nous voulons décomposer et illustrer l'effet
de chacun des moteurs à la ségrégation identifiés par Tréglia et Legrand
:
\begin{enumerate}
\item l'effet d'énergie de surface lié à la différence d'énergie de surface
des éléments purs,
\item l'effet d'alliage, lié au signe de l'énergie d'ordre, c'est-à-dire
à la tendance du système à former des liaisons homo- ou hétéro-atomiques.
\end{enumerate}
Nous présentons donc l'effet de chacun de ces moteurs à la ségrégation
sur un alliage A$_{1-x_{v}}$B$_{x_{v}}$. Nous illustrons et discutons
également l'effet de la dépendance en concentration locale sur chacun
des deux moteurs.

\subsection{Effet d'énergie de surface }

\begin{figure}[h]
\begin{centering}
\includegraphics[scale=0.4]{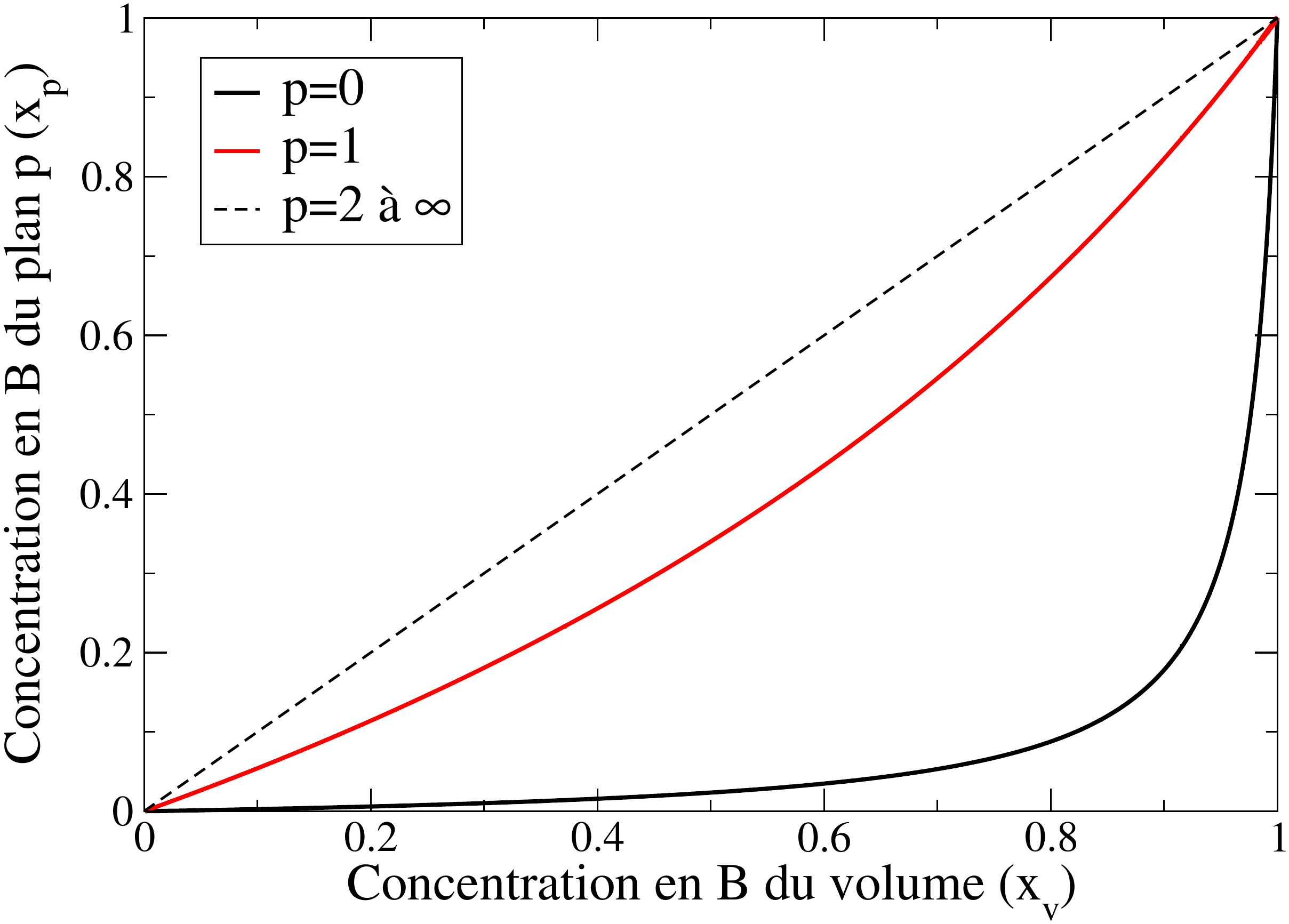}
\par\end{centering}

\caption{Isotherme de ségrégation des plans 0 (surface), 1 (sous-jacent à la
surface) et $\infty$ (volume) pour l'orientation $\left(100\right)$
à 1000\,K dans le cas où seul l'effet d'énergie de surface est moteur
à la ségrégation. Les $\epsilon_{AA}^{\left(i\right)}$ sont issus
du tableau \ref{tab:Eint_differentes_orientations}. $\epsilon_{AB}^{\left(2\right)}=\eta^{\left(2\right)}\epsilon_{AB}^{\left(1\right)}=\eta^{\left(2\right)}\frac{\epsilon_{AA}^{\left(1\right)}+\epsilon_{BB}^{\left(1\right)}}{2}$,
telle que $\Omega=0$.\label{fig:surface_seule}}
\end{figure}

On trace sur la figure \ref{fig:surface_seule} une isotherme de ségrégation
dans le cas où l'effet d'alliage est nul ($\Omega=0$) et où l'élément
B a une énergie de surface supérieure à l'élément A. On choisit les
énergies de paires homo-atomiques telles que les énergies de surface
des éléments purs soient les mêmes que celles de Fe et Cr ($\gamma_{A}=\gamma_{Fe}$
et $\gamma_{B}=\gamma_{Cr}$).

On représente les isothermes du plan de surface ($p=0$), du plan
sous-jacent et du plan de volume. Ce sont ici les isothermes pour
l'orientation (100). Cependant, quelle que soit l'orientation, les
isothermes de ségrégation ont même allure.

La différence d'énergie de surface entre fer et chrome est un moteur
fort à la ségrégation dans les deux premiers plans de surface. En
se référant au tableau \ref{tab:voisins_en_surface}, les deux premiers
plans sont les seuls qui ont des liaisons coupées. Pour toutes les
concentrations volumiques $x_{v}$ inférieures à 80\,\%\,Cr, le
plan de surface contient moins de 10\,\% en chrome. 

Les liaisons entre atomes de B sont plus fortes qu'entre atomes de
A. Il coûte donc moins d'énergie de couper des liaisons A--A que des
liaisons B--B : la surface s'appauvrit en B. Dans l'orientation (100),
le plan de surface ($p=0$) perd la moitié de ses 1$^{\text{ers}}$
voisins, et $\frac{1}{6}$ de ses 2$^{\text{e}}$ voisins : le premier
plan est très appauvri (trait noir continu sur la figure \ref{fig:surface_seule}).
Le plan suivant ($p=1$) perd seulement $\frac{1}{6}$ de ses 2$^{\text{e}}$
voisins. Il est donc également appauvri, mais dans une proportion
moindre (trait rouge continu sur la figure \ref{fig:surface_seule}).
Comme l'indique le tableau \ref{tab:voisins_en_surface}, les plans
suivants ($p>2$) ne perdent ni 1$^{\text{ers}}$ ni 2$^{\text{e}}$
voisins : la surface n'a plus d'effet sur eux en terme de liaisons
coupées. Ils ont en conséquence la même concentration que les plans
de volume (traits pointillés noirs sur la figure \ref{fig:surface_seule}).

C'est l'élément de plus basse énergie de surface qui ségrège en surface.
Les énergies de paires homo-atomiques ne dépendent pas de la concentration
locale, qui n'a donc pas d'effet sur ce moteur de la ségrégation.
Seuls les plans qui ont des liaisons coupées à cause de la surface
sont concernés par ce moteur.

\subsection{Effet d'alliage }

On ajuste maintenant les énergies de paires dans le système A$_{1-x_{v}}$B$_{x_{v}}$
pour que l'effet d'énergie de surface soit nul. Pour que les énergies
de surface des éléments A et B soient identiques, il faut que les
énergies de paires homo-atomiques soient égales ($\epsilon_{AA}^{\left(i\right)}=\epsilon_{BB}^{\left(i\right)},\forall i$
que l'on choisit arbitrairement nuls).

On choisit l'énergie de paire hétéro-atomique pour que l'énergie d'ordre
$\Omega$ soit égale à celle dans l'alliage fer--chrome. Afin de mettre
en évidence l'effet du changement de signe de l'énergie d'ordre sur
le moteur à la ségrégation, on considère les deux cas suivants :
\begin{itemize}
\item avec une énergie d'ordre constante $\Omega=-0.166$\,eV ajustée sur
la température critique expérimentale $T_{c}=965$\,K. Cela induit
une énergie de paire hétéro-atomique $\epsilon_{AB}^{\left(1\right)}$
égale à $0.0126$\,eV. L'énergie de paire hétéro-atomique aux deuxièmes
voisins $\epsilon_{AB}^{\left(2\right)}=\mbox{\ensuremath{\eta}}^{\left(2\right)}\epsilon_{AB}^{\left(1\right)}$
est égale à 0.0109\,eV.
\item avec une énergie d'ordre $\Omega$ dépendante de la concentration
locale ajustée sur les calculs \emph{ab initio} PWSCF-PAW et de la
température. Il s'agit des dépendances décrites dans le chapitre précédent
dans le tableau \ref{tab:parametres_Omega} et résumées dans la relation
\ref{eq:Omega_PAW_T}.
\end{itemize}
Dans ces deux cas, l'effet d'alliage est le seul moteur à la ségrégation.
Dans le cas d'un alliage à tendance à la démixtion comme l'alliage
Fe--Cr, le système tend à maximiser le nombre de liaisons homo-atomiques.
L'élément minoritaire tend donc à minimiser le nombre de liaisons
avec la matrice : il ségrège à la surface.

\subsection{Effet d'alliage au-dessus de la température critique}

\begin{figure}[H]
\begin{centering}
\includegraphics[scale=0.4]{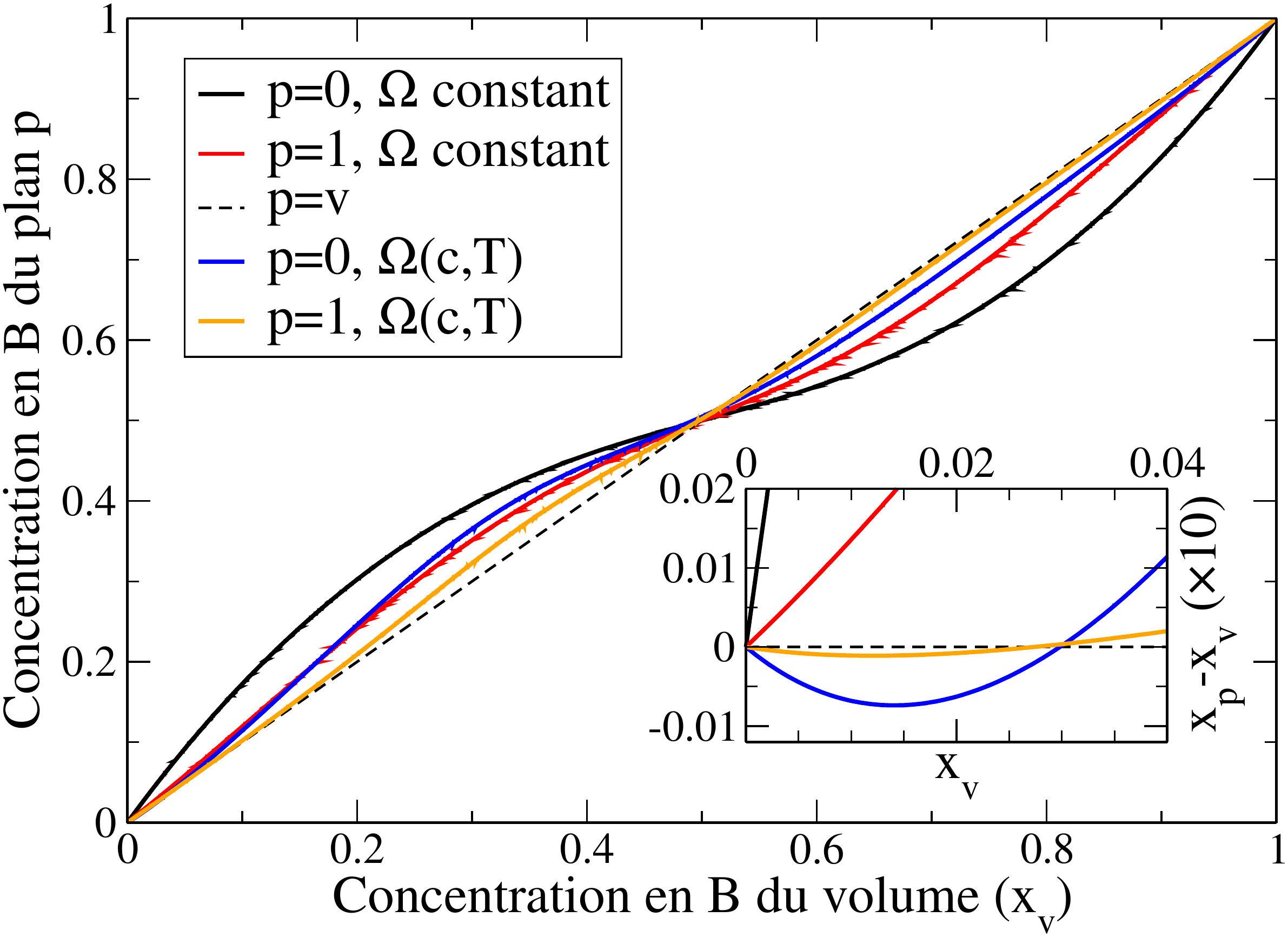}
\par\end{centering}

\caption{Isothermes de ségrégation d'un alliage modèle A$_{1-x}$B$_{x}$ piloté
uniquement par l'énergie d'ordre d'un alliage ayant tendance à la
démixtion à $T=1.1T_{c}$, pour une énergie d'ordre constante et une
énergie d'ordre dépendant de la concentration locale et de la température.\label{fig:Isotherme_alliage_seul}}
\end{figure}

On étudie d'abord le système fer--chrome au-dessus de la température
critique $T_{c}$. On est donc dans un régime dans lequel la solution
solide est stable quelle que soit la concentration. On représente
l'isotherme de ségrégation à la température $T=1.1T_{c}$ pour l'orientation
(100) sur la figure \ref{fig:Isotherme_alliage_seul}. Les deux cas
(énergie d'ordre constante et dépendante de la concentration et de
la température) sont indiqués pour comparaison.

À cette température, que l'énergie d'ordre soit constante ou pas,
l'effet d'alliage ajusté dans les conditions discutées ci-dessus est
un moteur à la ségrégation moins fort que l'effet d'énergie de surface.
C'est aux concentrations $x_{v}$ égales à 0.25 et à 0.75 que l'écart
entre concentration volumique et surface est le plus fort. La surface
y est respectivement enrichie (appauvrie) de 40\,\% par rapport au
volume.

Dans le cas où l'énergie d'ordre $\Omega$ est constante, les isothermes
de ségrégation de tous les plans ($\forall p$) sont symétriques par
rapport à la concentration $x_{v}=0.5$. Les concentrations surfaciques
et volumiques sont alors égales. À cette concentration précise, il
n'y a plus d'élément minoritaire ou majoritaire, ce qui annule le
moteur à la ségrégation. Dans tout le domaine de concentration, c'est
l'élément minoritaire qui ségrège en surface afin de maximiser le
nombre de liaisons homo-atomiques.

Dans le cas où l'énergie d'ordre $\Omega$ est dépendante de la concentration
et de la température, on observe la même tendance globale. Cependant,
trois phénomènes nouveaux apparaissent :
\begin{enumerate}
\item La tendance à l'enrichissement des plans de surface en élément minoritaire
est environ deux fois plus faible que dans le cas où l'énergie d'ordre
est constante.
\item La symétrie autour de A$_{\text{0.5}}$B$_{\text{0.5}}$ est brisée
(comme celle du diagramme de phases de volume). En particulier, l'enrichissement
en élément minoritaire A pour $x_{v}>0.5$ est quasiment négligeable
(traits bleu et orange sur la figure \ref{fig:Isotherme_alliage_seul}).
\item Enfin, on observe un changement de comportement de la surface du côté
très riche en A. Pour des concentrations $x_{v}$ inférieures à 3\,\%
on observe un très faible enrichissement de la surface en l'élément
majoritaire A. C'est une tendance inverse à l'enrichissement global
de la surface en élément minoritaire. Ce phénomène est très faible
($x_{0}-x_{v}<10^{-2}$) et la tendance change très vite avec la concentration. 
\end{enumerate}

\subsection{Effet d'alliage en-dessous de la température critique}

En-dessous de la température critique, à la température $T=0.9T_{c}$,
l'alliage est biphasé à l'intérieur des limites de solubilité.

\begin{figure}[h]
\begin{centering}
\includegraphics[scale=0.4]{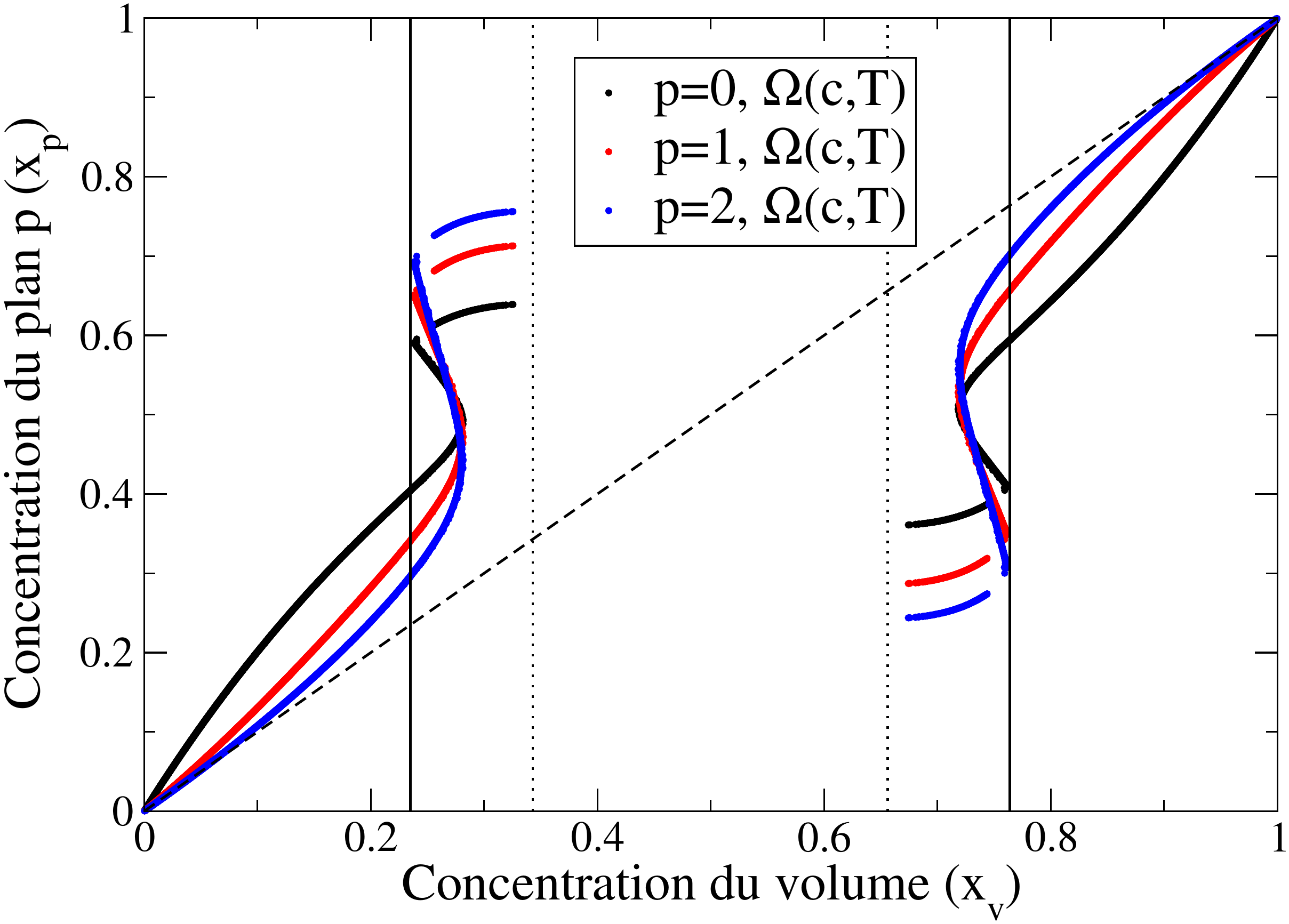}
\par\end{centering}

\caption{Isotherme de ségrégation des 3 premiers plans de surface d'un alliage
A$_{1-x_{v}}$B$_{x_{v}}$ à une température $T=0.9T_{c}$ et dont
les énergies d'interaction de paire homo-atomiques sont égales ($\gamma_{A}=\gamma_{B}$)
et arbitrairement nulles ($\epsilon_{kk}^{\left(i\right)}=0,\forall i=1,2;k=A,B$).
(lignes verticales continues noires) limites de solubilité de l'alliage
en volume, (pointillés verticaux noirs) limites de décomposition spinodale
en volume.\label{fig:iso_sous_tc}}
\end{figure}

Tant que la solution solide homogène est stable ou métastable (dérivée
seconde de l'énergie de mélange positive), on peut toujours trouver
une solution physique au système d'équations non-linéaires \ref{eq:44}.
Au contraire, sous la limite spinodale, la solution solide est instable.
On ne peut donc plus trouver de solution au système d'équations \ref{eq:44}.
C'est la raison pour laquelle les isothermes tracées sur la figure
\ref{fig:iso_sous_tc} ne peuvent être représentées pour des compositions
volumiques comprises entre 0.34 et 0.66.

On observe des boucles de Van der Waals annonçant des transformations
de phases en volume. Lorsque la composition d'un plan atteint la limite
de solubilité, une transition de phases a lieu en surface, induisant
un saut en concentration de la surface pour une très faible augmentation
de concentration en volume.

Ainsi, on observe une transition de phase en surface pour une concentration
volumique en chrome de 26\,\%, passant d'une concentration dans le
plan de surface de 0.44 à 0.62.

\section{Isothermes de ségrégation dans l'alliage Fe--Cr}

On intègre maintenant à la fois les effets de surface et d'alliage.
Les deux sont ajustées sur l'alliage Fe--Cr. Les énergies d'interactions
de paires homo-atomiques sont indiquées dans le tableau \ref{tab:Eint_differentes_orientations}.
Elles reproduisent les énergies de surface de Fe et Cr purs calculées
\emph{ab initio}. Elles dépendent pour cela de l'orientation de la
surface. La dépendance en concentration locale et en température de
l'énergie de paires hétéro-atomiques est toujours définie par les
équations \ref{eq:omega_PAWc} et \ref{eq:Omega_PAW_T} du chapitre
3. La portée des interactions est aux deuxièmes voisins.

\subsection{Au-dessus de la température critique}

\begin{figure}[H]
\begin{centering}
\includegraphics[scale=0.48]{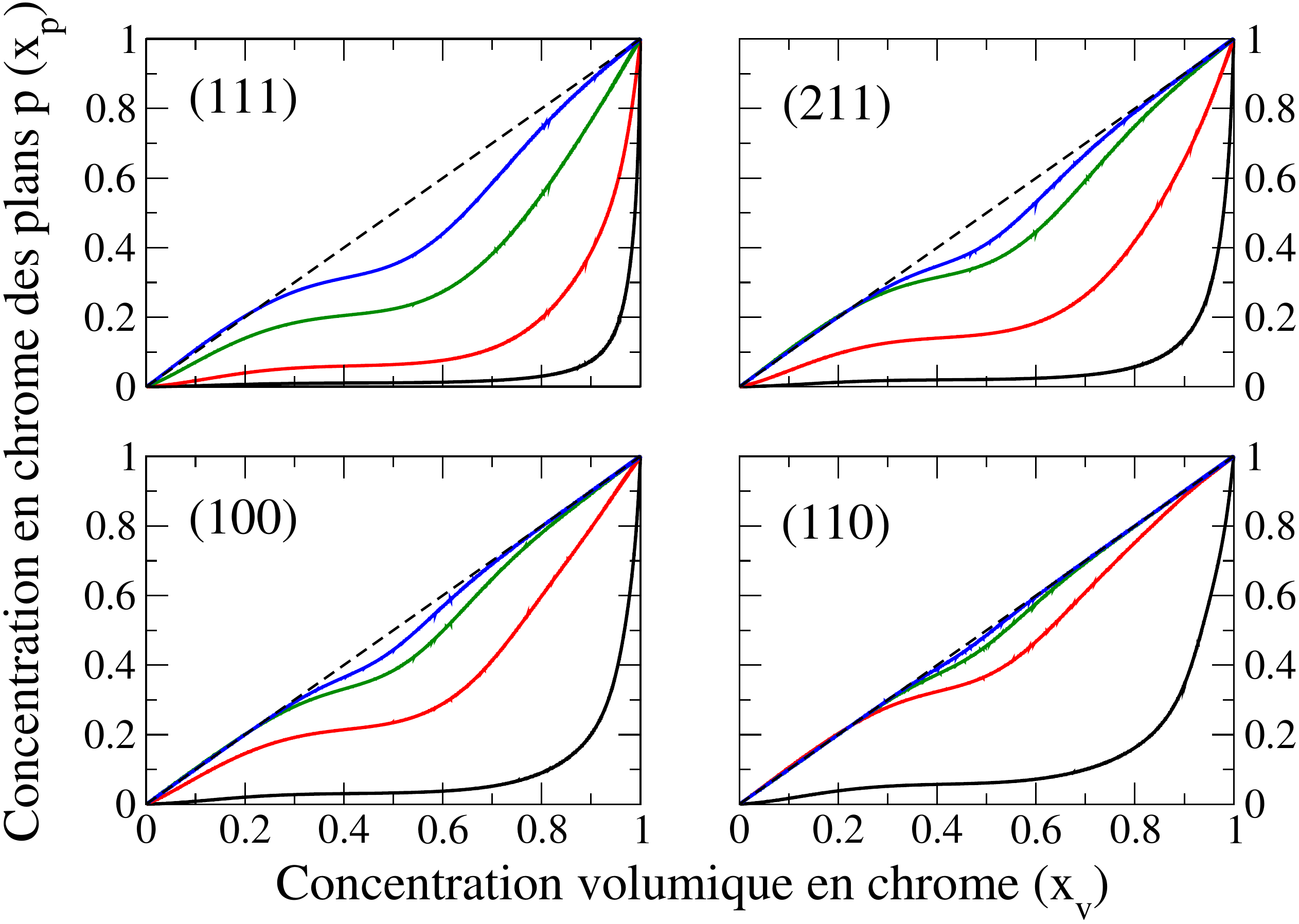}
\par\end{centering}

\caption{Isothermes de ségrégation du chrome dans le fer des quatre premiers
plans de surface ($p=$0, 1, 2 et 3) en noir, rouge, vert puis bleu,
à 1200\,K, c'est-à-dire à $1.1T_{c}$. Les tirets pointillés indiquent
un plan de concentration égale à la concentration de volume ($c_{p}=c_{v}$).
Les orientations sont classées de gauche à droite et de haut en bas
par densité croissante : $\rho_{\left(111\right)}<\rho_{\left(211\right)}<\rho_{\left(100\right)}<\rho_{(110)}$.\label{fig:iso1001200}}
\end{figure}

La figure \ref{fig:iso1001200} reproduit l'isotherme de ségrégation
des quatre premiers plans de surface ($p=$0, 1, 2 et 3) dans les
quatre orientations (111), (211), (100) et (110) à 1200\,K, c'est-à-dire
à $1.1T_{c}$.

Quel que soit le plan de surface, celui-ci est appauvrit en chrome
sur l'ensemble du domaine de concentration volumique. Plus la surface
est ouverte (moins elle est dense), plus l'appauvrissement du plan
de surface ($p=0$) est prononcé. Les surface sont ainsi composées
de moins de 5\,\%\,Cr jusqu'à environ 70\,\%\,Cr en volume. À
partir du cinquième plan sous la surface, les concentrations volumiques
et surfaciques sont indiscernables.

À 1200\,K, pour une composition volumique en chrome de 10\,\%, les
concentrations surfaciques sont comprises entre $10^{-4}$ et $3\cdot10^{-2}$
selon les orientations.

Plus la surface est dense, moins les effets sont prononcés. Ainsi,
le quatrième plan de la surface (111), la plus ouverte, a une concentration
encore largement différente de celle de volume, alors qu'il est indiscernable
du volume dans la surface (110).

Ces isothermes sont typiques d'isothermes de ségrégation dont la force
motrice principale est la différence d'énergies de surface. On reconnaît
le même type que sur la figure \ref{fig:surface_seule}. Notons l'épaulement
à 10\,\%\,Cr volumique environ. Équivalent à l'épaulement aux mêmes
concentrations sur la figure \ref{fig:Isotherme_alliage_seul}, il
est dû à la dépendance en concentration de l'énergie d'ordre.

\subsection{Descente en température}

\begin{figure}[H]
\begin{centering}
\includegraphics[scale=0.4]{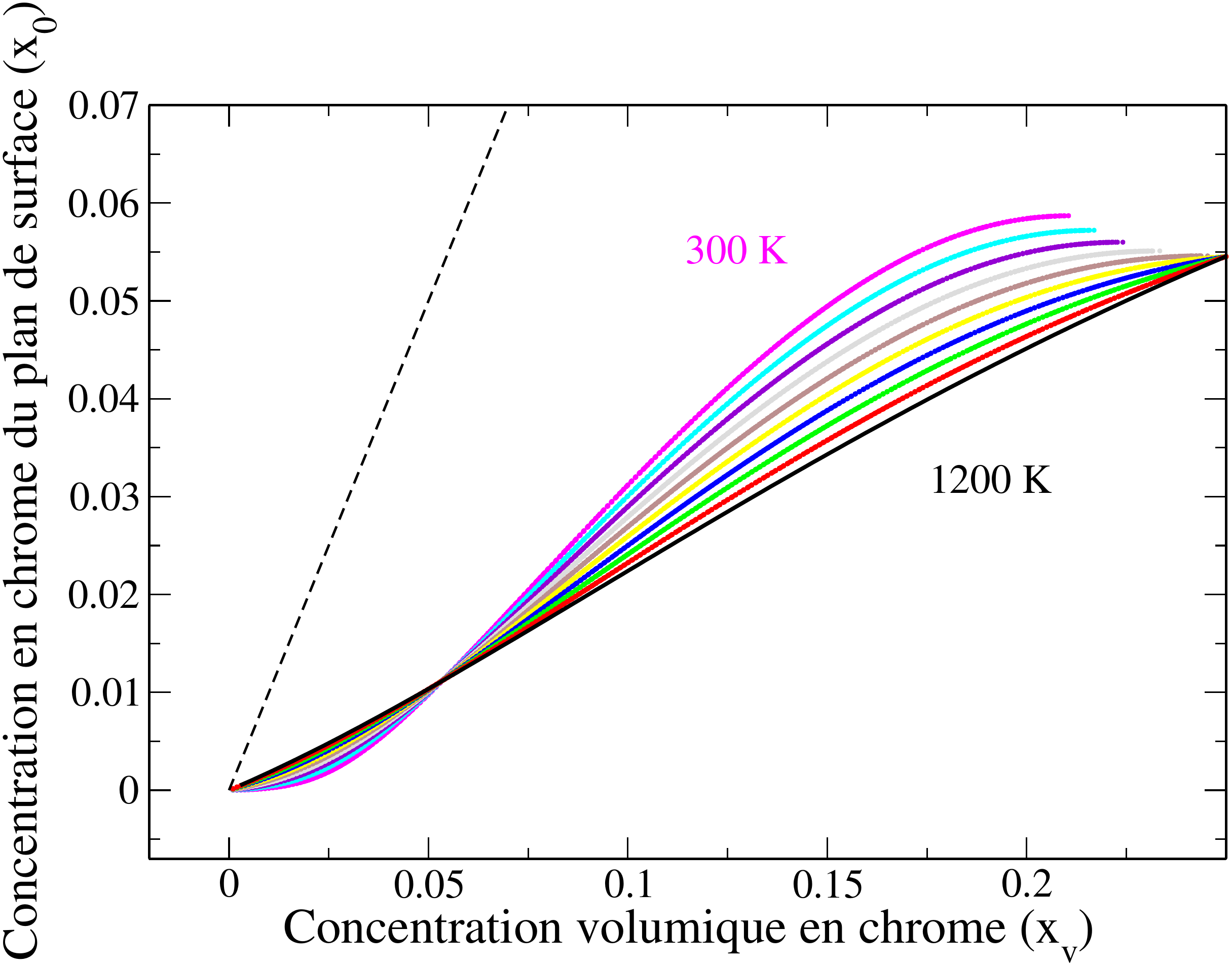}
\par\end{centering}

\caption{Évolution de la composition $x_{0}$ du plan de surface (100) en fonction
de la composition volumique $x_{v}$ pour des températures comprises
entre 1200\,K (en noir) et 300\,K (en magenta). Le pas en température
est de 100\,K. En tirets pointillés un repère pour $x_{0}=x_{v}$.\label{fig:evo_iso_T}}

\end{figure}

Sur la figure \ref{fig:evo_iso_T}, on représente l'évolution de la
concentration en chrome du plan de surface (100) lorsque la température
diminue de 1200\,K à 300\,K.

À l'intérieur des limites de décomposition spinodale, la solution
solide est instable. Il n'y a donc pas de solution aux systèmes d'équations
de champ moyen couplées \ref{eq:44}.

La concentration $x_{0}$ du plan de surface reste faible pour toutes
les températures et compositions volumiques. En accord avec les résultats
de Ropo et al. \cite{ropo_Crseg_2007}, la concentration surfacique
ne dépasse pas 10\,\%\,Cr. 

Au-dessous de 5\,\%\,Cr en volume, la concentration surfacique augmente
avec la température. Au-delà, le plan s'appauvrit en chrome quand
la température augmente. Pour toutes les températures, c'est à cette
composition volumique de 0.05 que se produit ce changement de régime.
Ce phénomène est induit par le changement de signe de l'énergie de
mélange à ces compositions.

Nous n'observons pas de phénomènes qui pourraient expliquer ce qui
est observer expérimentalement et que nous décrivions en début de
ce chapitre. Il est observé que si l'on dépose moins d'une mono-couche
de Cr sur du fer, une partie du chrome diffuse vers le volume. Si
on dépose plus d'une mono-couche, il n'y a plus de diffusion. Il nous
semble que ces phénomènes sont loin de l'équilibre thermodynamique
et qu'il est difficile de découpler ce qui relève ici de propriétés
d'équilibre de phénomènes cinétiques. Les résultats \emph{ab initio
}montrent justement que la présence d'une monocouche de chrome sur
Fe (100) et (110) est très défavorable. La présence de chrome dans
le plan sous-jacent de la surface est également défavorable. Il s'agit
donc là peut-être de barrières énergétiques à franchir. Il nous semble
donc difficile de conclure quant aux observations expérimentales.

\section{Conclusion}

\malettrine{D}{ans} ce dernier chapitre, nous avons commencé par
présenter les observations expérimentales concernant la ségrégation
du chrome dans le fer. Il semblerait que peu d'atomes de Cr déposés
sur Fe (100) diffusent vers le volume alors que plusieurs mono-couches
de Cr déposées ne diffuseraient pas. Il est probable que ces observations
soient des conséquences de phénomènes cinétiques, et non de phénomènes
d'équilibre.

Nous discutons ensuite des modèles proposés dans la littérature. Les
modèles d'Ackland \cite{ackland_orderedsigma_2009} et de Ropo et
al. \cite{ropo_Crseg_2007} sont plus spécifiques aux alliages fer--chrome.
Il s'agit dans le modèle d'Ackland de prendre en compte des interactions
de paires chimiques à courte portée et magnétiques à longue portée.
Les conclusions sont faussées par des énergies de surface erronées.
Ropo et al. font des calculs DFT-EMTO désordonnés dans l'approximation
CPA de l'évolution de la concentration d'un plan de surface (100)
en équilibre avec une infinité de plans de composition homogène. Les
calculs CPA permettent de tracer les isothermes dans le domaine monophasé
du diagramme de phases, c'est-à-dire en-dessous d'environ 5\,\%\,Cr
seulement. Ropo et al. ainsi qu'Ackland sont en accord pour noter
l'importance de l'effet d'énergie de surface sur la ségrégation, bien
que cela mène les auteurs à des conclusions contradictoires.

Bien qu'il existe de nombreux modèles généraux pour les alliages binaires,
celui de Tréglia et Legrand permet de mettre en évidence deux moteurs
principaux à la ségrégation : l'effet d'énergie de surface et l'effet
d'alliage. Le premier est lié à la différence d'énergie de surface
des éléments purs. C'est l'élément de plus basse énergie de surface
qui tend à ségréger. Le deuxième est lié au signe de l'énergie d'ordre
de l'alliage, c'est-à-dire à la tendance à former des liaisons homo-
ou hétéro-atomiques. Un alliage ayant tendance à la démixtion tend
à ségréger l'élément minoritaire en surface.

Pour qu'un modèle énergétique puisse capter les propriétés de ségrégation
d'équilibre d'un alliage, il est donc important qu'il reproduise les
énergies de surface des éléments purs ainsi que les propriétés de
mélange en volume. Ce n'est ni le cas du modèle d'Ackland qui produit
une énergie de surface du fer supérieure à celle du chrome, ni celui
du modèle de Ropo et al. dont les calculs ne tiennent pas compte de
la tendance à la démixtion.

Nous proposons un modèle énergétique qui reproduit à la fois les énergies
de surface calculées \emph{ab initio}, l'énergie de ségrégation dans
le plan de surface à dilution infinie calculée également \emph{ab
initio }et l'effet d'alliage avec le changement de signe de l'énergie
d'ordre qui donne sa large solubilité au chrome dans le fer. Il s'agit
du modèle d'interactions de paires aux deuxièmes voisins dépendant
de la concentration locale et de la température présenté au chapitre
3. Afin de reproduire les énergies de surface dans toutes les orientations,
les interactions dépendent de l'orientation. L'énergie d'ordre n'est
pas modifiée par rapport au modèle volumique du chapitre 3. Le diagramme
de phases, déduit des simulations Monte Carlo ou du traitement de
champ moyen, est donc le même qu'au chapitre précédent.

Nous traitons ce modèle énergétique en champ moyen de plan. Nous reproduisons
séparément les effets d'alliage et de surface pour le binaire Fe--Cr.
Ainsi, un alliage A--B dont les éléments A et B ont les mêmes énergies
de surface que Fe et Cr mais dont l'énergie d'ordre est nulle tend
fortement à faire ségréger l'élément de plus basse énergie de surface.
C'est une tendance très forte, pouvant induire des surfaces jusqu'à
neuf fois plus pauvres que le volume. Lorsque les éléments ont les
mêmes énergies de surface mais que l'énergie d'ordre est ajustée sur
les calculs \emph{ab initio} d'énergie de mélange, la composition
de surface reste proche de la composition volumique. Le changement
de signe de l'énergie de mélange a peu de conséquence sur les isothermes
de ségrégation. Lorsque l'énergie d'ordre dépend de la concentration,
on observe une légère tendance à la ségrégation de l'élément majoritaire
dans le domaine de concentration où il y a tendance à l'ordre. On
retrouve une tendance classique à la ségrégation du minoritaire au-dessus
de 3\,\%\,B en volume. En-dessous de la température critique, le
changement de signe de l'énergie de mélange a un effet négligeable.

Dans l'alliage Fe--Cr, en tenant compte à la fois de l'effet d'alliage
et d'énergie de surface, on observe que c'est ce dernier qui est prépondérant
la ségrégation. Dans l'ensemble du domaine de concentration dans lequel
la solution solide est stable ou métastable, la concentration surfacique
est très inférieure à la concentration volumique.

Nous avons donc présenté un modèle énergétique pour la ségrégation
de surfaces qui reproduit les résultats \emph{ab initio} qui montre
que c'est l'effet d'énergie de surface qui est le moteur principal
de la ségrégation, et que le changement de signe de l'énergie de mélange
récemment mis en évidence par les calculs \emph{ab initio} n'a pas
de conséquence importante sur ségrégation surfacique.\vspace{1cm}

\chapter*{Conclusion générale et perspectives}

\addcontentsline{toc}{chapter}{Conclusion générale et perspectives}
\markboth{Conclusion générale}{Conclusion générale et perspectives}

Dans cette étude, nous avons étudié l'alliage binaire fer--chrome
comme modèle de base des aciers ferritiques envisagés comme matériaux
de structure pour les réacteurs nucléaires du futur.

Nous commençons par des calculs dans la théorie de la fonctionnelle
de la densité avec les codes SIESTA et PWSCF. Ces deux codes utilisent
des approximations de pseudo-potentiels et de bases différentes. SIESTA
a l'avantage de permettre l'étude systématique de très grands systèmes,
au prix de résultats mois robustes. Les résultats clés sont donc vérifiés
avec le code PWSCF qui utilise des pseudo-potentiels de type PAW.
Avec ces deux méthodes de calcul, nous reproduisons les résultats
issus de la littérature comme la mise en solution favorable du chrome
dans le fer et défavorable du fer dans le chrome. Nous reproduisons
également le changement de signe de l'énergie de mélange avec la composition.
Ces deux propriétés sont intimement corrélées aux moments magnétiques
des atomes, en particulier de chrome. Nous montrons que le moment
magnétique des atomes de Cr est particulièrement sensible à leur environnement
chimique local. Ce moment magnétique est fortement augmenté lorsque
les Cr sont dilués dans une matrice de fer. Les pseudo-potentiels
à norme conservée utilisés par SIESTA surestiment les effets du magnétisme
et, en conséquence, les énergies de mise en solution et de mélange.
Ces tendances sont confirmées par les calculs plus robustes en pseudo-potentiels
PAW. Selon l'approximation utilisée, le changement de signe de l'énergie
de mélange a lieu pour une concentration en chrome comprise entre
7 et 16\,\%. Les paramètres de maille calculés et expérimentaux des
deux éléments sont similaires, ce qui rend négligeable l'effet des
relaxations atomiques.

Les tensions de surface expérimentales dans la phase liquide du fer
et du chrome sont approximativement égales. Extrapolées depuis la
phase liquide aux faibles températures, elles conduiraient à des énergies
de surface également proches. Afin de vérifier la validité de cette
extrapolation, nous avons calculé \emph{ab initio} les énergies de
surfaces des quatre orientations (100), (110), (111) et (211) des
éléments purs. On observe une nette différence entre les énergies
de surface du fer et du chrome. Celles du chrome sont supérieures
de 0.4\,J/m$^{\text{2}}$ en moyenne sur les quatre orientations
à celles du fer. C'est de nouveau la plus grande sensibilité électronique
du chrome à son environnement, ici aux surfaces libres, qui induit
cette différence. Le moment magnétique des atomes de surface est très
largement augmenté, parfois de plus de 100\,\% pour Cr. Alors que
l'énergie de surface varie linéairement avec la densité surfacique,
l'orientation (100) de Fe et Cr est singulière. Dans cette orientation,
tous les premiers voisins sont dans les plans adjacents et il n'y
a aucun premier voisin dans le plan, ce qui augmente davantage le
moment magnétique atomique. Cet effet est plus prononcé dans le chrome
qui présente un ferromagnétisme par plan.

La ségrégation d'une impureté de chrome dans des surfaces de fer est
également intimement liée aux effets du magnétisme. L'ensemble des
calculs \emph{ab initio} antérieurs à ce travail se limitait à des
alliages très concentrés de concentration surfacique supérieure à
25\,\% dans lesquels les interactions entre impuretés en surface
ne peuvent être négligées. Les calculs SIESTA nous permettent d'étudier
les surfaces très diluées. L'énergie de ségrégation dépend cependant
fortement des approximations de calcul. L'amplitude du moment magnétique
du chrome en impureté dans le fer, qui est déjà augmentée en volume,
est encore accrue en présence de surface. Cela induit, quelle que
soit la méthode de calcul, des répulsions fortes entre impuretés dans
la surface.

Dans un deuxième temps, nous développons un hamiltonien pour les calculs
thermodynamiques. L'hamiltonien le plus physique pour l'alliage fer--chrome
devrait tenir compte de l'effet de l'environnement chimique local
sur l'amplitude du moment magnétique des atomes de Cr. C'est une voie
riche d'un point de vue physique, mais tellement lourde numériquement
qu'il ne semble pas possible de réaliser des calculs d'équilibre ou
cinétiques avec un tel modèle. Avec un point de vue pragmatique, nous
choisissons d'étendre le modèle le plus simple qui puisse reproduire
une lacune de miscibilité : le modèle d'Ising. Il s'agit donc d'un
modèle d'interactions de paires sur réseau rigide. Afin de reproduire
les énergies de mélange calculées \emph{ab initio}, on introduit une
dépendance en concentration chimique locale des énergies de paires
hétéro-atomiques, tandis que les énergies de paires homo-atomiques
sont ajustées sur les énergies de cohésion des éléments purs. Ne tenant
pas explicitement compte du moment magnétique des atomes, cet hamiltonien
ne capte pas les phénomènes ayant lieu aux hautes températures, à
proximité des transitions de phases magnétiques. On le corrige donc
par une dépendance simple en température de l'énergie d'ordre. On
reproduit ainsi indirectement les phénomènes entropiques non-configurationnels
que sont l'entropie de vibration et l'entropie magnétique. Ce modèle
très efficace numériquement permet le calcul du diagramme de phases
du système dans l'approximation de champ moyen de site et par simulations
Monte Carlo. Ces dernières permettent d'évaluer l'effet de la zone
dans laquelle on calcule la concentration locale qui est négligeable.
Les limites de solubilité déduites à partir de notre modèle sont en
accord quantitatif avec les compilations expérimentales les plus récentes.

Enfin, nous étendons l'utilisation de ce modèle au calcul d'isothermes
de ségrégation dans les surfaces libres, c'est-à-dire au calcul des
concentrations des plans de surface en fonction de la composition
volumique. Sans rien modifier à l'hamiltonien pour les concentrations
volumiques, c'est-à-dire en conservant le même diagramme de phases,
on reproduit exactement les énergies de surface des éléments purs.
On reproduit également exactement les énergies de ségrégation dans
le plan de surface calculées \emph{ab initio} dans l'approximation
des PAW en ajoutant au modèle des énergies de site, ou interactions
de paires avec des lacunes au-dessus du plan de surface. Ce modèle
ne reproduit cependant pas le profil complet de ségrégation. En particulier,
les calculs \emph{ab initio} ont mis en évidence que l'énergie de
ségrégation dans le deuxième plan est inhabituellement positive. Le
traitement statistique en champ moyen de plan dans lequel la concentration
est homogène dans le plan permet de décomposer les moteurs à la ségrégation
en deux contributions : l'effet d'énergie de surface, lié à la différence
d'énergies de surface entre les éléments purs Fe et Cr, et l'effet
d'alliage lié à l'énergie d'ordre et à la tendance à l'ordre ou à
la démixtion en volume. C'est la première fois qu'un modèle énergétique
de ségrégation reproduit les résultats \emph{ab initio} d'énergies
de surface des éléments purs, d'énergie d'ordre et d'énergie de ségrégation
dans le plan de surface. On montre ainsi que l'effet d'énergie de
surface est très fort pour ce système. Au contraire, l'effet d'alliage,
déjà faible, n'est pas nettement modifié par le changement de signe
de l'énergie de mélange. C'est donc l'effet d'énergie de surface qui
est le moteur principal à la ségrégation. Celui-ci tend à faire ségréger
l'élément de plus faible énergie de surface, c'est-à-dire le fer.
En dehors des limites spinodales inaccessibles au traitement de champ
moyen par plan, les surfaces libres de l'alliage fer-chrome sont en
conséquence appauvries en chrome dans tout le domaine de concentration.\\

Nous terminons ce mémoire par quelques perspectives.

Il nous semblerait intéressant de combler le manque de données expérimentales
à basses températures. Cela demande bien sûr un grand effort de recherche
du fait des temps de mise à l'équilibre relativement longs. Cependant,
les limites de solubilité déduites à basse température ne sont actuellement
que des propositions basées pour la plupart sur des calculs \emph{ab
initio} à température nulle.

Nous avons montré que les résultats \emph{ab initio} pour le système
FeCr sont très sensibles aux approximations de calcul. Les pseudo-potentiels
PAW confirment leur robustesse, mais les bases d'onde planes ne sont
pas assez efficaces numériquement pour des études systématiques, en
particulier dans les domaines de compositions très diluées et en présence
de surfaces. Afin d'associer l'avantage numérique des bases localisées
et la robustesse des calculs PAW, Tristana Sondon, en post-doctorat
au SRMP, implémente actuellement le formalisme PAW dans le code SIESTA.

Enrique Martinez utilise actuellement le modèle énergétique présenté
au chapitre 3 pour le calcul de cinétiques de précipitation par simulations
Monte Carlo cinétiques. Le pragmatisme de l'approche rend les calculs
très efficaces numériquement, et les premiers résultats semblent en
accord avec les expériences de diffusion de neutrons aux petits angles
et de sonde tomographique.

Nous avons également commencé les simulations Monte Carlo pour la
ségrégation de surface. Les calculs de champ moyen sont beaucoup plus
rapides que les simulations Monte Carlo car ils permettent parfois
des études analytiques et autorisent une étude systématique de l'influence
de chacun des paramètres du modèle énergétique. Nous continuons de
développer ce modèle de champ moyen pour qu'il reproduise le profil
complet de ségrégation calculé \emph{ab initio}. Les simulations Monte
Carlo permettent cependant de travailler dans l'ensemble du domaine
de concentration et d'étudier les cinétiques de précipitation et de
ségrégation. Les limites du champ moyen de concentrations homogènes
dans le plan sont ainsi dépassées.

L'extension de l'étude aux joints de grains est également en cours.
Ceux-ci sont technologiquement très importants car la ségrégation
intergranulaire est mise en cause dans de nombreux phénomènes dans
les alliages : corrosion, accélération des cinétiques de diffusion
\ldots{} Les joints de grains ont également l'avantage d'être plus
facilement observables que les surfaces libres car ils sont à l'abri
de l'environnement. On pourrait ainsi comparer de façon plus rigoureuse
les isothermes de ségrégation expérimentales et théoriques.

Dans une perspective plus éloignée, nous pourrions utiliser notre
modèle pour étudier les phénomènes d'oxydation aux surfaces libres.
Cela nécessiterait de prendre en compte la présence d'oxygène. Notre
modèle tient compte de la présence de sites en surface occupés par
des lacunes. On pourrait ainsi imaginer occuper ces sites par des
atomes d'oxygène. Même si ce traitement statistique de champ moyen
de l'oxydation présente des limitations claires comme par exemple
les symétries imposées, les simulations Monte Carlo cinétiques semblent
bien adaptées.

\cleardoublepage


\appendix

\chapter{Paramètres d'ordre local de Warren-Cowley\label{sec:SRO}}

\malettrine{L}{}e paramètre d'ordre à courte distance (SRO) mesure
les corrélations entre les probabilités d'occupations de sites. La
définition la plus classique du SRO est donnée par Cowley \cite{Cowley_SRO_1950}.
Elle relie la probabilité $p_{Fe}^{lmn}$ d'occupation par un atome
de Fe du site de coordonnées $\left\{ l,m,n\right\} $ à la présence
en $\left\{ 0,0,0\right\} $ d'un atome de Cr. Le paramètre $\alpha_{Cr}^{lmn}$
dans un alliage Fe$_{x_{Fe}}$Cr$_{x_{Cr}}$ est défini par :

\begin{equation}
\alpha_{Cr}^{lmn}=1-\frac{p_{Fe}^{lmn}}{x_{Fe}}.
\end{equation}
Si le site $\left\{ l,m,n\right\} $ appartient à la $i$$^{\text{e}}$
sphère de coordination de Cr pris pour origine, alors 
\begin{equation}
\alpha_{Cr}^{\left(i\right)}=1-\dfrac{n_{Fe}^{\left(i\right)}}{x_{Fe}N^{\left(i\right)}}
\end{equation}
où $n_{\left(i\right)}$ est le nombre de Fe parmi les $N^{\left(i\right)}$
atomes de la $i$$^{\text{e}}$ sphère de coordination. $n_{Fe}^{\left(i\right)}/N^{\left(i\right)}=p_{Fe}^{\left(i\right)}$
est donc la probabilité de trouver un atome de Fe dans la $i$$^{\text{e}}$
sphère de coordination de l'atome de Cr pris pour origine. Si $p_{Fe}^{\left(i\right)}$
est égal à la probabilité moyenne de trouver un atome de fer dans
l'alliage $x_{Fe}$, il n'y a pas de corrélation entre Fe et Cr à
la portée $i$ ($\alpha_{Cr}^{\left(i\right)}=0$).

Si le SRO est négatif, la présence de Cr augmente la probabilité de
trouver un atome de Fe dans sa $i$$^{\text{e}}$ sphère de coordination.
Il y a donc un ordre local qui tend à former des paires Fe--Cr $i$$^{\text{e}}$
voisins. Au contraire, un SRO positif indique une tendance à former
des paires homo-atomiques aux $i$$^{\text{e}}$ voisins.

Certains auteurs dont Mirebeau et al. \cite{mirebeau_PRL_1984} utilisent
un paramètre d'ordre local $\beta$ spécifique aux alliages cubiques
centrés :
\begin{eqnarray}
\beta & = & \frac{8\alpha_{Cr}^{\left(1\right)}+6\alpha_{Cr}^{\left(2\right)}}{14}
\end{eqnarray}
 où $\beta$ est une moyenne pondérée%
\footnote{Il y a 8 premiers voisins et 6 deuxièmes voisins dans un réseau cubique
centré.%
} des paramètres de Warren-Cowley des deux premières sphères de coordination.

Si les SRO tendent vers une limite finie non-nulle quand $i$ devient
grand, alors il existe un ordre à longue distance.

\newpage{}

\chapter{Les structures quasi-aléatoires (SQS)\label{sec:annexe_SQS}}

\malettrine{L}{}es structures désordonnées d'un alliage A$_{1-x}$B$_{x}$
sont généralement générée par tirage aléatoire. Chaque site du réseau
a la probabilité $1-x$ d'être occupé par un atome A et $x$ d'être
occupé par un atome B. Il existe une méthode permettant de générer
des structures plus proches des structures parfaitement aléatoires
\cite{zunger_SQS_PRL_1990}. Pour cela, il faut définir des fonctions
de corrélations.

Une configuration $\sigma$ est discrétisée en figures d'interactions
$f=\left(k,m\right)$ de $k$ sites $m$$^{\text{e}}$ voisins. Sur
chaque site $i$, la variable d'occupation $\hat{S_{i}}$ prend la
valeur $-1$ ($+1$) si le site est occupé par l'espèce A (B). Pour
une solution solide A$_{1-x}$B$_{x}$ \emph{parfaitement aléatoire},
la fonction de corrélation pour la figure de type $f$ est
\begin{eqnarray}
\left\langle \overline{\Pi}_{f}\right\rangle _{al\acute{e}atoire} & = & \left\langle \overline{\Pi}_{k,m}\right\rangle _{al\acute{e}atoire}\\
 & = & \left\langle \hat{S}_{i}\right\rangle ^{k}\\
 & = & \left[\left(+1\right)\left(x\right)+\left(-1\right)\left(1-x\right)\right]^{k}\\
 & = & \left(2x-1\right)^{k}.\label{eq:sqs32}
\end{eqnarray}

Le principe de la SQS est de construire une configuration telle que
sa fonction de corrélation $\left\langle \overline{\Pi}_{k,m}\right\rangle _{SQS}$
tende vers la fonction de corrélation de la solution solide parfaitement
aléatoire $\left\langle \overline{\Pi}_{k,m}\right\rangle _{al\acute{e}atoire}$
pour le maximum de figures $f$ : 
\begin{equation}
\left\langle \overline{\Pi}_{k,m}\right\rangle _{SQS}\cong\left\langle \overline{\Pi}_{k,m}\right\rangle _{al\acute{e}atoire}.
\end{equation}

Pratiquement, on génère un grand nombre de structures de concentration
$x$ dont on calcule les fonctions de corrélations $\overline{\Pi}_{k,m}$.
On sélectionne ensuite la structure ayant les fonctions de corrélation
les plus proches de la fonction de corrélation \ref{eq:sqs32} d'une
solution solide parfaitement aléatoire.

\chapter{Les simulations Monte Carlo\label{sec:Principe-des-simulations_monte_carlo}}

\section{Description théorique succincte\label{sec:metro}}

\malettrine{C}{}haque répartition des atomes Fe et Cr sur le réseau
cubique centré représente une configuration $\sigma_{i}$ du système.
Au cours de la simulation numérique, des permutations entre atomes
ou des transmutations (selon l'ensemble statistique dans lequel on
travaille) font transiter le système d'une configuration vers une
autre. Pour l'étude des propriétés d'équilibre d'un système on définit
le \emph{bilan détaillé }suivant : 
\begin{equation}
P\left(\sigma_{i}\right)W_{\sigma_{i}\rightarrow\sigma_{j}}=P\left(\sigma_{j}\right)W_{\sigma_{j}\rightarrow\sigma_{i}}\label{eq:bilandetaille}
\end{equation}
où $P\left(\sigma_{i}\right)$ est la probabilité que le système soit
dans la configuration $\sigma_{i}$ et $W_{\sigma_{i}\rightarrow\sigma_{j}}$
est la probabilité de transiter de la configuration $\sigma_{i}$
vers la configuration $\sigma_{j}$. Si le bilan détaillé \ref{eq:bilandetaille}
est vérifié, le système converge vers l'état d'équilibre stationnaire,
où les probabilités $P\left(\sigma_{i}\right)$ sont données par :
\begin{equation}
P\left(\sigma_{i}\right)=\dfrac{1}{Z}\exp\left(-\dfrac{E\left(\sigma_{i}\right)}{k_{B}T}\right).
\end{equation}
$E\left(\sigma_{i}\right)$ est l'énergie totale de la configuration
$\sigma_{i}$ et $Z$ est la fonction de partition du système définie
par : 
\begin{equation}
Z=\sum_{i}\exp\left(-\dfrac{E\left(\sigma_{i}\right)}{k_{B}T}\right).
\end{equation}

L'algorithme que nous utilisons pour le choix des transitions est
l'algorithme de Metropolis \cite{metropolis__1953} On peut le résumer
ainsi :
\begin{enumerate}
\item À partir de la configuration $\sigma_{ini}$, on tire aléatoirement
deux atomes qu'on échange ou un atome qu'on transmute selon qu'on
réalise ces simulations dans l'ensemble canonique ou semi-grand-canonique
(voir annexe \ref{sec:Ensembles_canoniques_semi-grand-canonique}).
On définit ainsi une configuration $\sigma_{fin}$.
\item On calcule la différence d'énergie $\Delta E=E\left(\sigma_{fin}\right)-E\left(\sigma_{ini}\right)$.
\item Si $\Delta E<0$, alors on réalise la transition de $\sigma_{ini}$
vers $\sigma_{fin}$. Si $\Delta E>0$, alors on tire un nombre aléatoire
$\varsigma\in\left[0;1\right]$. Si $\varsigma<\exp\left(-\dfrac{\Delta E}{k_{B}T}\right)$,
alors on réalise également la transition.
\item Retour à l'étape 1.
\end{enumerate}
Cet algorithme assure d'échantillonner l'espace des phases de façon
homogène. Ses limitations sont la précision à haute température nécessitant
un échantillonnage très précis et le temps nécessaire à l'échantillonnage
à basse température puisque la transition de $\sigma_{ini}$ vers
$\sigma_{fin}$ est la plupart du temps refusée.

\section{Déterminer des limites de solubilité à partir de simulations Monte
Carlo\label{sec:Ensembles_canoniques_semi-grand-canonique}}

\malettrine{L}{}es simulations Monte Carlo décrites dans ce manuscrit
sont réalisées dans les ensembles statistiques canonique (nombre de
sites%
\footnote{La concentration de l'alliage est également constante dans l'ensemble
canonique.%
}, volume et température constants) ou semi-grand-canonique (nombre
de sites%
\footnote{Alors que le nombre de sites n'est pas constant dans l'ensemble grand-canonique,
il est imposé dans l'ensemble semi-grand-canonique. La concentration
globale du système reste cependant libre.%
}, potentiel chimique, volume et température constants).

Les limites de solubilité d'un alliage A--B peuvent être déterminées
dans l'ensemble canonique par plusieurs méthodes :
\begin{itemize}
\item en observant la présence de précipités à l'issue de la simulation
Monte Carlo réalisée soit avec un algorithme cinétique, soit avec
un algorithme Metropolis (voir section \ref{sec:metro} de cette annexe).
\item en observant l'évolution avec le temps (algorithme cinétique) du profil
de concentration dans un système initialement totalement démixé (tous
les atomes A d'un côté, tous les atomes B de l'autre). La présence
de l'interface $A/B$ a cependant un coût énergétique qui biaise la
simulation.
\end{itemize}
Le défaut majeur de cet ensemble pour déterminer des limites de solubilité
est qu'on ne peut vérifier si la simulation a atteint la solution
d'équilibre stationnaire. Pour s'en convaincre, on ne peut que faire
réaliser au système de nouveaux pas Monte Carlo et en observer l'effet.
On peut également estimer l'évolution de paramètres d'ordre locaux
au cours de la simulation, mais ils indiquent seulement la proximité
de la limite de solubilité \cite{xiong_grrrrbonny_2010}.

L'ensemble semi-grand-canonique est beaucoup plus pratique pour déterminer
des limites de solubilité. En imposant un potentiel chimique $\Delta\mu$,
le système adapte la proportion entre éléments A et B, c'est-à-dire
la concentration globale $x$. On boucle sur les paramètres chimiques
afin de pouvoir tracer la concentration globale du système en fonction
du potentiel chimique $x\left(\Delta\mu\right)$. La présence d'une
boucle d'hystérésis permet de déterminer les limites de solubilité.

L'ensemble semi-grand-canonique est donc le plus adapté pour déterminer
des limites de solubilité ou plus généralement pour observer des transformations
de phases thermodynamiques.

\chapter{Notion de concentration locale dans les simulations Monte Carlo\label{cha:def_clocal_methodo_annexe}}

\section{Définition de la concentration locale}

Dans notre modèle énergétique, l'énergie de paire hétéro-atomique
$\epsilon_{FeCr}^{\left(i\right)}$ dépend de la concentration locale
autour des deux atomes Fe et Cr. On définit la concentration locale
$x_{local}$ comme le nombre de Cr présents dans les $i$ premières
sphères de coordination. 

Il est important de noter la différence entre la concentration locale
définie ci-dessus, et la portée des interactions inter-atomiques.
Si notre modèle est aux deuxièmes voisins, cela signifie que seules
les énergies de paires entre premiers et deuxièmes voisins sont non-nulles
($\epsilon_{FeFe}^{\left(1\right)}$, $\epsilon_{FeFe}^{\left(2\right)}$,
$\epsilon_{CrCr}^{\left(1\right)}$, $\epsilon_{CrCr}^{\left(2\right)}$,
$\epsilon_{FeCr}^{\left(1\right)}$ et $\epsilon_{FeCr}^{\left(2\right)}$
seulement sont différents de zéro). La concentration locale dépend
de la zone prise en compte pour le calcul de $\epsilon_{FeCr}^{\left(1\right)}$
et $\epsilon_{FeCr}^{\left(2\right)}$. On peut donc avoir une portée
aux deuxièmes voisins, et une dépendance en concentration locale s'étendant
jusqu'aux dixièmes voisins de Fe et Cr.

\begin{figure}[h]
\begin{centering}
\subfloat[Premiers voisins : $x_{local}=\nicefrac{2}{22}$.\label{fig:clocal1}]{\begin{centering}
\includegraphics[scale=0.25]{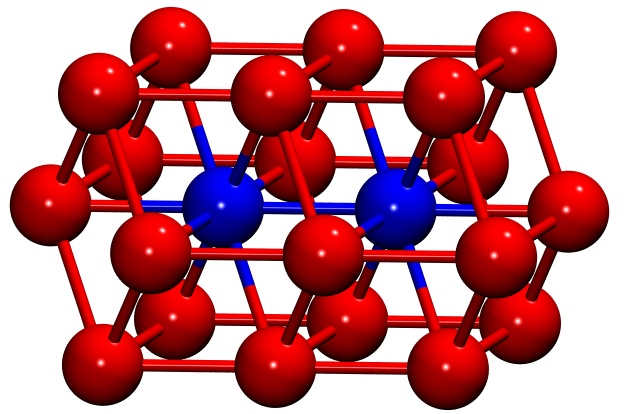}
\par\end{centering}

}\subfloat[Deuxièmes voisins : $x_{local}=\nicefrac{2}{24}$.\label{fig:clocal2}]{\begin{centering}
\includegraphics[scale=0.3]{figures_thermo/cell_2at_2nn_et_leurs_1er_et_2nd_voisins}
\par\end{centering}

}
\par\end{centering}

\caption{Illustration du calcul de la concentration locale jusqu'à deux sphères
de coordination autour de la paire d'atomes bleus (a) lorsqu'ils sont
premiers (b) et deuxièmes voisins entre eux. En plus d'illustrer le
calcul de la concentration locale, cette figure illustre la différence
entre portée des interactions (premiers ou deuxièmes voisins entre
deux atomes) et sphères de coordination sur lesquelles est calculée
la concentration locale.\label{fig:Sch=0000E9ma-c_local}}
\end{figure}

La figure \ref{fig:Sch=0000E9ma-c_local} illustre le calcul de la
concentration locale autour des deux atomes bleus :
\begin{enumerate}
\item On calcule d'abord le nombre total d'atomes $k$$^{\mbox{e}}$ voisins
de l'un ou l'autre des atomes bleus ($k\leq i$) : les atomes voisins
des deux atomes bleus ne sont comptabilisés qu'une fois.
\item On calcule ensuite le rapport du nombre d'atomes bleus sur le nombre
total d'atomes : c'est la concentration locale aux $i$$^{\mbox{e}}$
voisins.
\end{enumerate}
Si la portée des interactions est aux deuxièmes voisins, il faut considérer
les interactions aux premiers (cas \ref{fig:clocal1} de la figure
\ref{fig:Sch=0000E9ma-c_local}) et aux deuxièmes voisins (cas \ref{fig:clocal2}
de la figure \ref{fig:Sch=0000E9ma-c_local}). Dans ces deux cas,
l'étendue de la zone de dépendance en concentration n'est pas la même.

Dans les deux cas \ref{fig:clocal1} et \ref{fig:clocal2} illustrés
sur la figure \ref{fig:Sch=0000E9ma-c_local}, la concentration locale
est calculée sur une zone équivalente aux deux premières sphères de
coordination des atomes bleus ($i=2$).

\begin{itemize}
\item Si les deux atomes bleus sont premiers voisins (cas \ref{fig:clocal1}),
il y a au total 22 atomes premiers ou deuxièmes voisins de l'un ou
l'autre des atomes bleus. La concentration locale est $x_{local}=\frac{2}{22}$.
\item Si les deux atomes bleus sont deuxièmes voisins (cas \ref{fig:clocal2}),
il y a au total 24 atomes premiers ou deuxièmes voisins de l'un ou
l'autre des atomes bleus. La concentration locale est $x_{local}=\frac{2}{24}$.
\end{itemize}
Le nombre total de voisins s'accroît très rapidement avec la portée
maximale des interactions $i$. Les éléments permettant de choisir
$i$ sont :
\begin{itemize}
\item une discrétisation suffisante de la concentration locale (voir annexe
\ref{sec:discretisation_suffisante_clocal}),
\item un temps de calcul raisonnable (voir annexe \ref{sub:influence_portee_CPU}),
\item et une répartition cohérente de l'énergie sur les différentes sphères
de coordination (voir annexe \ref{sub:Mod=0000E8le-d'int=0000E9raction}).
\end{itemize}

\section{La discrétisation suffisante de la concentration locale\label{sec:discretisation_suffisante_clocal}}

La définition de la concentration locale $x_{local}$ doit permettre
une discrétisation suffisante de cette dernière pour reproduire les
variations de l'énergie de mélange observées grâce aux calculs \emph{ab
initio}. Si la zone de définition de $x_{local}$ pour le calcul de
$\epsilon_{FeCr}^{\left(i\right)}$ s'étend seulement aux premiers
voisins de Fe et Cr, la concentration locale minimale est $x_{local}=\nicefrac{1}{15}\approx0.07$.
Il est alors impossible d'observer une concentration locale intermédiaire
entre 0 et 0.07. L'énergie de mélange de l'alliage est donc systématiquement
positive.

La définition de la concentration locale s'étend donc au minimum aux
deuxièmes voisins.

Mais augmenter la zone de calcul de $x_{local}$ a une incidence forte
sur les temps de calcul. Les problèmes algorithmiques liés à l'étendue
de la zone de calcul de la concentration locale et son influence sur
les temps de calculs sont discutés dans les paragraphes \ref{sub:influence_portee_CPU}
et \ref{sub:Influencede-la-port=0000E9e_des_interactions}.

L'influence de la portée des interactions sur le diagramme de phases
est présentée dans le paragraphe \ref{sub:Influencede-la-port=0000E9e_des_interactions}.

\section{Effet de l'étendue de la concentration locale sur le temps de calcul\label{sub:influence_portee_CPU}}

Le temps de calcul de chaque pas Monte Carlo dépend de la définition
choisie pour la concentration locale $x_{local}$. 

\begin{figure}[h]
\begin{centering}
\subfloat[$x_{local}$ s'étend jusqu'aux premiers voisins.\label{fig:xlocal1}]{\begin{centering}
\includegraphics[scale=0.25]{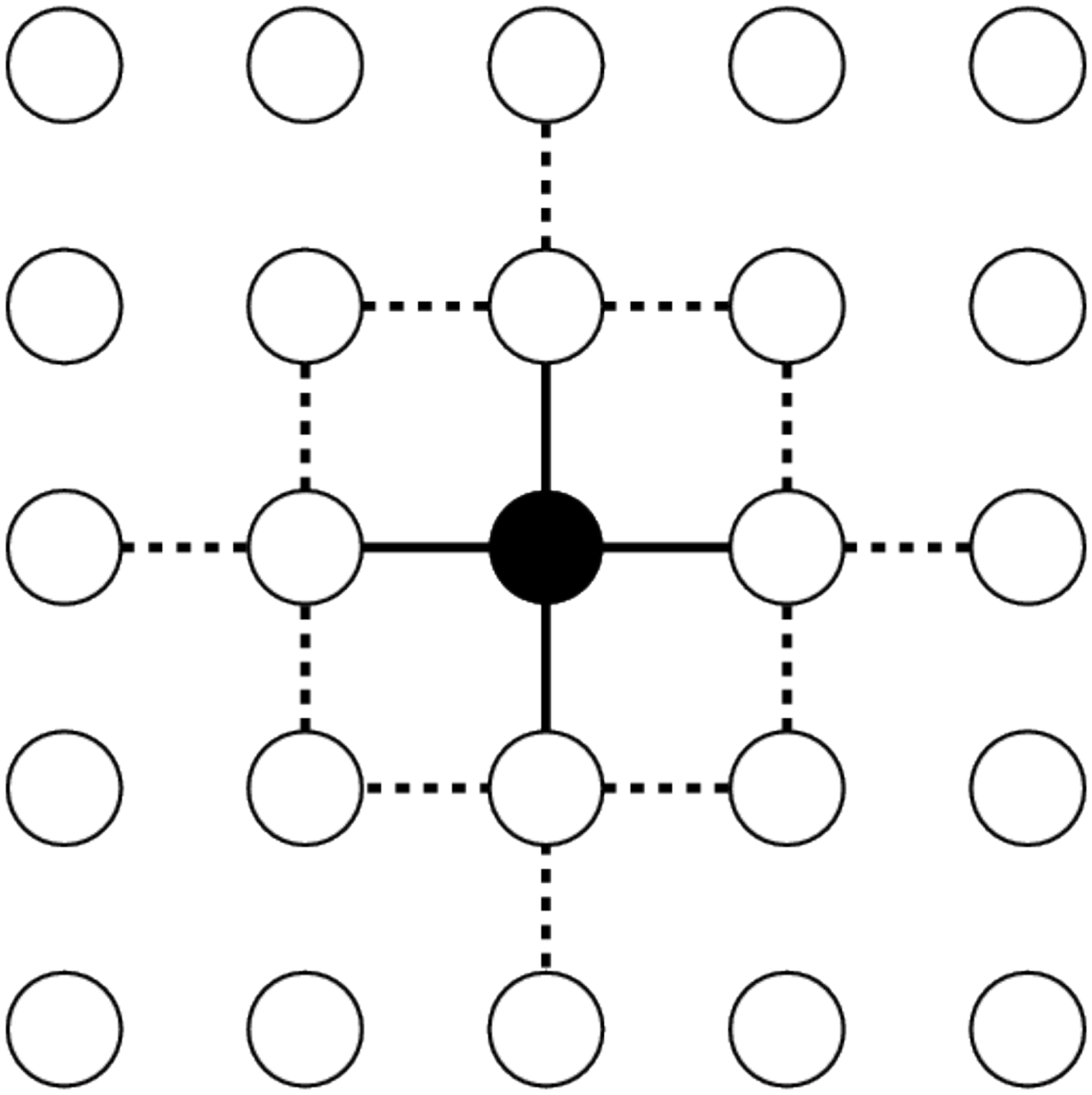}
\par\end{centering}

} \subfloat[$x_{local}$ s'étend jusqu'aux deuxièmes voisins.\label{fig:xlocal2}]{\begin{centering}
\includegraphics[scale=0.25]{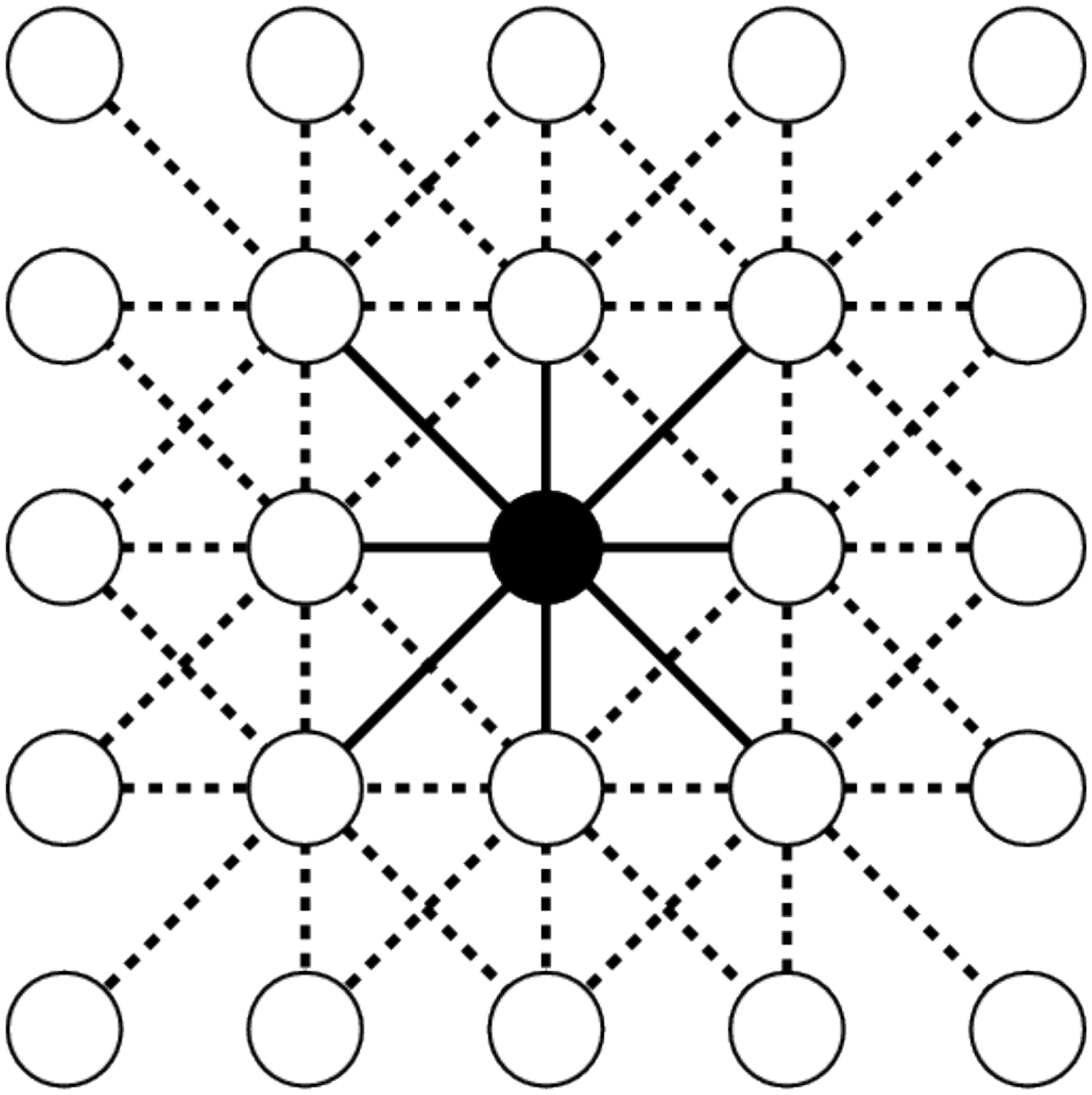}
\par\end{centering}

}
\par\end{centering}

\caption{Illustration du nombre de liaisons dont il faut tenir compte dans
le bilan énergétique d'une transmutation d'un atome blanc en atome
noir dans un réseau carré. Si les énergies de paires sont indépendantes
de la concentration locale, seules les énergies de paires en trait
plein sont à calculer. Si les énergies de paires sont dépendantes
de la concentration locale, toutes les énergies de paires en traits
pleins et traits pointillés sont à re-calculer.\label{fig:algo_ecal}}
\end{figure}

Le schéma \ref{fig:algo_ecal} illustre l'effet de la dépendance en
concentration locale sur le nombre d'énergies de paires à re-calculer
après la transmutation d'un atome blanc en atome noir. On considère
deux extensions de la zone de calcul de $x_{local}$ : jusqu'aux premiers
voisins seulement ($i=1$) et jusqu'aux deuxièmes voisins ($i=2$).
On se limite à un réseau carré pour l'illustration.
\begin{itemize}
\item Si les énergies de paires sont constantes, dans le cas d'une permutation
d'un atome \og blanc \fg{} en atome \og noir \fg{}, seules les
énergies des paires contenant cet atome interviennent dans le bilan
énergétique (paires en traits pleins).
\item Si les interactions dépendent de la concentration locale, toutes les
paires pour lesquelles l'atome permuté intervient dans le calcul de
la concentration locale interviennent dans le bilan énergétique (paires
en traits pleins et en traits pointillés).
\end{itemize}

Sur la figure \ref{fig:xlocal1}, la zone de calcul de $x_{local}$
s'étend jusqu'aux premiers voisins seulement ($i=1$). Il y a 4 liaisons
directes avec l'atome noir et 16 liaisons directes ou indirectes dont
il faut recalculer l'énergie la concentration locale et l'énergie
de paire correspondante. Sur la figure \ref{fig:xlocal2}, la zone
de calcul de $x_{local}$ s'étend jusqu'aux deuxièmes voisins ($i=2$).
Huit énergies de liaisons sont à calculer dans un modèle de paires
constantes et 52 dans un modèle dépendant de $x_{local}$.

\begin{figure}[h]
\begin{centering}
\includegraphics[scale=0.4]{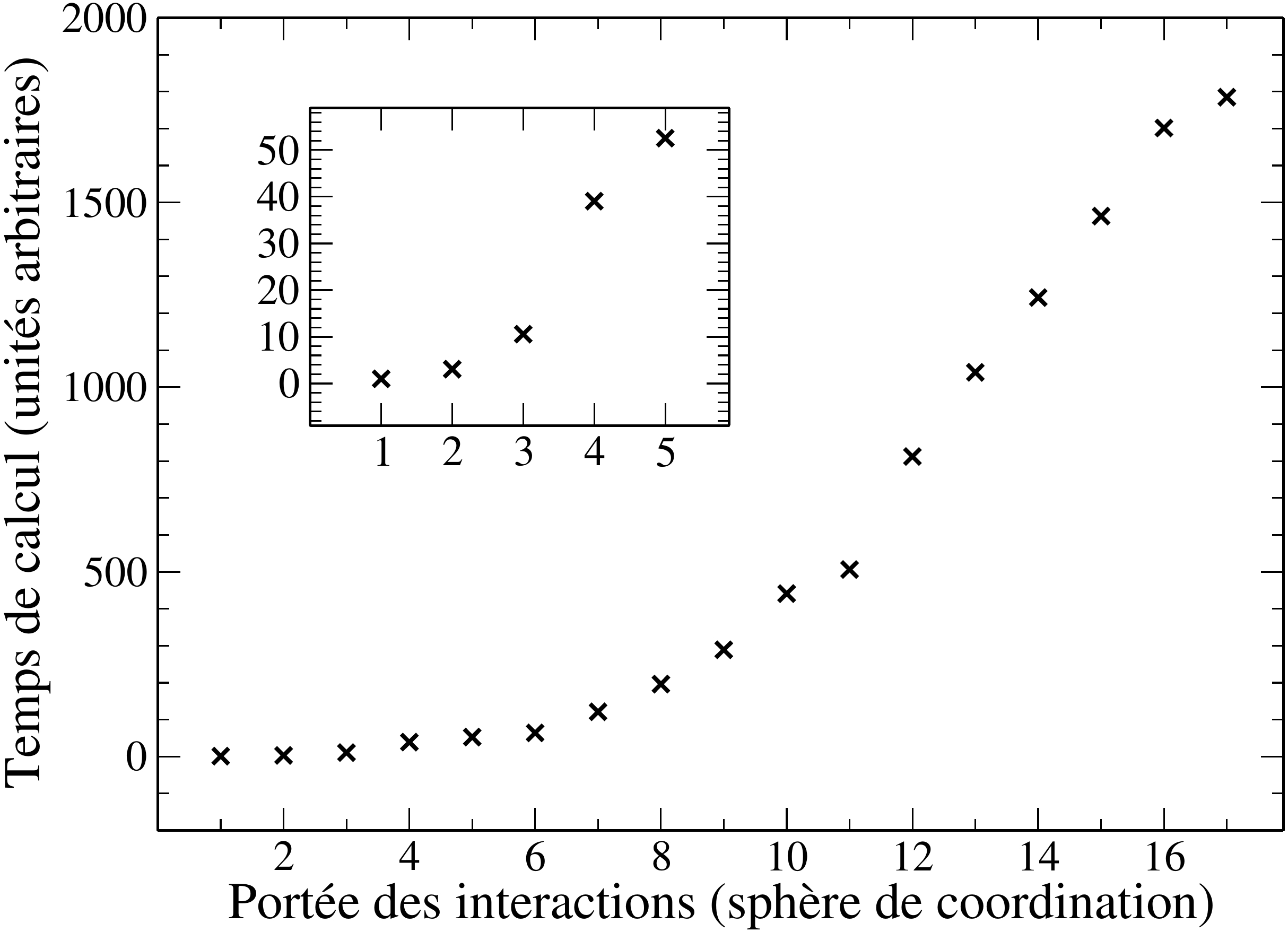}
\par\end{centering}

\caption{Évolution du temps de calcul avec la portée des interactions considérées.
En encart, les cinq premières sphères de coordination.\label{fig:evo_tCPU_portee}}
\end{figure}

Ainsi, on montre que le temps de calcul de chaque pas Monte Carlo
varie en $N^{2}$ (aux quelques voisins de voisins communs près),
où $N$ est le nombre d'atomes dans la zone de calcul de $x_{local}$.
L'évolution du temps de calcul avec la portée des interactions est
représentée sur la figure \ref{fig:evo_tCPU_portee}.

On rappelle que $N=8$ aux premiers voisins, 14 aux deuxièmes voisins,
26 aux troisièmes \ldots{} 

On observe sur la figure \ref{fig:evo_tCPU_portee} que le temps de
calcul est triplé quand l'étendue de la zone de calcul de la concentration
locale passe des premiers aux deuxièmes voisins. Il est multiplié
par 11 quand elle s'étend aux troisièmes voisins ! On souhaitera ainsi
limiter la portée des interactions pour limiter la lourdeur des calculs.

\chapter{Répartition de l'énergie sur les différentes sphères de coordination\label{sub:Mod=0000E8le-d'int=0000E9raction}}

\malettrine{O}{}n impose un rapport fixe entre les énergies de paires
aux premiers et aux $i$$^{\mbox{e}}$ voisins. On définit ce rapport
entre deux atomes $n$ et $m$ :
\begin{equation}
\eta^{\left(i\right)}=\frac{\epsilon_{nm}^{\left(i\right)}}{\epsilon_{nm}^{\left(1\right)}}
\end{equation}
où $\epsilon_{nm}^{\left(i\right)}$ est l'énergie de paire entre
deux atomes $n$ et $m$ $i$$^{\mbox{e}}$ voisins. Le choix des
$\eta^{\left(i\right)}$ est arbitraire. Nous avons testé trois possibilités
:
\begin{enumerate}
\item l'énergie de paires entre premiers et deuxièmes voisins peut être
égale ($\epsilon_{nm}^{\left(1\right)}=\epsilon_{nm}^{\left(2\right)}$,
soit $\eta^{\left(2\right)}=1$). C'est l'hypothèse utilisée par exemple
par Ackland et al. \cite{ackland_magnetically_2006,ackland_orderedsigma_2009}.
\item on peut également prendre une énergie de paire deux fois plus faible
aux deuxièmes voisins ($\eta^{\left(2\right)}=\frac{1}{2}$).
\item dans le cas de portées plus importantes (jusqu'aux 5$^{\text{e}}$
voisins dans notre cas) ($i=1$ à 5), on considère que l'énergie de
paire est inversement proportionnelle à la distance entre atomes de
la paire :
\end{enumerate}
\begin{equation}
\eta^{\left(i\right)}=\dfrac{\epsilon_{nm}^{\left(i\right)}}{\epsilon_{nm}^{\left(1\right)}}=\dfrac{d^{\left(1\right)}}{d^{\left(i\right)}}.\label{eq:eta}
\end{equation}

Pour cette dernière situation, le tableau \ref{tab:coordination_en_volume}
présente les valeurs de $\eta^{\left(i\right)}$ pour les différentes
sphères de coordination et le nombre total de voisins dans chaque
sphère $Z^{\left(i\right)}$.

\begin{table}[h]
\begin{centering}
\begin{tabular}{|c|c|c|}
\hline 
portée ($i$) & $\eta^{\left(i\right)}=\dfrac{d^{\left(1\right)}}{d^{\left(i\right)}}$ & $Z^{\left(i\right)}$ \\
\hline 
\hline 
1 & 1 & 8 \\
\hline 
2 & $\frac{\sqrt{3}}{2}\approx0.866$ & 6 \\
\hline 
3 & $\frac{\sqrt{3}}{2\sqrt{2}}\approx0.612$ & 12 \\
\hline 
4 & $\frac{\sqrt{3}}{\sqrt{11}}\approx0.522$ & 24 \\
\hline 
5 & $\frac{1}{2}$ & 8 \\
\hline 
6 & $\frac{\sqrt{3}}{4}\approx0.433$ & 6 \\
\hline 
\end{tabular}
\par\end{centering}

\caption{Portée de la liaison et poids de celle-ci en comparaison à une liaison
équivalente aux premiers voisins, pour les 6 premières sphères de
coordination pour le calcul de l'énergie de liaison dans le cas où
la portée est supérieure aux deuxièmes voisins, et nombre total de
voisins dans chaque sphère. \label{tab:coordination_en_volume}}
\end{table}

\chapter{Algorithme de minimisation du système d'équations non-linéaires}

Les systèmes d'équations couplées \ref{eq:couple1} et \ref{eq:44}
sont non-linéaires. La concentration dans chacun des plans $p$ dépend
de la concentration dans tous les autres plans. L'objectif est de
déterminer la configuration $\left\{ x_{p}\right\} $ minimisant l'énergie
totale du système traité en champ moyen de plans.

La solution solide est instable à l'intérieur des limites de décomposition
spinodale. Il n'existe donc pas de configuration minimisant l'énergie
du système à ces concentrations.

Deux algorithmes différents ont été testés pour déterminer numériquement
la configuration d'équilibre $\left\{ x_{p}\right\} $ à partir d'une
configuration $\left\{ x_{p}\right\} _{0}$ choisie par l'utilisateur
: les méthodes de Newton-Raphson \cite{Numerical_recipies_fortran_1996}
et d'itérations naturelles \cite{kikuchi_natural_iteration_method_1974}. 

C'est la méthode d'itérations naturelles de Kikuchi qui s'est révélée
la plus légère et efficace numériquement. Elle consiste en la résolution
de la boucle auto-cohérente suivante :
\begin{enumerate}
\item Initialiser le système $\left\{ x_{p}\right\} _{0},\forall p=1,v$,
où le $v$$^{\text{e}}$ plan est le plan de raccord avec le volume
: $x_{p}=x_{v},\forall p\geq v$.
\item Calculer $\Delta E_{seg}\left(p=0\right)$ et en déduire un nouveau
$x_{v}$.
\item Utiliser ce nouveau $x_{v}$ pour recalculer les nouveaux $\Delta E_{seg}\left(p\right)=f\left(x_{v},x_{p\pm i}\right)$,
pour $i=1$ à $v$. On a un nouvelle configuration $\left\{ x_{p}\right\} _{0}^{'}$.
\item Comparer $x_{v}$ à $x_{v-1}$. Si la différence est supérieure à
un critère de convergence, alors on définit $\left\{ x_{p}\right\} _{1}=\left\{ x_{p}\right\} _{0}^{'}$
et la boucle auto-cohérente reprend à l'étape n\textdegree{}\,2.
Si le critère de convergence est vérifié, la configuration $\left\{ x_{p}\right\} _{i}^{'}$
est la configuration d'équilibre.
\item Incrémenter $x_{v}$ et retourner à l'étape n\textdegree{}2.
\end{enumerate}
L'objectif est de déterminer la concentration de chaque plan $x_{p}$
en fonction de la concentration en volume $x_{v}$. Le choix du plan
sur lequel la concentration est fixée peut paraître naturellement
le plan de volume. Cependant, pour des raisons de convergence, nous
observons qu'il est parfois préférable de boucler sur un autre plan
qui peut par exemple être le plan de surface.

\newpage{}

Cette thèse et du travail post-doctoral personnel ont donné lieu aux publications suivantes :
\begin{enumerate}
\item Levesque et al., D{\'e}mixtion et s{\'e}gr{\'e}gation superficielle dans les alliages fer-chrome: de la structure {\'e}lectronique aux mod{\`e}les thermodynamiques \cite{levesque2010demixtion}
\item Olsson et al., DFT data on defect energies in Fe-Cr alloys, including magnetic configurations and comparison with XMCD \& EXAFS results \cite{olsson2010dft}
\item Levesque et al., Simple concentration-dependent pair interaction model for large-scale simulations of Fe-Cr alloys \cite{levesque2011simple}
\item Martinez et al., Simulations of decomposition kinetics of Fe-Cr solid solutions during thermal aging \cite{martinez2011simulations}
\item Levesque et al., levesque2011iron \cite{levesque2011iron}
\item Levesque et al., Electronic origin of the anomalous segregation behavior of Cr in Fe-rich Fe-Cr alloys \cite{levesque2012electronic}
\item Gupta et al., Effect of surface hydrogen on the anomalous surface segregation behavior of Cr in Fe-rich Fe-Cr alloys \cite{gupta2012effect}
\item Levesque, Anomalous surface segregation profiles in ferritic Fe-Cr stainless steel \cite{levesque2013anomalous}
\end{enumerate}


\colophon{Ce document a été préparé sous Ubuntu Linux grâce aux logiciels de sources libres \LaTeX{}, LyX, Grace, Xfig et VMD. La classe de document \LaTeX{} est une version personnalisée de \textit{these$\_$gi} développée par J. Chiquet.}
\end{document}